\documentclass[sigconf,edbt]{acmart-edbt2025}

\def\BibTeX{{\rm B\kern-.05em{\sc i\kern-.025em b}\kern-.08em
    T\kern-.1667em\lower.7ex\hbox{E}\kern-.125emX}}

\usepackage{booktabs} 
\usepackage{balance}

\setcopyright{rightsretained}

\acmDOI{}

\acmISBN{978-3-89318-099-8}

\acmConference[EDBT 2025]{28th International Conference on Extending Database Technology (EDBT)}{25th March-28th March, 2025}{Barcelona, Spain}
\acmYear{2025}

\usepackage{xspace}
\usepackage{soul}
\usepackage{multirow}
\usepackage{enumitem}
\newcommand{\systemname}{\textsc{Co\-met}\xspace}
\newcommand{\mypara}[1]{\smallskip\noindent#1.}
\newcommand{\smallsection}[1]{\smallskip\noindent\textbf{#1}}
\usepackage{subcaption}
\usepackage{dcolumn}
\usepackage{xcolor}
\usepackage{colortbl}
\definecolor{c_blue}{HTML}{0173B2}

\newcommand{\pt}{\%\textit{pt}\xspace}
\usepackage{placeins}
\newcommand{\revision}[1]{\textcolor{black}{#1}}

\begin{document}
\title[Step-by-Step Data Cleaning Recommendations to Improve ML Prediction Accuracy]{Step-by-Step Data Cleaning Recommendations\\ to Improve ML Prediction Accuracy}

\author{Sedir Mohammed}
\affiliation{%
  \institution{Hasso Plattner Institute\\ University of Potsdam}
  \country{Germany}
  \postcode{43017-6221}
}
\email{Sedir.Mohammed@hpi.de}

\author{Felix Naumann}
\affiliation{%
  \institution{Hasso Plattner Institute\\ University of Potsdam}
  \country{Germany}
  \postcode{43017-6221}
}
\email{Felix.Naumann@hpi.de}

\author{Hazar Harmouch}
\affiliation{%
  \institution{University of Amsterdam}
  \country{The Netherlands}
  \postcode{43017-6221}
}
\email{h.harmouch@uva.nl}
\renewcommand{\shortauthors}{}

\begin{abstract}
Data quality is crucial in \textit{machine learning}~(ML) applications, as errors in the data can significantly impact the prediction accuracy of the underlying ML model.
Therefore, data cleaning is an integral component of any ML pipeline.
However, in practical scenarios, data cleaning incurs significant costs, as it often involves domain experts for configuring and executing the cleaning process. 
Thus, efficient resource allocation during data cleaning can enhance ML prediction accuracy while controlling expenses.

This paper presents \systemname, a system designed to optimize data cleaning efforts for ML tasks. 
\systemname gives step-by-step recommendations on which feature to clean next, maximizing the efficiency of data cleaning under resource constraints.
We evaluated \systemname across various datasets, ML algorithms, and data error types, demonstrating its robustness and adaptability.
Our results show that \systemname consistently outperforms feature importance-based, random, and another  well-known cleaning method, achieving up to 52 and on average 5~percentage points higher ML prediction accuracy than the proposed baselines.

\end{abstract}

\maketitle

\section{Data Cleaning for ML}
\label{sec:intro}
In an era dominated by data-driven strategies and rapid development in \textit{machine learning}~(ML), \emph{data quality} has become a significant factor in the success of ML applications.
The availability of vast and diverse datasets has empowered various domains, such as healthcare, biology and finance, to leverage the capabilities of ML algorithms, fostering significant improvements~\cite{zha2023data, polyzotis2021can}. 
However, the effectiveness of these algorithms depends on the quality of the input data during training and testing phases~\cite{DBLP:conf/icde/ForoniLV21}.
Real-world datasets often contain imperfections, inconsistencies, and inaccuracies that can significantly impact the prediction accuracy of ML models.
Thus, data cleaning, which entails identifying and correcting data errors, is essential for reliable and accurate ML predictions.
\begin{figure}[!tb]
\centering
\includegraphics[width=.85\linewidth]{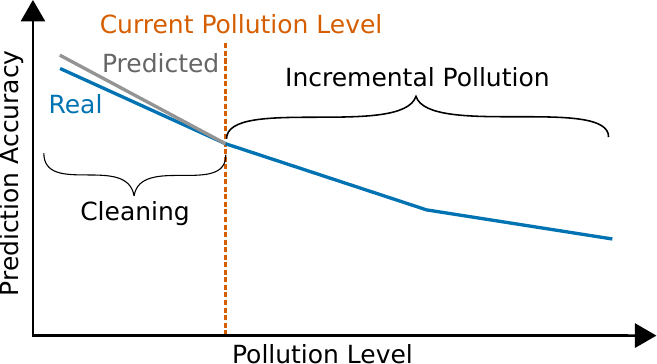}
\caption{\systemname incrementally pollutes features and based on the observed negative effect of the pollution on the prediction accuracy, it estimates the positive impact of data cleaning.}
\label{fig:main_macro_example}
\end{figure}

Traditionally, data scientists performed cleaning during data acquisition, often without considering the underlying ML task~\cite{DBLP:journals/jdiq/IlyasR22}.
However, the focus has shifted from isolated data cleaning strategies preceding the ML task~(``Cleaning before ML'') to an integrated perspective that views data cleaning and the underlying ML task as a cohesive entity~(``Cleaning for ML'')~\cite{neutatz2021cleaning}.
This development marks a new era in which data quality and ML outcomes are symbiotic components within the ML pipeline.

The growing paradigm of data-centric \textit{Artificial Intelligence}~(AI) emphasizes the role of data quality throughout the ML pipeline~\cite{whang2023data, zha2023data}.
Initially, developers and researchers focused on models to improve prediction accuracy.
With data-centric AI, the focus is on the data and its quality instead of the ML model, potentially leading to more robust and generalizable AI applications.
However, this shift also presents challenges, like ensuring data quality, obtaining reliable annotations, handling missing values, and fostering data diversity.
Especially in real-world scenarios where data cleaning is often associated with costs as this task is typically performed by users, respectively domain experts.
These challenges motivate the need for strategic data preparation methods to ensure effective ML models and an efficient data cleaning process~\cite{polyzotis2021can, zha2023data}.

Given the common scenario where a user possesses a~(dirty) dataset and aims to deploy an ML model within a limited resource allocation for data cleaning~\cite{mazumder2022dataperf}, it is beneficial to determine the order of clean operations, and thus to maximize the improvement of the downstream ML task within a budget.
To determine this order, we introduce the \underline{C}leaning \underline{O}ptimization and \underline{M}odel \underline{E}nhancement \underline{T}oolkit (\systemname), a progressive approach for cleaning recommendations in such scenarios.
Our approach provides the user with a step-by-step recommendation on which feature to clean next.

By analyzing the relationships between feature-wise data quality, cleaning costs, and their impact on ML outcomes, \systemname allocates the cleaning budget to maximize improvements in prediction accuracy.
To estimate how feature-wise data cleaning affects prediction accuracy, \systemname incrementally injects additional errors into features~(see Figure~\ref{fig:main_macro_example}) and measures prediction accuracy after each pollution.
It then interpolates a trend from these measurements to predict the effect of cleaning the respective feature. 
Considering the cleaning costs, \systemname recommends the next feature to clean.
The principle of incremental data pollution has shown promise in other fields~\cite{chang_quantum_2023}:
In quantum computing, researchers amplify noise during calculations to back-calculate results in a noise-free environment.

The cleaning process within~\systemname can be executed manually or through automatic error correction algorithms~\cite{DBLP:journals/jdiq/IlyasR22}.
Recently, \textit{Automated Machine Learning}~(AutoML) emerged to streamline and automate various ML stages, from data preprocessing, including data cleaning, to feature engineering, algorithm selection, and hyperparameter tuning~\cite{DBLP:journals/kbs/HeZC21}.
The flexibility of \systemname allows it to fit seamlessly into an AutoML pipeline. 
In critical domains, where the data cleaning process might require expert validation even when it is done automatically, \systemname's step-wise feature cleaning recommendations provide a controlled and transparent approach.

In this paper, ``prediction accuracy'' refers to metrics like accuracy or F1 score, even though \systemname can also optimize other AI-related metrics, such as fairness or bias~\cite{whang2023data, zha2023data}.

To evaluate our approach, we introduce \revision{three} baselines and a related approach.
The first baseline uses feature importance from the~(dirty) input data to guide feature-wise cleaning recommendations~\cite{lundberg2017unified}.
The second baseline acts on random feature selection -- a contrastingly non-strategic method.
\revision{The third baseline breaks \systemname's step-wise approach, using a ranked feature list from the first step for all subsequent cleaning operations.}
The fourth contender is \textit{ActiveClean}~\cite{krishnan2016activeclean}, a well-known method aiming to optimize the cleaning process like \systemname.
Our empirical results show that \systemname, on average, outperforms the baselines across diverse datasets, error types and ML algorithms. 
The results show the potential for improved model prediction, emphasizing the importance of an informed data cleaning strategy in resource-constrained scenarios.

\smallsection{Contribution.} In summary, our research advances the understanding of data cleaning as a key facet of data-centric AI\@.
It offers practitioners a principled approach to optimize data-cleaning efforts within budget constraints. 
As ML increasingly permeates various domains, insights from this study promise to enable more accurate and cost-efficient deployment of ML models on real-world, imperfect data. 
We specifically make the following main contributions:
\begin{enumerate}
    \item We present \systemname, an innovative approach that incrementally pollutes data to assess the effects of data cleaning on prediction accuracy and iteratively makes cleaning recommendations while considering cleaning costs.
    \item We evaluate the performance of \systemname by simulating the cleaning of various datasets and by comparing the prediction accuracies against two cleaning baselines and ActiveClean~\cite{krishnan2016activeclean}.
    \item We further evaluate \systemname using the CleanML benchmark~\cite{DBLP:conf/icde/LiRBZCZ21, cleanmldatasetdescriptionspdf}, which provides both dirty and clean versions of real-life datasets for comprehensive analysis.
\end{enumerate}

\smallsection{Outline.}
Section~\ref{sec:related_work} presents relevant related work.
Section~\ref{sec:system} details the functionality and various components of \systemname.
We present the data considered in the experiments, ML algorithms, error types, cleaning cost models, and other implementation details of \systemname in Section~\ref{sec:experimental_setup}.
Section~\ref{sec:results} shows the results of the experiments in which we compare \systemname to the baselines and related work.
Finally, Section~\ref{sec:conclusion} summarizes our findings and outlines future research directions.

\section{Related Work} \label{sec:related_work}
Traditional error detection and cleaning methods, such as NA\-DEEF~\cite{DBLP:conf/sigmod/DallachiesaEEEIOT13} and KATARA~\cite{DBLP:journals/pvldb/ChuOMIP0Y15}, rely on user-defined rules or domain expertise, which can be costly and labor-intensive.
As the complexity and volume of data increased, and ML methods became more popular, the need for automated and adaptive data cleaning methods became more pressing.
An example of ML-based \emph{error detection} is HoloDetect~\cite{DBLP:conf/sigmod/HeidariMIR19}, which learns error patterns in the data to generate further training data synthetically and trains an error detection model.
Another approach is ED2~\cite{DBLP:conf/cikm/NeutatzMA19} which uses active learning and selects the entries for which the ML model is currently unsure for labelling by the user.
There exist also dedicated approaches for automatic \emph{data cleaning}.
HoloClean~\cite{DBLP:journals/corr/RekatsinasCIR17} processes different cleaning signals in data, using probability theory to reconcile inconsistencies across various signals and repair data.
Similarly, SCARE~\cite{DBLP:conf/sigmod/YakoutBE13} uses ML techniques to repair data by maximizing data likelihood while introducing minimal changes.

The presented methods and a large amount of similar research focus solely on data without considering its downstream application.
However, erroneous data may not be equally disruptive, as~\citet{DBLP:conf/icde/ForoniLV21} have explicitly demonstrated for the ML context as a downstream task.
This aligns with our intuition in this paper.  
In the following, we discuss existing approaches that consider data cleaning in the context of ML applications.

\smallsection{ActiveClean} treats data cleaning as a stochastic gradient descent problem~\cite{krishnan2016activeclean}.
It selects iteratively records for cleaning that are estimated to have the highest gradient of the loss function after cleaning the respective records.
The gradient per record is estimated from the gradient of the dirty record and previous cleaning procedures.
In contrast, \systemname predicts the cleaning effect in each iteration by introducing additional errors into the current pollution state, making it adaptive by always estimating cleaning impacts based on the current state of the data.
Unlike ActiveClean, which starts with a random sample due to a lack of gradient information, \systemname provides well-founded recommendations from the first iteration.
Despite these differences, ActiveClean is similar enough to \systemname in its approach to serve as a baseline in our experiments.

\smallsection{CPClean} adopts an incremental cleaning strategy similar to \systemname, using a step-by-step process primary informed by validation sets and counting queries~\cite{karlavs2020nearest}.
This process persists until it determines that additional cleaning will no longer affect the prediction accuracy.
However, CPClean is optimized specifically for nearest-neighbor models, lacking the model-agnostic flexibility that \systemname offers.

\smallsection{BoostClean} approaches data cleaning as a boosting problem, generating a cleaning program applicable to training or test records, while actively conducting the cleaning~\cite{boostclean}.
The primary criterion for cleaning decisions is the model's test accuracy.
Although BoostClean estimates the impact of data cleaning similarly to our approach, it requires fully clean labels in the test data and pre-defined detection and cleaning functions by a domain expert.
In contrast, \systemname operates without prerequisites on the cleaning state of training and test data and autonomously estimates the impact of cleaning without relying on pre-defined rules or functions.


\smallsection{AutoSklearn} is a complete AutoML system that automates data cleaning by constructing and evaluating ML pipelines with model validation accuracy~\cite{feurer2015efficient}. 
It focuses on imputation when cleaning errors based on mean, median or most frequent values.
~\citet{neutatz2022data} extended the system with additional data cleaning methods for outlier detection and advanced imputation strategies.
Although these AutoML frameworks offer comprehensive solutions, their data cleaning functionalities are tailored to handle specific error types and require pre-defined detection and cleaning routines.

\smallsection{DiffPrep}, similar to \systemname aims to optimize data cleaning but focuses on feature-wise transformations during pre-processing while considering the impact on the underlying ML model~\cite{diffprep}.
DiffPrep determines the order of transformation types~(e.g., cleaning missing values first, then outliers) and selects the appropriate cleaning method~(e.g., mean imputation for missing values, Z-score for outliers) for each feature, tailored to the ML model.
Unlike \systemname, DiffPrep does not clean incrementally; it establishes a final transformation pipeline for all features without guiding the user step-by-step, which limits direct comparability with \systemname.
\section{Cleaning Recommendation} \label{sec:system}

\begin{figure*}[tb]
\centering
\includegraphics[width=1.0\linewidth]{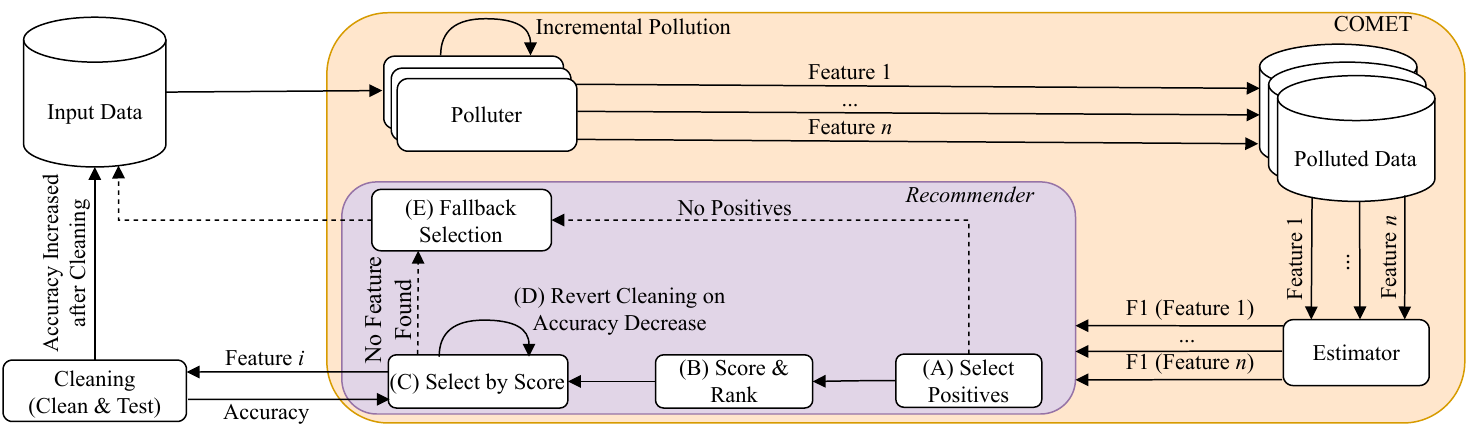}
\caption{\revision{\systemname workflow for an individual error type: (1)~\emph{Polluter}: Introduce further pollution; (2)~\emph{Estimator}: Evaluate pollution/cleaning effects on ML model accuracy; (3)~\emph{Recommender}: Propose feature-wise cleaning strategies based on scoring.}}
\label{fig:architecture}
\end{figure*}

\systemname provides cost-aware feature-wise cleaning recommendations in a progressive fashion, i.e., offering step-by-step guidance on which features should the next cleaning effort be spent on.
These cleaning recommendations are always error type and ML algorithm-specific.
However, \systemname can handle arbitrary error types and ML algorithms\revision{, without being restricted to a specific ML task.
In this manuscript, we configured \systemname for four specific error types (see Section~\ref{sec:error_types}) and focused on classification.}
The actual cleaning can be performed by either a data cleaning algorithm, such as HoloClean~\cite{DBLP:journals/corr/RekatsinasCIR17}, SCARE~\cite{DBLP:conf/sigmod/YakoutBE13}, or, more importantly, a dedicated domain expert.
At no point during the process does \systemname require information about the actual pollution level of the individual features, nor which entries are actually erroneous.
Figure~\ref{fig:architecture} shows the architecture of \systemname, consisting of three modules:~\emph{Polluter} (Section~\ref{polluter}),~\emph{Estimator} (Section~\ref{estimator}) and~\emph{Recommender} (Section~\ref{recommender}).
\systemname initially requires~(dirty) data as input, which serves as the basis for generating cleaning recommendations for the subsequent cleaning step. In the following subsections, we provide a detailed description of each module and its functionality.

In the rest of the paper, we refer to the number of data entries to be cleaned in each iteration as the~\emph{cleaning step}.
We also use the term \revision{\textit{Cleaner}} to denote the data cleaning methods, encompassing both the algorithm-based and the human-based techniques.

In the description of each module, we assume a specific error type. 
However, the process can be applied separately to different error types to identify the optimal feature and error type combination.

\subsection{Incremental data pollution}
\label{polluter}
The first module of \systemname, the~\emph{Polluter}, incrementally pollutes the input data given a specific error type for which \systemname should make a cleaning recommendation in the current iteration.
The \emph{Polluter} introduces additional erroneous entries for each feature, quantified by a \textit{pollution step}, which is the counterpart to a cleaning step.
Additionally, we use the notion of the pollution level, which denotes the percentage of cells that should be additionally polluted.
Since selecting the entries to be polluted may also have a potentially different effect on the prediction accuracy, the \emph{Polluter} selects randomly multiple combinations of entries per feature to be polluted.
So, the functionality of the~\emph{Polluter} is represented mathematically by the function:
\begin{equation}
    \revision{\textit{Polluter}}(d,f,\textit{Err},\rho) = d'_{f,\rho,c}
\end{equation}
where~$d$ represents the input data,~$f$ is the considered feature to be polluted,~$\textit{Err}$ is the error type,~$\rho$ signifies the pollution level for the feature~$f$, and~$d'_{f,\rho,c}$ represents a polluted data state for the pollution level~$\rho$ and the combination~$c$.

\mypara{\revision{Example}}
\revision{Suppose we have a feature ``Income'' in a dataset $d$.
If the error type \textit{Err} is missing values and the pollution level $\rho$ is set to 1\%, the \emph{Polluter} introduces missing values to 1\% of the entries in the ``Income'' column.
This results in a polluted data state $d'_{\text{Income},1\%,c}$, where $c$ represents one specific combination of polluted entries.}

By default, \systemname assumes all features contain dirty entries and must be cleaned until the \emph{\revision{\textit{Cleaner}}} marks them as clean.
To produce the next cleaning recommendation, the \emph{Polluter} performs two further pollution \revision{steps} for each feature~(see Figure~\ref{fig:main_macro_example}). 
Each new pollution level corresponds to an additionally polluted input data state.
Pollution is applied separately to the train and the test data to prevent information leakage between them.

It is worth noting that, given that the~\emph{Polluter} lacks knowledge about which specific entries in a feature are erroneous, it may overwrite already dirty entries with artificial errors.
Thus, the targeted pollution level might not be fully achieved after a pollution step.
Using the hypergeometric distribution, we estimate that the probability of selecting clean entries to pollute remains high when the number of already dirty entries is small relative to the total.

\subsection{Cleaning impact estimation}
\label{estimator} 
The~\emph{Estimator} receives from the~\emph{Polluter} feature by feature the different versions of the data, including~$d$ and a set~$D'_{f}$ of~$d'_{f,\rho,c}$.
The~\emph{Estimator} operates in two steps:

\paragraph{Step~1: Pollution Effect Measurement}
We evaluate the influence of incremental feature-wise pollution on the prediction accuracy for the considered pollution levels~$\rho$.
Since the prediction accuracy depends on the used ML algorithm, \systemname expects an ML algorithm as an additional input in the~\textit{Estimator}.
This phase is expressed as
\begin{equation}
    E_1(d, D'_f, f, MLA) = A(f)
\end{equation}
where~$MLA$ is the chosen ML algorithm, and~$A(f)$ is a set of measured prediction accuracies per pollution step for the feature~$f$.

\mypara{\revision{Example}}
\revision{Given the data states $D'_{f} = \{d'_{\text{Income},1\%,c}, d'_{\text{Income},2\%,c}\}$ and the ML algorithm $MLA$ (e.g., SVM), the Estimator computes the prediction accuracy $A(\text{Income}) = \{0.85, 0.82\}$ for the respective pollution levels $\rho = 1\%, 2\%$.}

\paragraph{Step~2: Predictive Model Construction}
In the second step, denoted as~$E_2$, the~\emph{Estimator} utilizes the prediction accuracies measured from the incrementally polluted input data per feature.
Based on the measured prediction accuracies, the~\emph{Estimator} trains a Bayesian regression model to predict the prediction accuracy for the next cleaning step~(see the line denoted as ``predicted'' in Figure~\ref{fig:main_macro_example}).
The regression-based prediction is formulated as
\begin{equation}
    E_2(A(f)) = (P_{\textit{next}}(f)\revision{, U(f)})
\end{equation}
where~$P_{\textit{next}}(f)$ is the predicted accuracy for the feature~$f$ and $U(f)$ is the uncertainty of the prediction.
\revision{The Bayesian regression model quantifies the prediction uncertainty, which the \emph{Recommender} uses later for feature ranking and selection.}

\subsection{Optimal feature selection}
\label{recommender}
The last module, the~\emph{Recommender}, selects the next feature for the \emph{\revision{\textit{Cleaner}}} to clean~(see Figure~\ref{fig:architecture}).
This decision is based on the predicted accuracy per feature, factoring in the cleaning costs.
Initially, the \emph{Recommender} considers only features with a positive predicted accuracy~(Figure~\ref{fig:architecture} -- \revision{(A)} Select Positives).
These are features where the~\emph{Estimator} predicts that the prediction accuracy improves after cleaning by one cleaning step.
It then scores and ranks these features, using a score that balances the potential accuracy gain against its associated uncertainty and cleaning cost~(Figure~\ref{fig:architecture} -- \revision{(B)} Score \& Rank).
We define this score for a feature~$f$ as
\begin{equation}
    \textit{Score}(f) = \frac{P_{\textit{next}}(f) - U(f)}{C(f)}
\end{equation}
where~$P_{\textit{next}}(f)$ is the predicted accuracy gain for a feature~$f$,~$U(f)$ is the uncertainty associated with this prediction, and~$C(f)$ is the cleaning cost function for the feature~$f$.
The uncertainty~$U(f)$ is calculated as the difference between the upper and lower bounds of the confidence interval of the Bayesian regression model.
This ranking prioritizes features that provide the most significant prediction accuracy improvement per unit of cleaning cost, while also considering the confidence level of these predictions.
Thus, \systemname recommends the most cost-effective feature\revision{~$f$} for cleaning in the current iteration (Figure~\ref{fig:architecture} -- \revision{(C)} Select by Score).

\mypara{\revision{Example}}
\revision{For the feature ``Income'', the predicted accuracy improvement is $P_{\textit{next}}(\text{Income})=0.88$, the uncertainty is $U(\text{Income}) \\= 0.02$, and the cleaning cost is $C(\text{Income}) = 1$.
The score is computed as: $\textit{Score}(\text{Income}) = \frac{0.88 - 0.02}{1} = 0.86$.
}

As part of the cleaning recommendation, \systemname provides the \emph{\revision{\textit{Cleaner}}} with details on which entries of the recommended feature were temporarily polluted by the \emph{Polluter}~($D'_{f}$).
These entries, which have led to the recommendation, are potentially partially dirty \revision{in the input data $d_{f}$} \revision{(feature $f$ of the input data)} -- the \revision{\textit{Cleaner}} should first clean them.
If the already dirty entries in~$d_{f}$  are fewer than required for a cleaning step, the \emph{\revision{\textit{Cleaner}}} must clean additional random entries from this feature.
After cleaning, the \emph{Recommender} evaluates the impact on prediction accuracy and considers the cleaning step as successful if the accuracy increases \revision{-- a new iteration of \systemname starts}~(Figure~\ref{fig:architecture} -- \revision{\textit{Cleaner},} Clean~\&~Test).  
If the prediction accuracy decreases after cleaning, \systemname reverts the data to its pre-cleaning state, retaining the cleaned data in a \textit{cleaning buffer} \revision{(step (D) in Figure~\ref{fig:architecture})}.
The~\emph{Recommender} then moves to the next highest-ranked feature for cleaning~(Figure~\ref{fig:architecture} -- (C) Select by Score).
If the feature to be cleaned is in the cleaning buffer, the~\emph{Recommender} removes it and applies the changes to data.

In rare cases where all ranked features result in a decrease in prediction accuracy after cleaning, or none are predicted to improve accuracy, the~\emph{Recommender} switches to a fallback strategy~(Figure~\ref{fig:architecture} -- (E) Fallback Selection).
This strategy focuses on cleaning the feature that previously archived highest F1 score after cleaning.
If a feature is deemed completely cleaned, the~\emph{Recommender} moves on to the next most important feature in the ranking.

Once a feature is cleaned by one cleaning step, the~\emph{Recommender} compares the actual increase in prediction accuracy to the one predicted by \systemname, identifying any discrepancies in the predictive modelling for the considered feature.
The \emph{Estimator} then adjusts the prediction to improve its accuracy, which involves calculating the mean of the measured discrepancies and then modifying the prediction by adding or subtracting this mean value.
Even if the \emph{Recommender} evaluates the cleaning as inefficient, and \systemname restores the pre-cleaning state, or activates the fallback strategy, the~\emph{Estimator} adjusts the prediction model for that feature.

The~\emph{Recommender} based decision-making, denoted as~$R$, is formalized as
\begin{equation} 
    R\left(\{P_{\textit{next}}(f_i),\, C(f_i),\, U(f_i)\}_{i=1}^{n}\right) = f_{\textit{rec}}
\end{equation} 
where the~\emph{Recommender}~$R$ expects the predicted accuracy, the cleaning cost, and the uncertainty for each considered feature $f_1\ldots f_n$ to recommend a feature~$f_{\textit{rec}}$ for cleaning the considered error type by one cleaning step.

\subsection{\revision{Error types}} \label{sec:error_types}
\revision{\systemname is conceptually an error type-agnostic approach.
However, we configured \systemname for a specific set of error types \revision{$Err$}, which we detail in the following.}
We consider error types that emulate common data errors, such as incomplete or incorrectly data, scale inconsistencies, and random noise.
For the data pollution, \revision{the \textit{Polluter} uses} the JENGA framework~\cite{Jenga} to randomly sample a specific number of records~$x$ from a feature, \revision{based on the pollution level}, and introduce the respective error type.
The following presents the considered error types, including the respective pollution process.

\smallsection{Missing values.}
Missing values are a common problem in many real-world data~\cite{rubin1976inference, emmanuel2021survey}.
These can appear as empty entries or be represented by placeholders, such as ``NaN''.
However, hidden missing values may use unconventional placeholders~\cite{FAHES2018}, such as representing a missing date with \texttt{1900-01-01}.
To pollute the data, \revision{the \textit{Polluter}} replaces the randomly selected \revision{records} of a feature with a placeholder.

\smallsection{Gaussian noise.}
Numerical values can be noisy, e.g., due to erroneous sensors, external interference, or incorrect user input.
To pollute the data for a given feature, the \textit{Polluter} randomly samples~$x$~records and adds Gaussian noise to each selected data point, generated with a mean of zero and a randomly chosen standard deviation within~[1-5].

\smallsection{Categorical shift.}
Categorical shift is the counterpart to Gaussian noise for categorical variables, where incorrect categories are assigned values.
During the pollution, \revision{the \textit{Polluter}} samples $x$~records from a feature and swap their categories with other categories in the feature.

\smallsection{Scaling.} 
We assume scaling in numerical values could occur due to incorrect conversion of units, such as from \textit{cm} to \textit{m}.
To pollute the data, \revision{the \textit{Polluter}} randomly increases the scaling by~10,~100, or~1000 in the selected rows of the considered feature.
\section{Experimental setup} \label{sec:experimental_setup}
This section details the methods and datasets we use to demonstrate the performance of \systemname in giving valuable recommendations for iterative feature-wise cleaning.
In the rest of the paper, we refer to a unique combination of a dataset, an ML algorithm, and a specified error type in the data as a~\emph{configuration}.
Each configuration is a distinct experimental scenario, allowing us to study the effects of various data conditions, algorithms, and errors on \systemname.

\subsection{Pollution and cleaning setup}
To validate the recommendation-based cleaning, we use datasets that have both dirty and cleaned versions (ground truth) and other datasets that we artificially pre-polluted to establish a ground truth.
We introduce the datasets in Section~\ref{sec:datasets}.
Each setting of pollution levels across features is referred to as a \emph{pre-pollution setting}.
These settings affect only features, so we do not add any errors to the labels in our experiments.
Since the pre-pollution setting is at dataset level, we keep the same pre-pollution settings across different configurations involving the same dataset. 
Based on discussions with ML experts, both training and test data are equally polluted, reflecting common real-world scenarios.

Given the potential variations in data quality per feature in real-world datasets and the significant influence of the pre-pollution level distribution on the prediction quality~\cite{DBLP:conf/icde/ForoniLV21}, it is essential to consider diverse pre-pollution settings.
Thus, we sample for each pre-pollution setting the pollution level per feature using an exponential distribution to ensure a wide-ranging representation of pollution level distribution.
The cleaning and pollution step is set at 1\% of the total data size of the train or test data to ensure a consistent impact across experiments.
\revision{We consider two scenarios for the pre-pollution: In the first scenario, we pollute the data according to the pre-pollution level, with only one error type.
In the second scenario, we randomly select an error type for each pollution step of a feature during pre-pollution.
To conduct the pre-pollution, we use the same pollution methods used by the \textit{Polluter}.}

\subsection{Cleaning cost models} \label{sec:cost_models}
For the two scenarios, where we consider either single or multiple errors, we use different cleaning cost models \revision{per feature}. 
In the first scenario, applied within the single-error context, we assume a \textit{constant} cost function for each feature: each cleaning step incurs a uniform cost of one unit. 
In this way, we maintain comparability and avoid favoring any particular feature.
However, in practice, different error types might have varying cleaning costs, thus in the multi-error scenario, we go beyond \textit{constant} cost function by incorporating \textit{linear} and \textit{\revision{one-shot}} cost functions associated with different error types. 
Note that our assignment of cost functions to error types is an example and can be adapted as needed.

\smallsection{Constant cost.}
We apply this cost to both \textit{categorical shifts} and \textit{scaling} errors.
Categorical shifts can be identified by detecting frequent correlations through FD discovery algorithms or association rule mining, with deviations corrected via imputation. 
Similarly, scaling errors identified by outlier detection are also corrected using imputation methods.
In the experiments, we assign the cost of one unit per cleaning step.

\smallsection{\revision{One-shot} cost.}
A \revision{one-shot} cost function implies a higher initial cleaning cost that does not recur in subsequent steps. 
This model is used for cleaning \textit{missing values}, starting with identifying missing values and then performing data imputation once for the entire column. 
In the experiments, we assign costs of two units for the first cleaning step and zero for all further cleaning steps.

\smallsection{Linear cost.}
In this model, each cleaning step of a particular error type costs incrementally more than the previous one. We assume a linear cost function for the cleaning of \textit{Gaussian noise}.
Initial detection involves estimating noise distribution and identifying strong outliers, which becomes gradually harder as subtle deviations are harder to detect.
In the experiments, we assume an increase of one unit per cleaning step performed, with initial costs of one unit.

\smallskip For both scenarios, we limited the cleaning budget to~50~units, balancing practicality and experimental rigor.
We are aware that the effectiveness of such budgets differs notably across various pre-pollution settings: it may be adequate for thoroughly cleaning datasets with fewer dirty features or lower pollution levels, or, 
conversely, it may only allow partial cleaning for more dirty features or overall higher pollution levels.
This variation mirrors real-world data cleaning challenges, where the extent of pollution significantly impacts the resources needed for effective cleaning.

\subsection{Datasets} \label{sec:datasets}
We use seven different datasets commonly used for classification tasks.
Three of them have both clean and dirty versions available and include at least one of the error types considered in this paper.
These datasets are part of the benchmark provided by CleanML~\cite{DBLP:conf/icde/LiRBZCZ21, cleanmldatasetdescriptionspdf}.
Additionally, we established three distinct pre-pollution settings for four further datasets from the UCI Machine Learning Repository and Kaggle.
We present the used datasets in the following; in addition, Table~\ref{tbl:datasets} shows an overview of the characteristics of these datasets.

\smallsection{Datasets used with pre-pollution.}
The \textit{Contraceptive Method Choice}~(CMC) dataset contains a subset of the 1987 Indonesian National Contraceptive Prevalence Survey~\cite{misc_contraceptive_method_choice_30}.
The classification task is to predict the current contraceptive method of the women surveyed.
The \textit{EEG Eye State}~(EEG) dataset contains data from EEG measurements with the Emotiv EEG neuroheadset
~\cite{misc_eeg_eye_state_264}.
The classification task is to predict the eye state (closed or open). The ``categorical shift'' is not applicable to this dataset, since it contains only numerical variables.
The \textit{Telco customer churn}~(Churn) dataset from IBM consists of fictional customer data from a telecommunications company~\cite{noauthor_telco_nodate}.
The classification task is to predict whether the customer terminated their relationship with the company as an active customer in the last month.
Finally, the \textit{South German Credit}~(S-Credit) dataset contains bank data from a southern German bank from~1973 to~1975~\cite{statlog_dataset, groemping2019south, budach2022effects}.
we use in our experiments the version from~\cite{groemping2019south}.
The classification task is to predict whether a bank customer will comply with the terms of the contract or not.

\smallsection{Datasets provided by CleanML.}
The~\textit{Airbnb} dataset contains hotel information from the top~10 tourist destinations and US metropolitan areas~\cite{DBLP:conf/icde/LiRBZCZ21, cleanmldatasetdescriptionspdf, airbnb_dataset}.
The CleanML authors provided the data by scraping the Airbnb.com website.
The classification task is to predict whether the rating of a hotel is~5 or not.
In the context of this dataset, we consider scaling errors.
The~\textit{Credit} dataset consists of credit data, whereby the classification task is to predict whether a client will be in financial distress in the next two years~\cite{DBLP:conf/icde/LiRBZCZ21, cleanmldatasetdescriptionspdf, credit_dataset}.
In the context of this dataset, we consider missing values and scaling errors.
The \textit{Titanic} dataset contains passenger records~\cite{DBLP:conf/icde/LiRBZCZ21, cleanmldatasetdescriptionspdf, titanic_dataset}.
The classification task is to determine whether a passenger survives.
In the context of this dataset, we consider missing values.

\begin{table}[!h]
\begin{center}
\begin{tabular}{lrrrr}
\hline
Name     & \# Rows & \# Cat. & \# Num. & \# Class \\ \hline
\multicolumn{5}{l}{\underline{Datasets used with Pre-pollution}} \\
CMC      & 1,473   & 7              & 2            & 3        \\
Churn    & 7,032   & 16             & 3            & 2        \\
EEG      & 14,980  & 0              & 14           & 2        \\
S-Credit & 1,000   & 17             & 3            & 2        \\
\hline
\multicolumn{5}{l}{\underline{Datasets provided by CleanML}} \\
Airbnb \revision{(scaling errors)}   & 26,288  & 3              & 37           & 2        \\
Credit \revision{(scaling errors)}   & 11,985  & 0              & 10           & 2        \\
Titanic \revision{(missing values)}  & 891     & 6              & 2            & 2        \\
\hline
\end{tabular}
\caption{Overview of our used datasets.}
\label{tbl:datasets}
\end{center}
\end{table}

\subsection{ML algorithms} \label{sec:ml_algorithms}
\revision{While \systemname is not limited to a specific ML task, here we focus on classification.
To demonstrate the superiority of \systemname over state-of-the-art baselines, we selected four diverse ML algorithms:} 
Support Vector Machine~(SVM)~\cite{cortes1995support}, a \textit{k}-nearest neighbors classifier~(KNN)~\cite{altman1992introduction}, a multi-layer perceptron~(MLP)~\cite{specht1991general} and a Gradient Boosting classifier~(GB)~\cite{10.1214/aos/1013203451}.
We performed a 10-sampled random hyperparameter optimization for each configuration and pre-pollution setting.
With this, we simulate a real-world scenario wherein users working with dirty data aim for the highest prediction accuracy given the dataset's current state.

\subsection{Evaluation metrics and baselines}
As \systemname makes recommendations for individual cleaning steps, we compare the prediction accuracy per step to evaluate the performance of \systemname against \revision{four} baselines: \textit{random recommendations}~(RR), \textit{feature importance-based recommendations}~(FIR)\revision{, light version of \systemname (CL)} and \textit{ActiveClean}~(AC).
As a metric for the prediction accuracy, we use the well-known F1 score.

RR adopts a randomized approach, randomly selecting a feature in each cleaning step among those that have been marked to be cleaned. 
For our evaluation, this process is repeated five times for each pre-pollution setting, and we averaged the resulting F1 scores per cleaning step. 
FIR uses Shapley values~\cite{lundberg2017unified} to rank features by importance within each configuration and pre-pollution setting, selecting the highest-ranked yet polluted feature for cleaning in each step until it is fully cleaned, then moving to the next.
\revision{CL is a simplified version of \systemname:
it applies the idea of \systemname once to generate a ranked list of features based on the \emph{Estimator}'s output.
Similar to FIR, the \emph{Recommender} selects the feature with the highest rank at each cleaning step and continues until that feature is fully cleaned.
Similar to \systemname, the \emph{Recommender} in CL uses the same cleaning step and reverting and fallback strategies.}
The authors of AC have focused on ML algorithms with a convex loss function.
Specifically, they used~SVM~(in the following denoted as~\textit{AC-SVM}), linear regression~(LIR) and logistic regression~(LOR)~\cite{krishnan2016activeclean}. 
Thus, we focus on these ML algorithms when comparing \systemname with AC.
\revision{AC relates the cleaning procedure to the number of records that can be cleaned. 
We adapt AC to align with our concept of cleaning on a feature level, ensuring a comparable experimental setting.}

Additionally, we introduce a \textit{local optimum}~(Oracle), which shows the optimal feature for cleaning in each step based on the gain in F1 score relative to the associated cleaning costs.
The cleaning order generated by the Oracle does not always guarantee the highest accuracy at every cleaning step compared to \systemname or the other baselines.
A single divergent decision on feature cleaning by \systemname or the baselines can lead to divergent cleaning orders that can reach, at some point, a higher accuracy than the Oracle.
However, Oracle performs much better than any other approach on average and thus can be treated as an upper bound.


\section{Experimental Results}
\label{sec:results}

\begin{figure*}[!tp]
    \centering
    \begin{subfigure}{0.24\textwidth}
        \includegraphics[width=\linewidth]{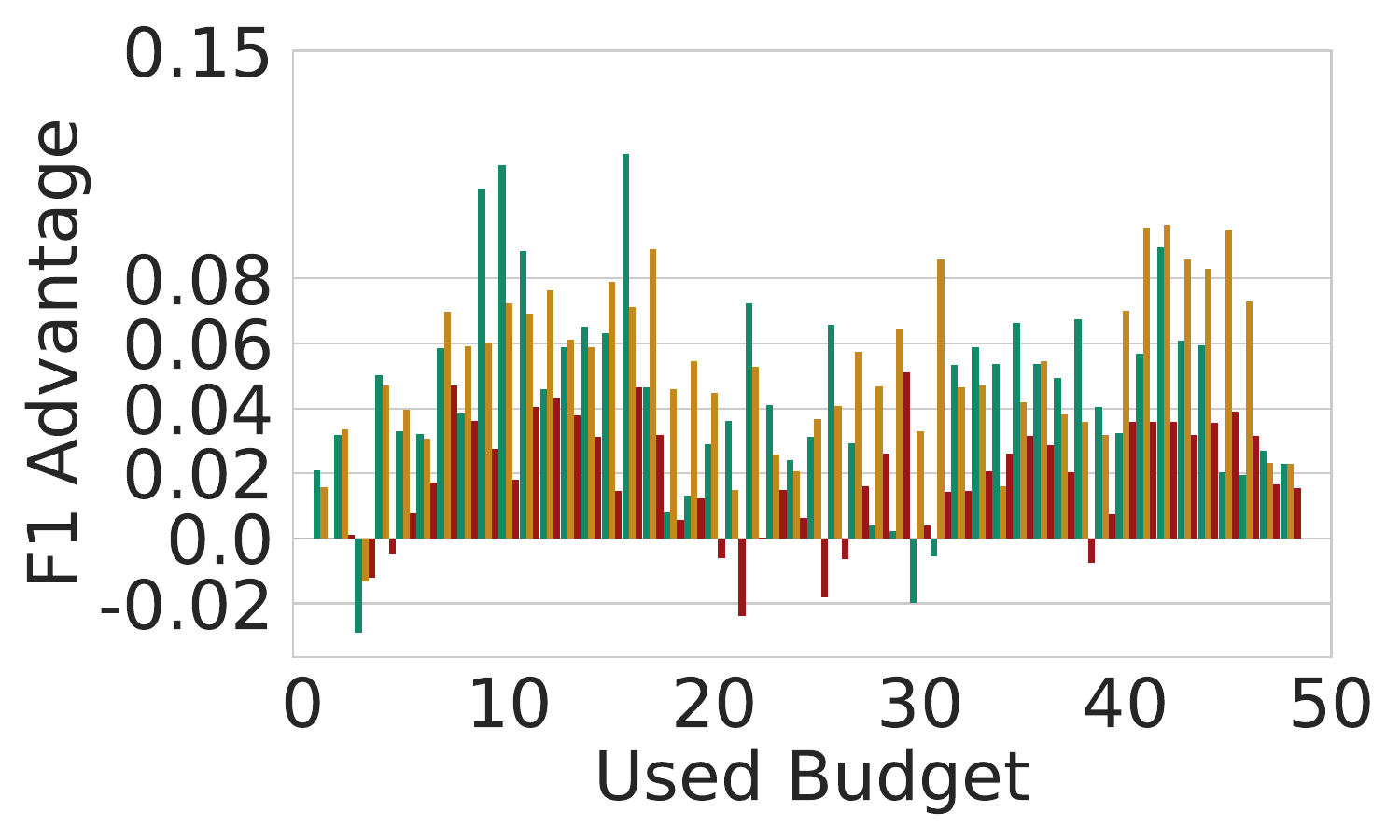}
        \caption{CMC}
    \end{subfigure}
    \begin{subfigure}{0.24\textwidth}
        \includegraphics[width=\linewidth]{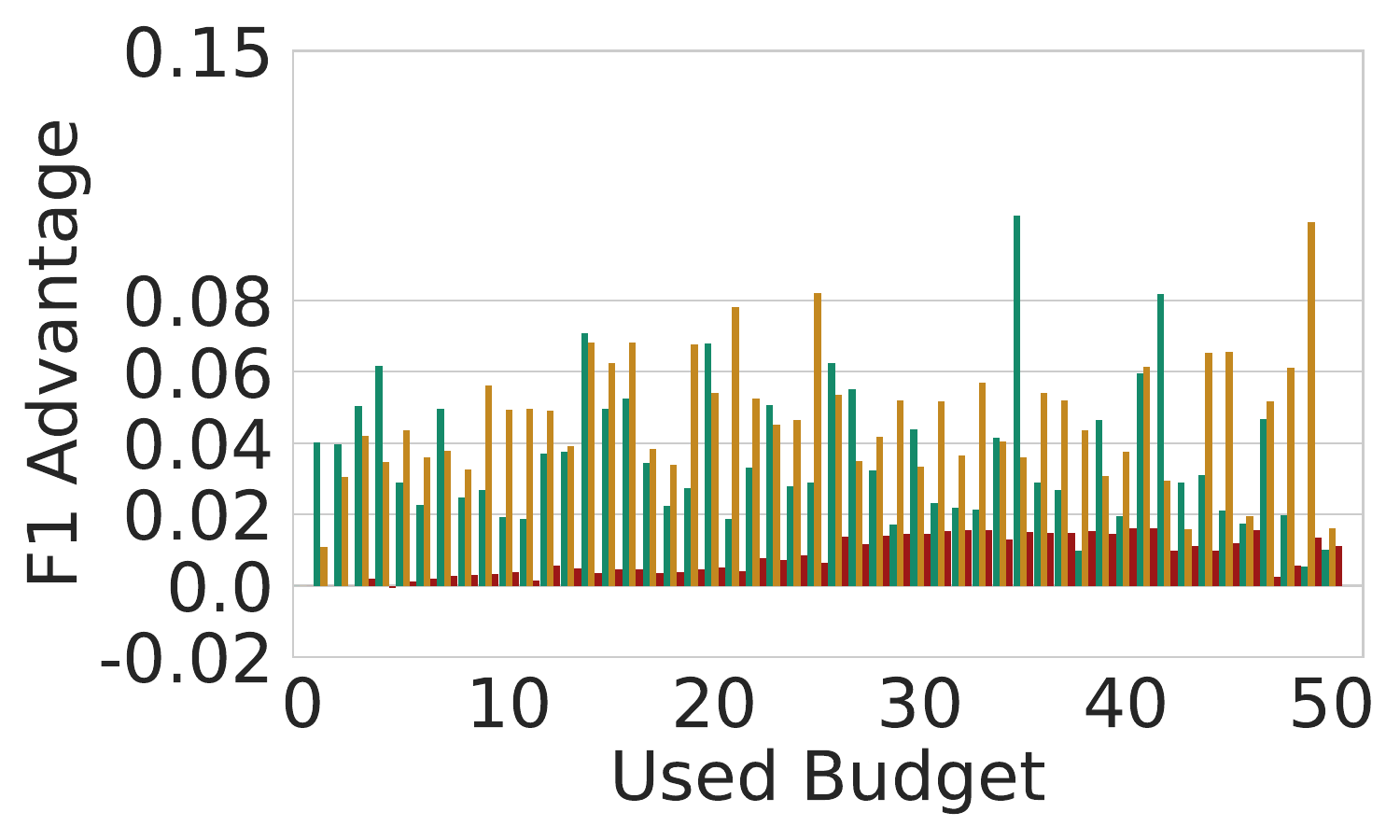}
        \caption{Churn}
    \end{subfigure}
    \begin{subfigure}{0.24\textwidth}
        \includegraphics[width=\linewidth]{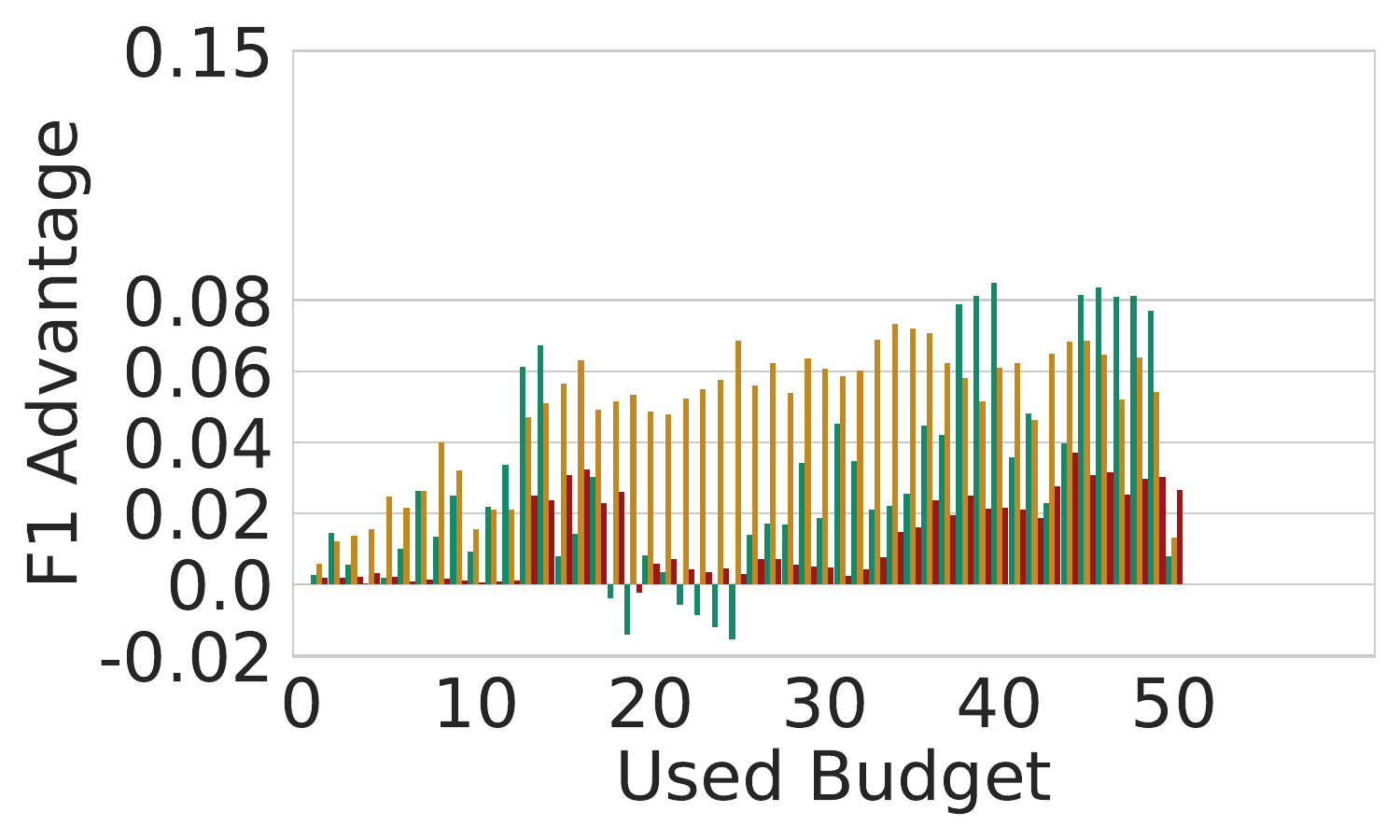}
        \caption{EEG}
    \end{subfigure}
    \begin{subfigure}{0.24\textwidth}
        \includegraphics[width=\linewidth]{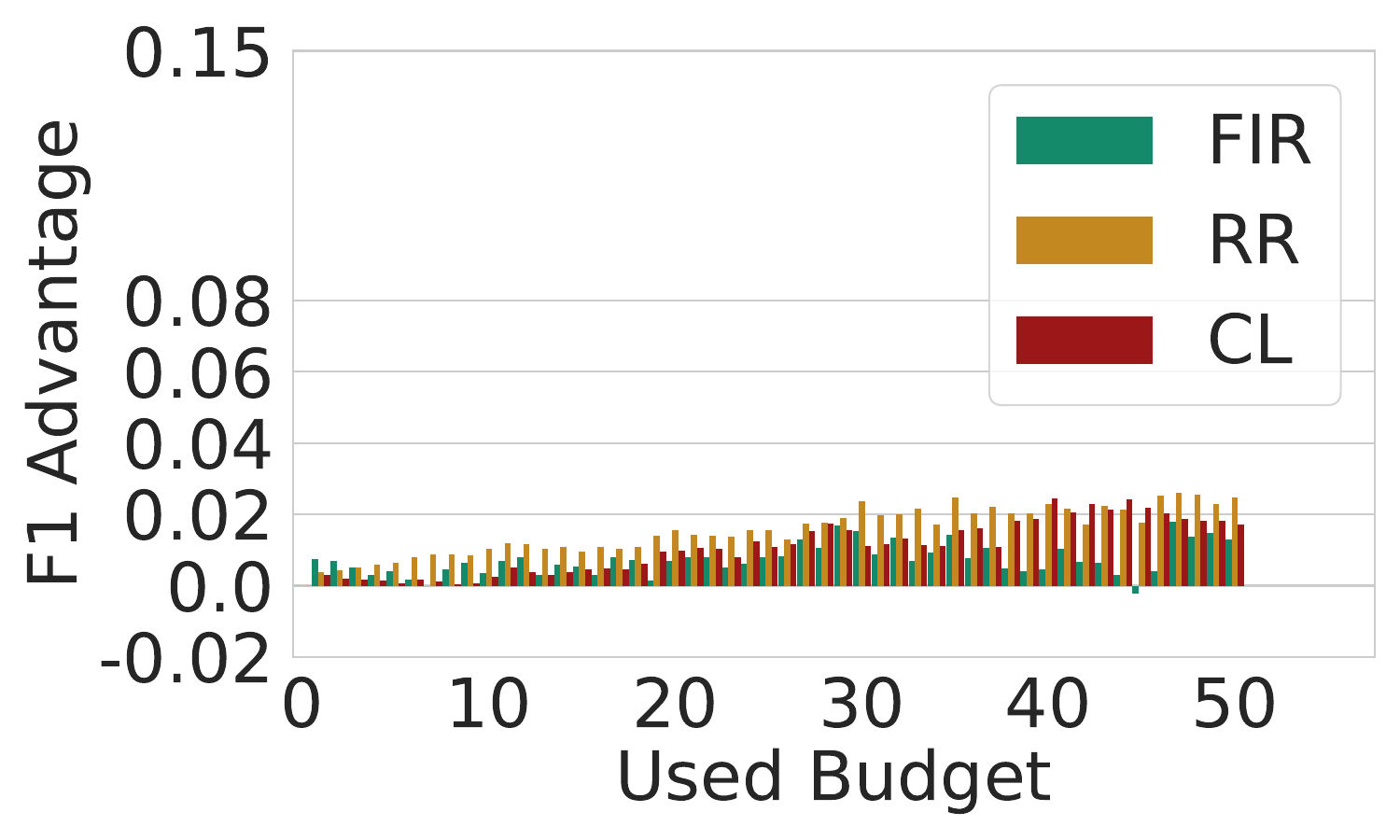}
        \caption{S-Credit}
    \end{subfigure}
    \caption{\revision{Comparison of~\systemname with the baselines for SVM across multiple error types and cost functions.}}
    \label{fig:agg_bl_multi_error_results}
\end{figure*}
\begin{figure*}[!tp]
    \centering
    \begin{subfigure}{0.24\textwidth}
        \includegraphics[width=\linewidth]{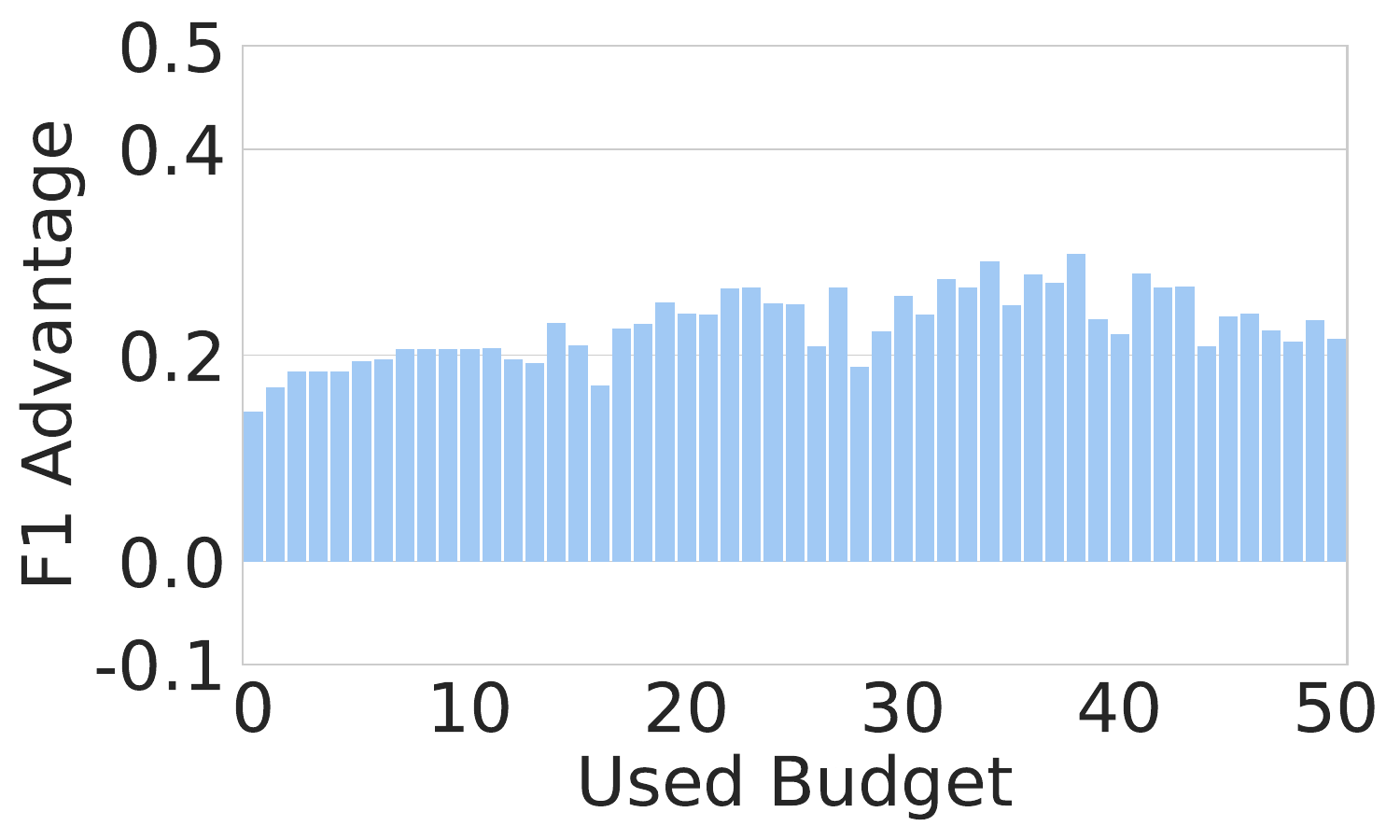}
        \caption{CMC}
    \end{subfigure}
    \begin{subfigure}{0.24\textwidth}
        \includegraphics[width=\linewidth]{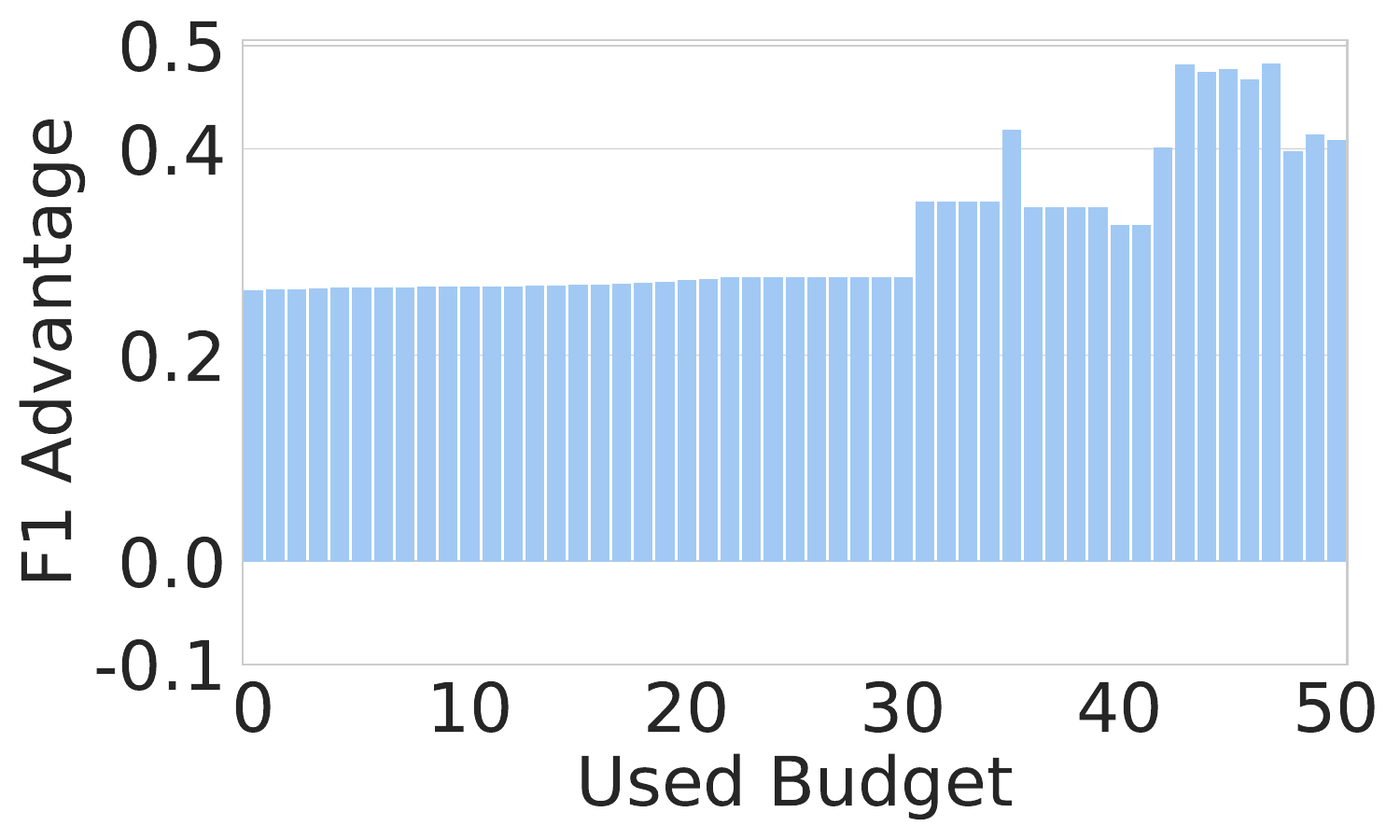}
        \caption{Churn}
    \end{subfigure}
    \begin{subfigure}{0.24\textwidth}
        \includegraphics[width=\linewidth]{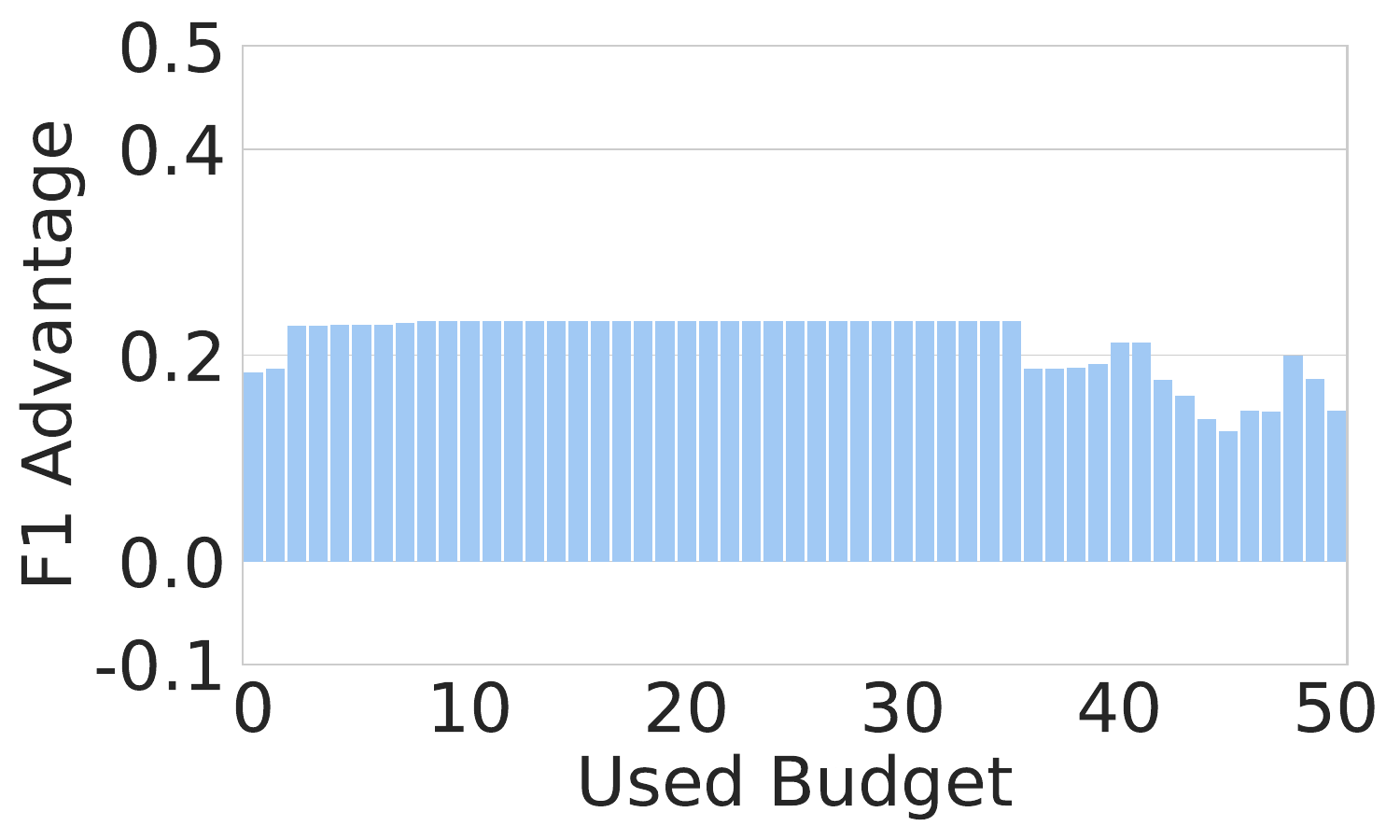}
        \caption{EEG}
    \end{subfigure}
    \begin{subfigure}{0.24\textwidth}
        \includegraphics[width=\linewidth]{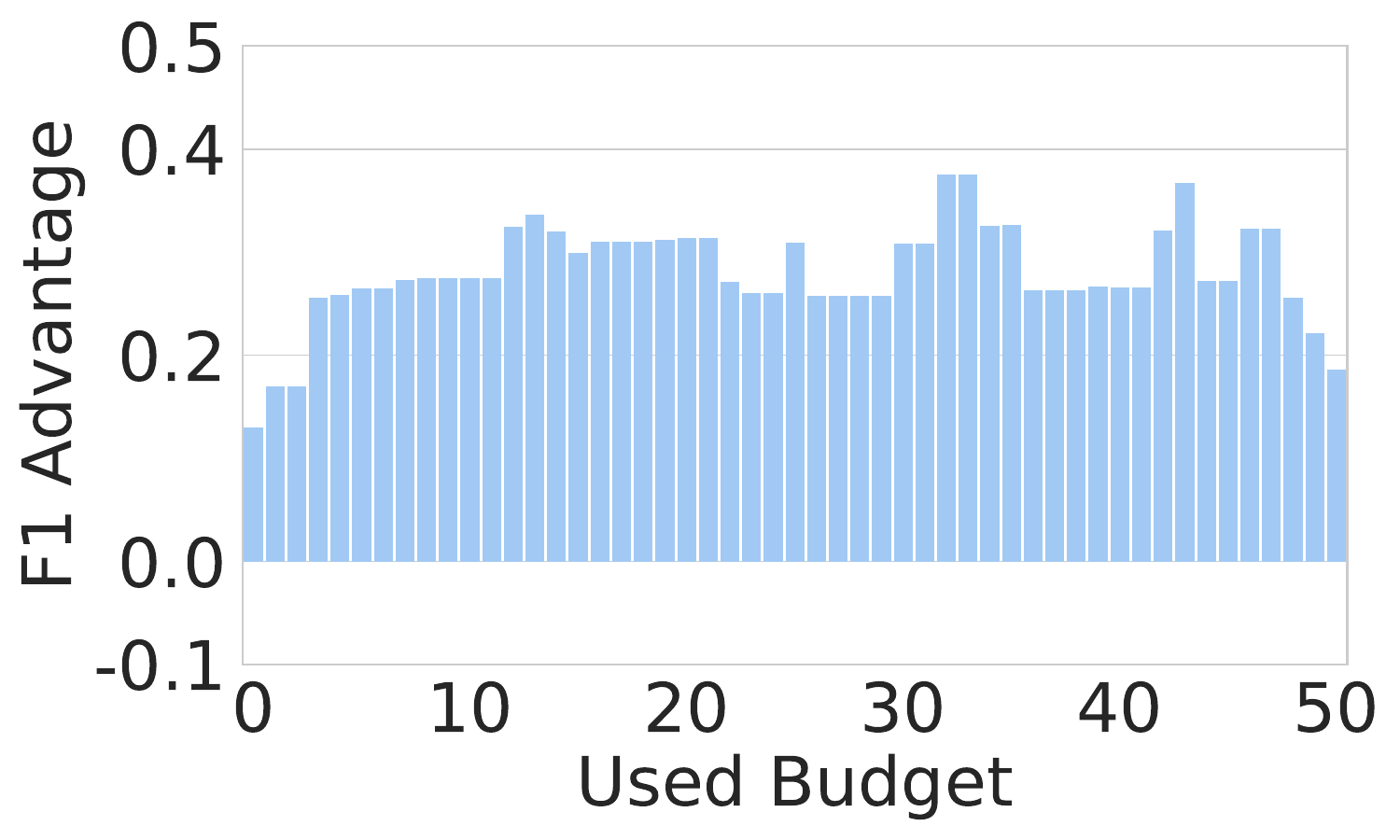}
        \caption{S-Credit}
    \end{subfigure}
    \caption{Comparison of~\systemname with AC for LIR across multiple error types and cost functions.}
    \label{fig:agg_ac_multi_error_results}
\end{figure*}

This section delves into the findings from our extensive evaluation of \systemname, conducted across various datasets, error types, ML algorithms, and baselines.
We designed these experiments to answer the following research questions~(RQ) under two scenarios:

{\setlength{\leftmargini}{13pt}
\begin{enumerate}
    \item \revision{How effective is \systemname in multi-error types and diverse cost functions scenario?}\label{rq:6}
    \item How effective is \systemname compared to baselines FIR, RR, and CL?\label{rq:1}
    \item How effective is \systemname compared to related work AC~\cite{krishnan2016activeclean}?\label{rq:2}
    \item How does \systemname performance vary among different error types and ML algorithms?\label{rq:3}
    \item How accurate are the predictions produced by the \emph{Estimator}?\label{rq:4}
    \item How efficient is \systemname in terms of runtime?\label{rq:5}
\end{enumerate}}
\revision{For the comparison with the respective baselines (RQ~\ref{rq:6}-RQ~\ref{rq:3}), we calculate the F1 score differences (shown as \textit{F1 advantage} in the plots) for each cleaning step and the considered baseline we average these differences across the pre-pollution setting (for the datasets used with pre-pollution).
A positive F1 advantage indicates that \systemname outperforms the corresponding baseline.}

\subsection{\revision{Comparison to baselines for multiple error types and diverse cost functions}}

For the first research question~(RQ~\ref{rq:6}), we compare the performance of \systemname with FIR\revision{, RR, CL (see Figure~\ref{fig:agg_bl_multi_error_results}) and} AC (see Figure~\ref{fig:agg_ac_multi_error_results}), considering multiple error types present in the data/per feature and various cost functions \revision{, as introduced in Section~\ref{sec:cost_models},} for cleaning.
As the CleanML datasets contain only single error types, we excluded them and focused on the datasets we pre-polluted~(see Table~\ref{tbl:datasets}).
\revision{Since our experimental setup consists of many configurations and pre-pollution settings, we focus on several specific configurations: those using SVM as the ML algorithm for FIR, RR, and CL, and those using LIR for the comparison with AC\@.
Our analysis showed that \systemname achieved the best results with SVM compared to FIR, RR, and CL, and the best results with LIR in comparison to AC.}


\smallsection{Comparison to FIR\revision{, RR and CL}.}
In the comparison with FIR, RR and CL, the results show that \systemname outperforms the baselines in this scenario.
\systemname achieves an F1 score difference of up to 11\pt for CMC\@.
\systemname is also consistently superior throughout the cleaning process for Churn, EEG and S-Credit, although the differences vary.
For example, \systemname achieves a difference of 10\pt in the F1 score compared to FIR in Churn with an invested budget of 35, while for the next cleaning step, the difference drops to 3\pt.
Overall, \systemname maintains a positive difference across all datasets.
\revision{The difference for S-Credit is smaller.
Our analysis shows that the average difference between the F1 score in dirty and fully cleaned state is -1.5\pt, limiting the potential for clear superiority.}

Considering different error types within a feature and introducing various cost functions expands the search space for potential cleaning steps.
FIR and RR more frequently make suboptimal decisions.
The different cost functions reinforce these findings, especially for linear cleaning costs, where repeated poor decisions are heavily penalized by consuming more budget.
In contrast, \systemname accurately estimates the impact of cleaning combinations of features and error types, maintaining its superiority.

\begin{figure*}
    \centering
    \raisebox{1.4\height}{\rotatebox{90}{\textbf{CMC}}}\hspace{0.3em}%
    \begin{subfigure}{0.24\textwidth}
        \includegraphics[width=\linewidth]{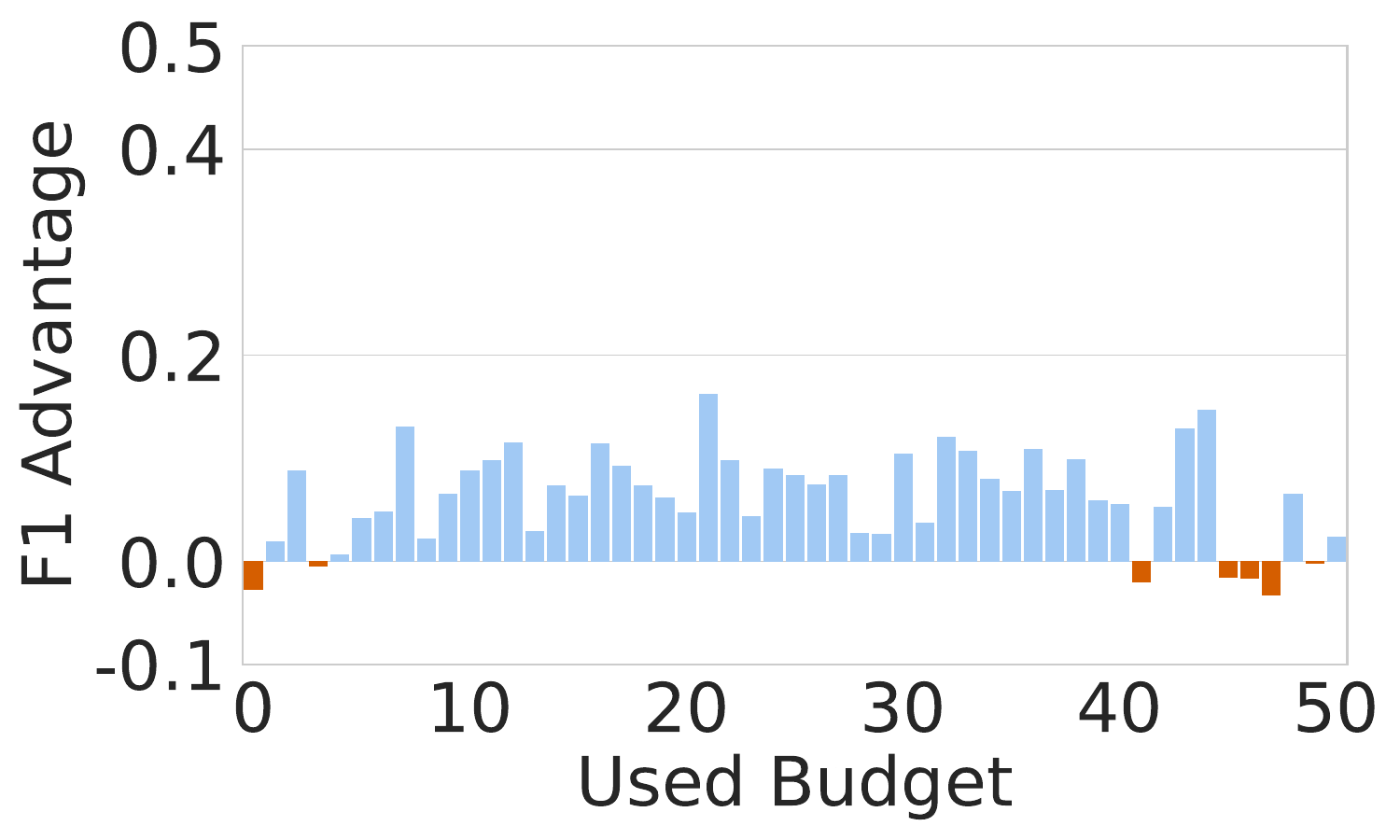}
    \end{subfigure}\hfill
    \begin{subfigure}{0.24\textwidth}
        \includegraphics[width=\linewidth]{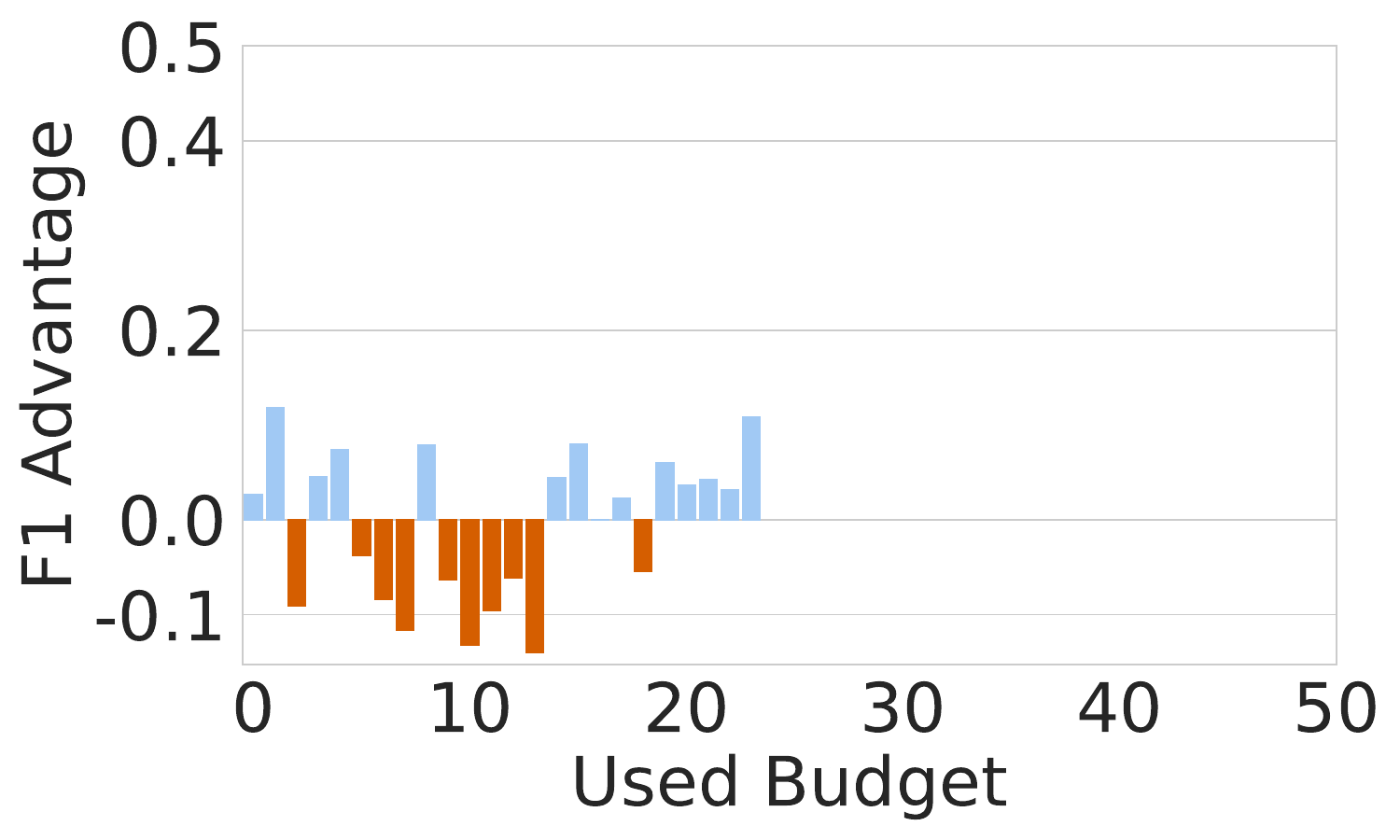}
    \end{subfigure}\hfill
    \begin{subfigure}{0.24\textwidth}
        \includegraphics[width=\linewidth]{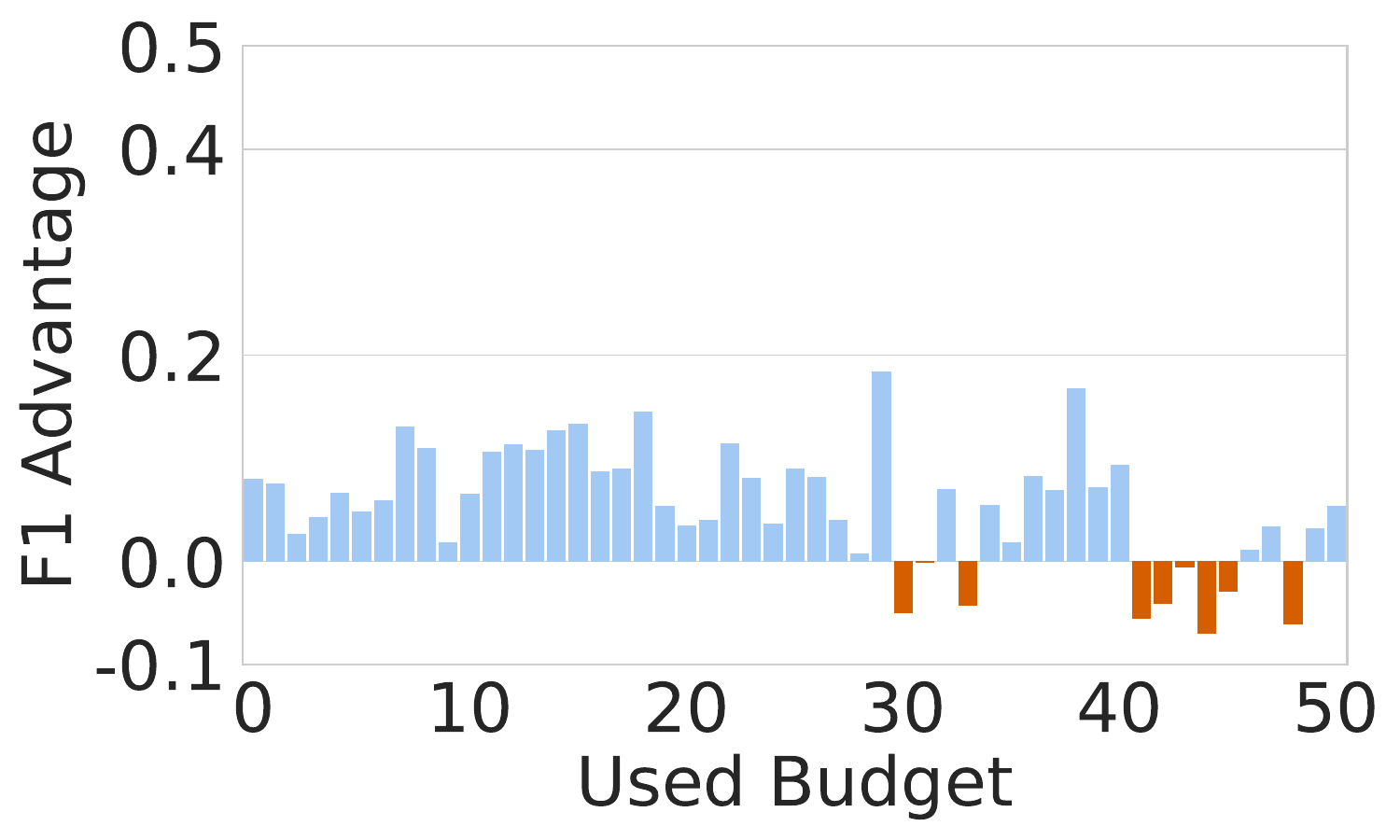}
    \end{subfigure}\hfill
    \begin{subfigure}{0.24\textwidth}
        \includegraphics[width=\linewidth]{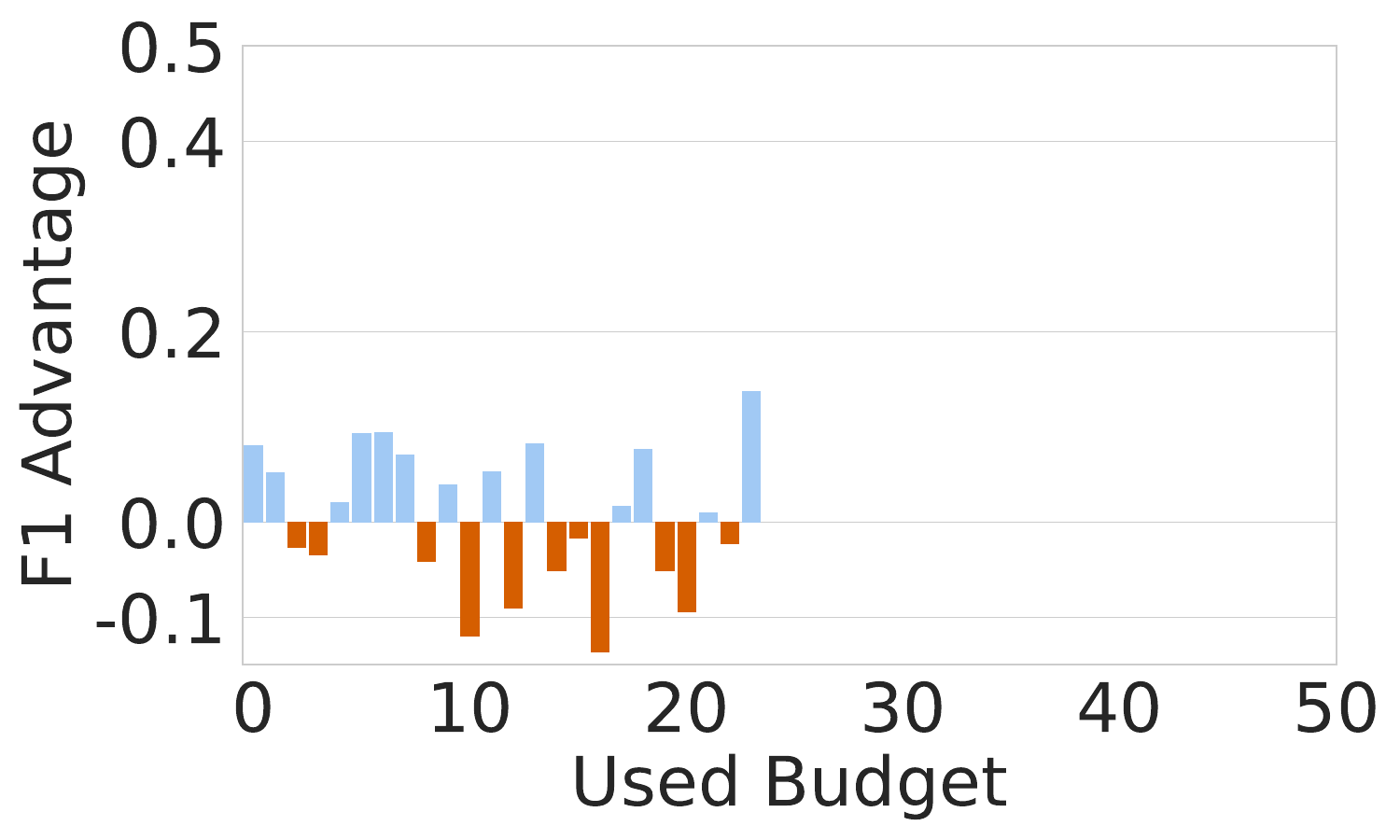}
    \end{subfigure}
    
    \vspace{-0.1em}

        \raisebox{1.2\height}{\rotatebox{90}{\textbf{Churn}}}\hspace{0.3em}%
    \begin{subfigure}{0.24\textwidth}
        \includegraphics[width=\linewidth]{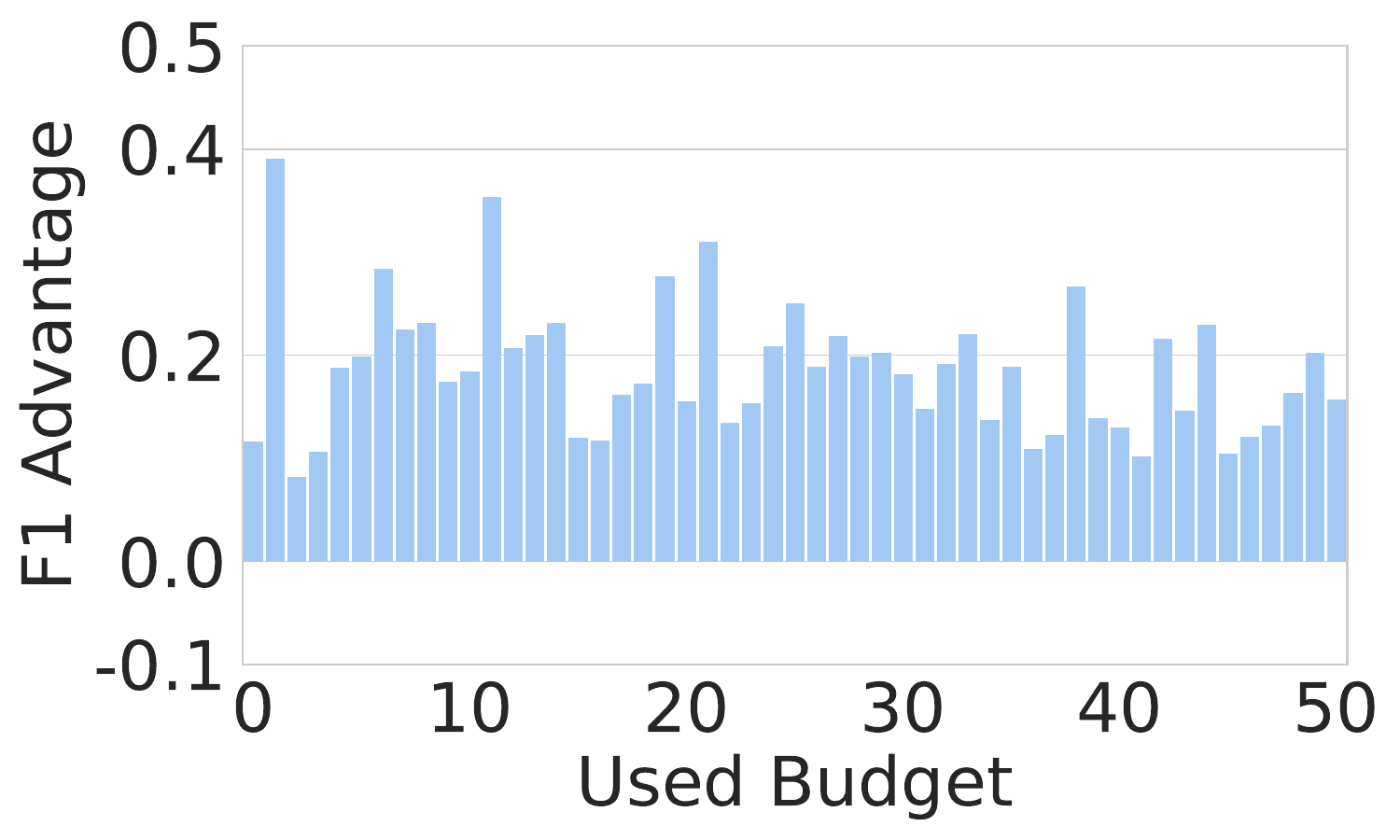}
    \end{subfigure}\hfill
    \begin{subfigure}{0.24\textwidth}
        \includegraphics[width=\linewidth]{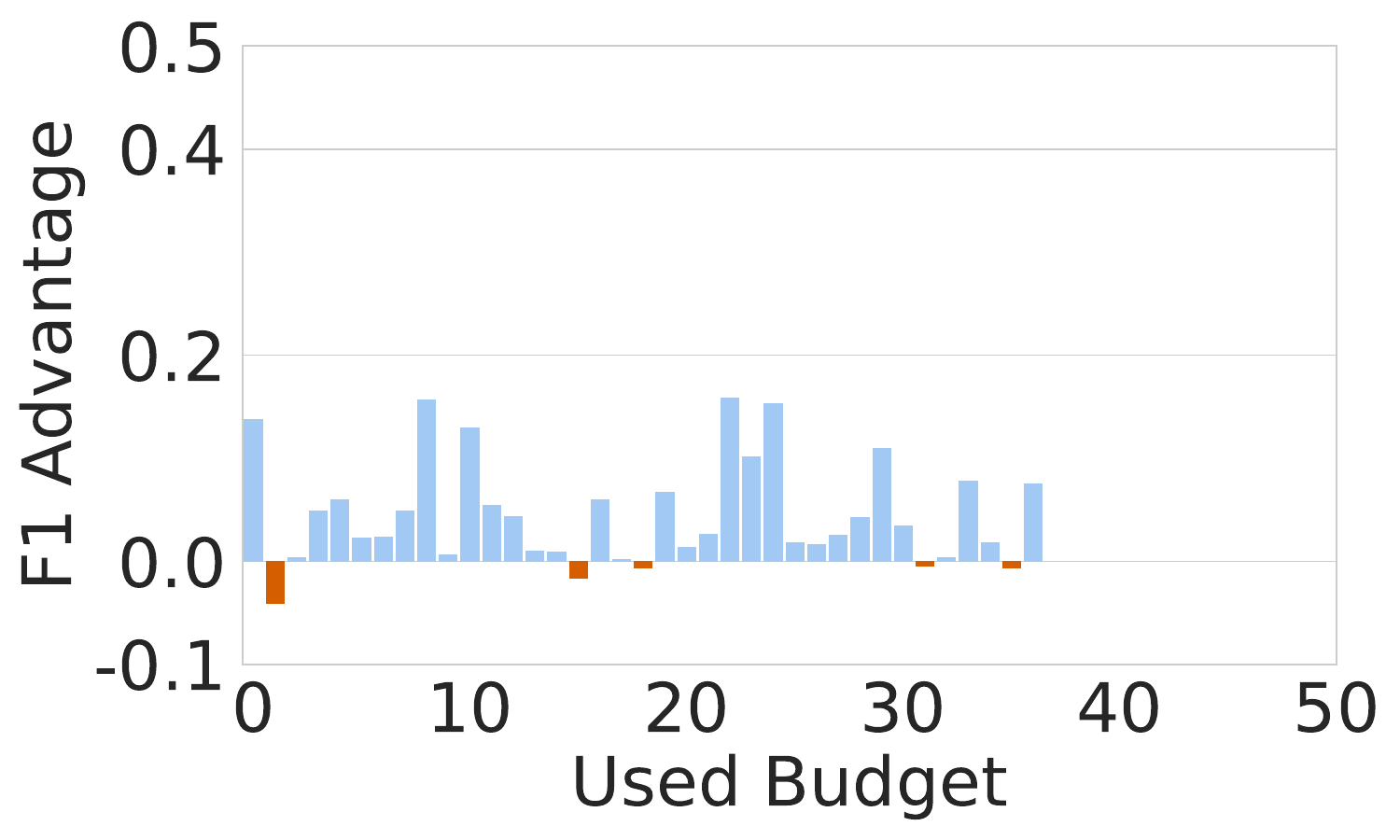}
    \end{subfigure}\hfill
    \begin{subfigure}{0.24\textwidth}
        \includegraphics[width=\linewidth]{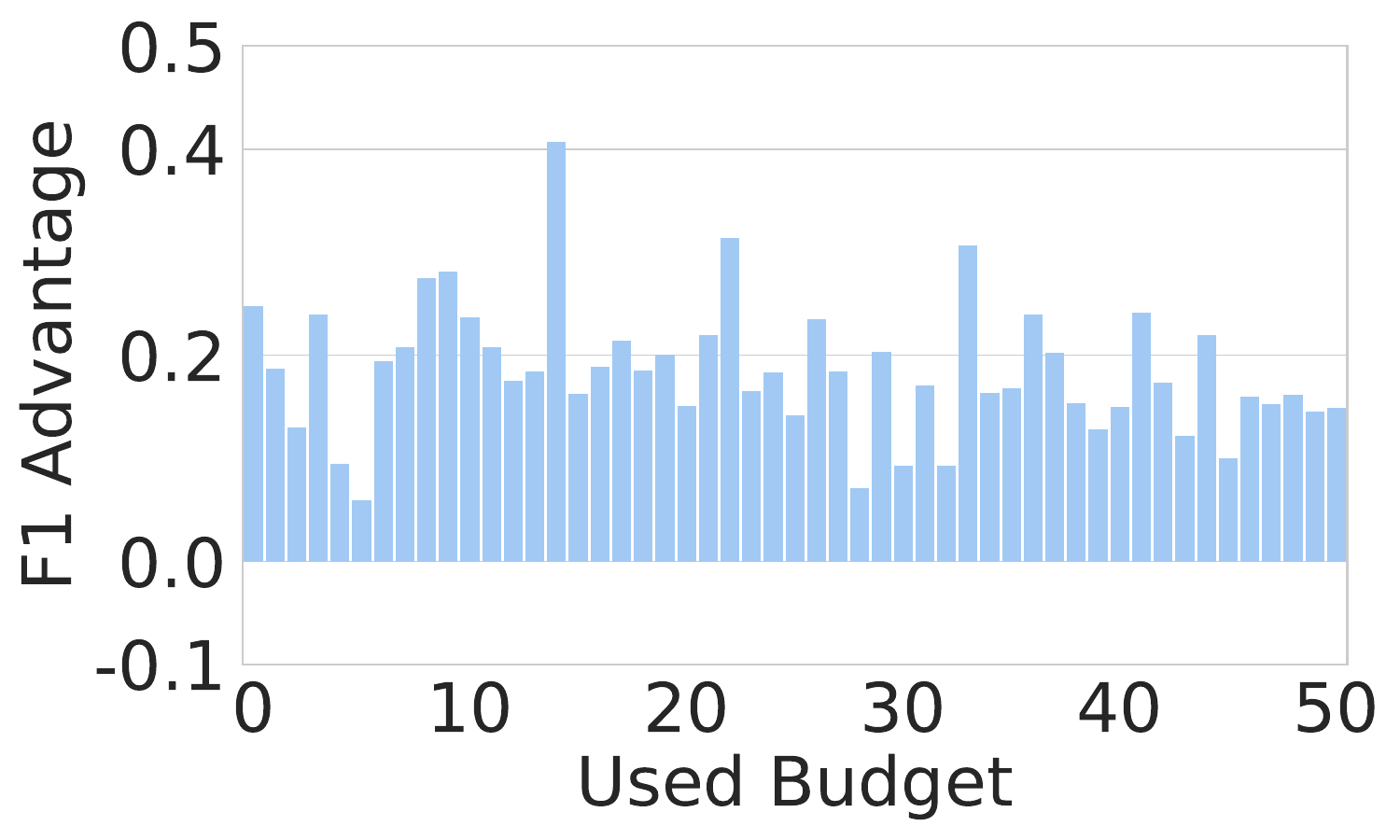}
    \end{subfigure}\hfill
    \begin{subfigure}{0.24\textwidth}
        \includegraphics[width=\linewidth]{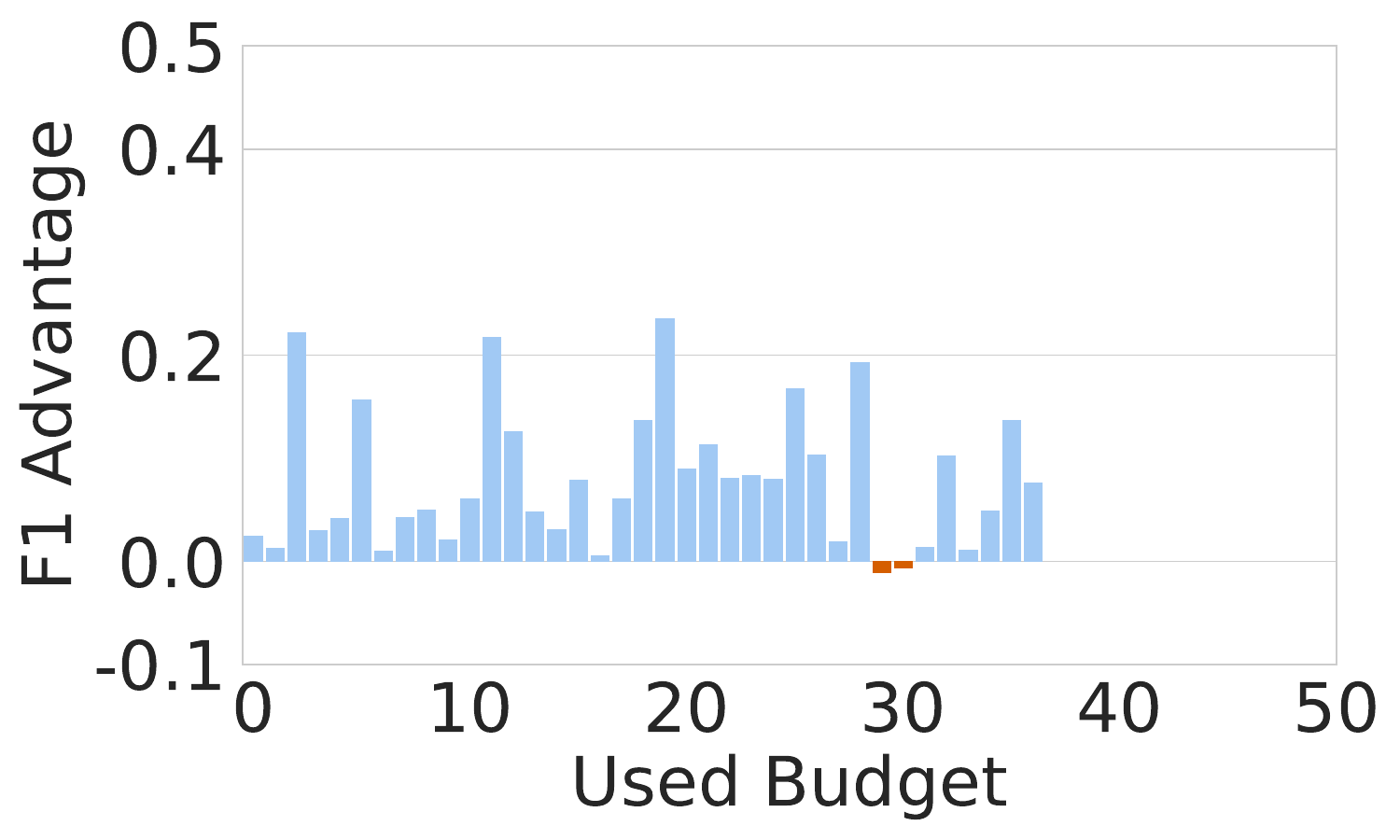}
    \end{subfigure}
    
    \vspace{-0.1em}

    \raisebox{2.\height}{\rotatebox{90}{\textbf{EEG}}}\hspace{0.3em}%
    \begin{subfigure}{0.24\textwidth}
        \centering\raisebox{3.85\height}{\parbox{0.75\linewidth}{\texttt{EEG only contains numerical features.}}}
    \end{subfigure}\hfill
    \begin{subfigure}{0.24\textwidth}
        \includegraphics[width=\linewidth]{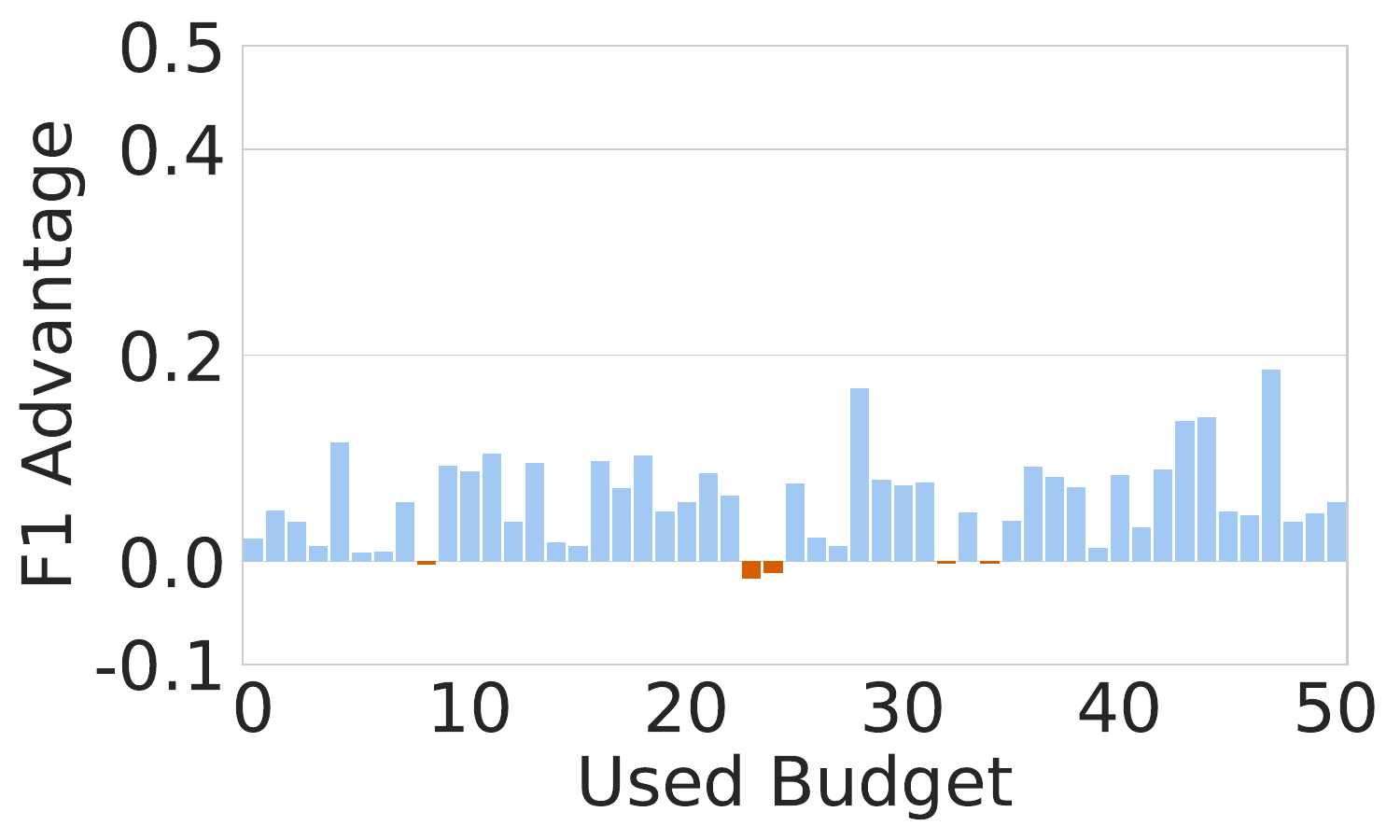}
    \end{subfigure}\hfill
    \begin{subfigure}{0.24\textwidth}
        \includegraphics[width=\linewidth]{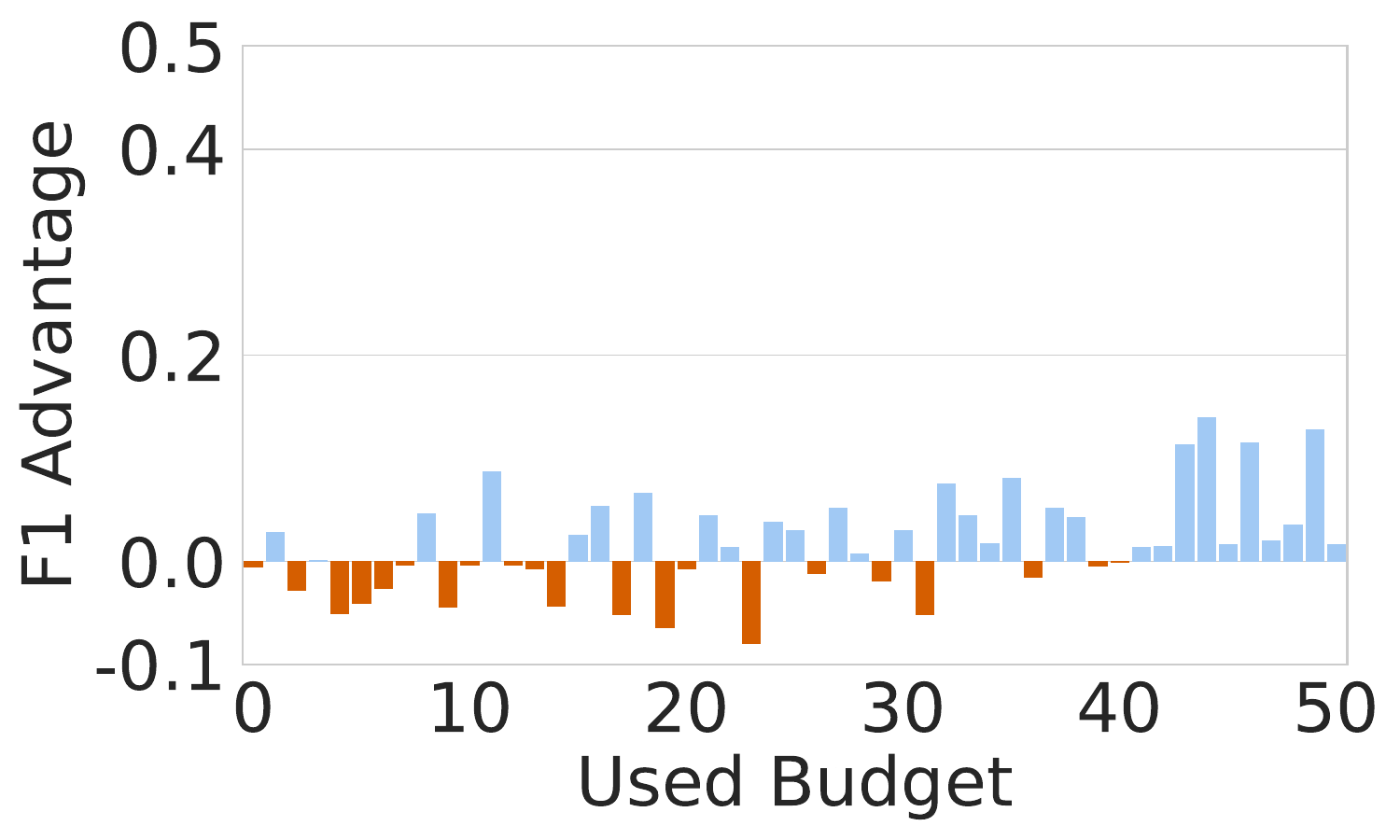}
    \end{subfigure}\hfill
    \begin{subfigure}{0.24\textwidth}
        \includegraphics[width=\linewidth]{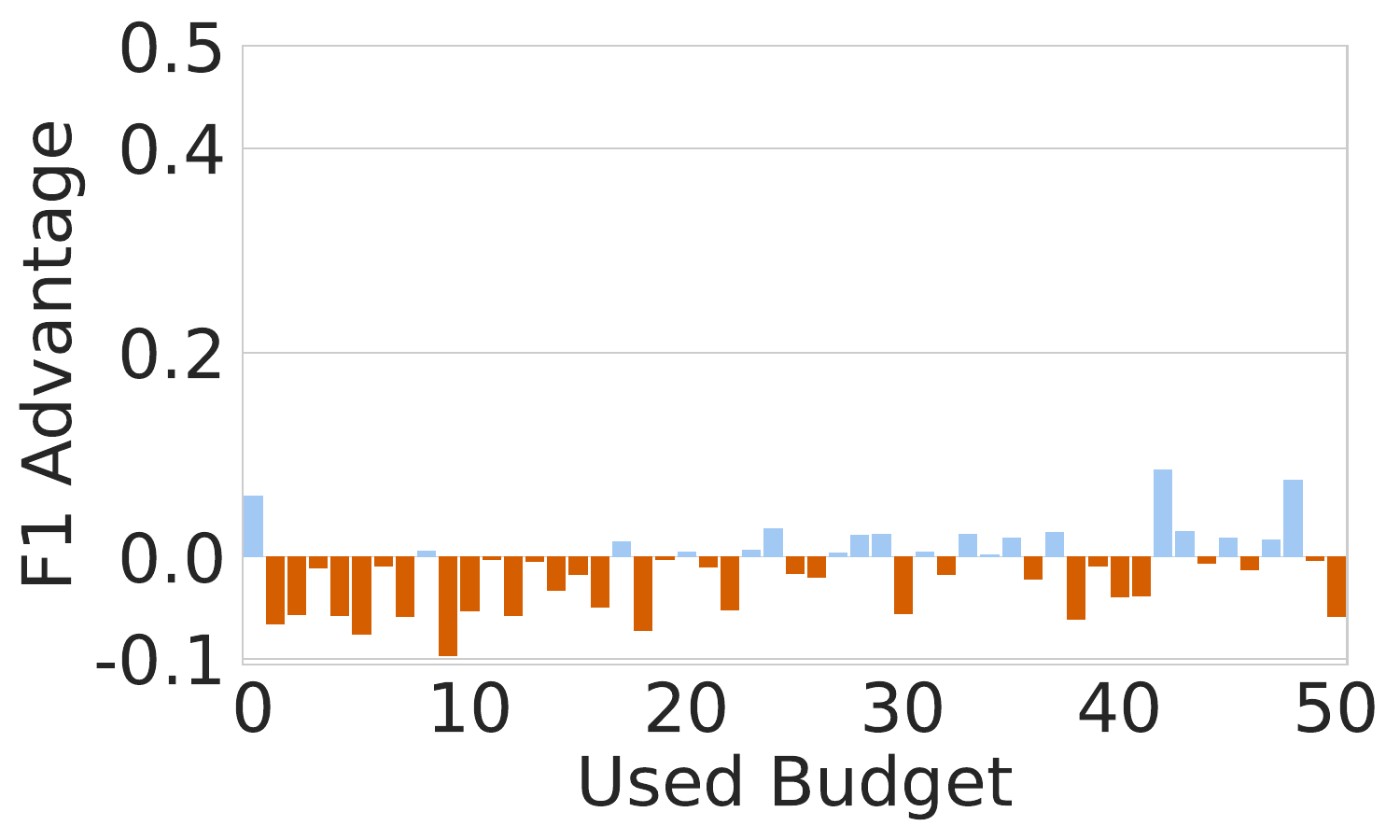}
    \end{subfigure}
    
    \vspace{-0.1em}
    
    \raisebox{1.2\height}{\rotatebox{90}{\textbf{S-Credit}}}\hspace{0.3em}%
    \begin{subfigure}{0.24\textwidth}
        \includegraphics[width=\linewidth]{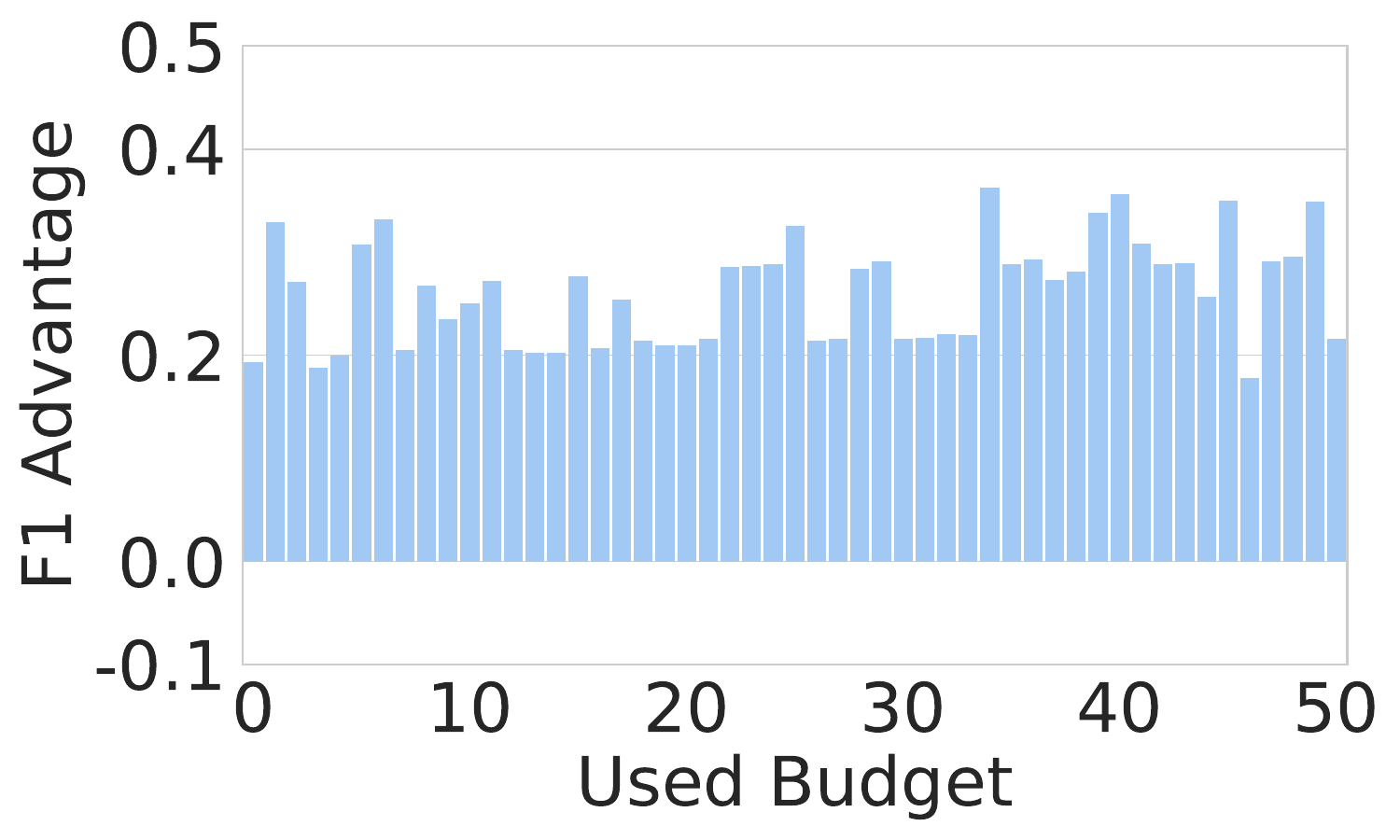}
        \caption{Categorical Shift}
    \end{subfigure}\hfill
    \begin{subfigure}{0.24\textwidth}
        \includegraphics[width=\linewidth]{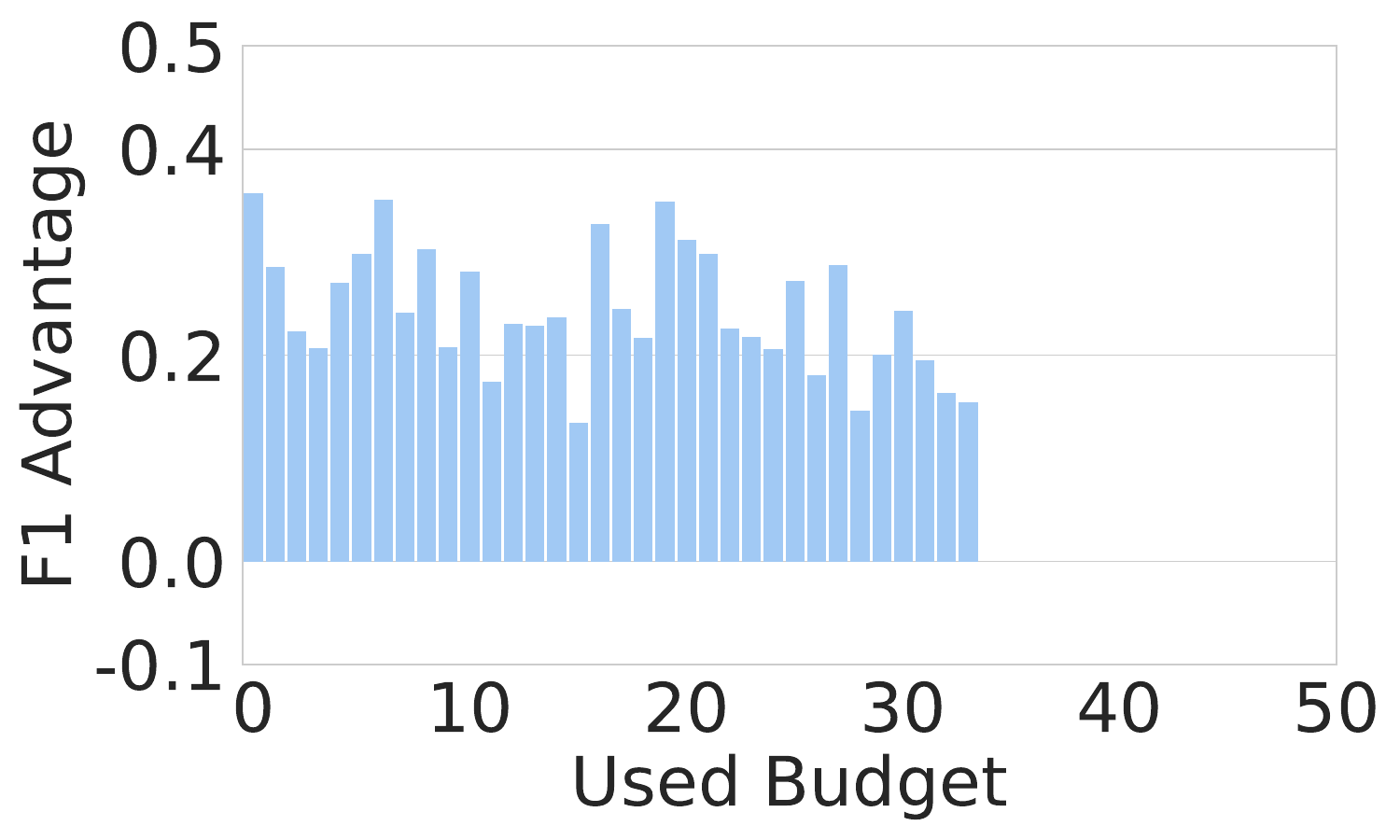}
        \caption{Gaussian Noise}
    \end{subfigure}\hfill
    \begin{subfigure}{0.24\textwidth}
        \includegraphics[width=\linewidth]{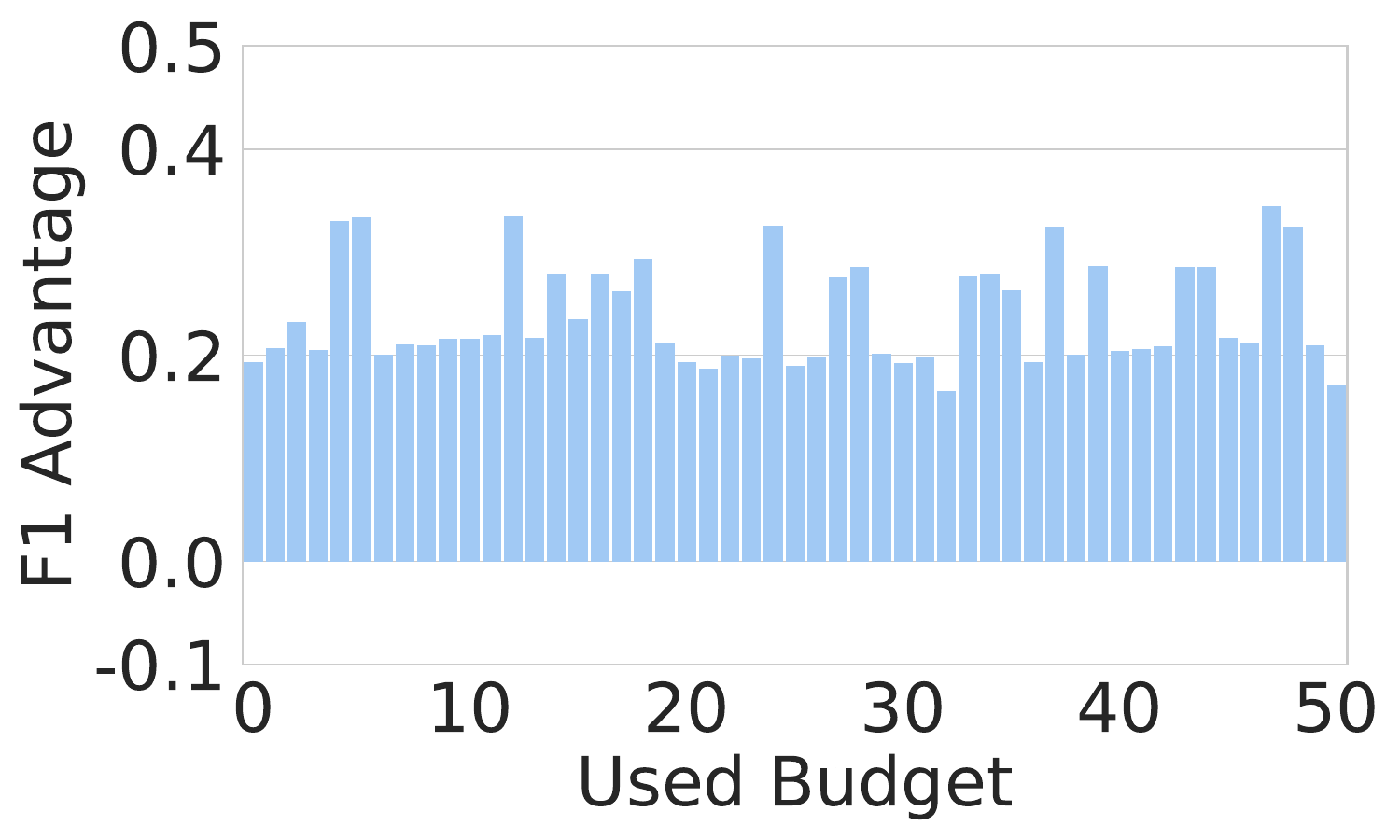}
        \caption{Missing Values}
    \end{subfigure}\hfill
    \begin{subfigure}{0.24\textwidth}
        \includegraphics[width=\linewidth]{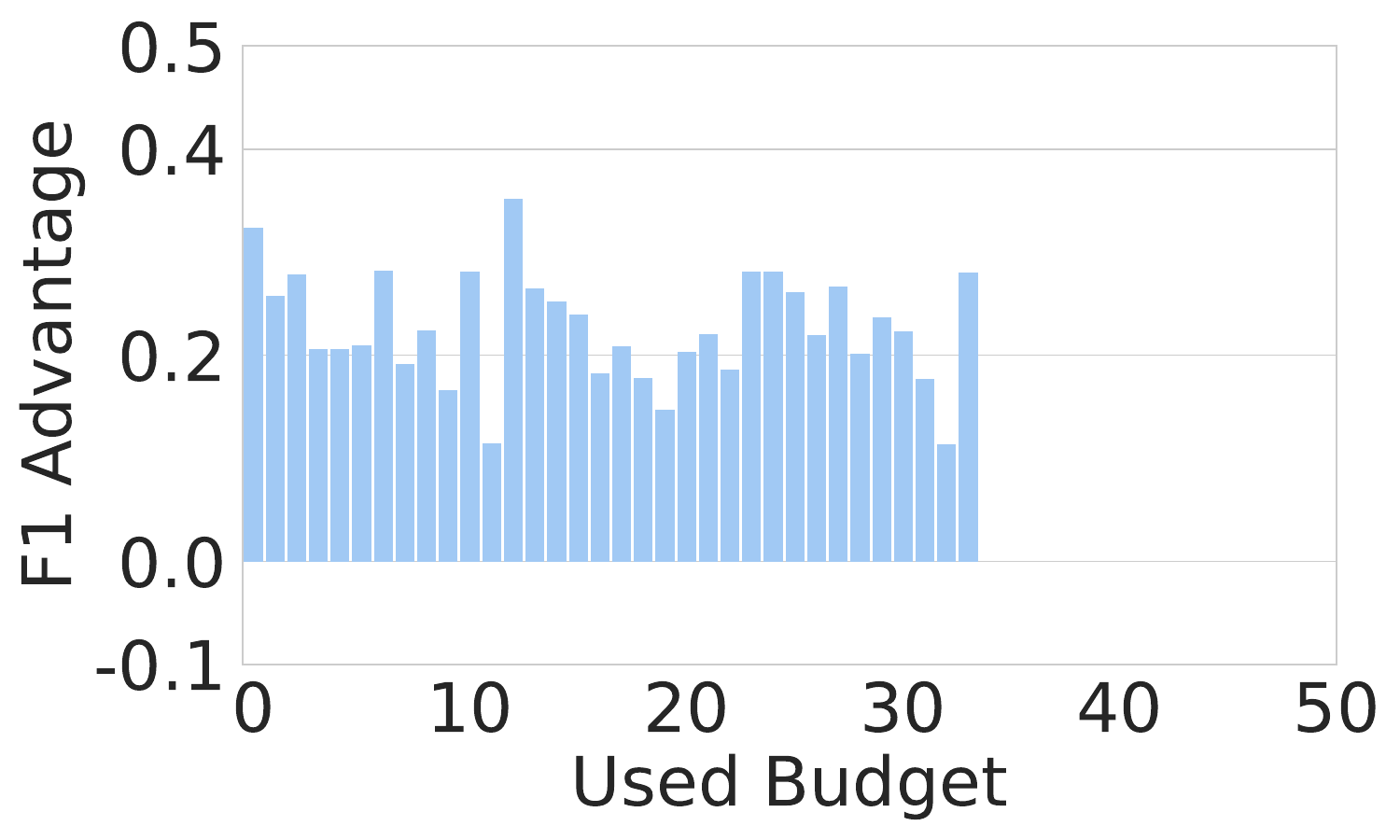}
        \caption{Scaling}
    \end{subfigure}
    \caption{Comparison of~\systemname with AC for AC-SVM across error types.}
    \label{fig:agg_ac_results}
\end{figure*}

\smallsection{Comparison to AC.}
In the comparison to AC, Figure~\ref{fig:agg_ac_multi_error_results} shows that \systemname consistently outperforms AC throughout the entire cleaning process.
In most cases, \systemname achieves a difference of 20\pt in the F1 score across all datasets.
For Churn~(Figure~\ref{fig:agg_ac_multi_error_results}(b)), \systemname even achieves a maximum F1 score difference of nearly 50\pt. 
Overall, the differences are less erratic than in the single-error and constant cost function comparisons~(Figure~\ref{fig:agg_ac_results}), which is attributed to the influence of the cost functions. 
Since AC cleans on a per-record basis, different error types are corrected across multiple features during each cleaning step. 
For comparison, we propagate for longer periods the F1 scores achieved from previously utilized budget units until an actual F1 score is measured for the current unit. 

\subsection{Comparison to FIR\revision{, RR, and CL for a single error type}}
\label{sec:rq1}

\begin{figure*}
    \centering
    \raisebox{1.4\height}{\rotatebox{90}{\textbf{CMC}}}\hspace{0.3em}%
    \begin{subfigure}{0.24\textwidth}
        \includegraphics[width=\linewidth]{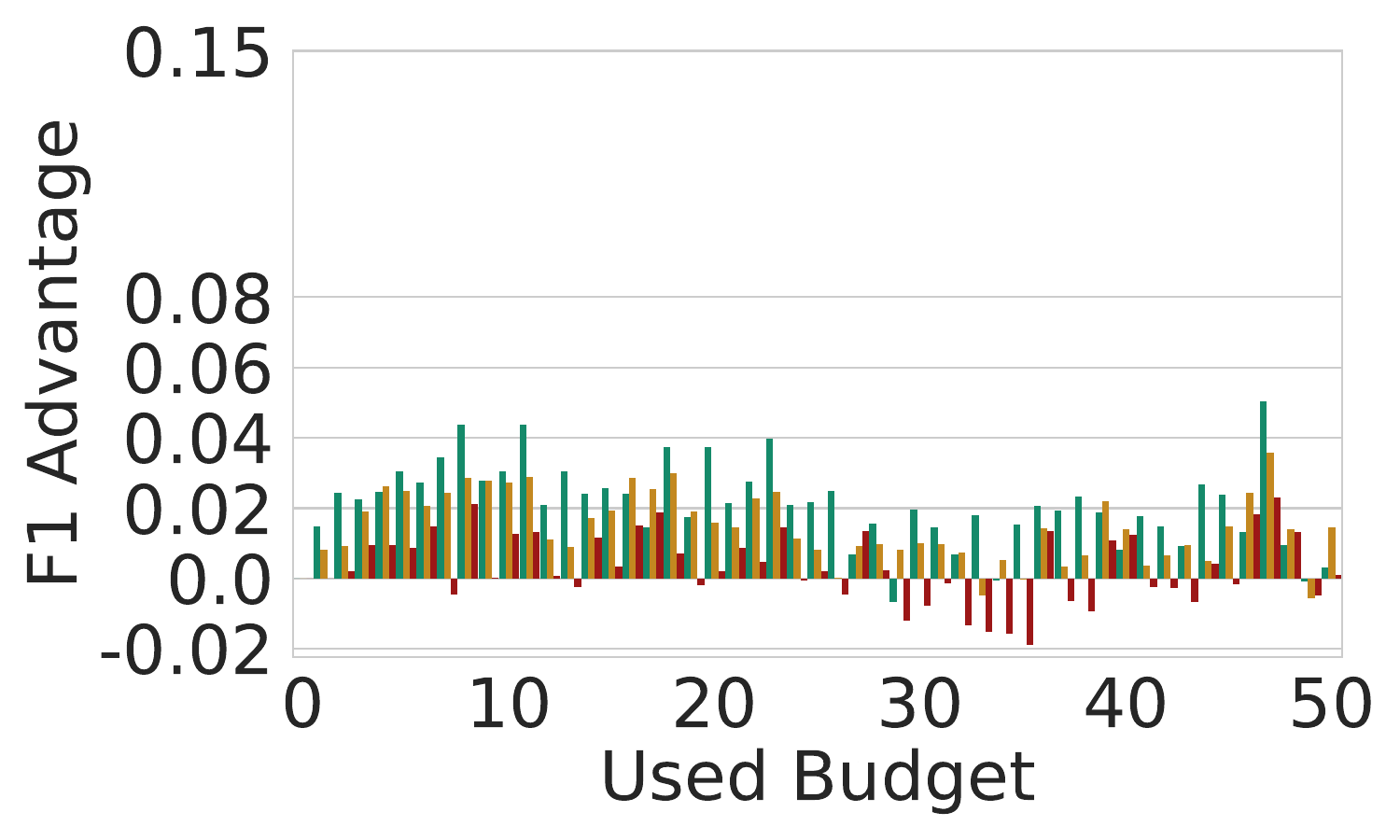}
    \end{subfigure}\hfill
    \begin{subfigure}{0.24\textwidth}
        \includegraphics[width=\linewidth]{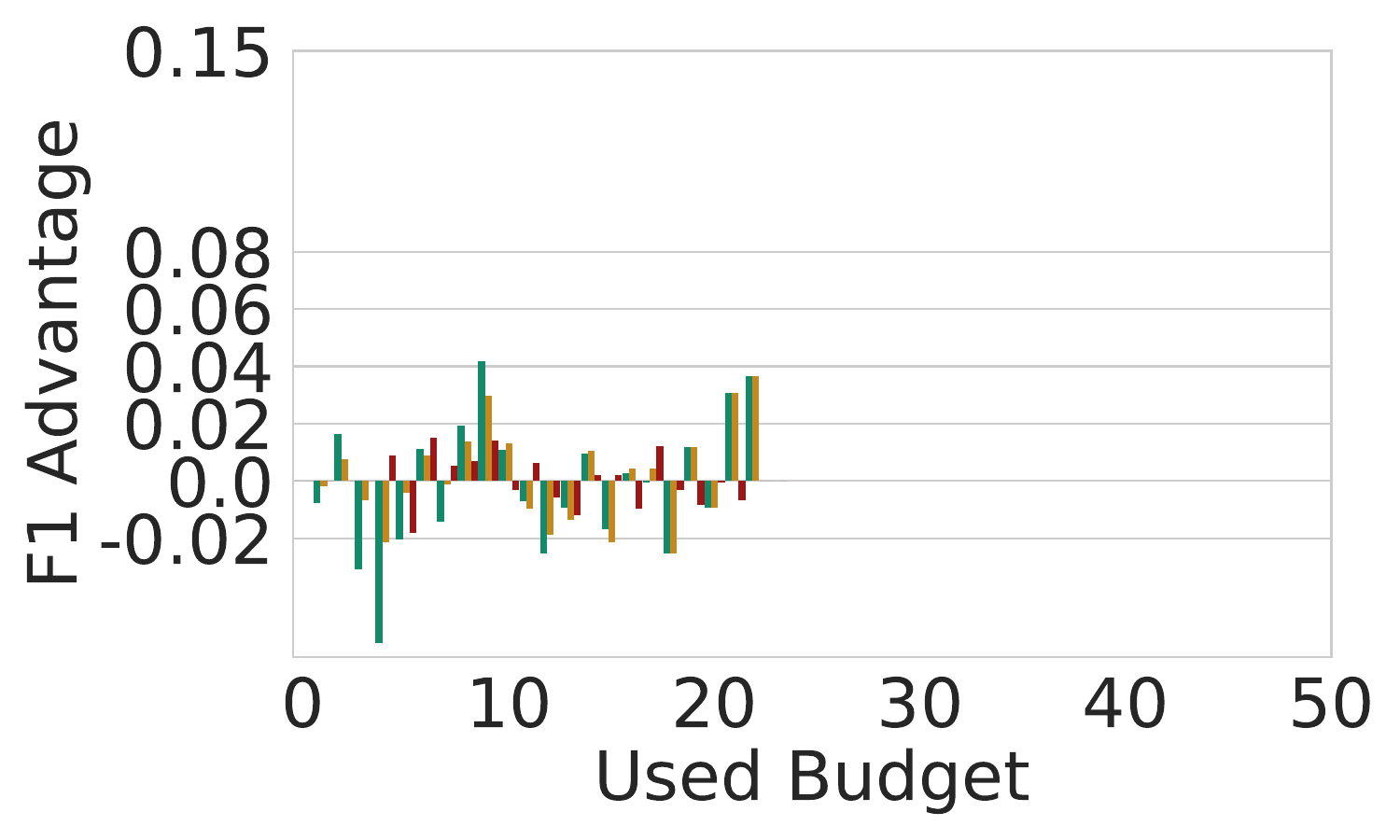}
    \end{subfigure}\hfill
    \begin{subfigure}{0.24\textwidth}
        \includegraphics[width=\linewidth]{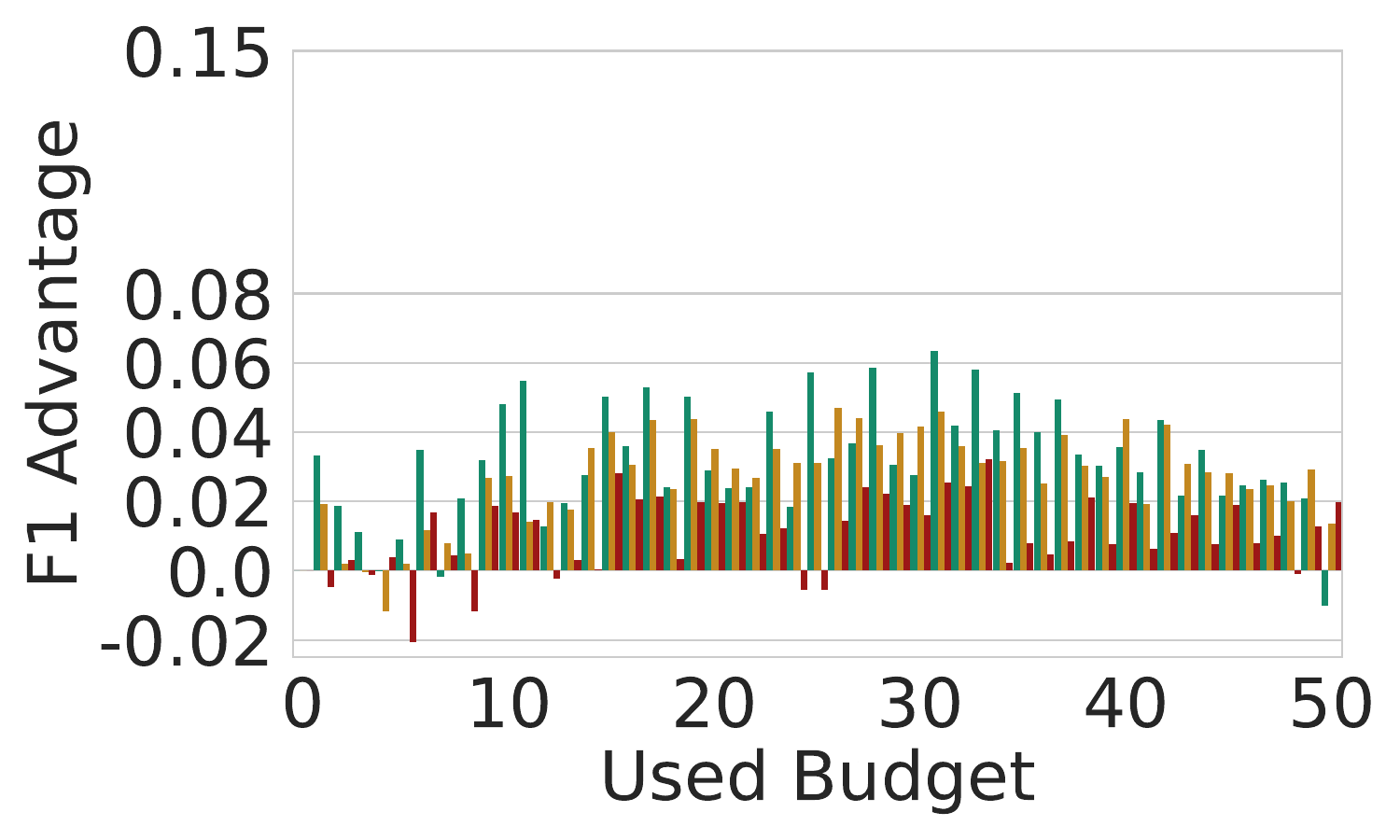}
    \end{subfigure}\hfill
    \begin{subfigure}{0.24\textwidth}
        \includegraphics[width=\linewidth]{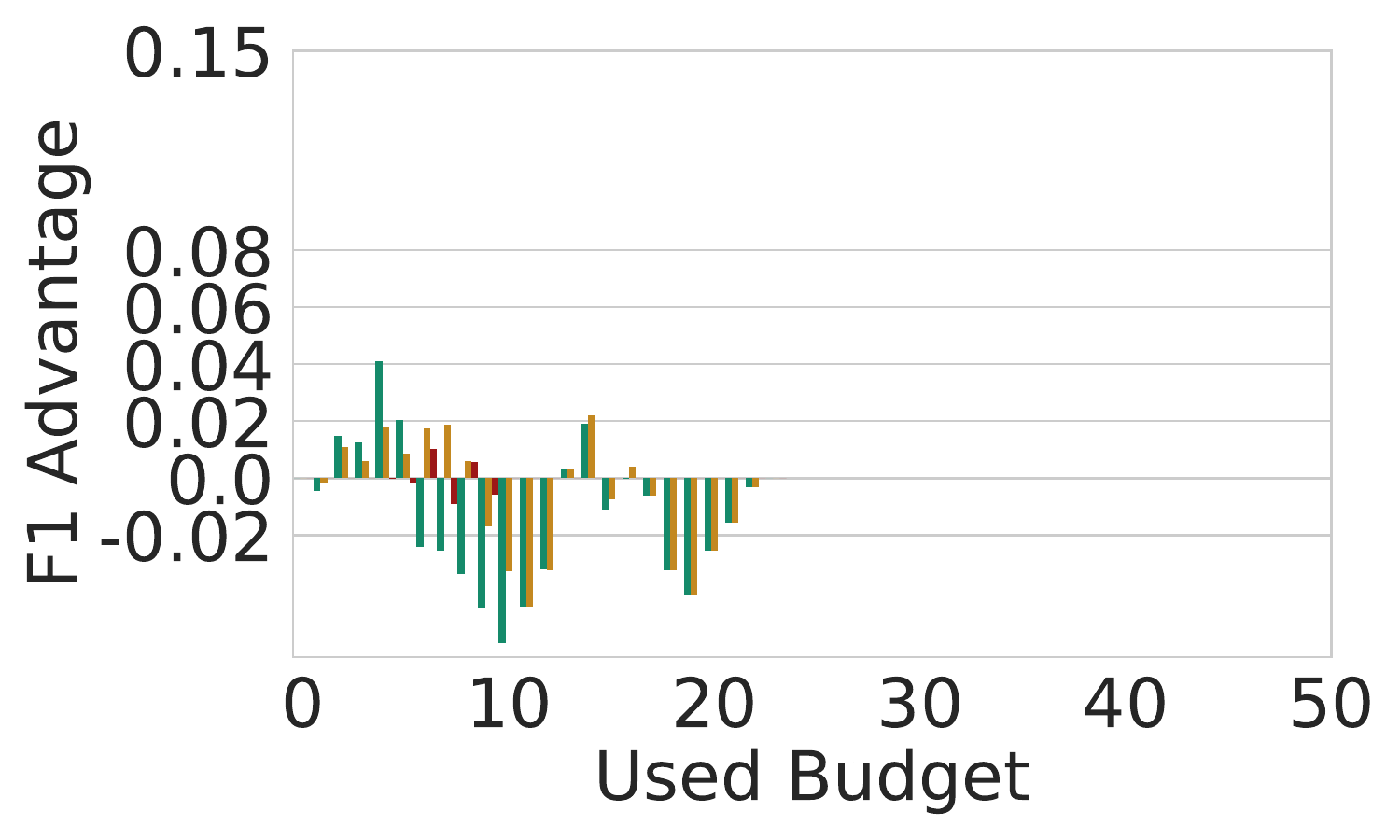}
    \end{subfigure}
    
    \vspace{-0.1em}

        \raisebox{1.2\height}{\rotatebox{90}{\textbf{Churn}}}\hspace{0.3em}%
    \begin{subfigure}{0.24\textwidth}
        \includegraphics[width=\linewidth]{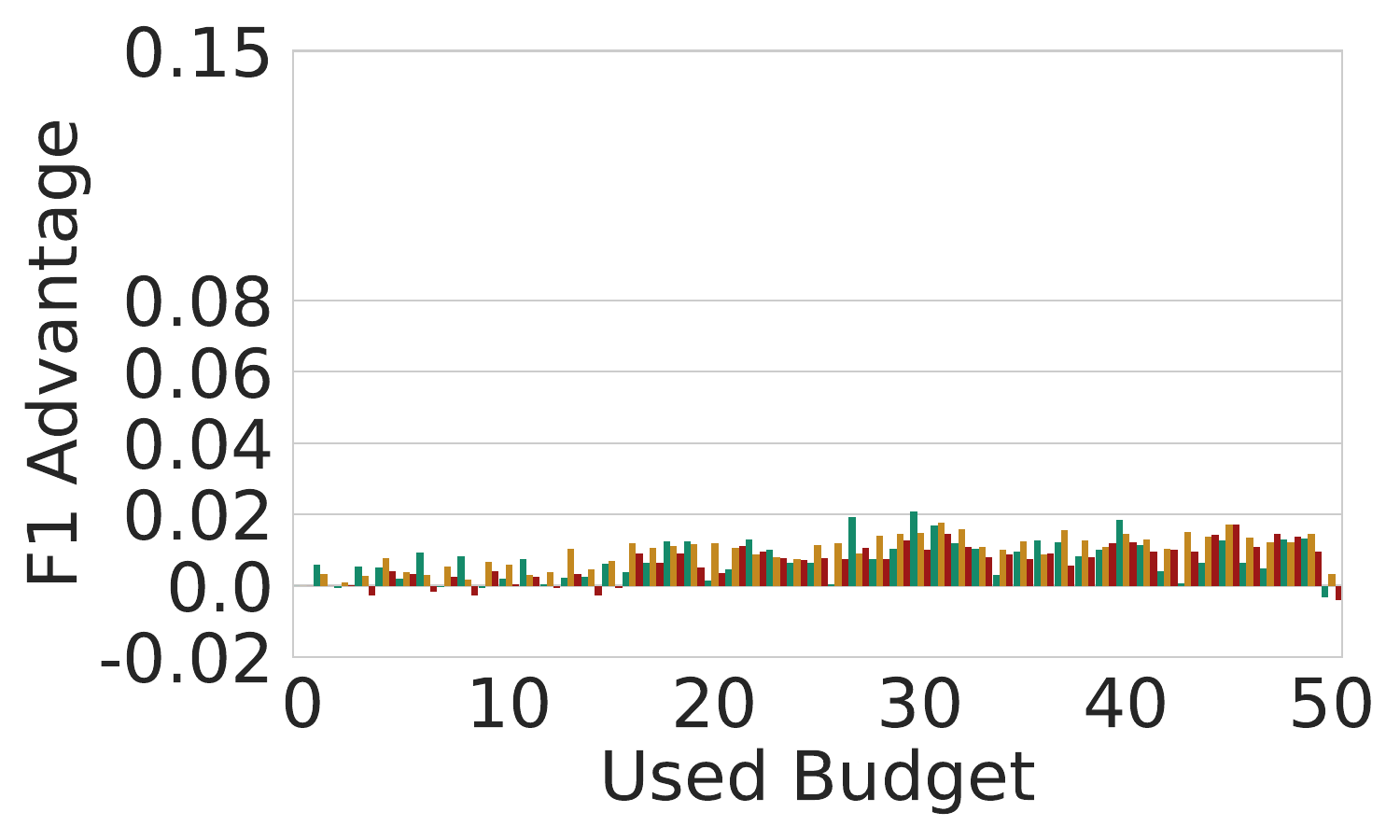}
    \end{subfigure}\hfill
    \begin{subfigure}{0.24\textwidth}
        \includegraphics[width=\linewidth]{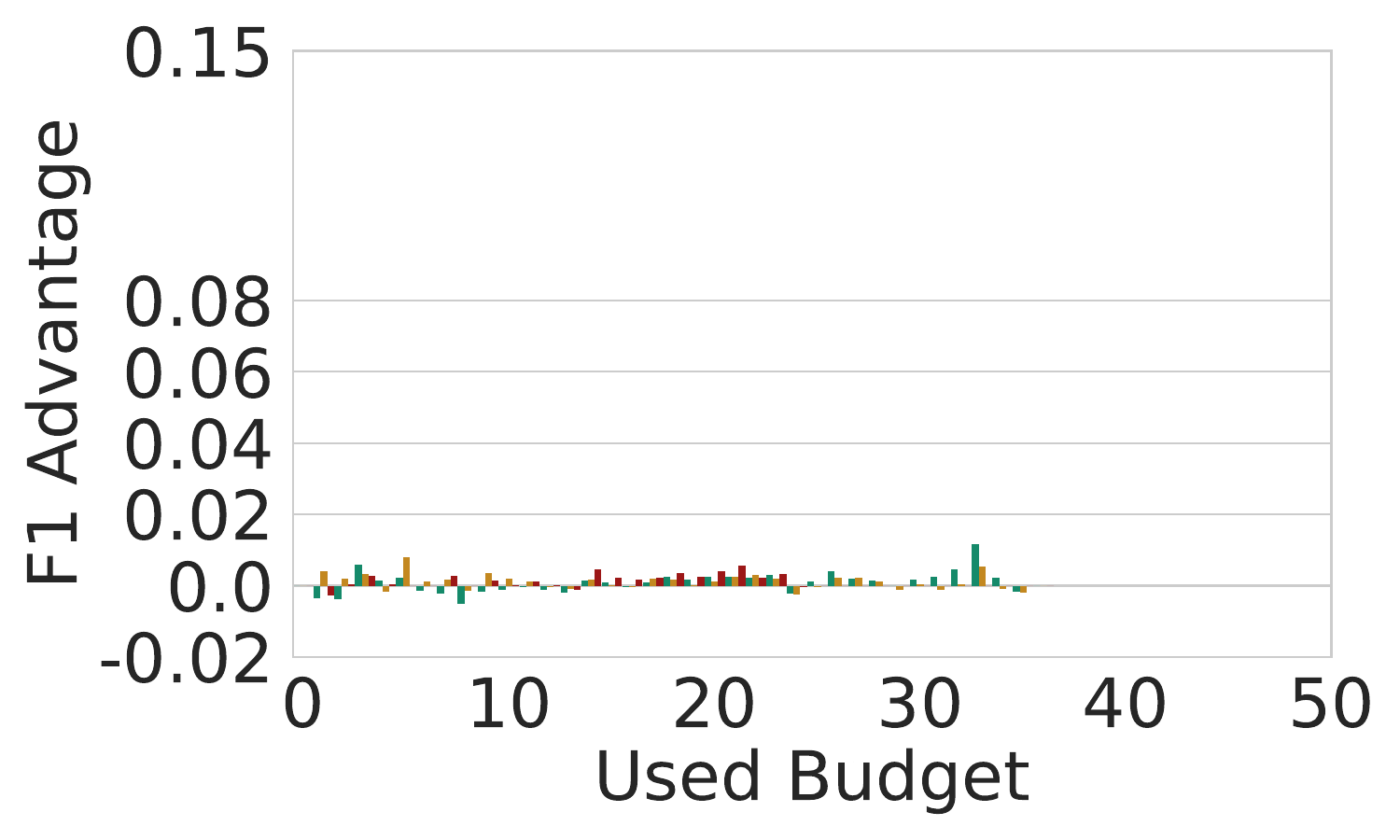}
    \end{subfigure}\hfill
    \begin{subfigure}{0.24\textwidth}
        \includegraphics[width=\linewidth]{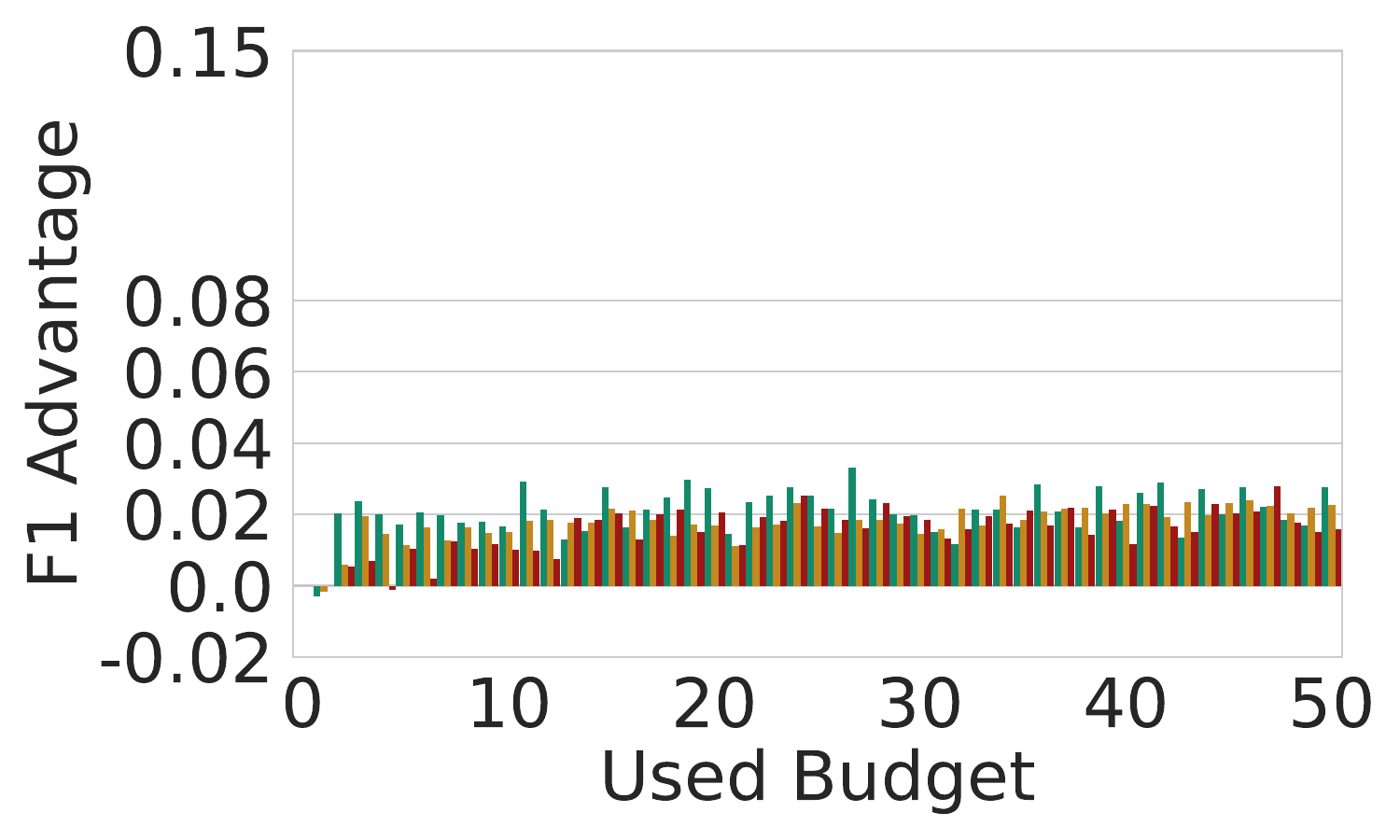}
    \end{subfigure}\hfill
    \begin{subfigure}{0.24\textwidth}
        \includegraphics[width=\linewidth]{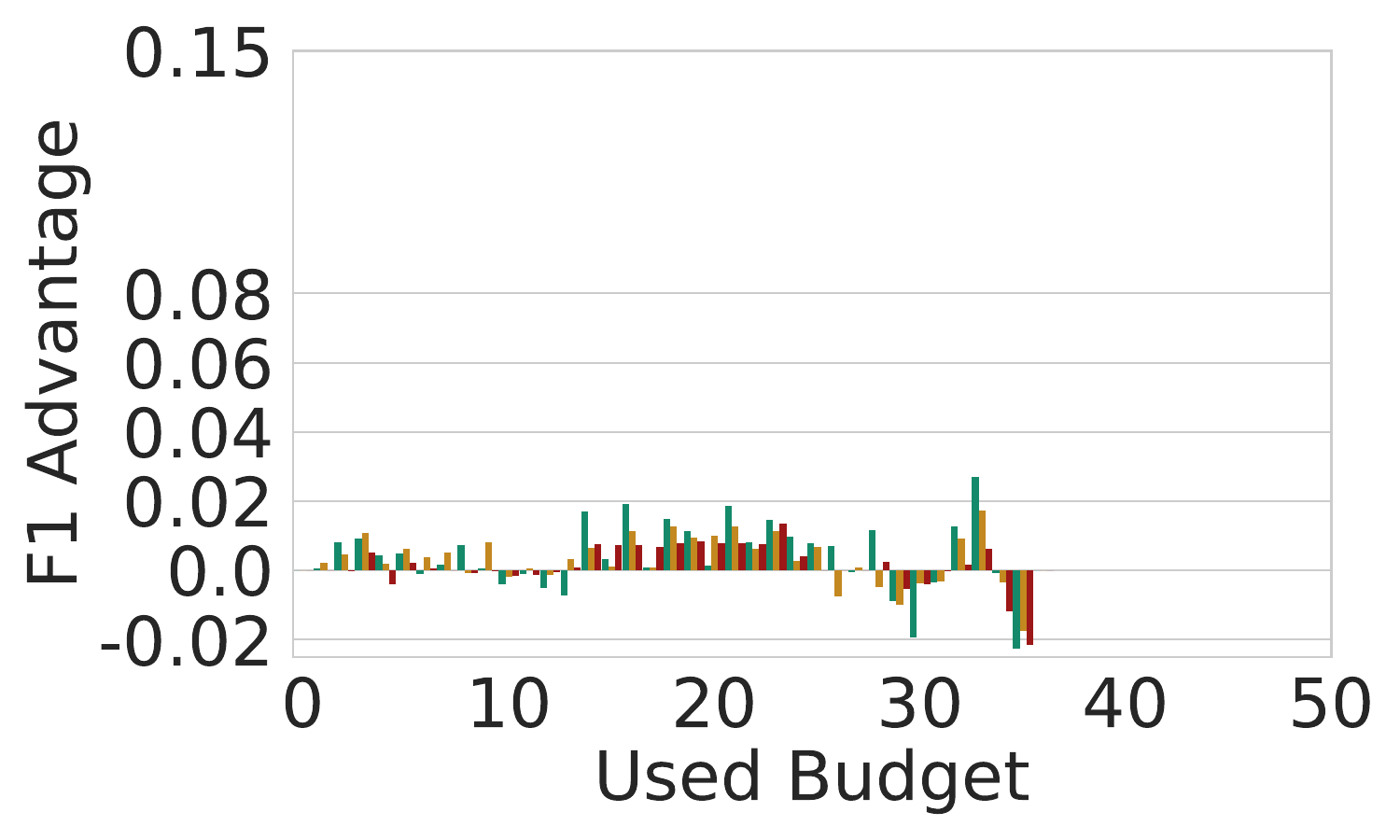}
    \end{subfigure}
    
    \vspace{-0.1em}

    \raisebox{2.\height}{\rotatebox{90}{\textbf{EEG}}}\hspace{0.3em}%
    \begin{subfigure}{0.24\textwidth}
        \centering\raisebox{3.85\height}{\parbox{0.75\linewidth}{\texttt{EEG contains only numerical features.}}}
    \end{subfigure}\hfill
    \begin{subfigure}{0.24\textwidth}
        \includegraphics[width=\linewidth]{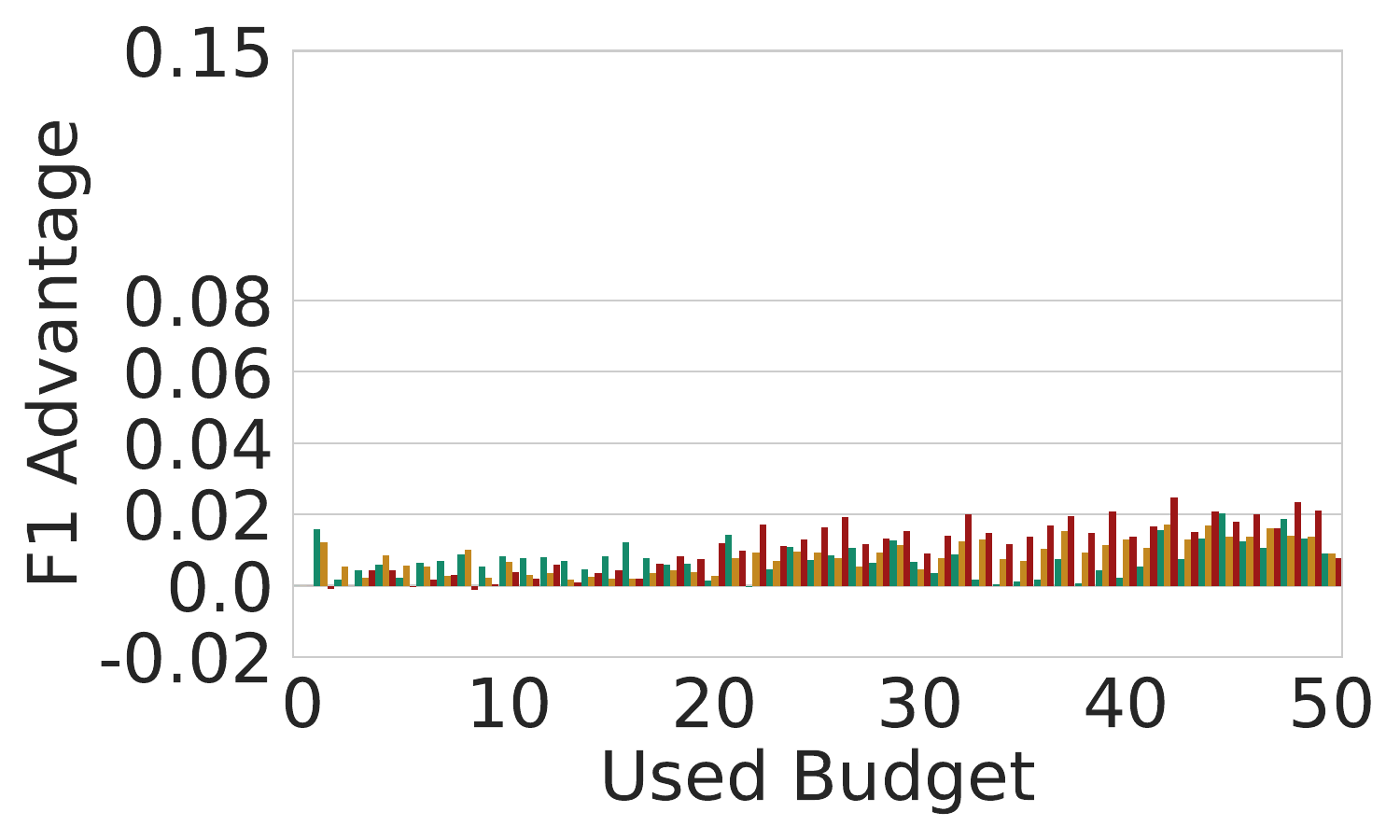}
    \end{subfigure}\hfill
    \begin{subfigure}{0.24\textwidth}
        \includegraphics[width=\linewidth]{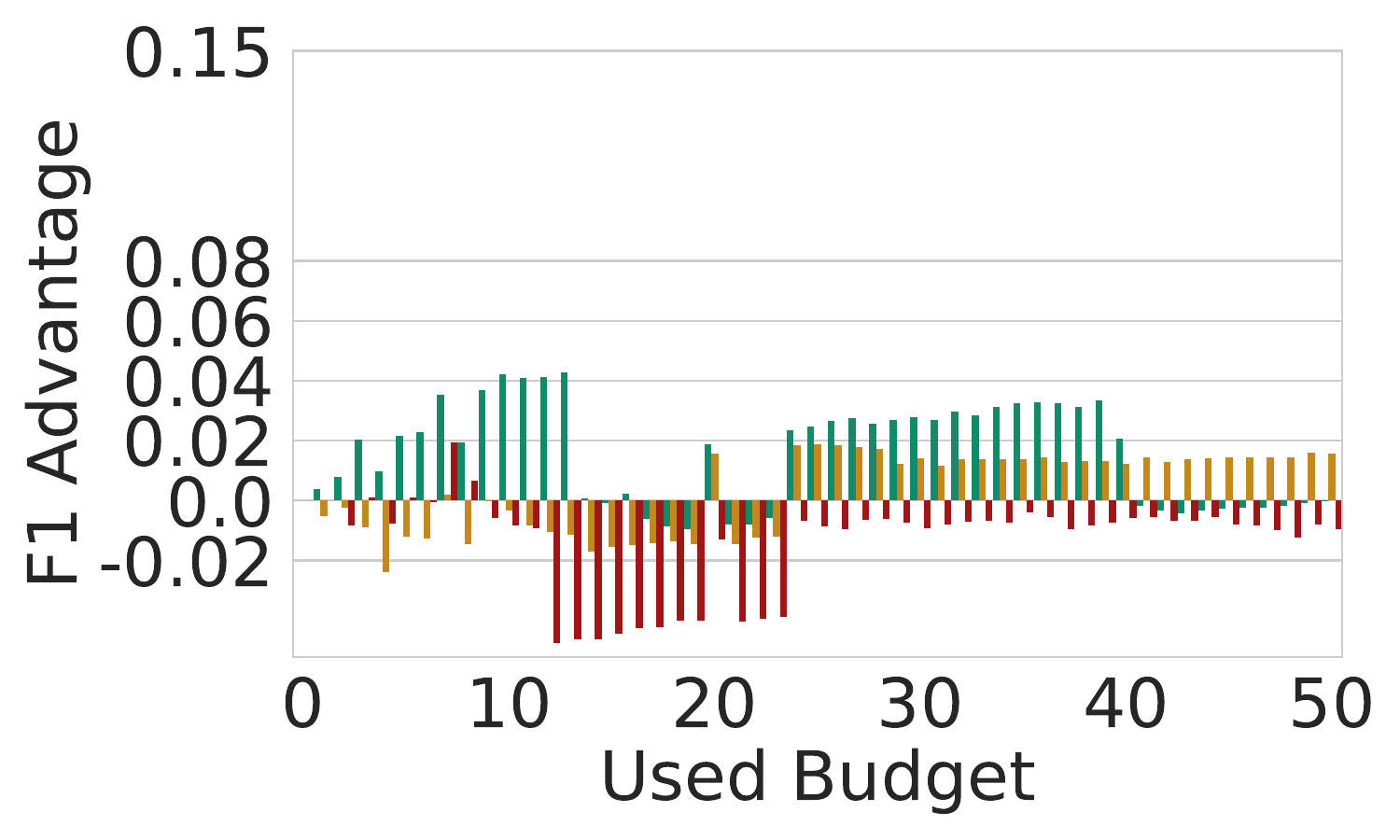}
    \end{subfigure}\hfill
    \begin{subfigure}{0.24\textwidth}
        \includegraphics[width=\linewidth]{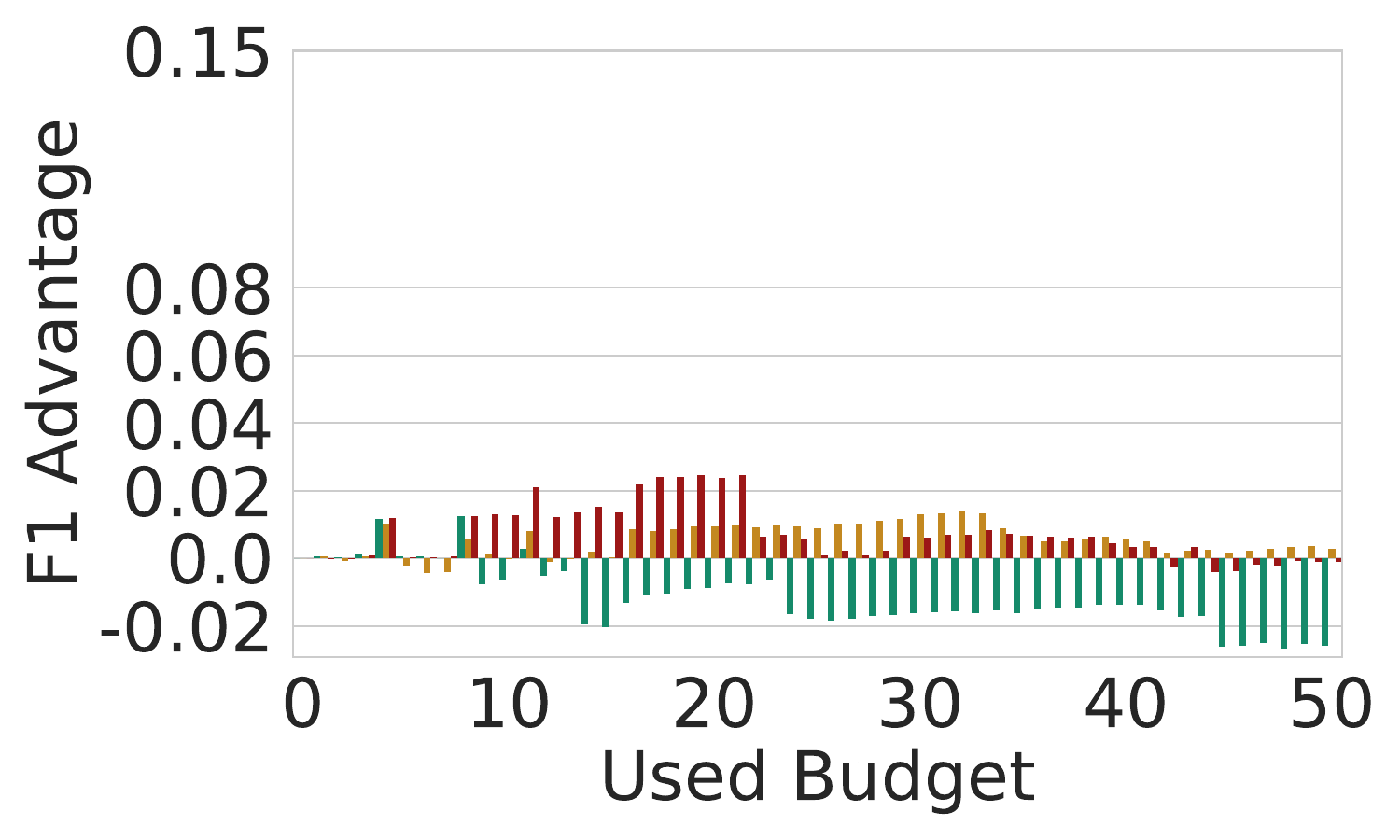}
    \end{subfigure}
    
    \vspace{-0.1em}
    
    \raisebox{1.2\height}{\rotatebox{90}{\textbf{S-Credit}}}\hspace{0.3em}%
    \begin{subfigure}{0.24\textwidth}
        \includegraphics[width=\linewidth]{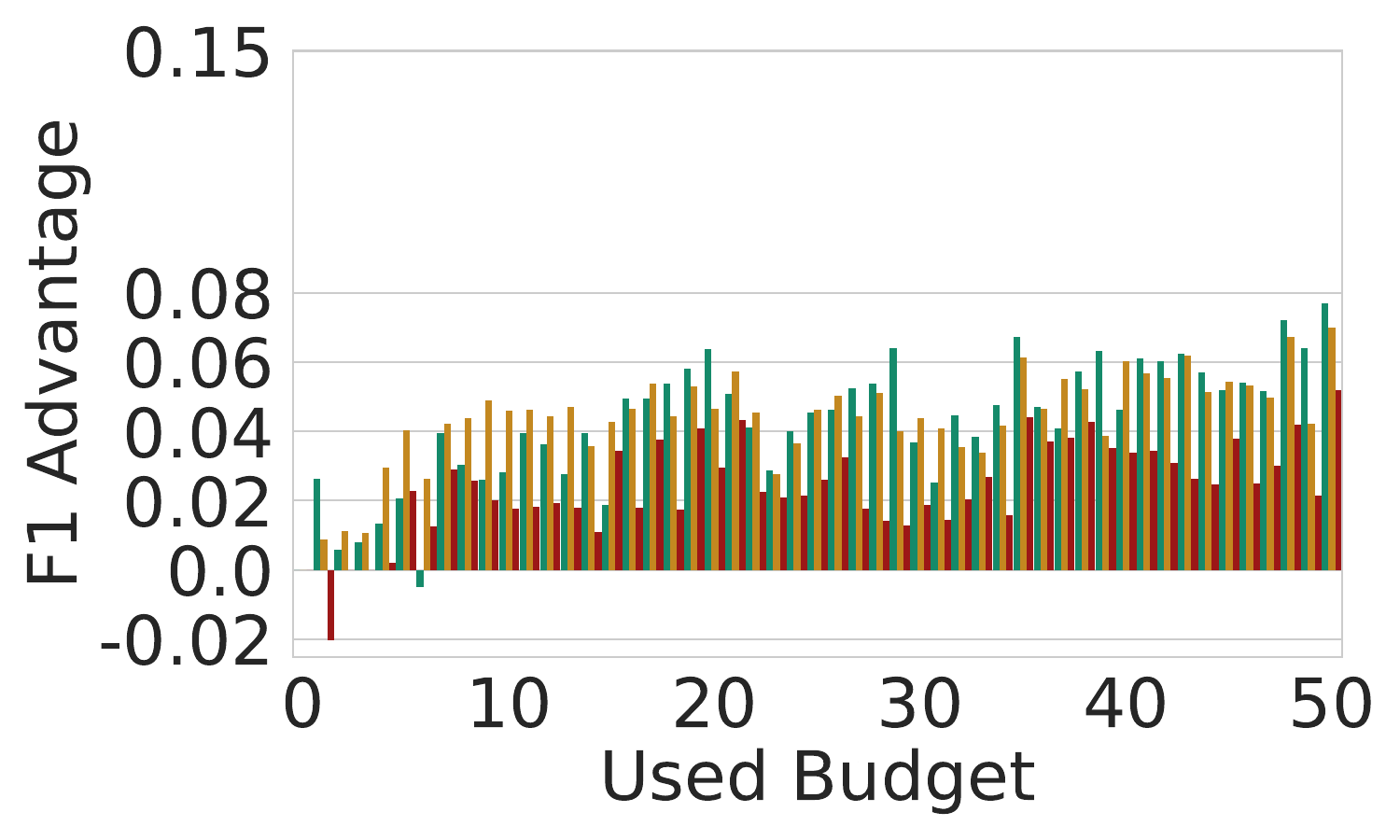}
        \caption{Categorical Shift}
    \end{subfigure}\hfill
    \begin{subfigure}{0.24\textwidth}
        \includegraphics[width=\linewidth]{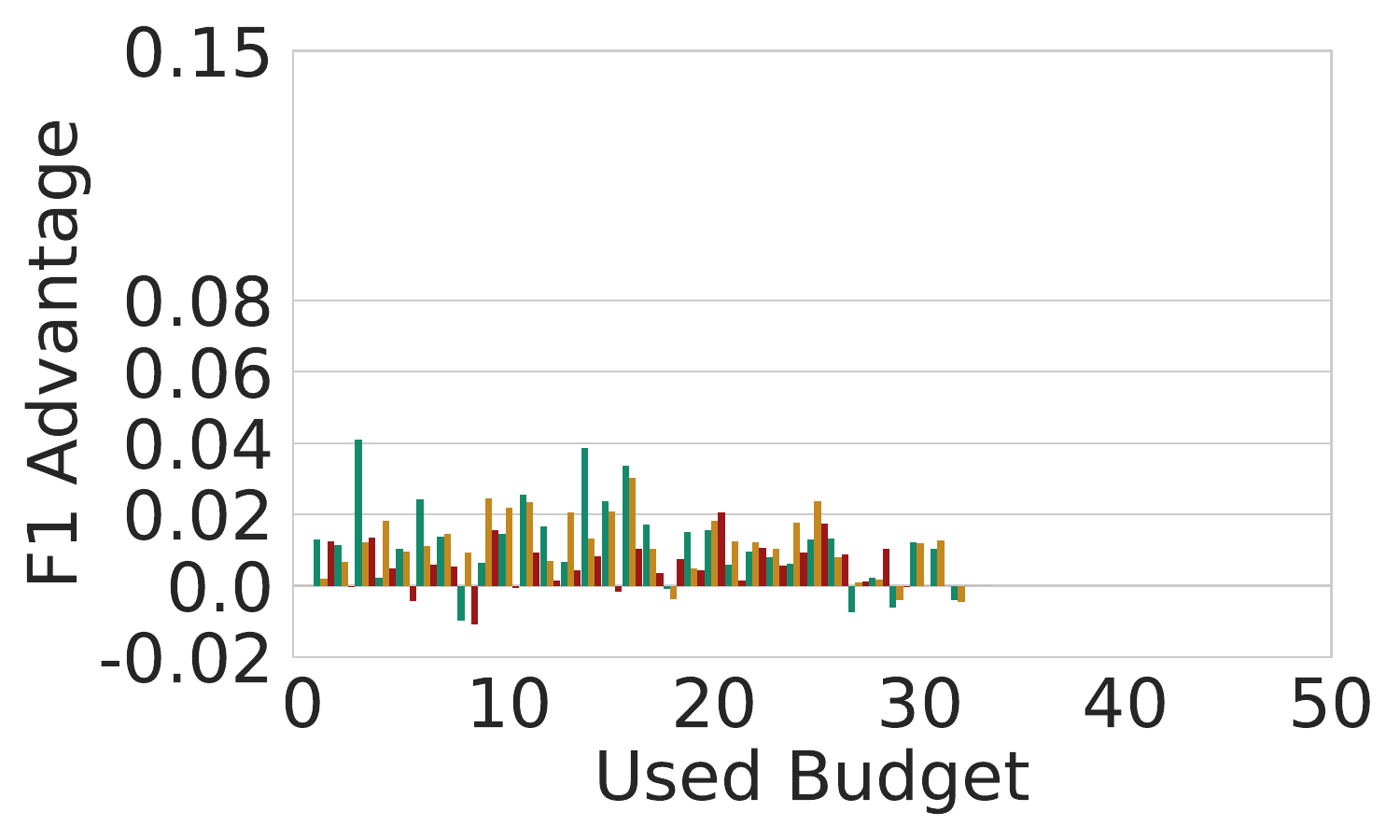}
        \caption{Gaussian Noise}
    \end{subfigure}\hfill
    \begin{subfigure}{0.24\textwidth}
        \includegraphics[width=\linewidth]{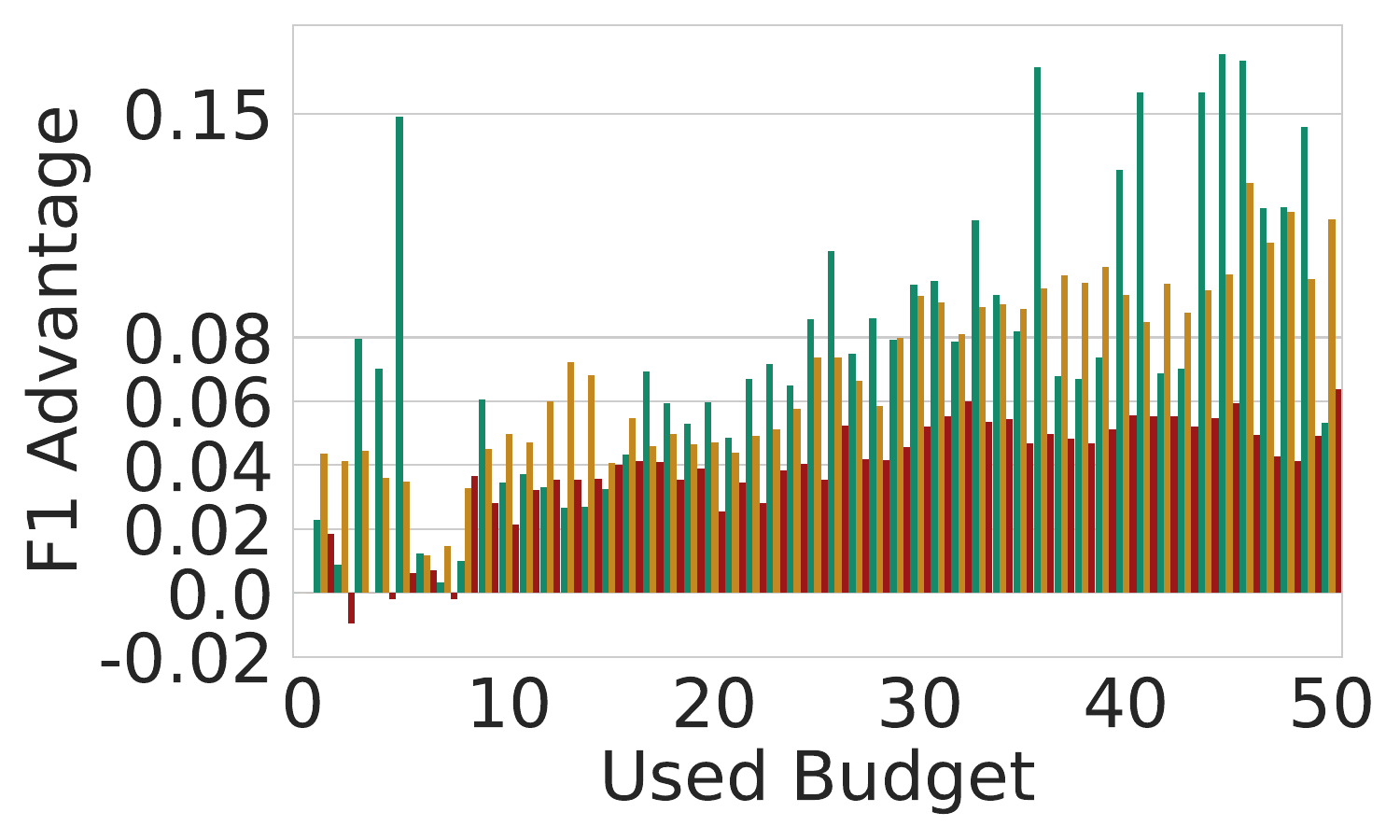}
        \caption{Missing Values}
    \end{subfigure}\hfill
    \begin{subfigure}{0.24\textwidth}
        \includegraphics[width=\linewidth]{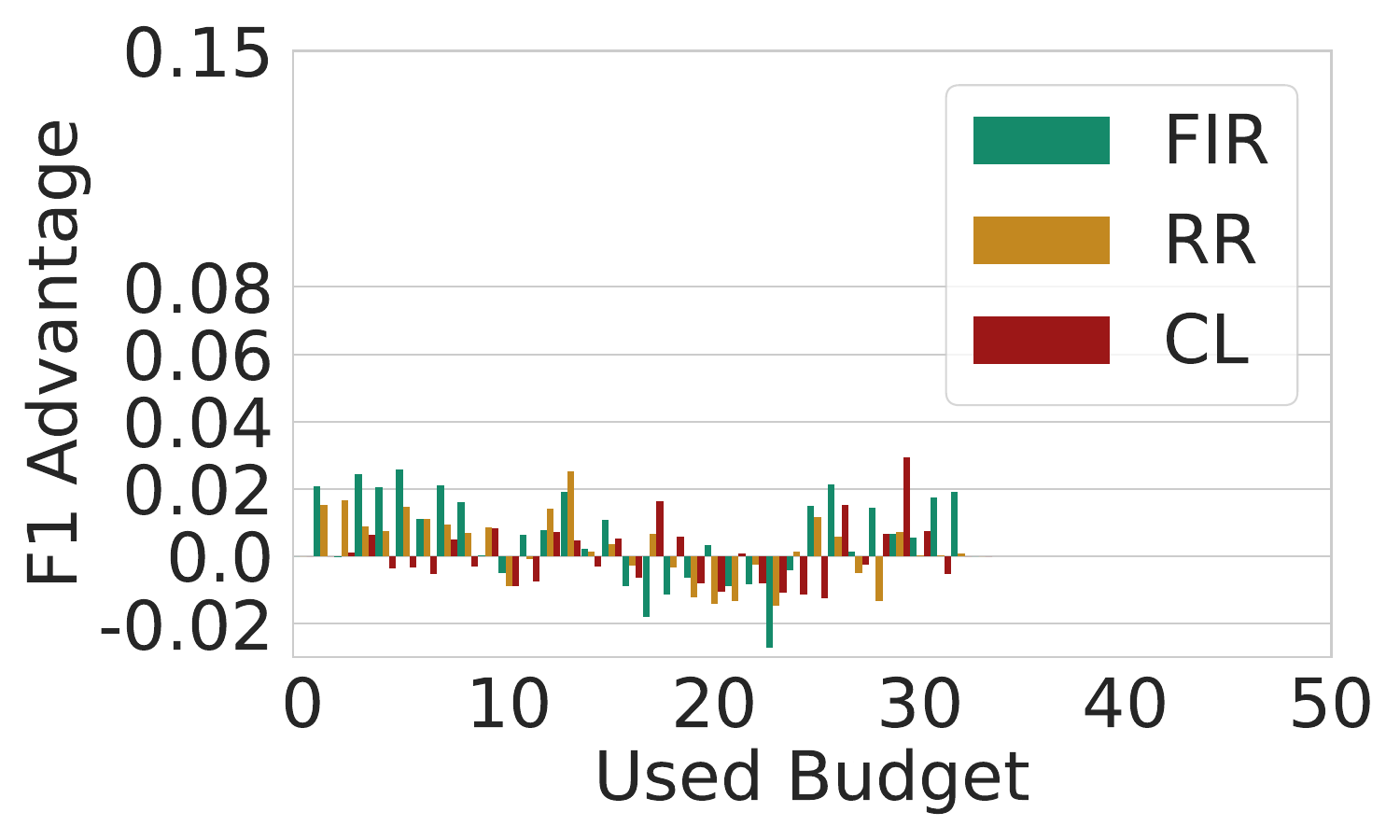}
        \caption{Scaling}
    \end{subfigure}
    \caption{\revision{Comparison of~\systemname with the baselines FIR, RR and CL for MLP across error types.}}
    \label{fig:agg_bl_results}
\end{figure*}

\begin{figure*}
    \centering
    \begin{subfigure}{0.24\textwidth}
        \includegraphics[width=\linewidth]{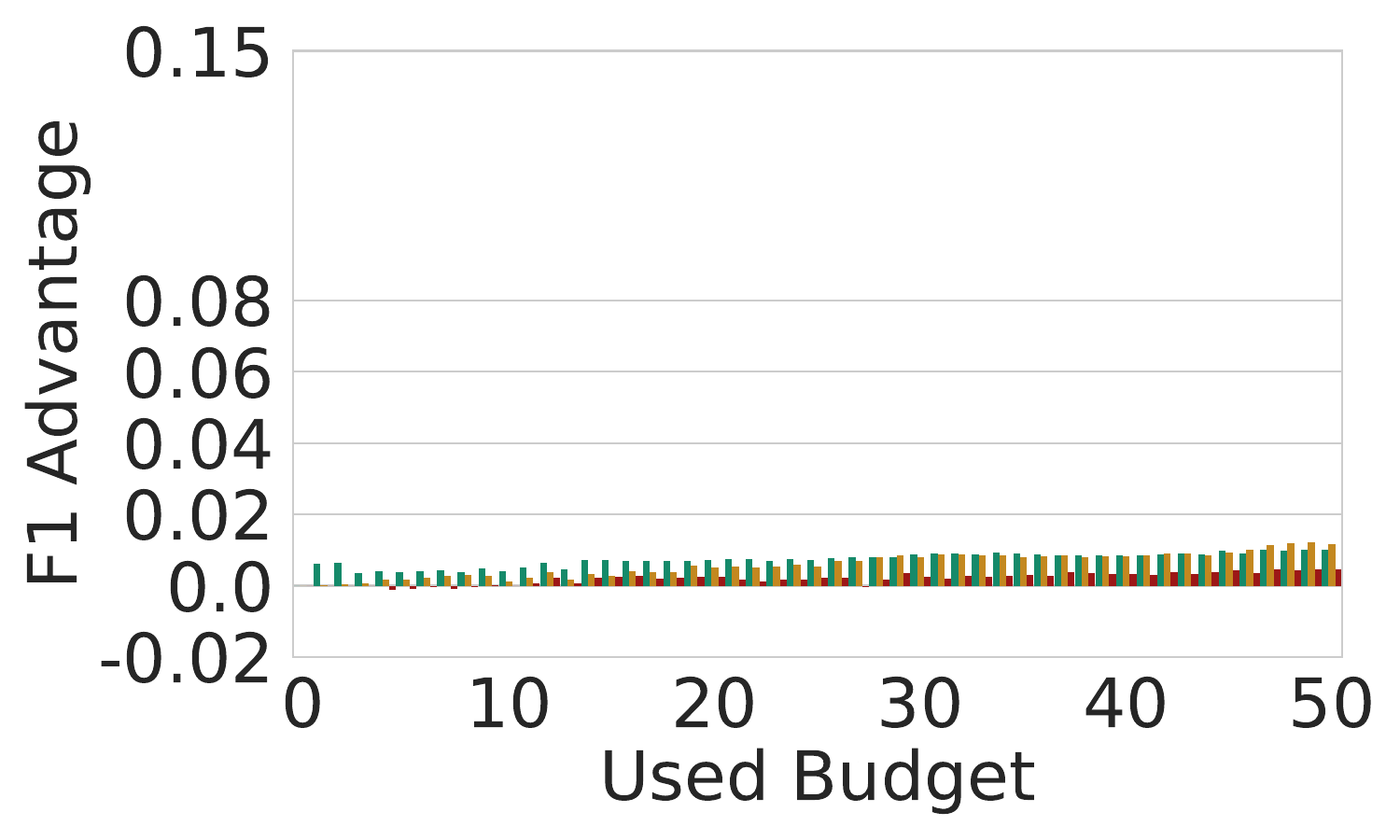}
        \caption{Airbnb - Scaling}
    \end{subfigure}
    \begin{subfigure}{0.24\textwidth}
        \includegraphics[width=\linewidth]{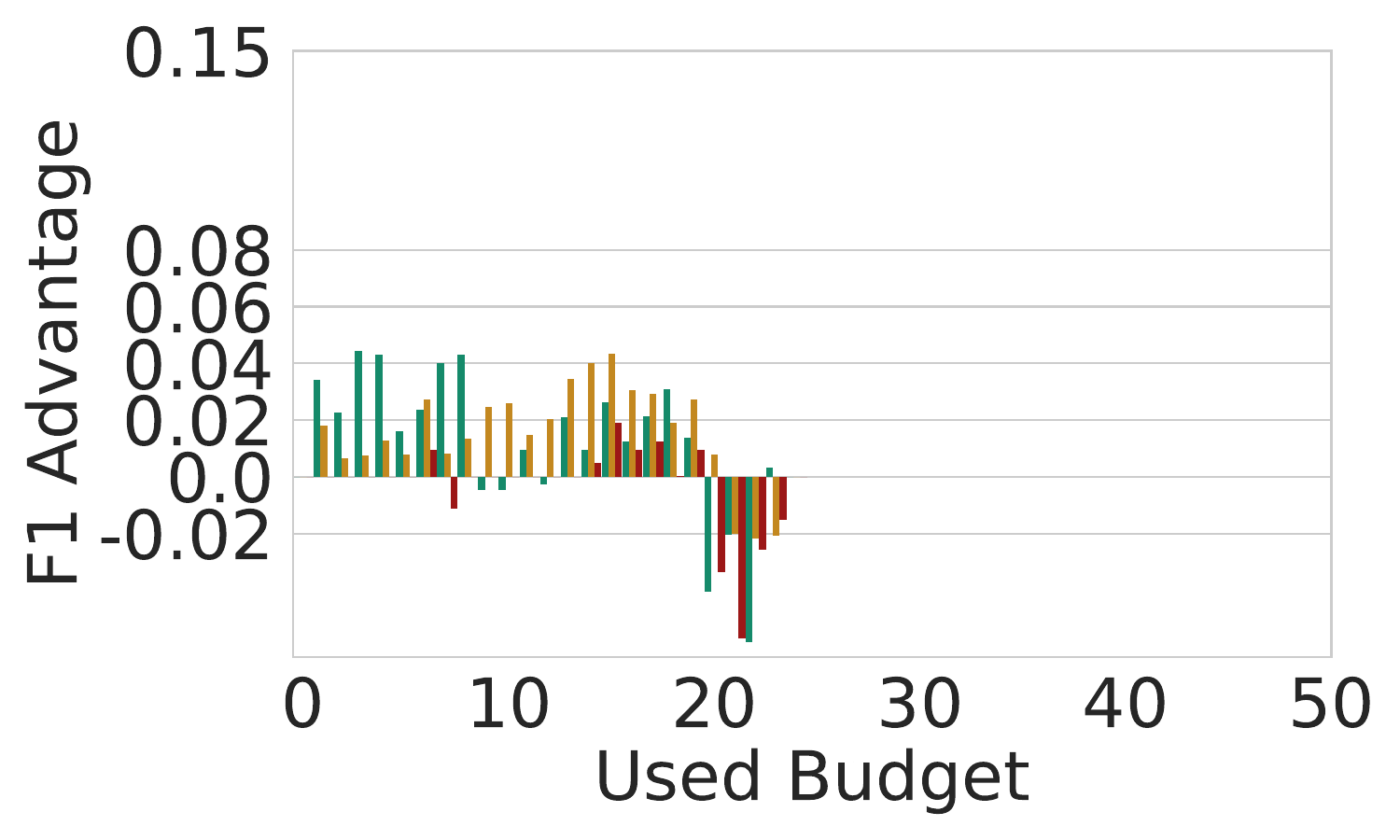}
        \caption{Credit - Scaling}
    \end{subfigure}
    \begin{subfigure}{0.24\textwidth}
        \includegraphics[width=\linewidth]{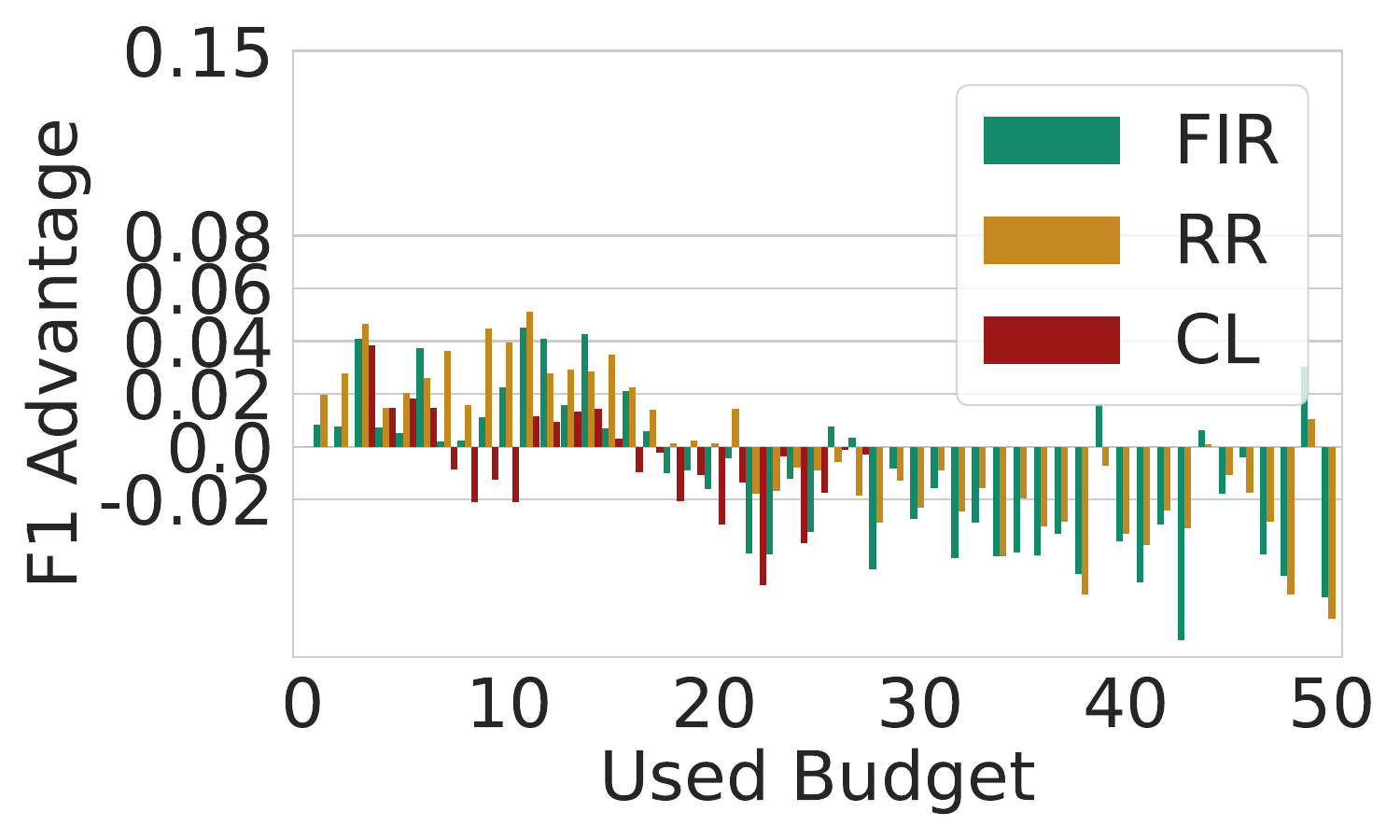}
        \caption{Titanic - Missing Values}
    \end{subfigure}
    \caption{\revision{Comparison of~\systemname with the baselines FIR, RR and CL for MLP across error types, for datasets from CleanML.}}
    \label{fig:agg_bl_results2}
\end{figure*}
\revision{In RQ~1, we examined the performance of \systemname in a multi-error setting, considering various cost functions.
To assess the impact of individual error types on \systemname's performance compared to FIR, RR, and CL, we now focus on single-error scenarios as part of the second research question~(RQ~\ref{rq:1}). The results are shown in Figures~\ref{fig:agg_bl_results} and~\ref{fig:agg_bl_results2}.}
Here, we assume constant costs for each cleaning step, regardless of the error type, to ensure comparability.
An early stop in the cleaning process indicates that the dataset is fully cleaned.
\revision{Again, due to the extensive configurations and pre-pollution settings, we focus on one ML algorithm: MLP\@.
As \systemname performs worse with MLP than other algorithms, this experiment highlights its worst-case performance.}
However, the results for the other algorithms are similar enough that we do not go into further detail for them.


\smallsection{Categorical shift.}
\label{sec:categorical_shift_results}
Considering categorical shift errors~(see Figure~\ref{fig:agg_bl_results}a), \systemname achieves a higher F1 score throughout the cleaning process compared to FIR, RR, and CL except for a few outliers~(EEG, being numerical, does not contain categorical shift errors).
\systemname shows the most significant advantage compared to the baselines in the S-Credit dataset.
Here, \systemname achieves a F1 score advantage for the baselines of up to 0.08 (which is equivalent to an advantage of 8 \textit{percentage points}~(\pt)).
The higher the budget invested, the greater the advantage between \systemname and the baselines.

\begin{figure}[tb]
\centering
\includegraphics[width=1.\linewidth]{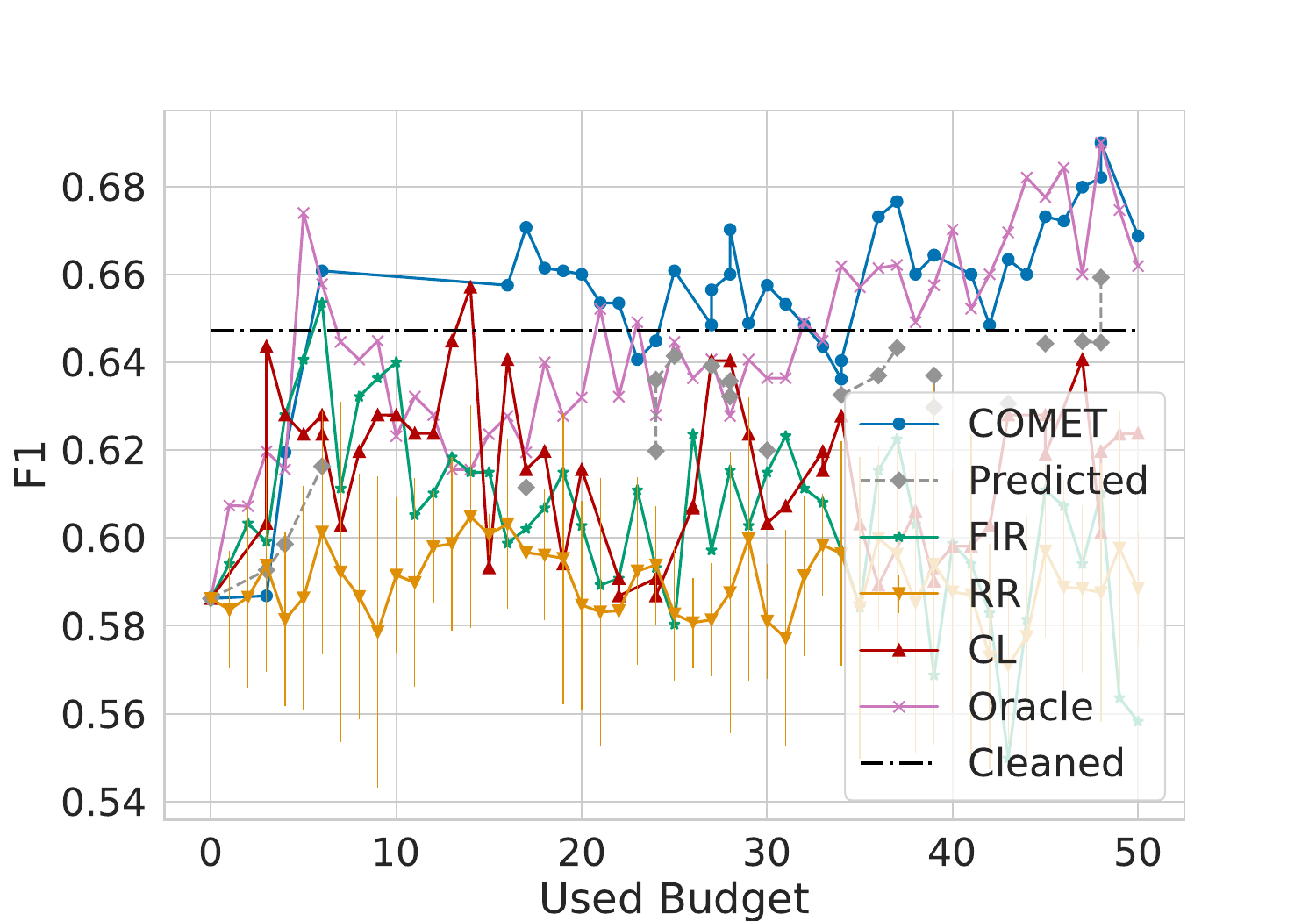}
\caption{\revision{Example of \systemname's advantage over FIR and RR, considering S-Credit with categorical shift errors and MLP.}}
\label{fig:bl_south_cat_pre2}
\end{figure}

Figure~\ref{fig:bl_south_cat_pre2} shows an example of the cleaning process of S-Credit considering one pre-pollution setting.
Throughout the cleaning budget range, \systemname maintains up to a~11\pt higher F1 score over the considered baselines\@.
Notably, between budgets 6 and 32, \systemname's F1 score even surpasses the Oracle, underscoring the effectiveness of its cleaning recommendations.
\revision{However, the baselines and even \systemname show a fluctuating behavior.
Cleaning categorical shift errors causes the cleaned feature distributions to become more aligned with their true categories, which might cause the model to adjust how it weighs other features.
This adjustment can result in unpredictable, temporary dips in performance.}

The horizontal~``cleaned'' line in the figure represents the scenario where the dataset is completely clean, independent of the budget, showing that Oracle and \systemname achieve higher F1 scores than complete cleaning for budgets \revision{exceeding}~6.

\smallsection{Gaussian noise.}
\systemname demonstrates superior performance compared to the baselines in handling Gaussian noise, albeit with a slightly less pronounced advantage than observed for categorical shifts~(see Figure~\ref{fig:agg_bl_results}b).
In most cases, \systemname's recommendations lead to F1 scores up to 4\pt higher than FIR, RR and CL\@.
EEG shows\revision{, especially for CL,} an increasing trend in the F1 score difference with a higher budget.
\revision{For CL, the initial feature ranking appears beneficial during the first few iterations, resulting in a smaller advantage for \systemname over CL\@.
However, this ranking becomes outdated in later iterations and no longer represents an optimized cleaning order -- the advantage from \systemname increases.}
In the Churn dataset, the difference between \systemname and the baselines is minimal, with \systemname achieving only slightly higher F1 scores due to the small difference (max 1.5\pt) between the dirty and cleaned states.
Therefore, no significant advantage of \systemname over the baselines can develop during the cleaning process.
The CMC dataset shows an outlier of \systemname.
Here, the baselines FIR and RR briefly achieve an advantage of up to~5\pt over \systemname.

\smallsection{Missing values.}
\systemname outperforms the baselines in most experiments, achieving \revision{a F1 score advantage} in most cleaning steps, as shown in Figure~\ref{fig:agg_bl_results}c and~\ref{fig:agg_bl_results2}c.
In S-Credit, the F1 score difference again increases with a higher budget, with \systemname occasionally leading by over 15\pt.
This advantage is also evident in other datasets.
In Churn, the advantage similarly increases, while CMC shows a bell curve trend: the F1 score advantage increases up to 6\pt for half the budget, then gradually decreases\revision{, meaning that the baselines compensate previous wrong decisions in later iterations.}
The EEG results show a sudden drop in the advantage between \systemname and FIR\revision{, respectively between \systemname and CL,} from a budget of~12.
\revision{FIR maintains its advantage until a budget of 22, after which \systemname recovers and surpasses FIR, while CL continues to maintain its superiority.}
However, from a budget of 40, the difference between \systemname and FIR narrows again, mainly due to a specific pre-pollution setting in EEG.
This pattern can be explained by the fact that each step of the cleaning process builds upon the previous one.
As a result, there is a risk that \systemname may recommend a suboptimal feature at certain stages, causing prediction accuracy to stagnate temporarily.
This stagnation can last for several cleaning steps before a noticeable improvement is observed. 
This phenomenon also occurs in other experiments, where the considered baseline or AC is superior, as the incremental nature of the cleaning process sometimes leads to delayed improvements in accuracy.
The results for Titanic represent an outlier in terms of \systemname's performance compared to the baselines.
\systemname outperforms the baselines up to a budget of 17, but the baselines are superior afterward, with an advantage of up to 5\pt.

While the baselines perform better in a few budget-cases, \systemname remains the superior choice in most cleaning scenarios.

\smallsection{Scaling.}
The trends of the individual methods for scaling errors are similar to those of Gaussian noise\revision{~(see Figure~\ref{fig:agg_bl_results}d)}.
Overall, \systemname again shows an advantage over the baselines, although this advantage is smaller than categorical shifts and missing values~(see  Figures~\ref{fig:agg_bl_results2}a and~\ref{fig:agg_bl_results2}b).
When considering scaling errors, \systemname achieves an F1 score advantage of up to~4\pt compared to FIR, RR and CL\@.
Similar to the previously considered results, a trend can be seen for Airbnb: the difference between \systemname and the baselines increases as the invested budget increases. 
However, there are also some fluctuations.
In CMC, the baselines FIR and RR generally outperform \systemname, and in EEG, FIR outperforms \systemname from a budget of~11, though \systemname \revision{keeps its superiority over RR and CL}.



\begin{figure*}
    \centering
    \begin{subfigure}{0.24\textwidth}
        \includegraphics[width=\linewidth]{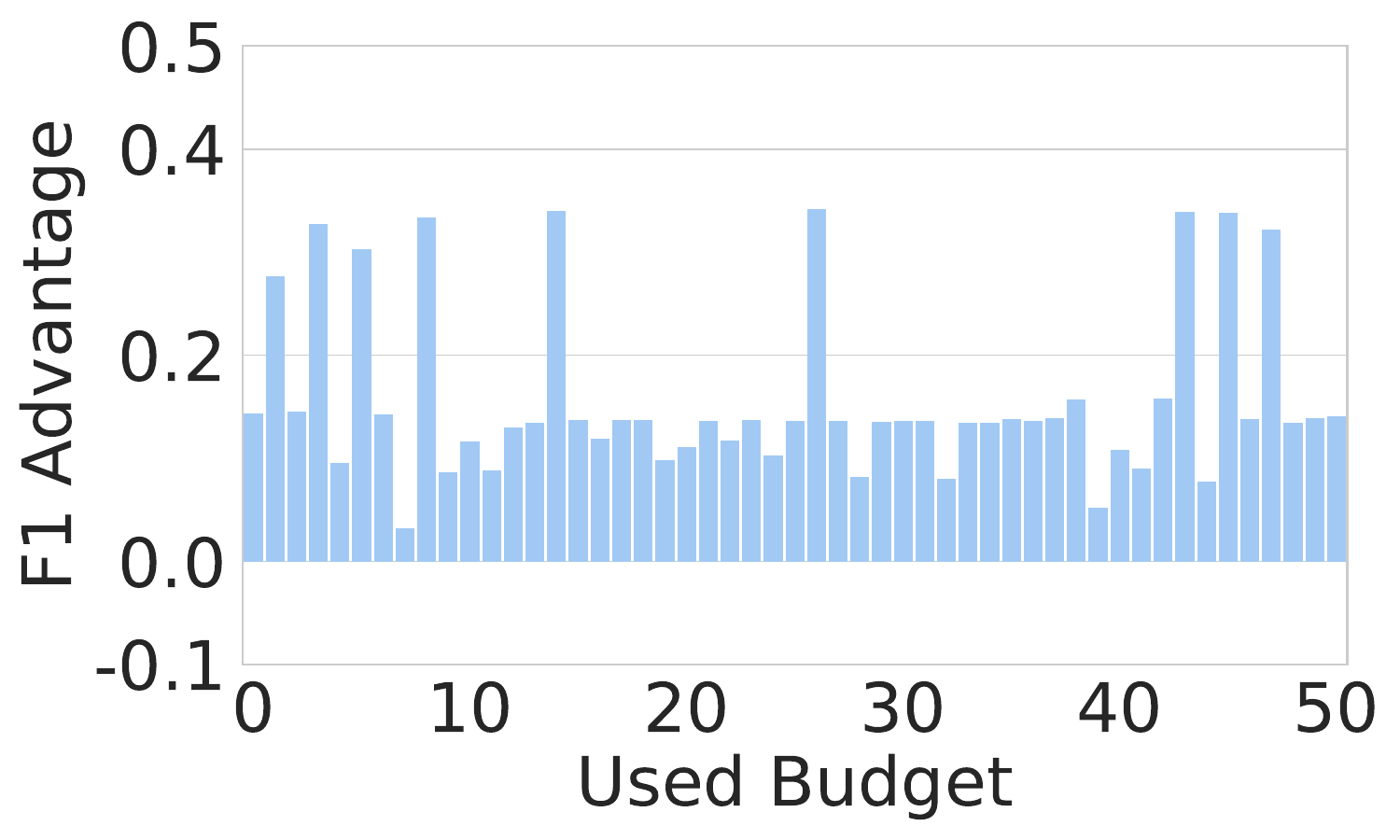}
        \caption{Airbnb - Scaling}
    \end{subfigure}
    \begin{subfigure}{0.24\textwidth}
        \includegraphics[width=\linewidth]{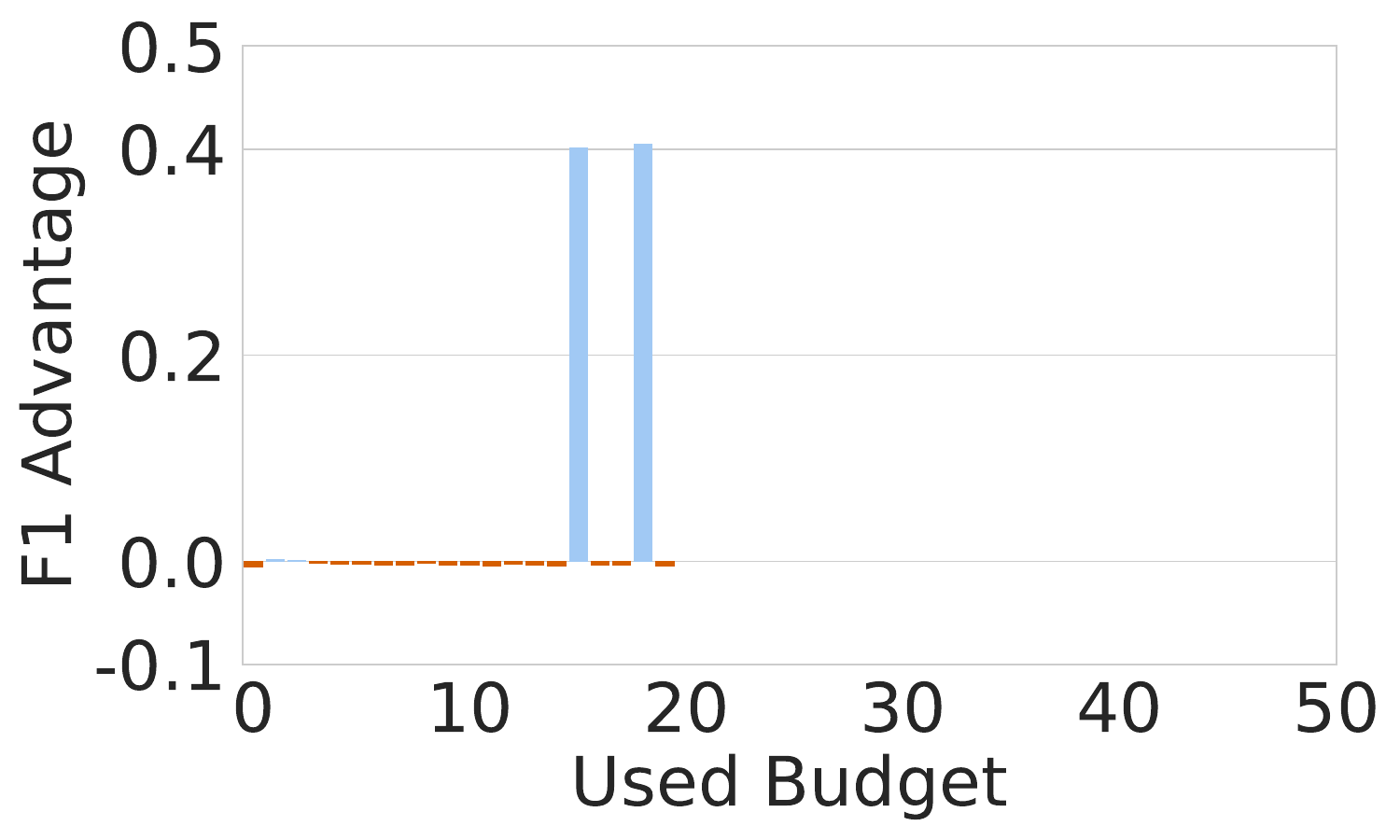}
        \caption{Credit - Scaling}
    \end{subfigure}
    \begin{subfigure}{0.24\textwidth}
        \includegraphics[width=\linewidth]{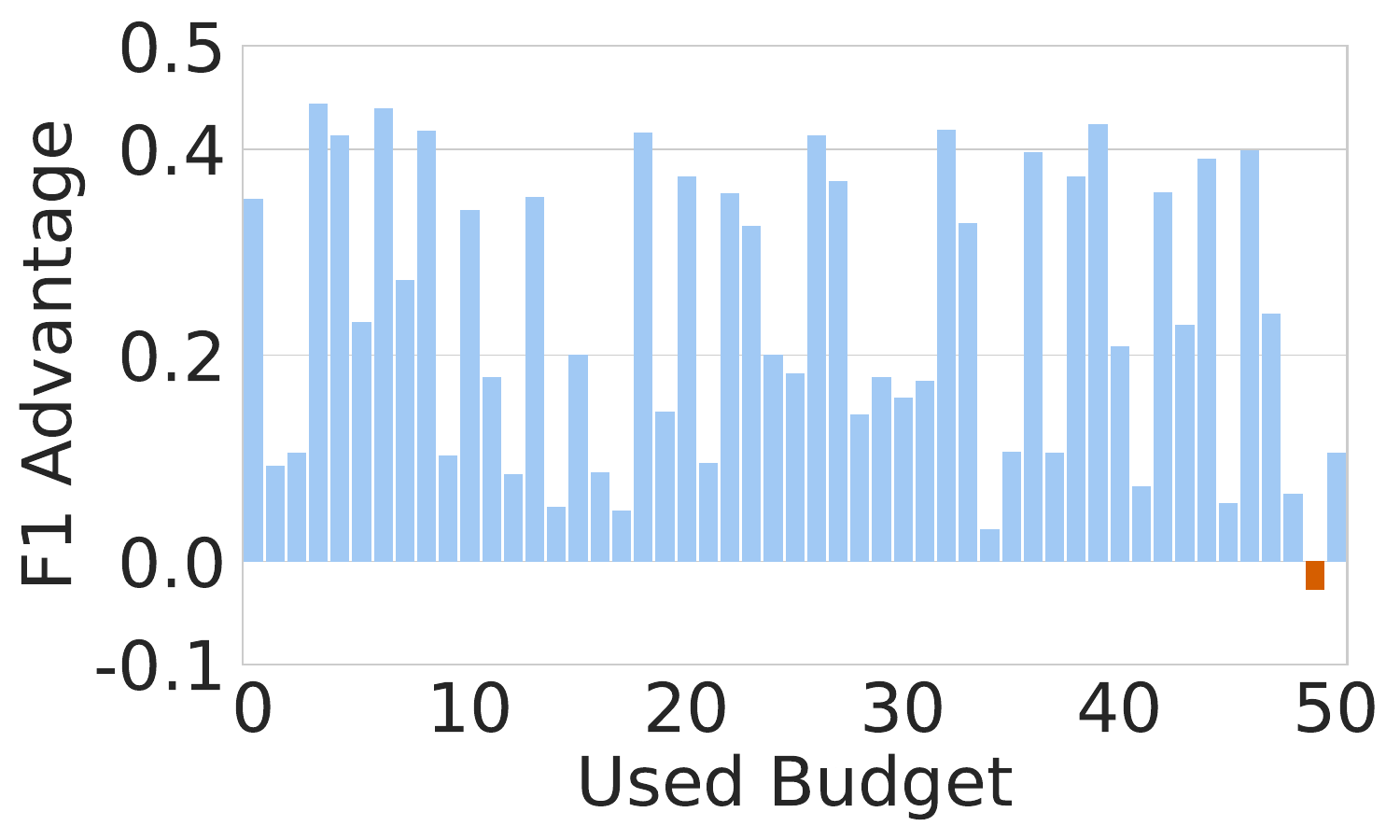}
        \caption{Titanic - Missing Values}
    \end{subfigure}
    \caption{Comparison of~\systemname with AC for AC-SVM across error types, for datasets from CleanML.}
    \label{fig:agg_ac_results2}
\end{figure*}

\subsection{Comparison to AC for a single error type}
\label{sec:activecleanexp}
We compared \systemname's performance with ActiveClean~(\cite{krishnan2016activeclean}, RQ~\ref{rq:2}), focusing on AC-SVM among the many configurations and pre-pollution settings considered\revision{, in a single-error scenario}.
\revision{Similar to RQ~\ref{rq:1}, we also assume constant costs for each cleaning step in RQ~\ref{rq:2}, regardless of the error type.}
Figures~\ref{fig:agg_ac_results} and~\ref{fig:agg_ac_results2} show \systemname's performance per budget, comparing it to AC for individual error types with AC-SVM\@.
The overall trends are consistent with those observed for other ML algorithms (LIR, LOR).

Our implementation of AC is based on the code published by the authors\footnote{\url{https://www.dropbox.com/sh/r2vv252m5lnqpmm/AAAMj0WRaZX9EKH_8dLOHQpIa?e=3&dl=0}}, which contains only the basic functions of AC; we extended it by the component of gradient-based selection of records.
AC's approach also includes an error detection component, which we skip in the experiments.
Furthermore, AC always assumes record-wise data cleaning, while \systemname assumes feature-wise cleaning.
To integrate AC into our setting, we first pre-train the respective ML model with the records that are already clean according to the pre-pollution settings.
We then follow AC's approach by selecting a sample of records according to the gradients and cleaning the records across all features.
The sample corresponds to the size of a cleaning step.
Though the number of cleaned entries per iteration or budget may differ, the discrepancies are minor due to our assumed equal error distribution.

\smallsection{Categorical shift.}
Except for a few minor deviations in CMC, \systemname generally outperforms AC \revision{for all considered datasets}, as Figure~\ref{fig:agg_ac_results}a shows.
The F1 score \textit{advantage} reaches up to 40\pt in certain datasets like Churn.
\revision{The fluctuating F1 score advantages highlight that the performance of AC is quite erratic.
Our analysis shows that, for example in S-Credit, the F1 score can drop by up to 30\pt after a cleaning step, only to recover after a few further cleaning steps.}
In contrast, \systemname shows a steady increase, with F1 scores improving by up to 1\pt, and far lower variance.

\smallsection{Gaussian noise.}
\systemname also outperforms AC when dealing with Gaussian noise, although the advantage per used budget are smaller (Figure~\ref{fig:agg_ac_results}b).
Apart from CMC, \systemname is almost always preferred over AC\@.
The highest difference can be seen with S-Credit, where \systemname achieves a gap of up to 40\pt, though this advantage narrows with a higher invested budget.
In CMC, however, AC periodically outperforms \systemname, particularly with a budget between 5 and 13, where AC holds a 10\pt advantage, mainly again due to sudden F1 score changes.

\smallsection{Missing values.}
When the data includes missing values, \systemname outperforms the AC in most cleaning steps (see Figures~\ref{fig:agg_ac_results}c and~\ref{fig:agg_ac_results2}c.
Notably, \systemname shows the most consistent performance across all cleaning steps in the S-Credit dataset, maintaining a performance gap of 20\pt to 35\pt throughout the cleaning process.
For the Churn dataset, \systemname leads AC by up to 40\pt.
However, in the EEG and CMC datasets, \systemname's advantage over AC is less pronounced, peaking at around 20\pt. 
In EEG, AC frequently outperforms \systemname due to one pre-pollution setting.
For the Titanic dataset, performance differences vary abruptly~(Figure~\ref{fig:agg_ac_results2}c).
Once more, no consistent pattern relative to the cleaning budget emerges, even when using \systemname throughout the cleaning steps.

\smallsection{Scaling.}
The performance of \systemname in addressing scaling errors mirrors its effectiveness with missing values.
In the EEG dataset, AC slightly outperforms \systemname, a trend also occasionally observed in the CMC dataset where AC demonstrates superiority. 
In the credit dataset, \systemname and AC perform similarly, except in two cases where \systemname excels.
However, \systemname distinctly outperforms AC in these outlier scenarios.
For other datasets, \systemname consistently maintains its advantage over AC throughout the cleaning steps. 
The Airbnb dataset shows a pattern similar to S-Credit, with \systemname steadily leading AC, except for a few exceptions.


\subsection{Overall performance per Baseline} \label{sec:performance_per_baseline}
\begin{figure}[tb]
\centering
\includegraphics[width=1.\linewidth]
{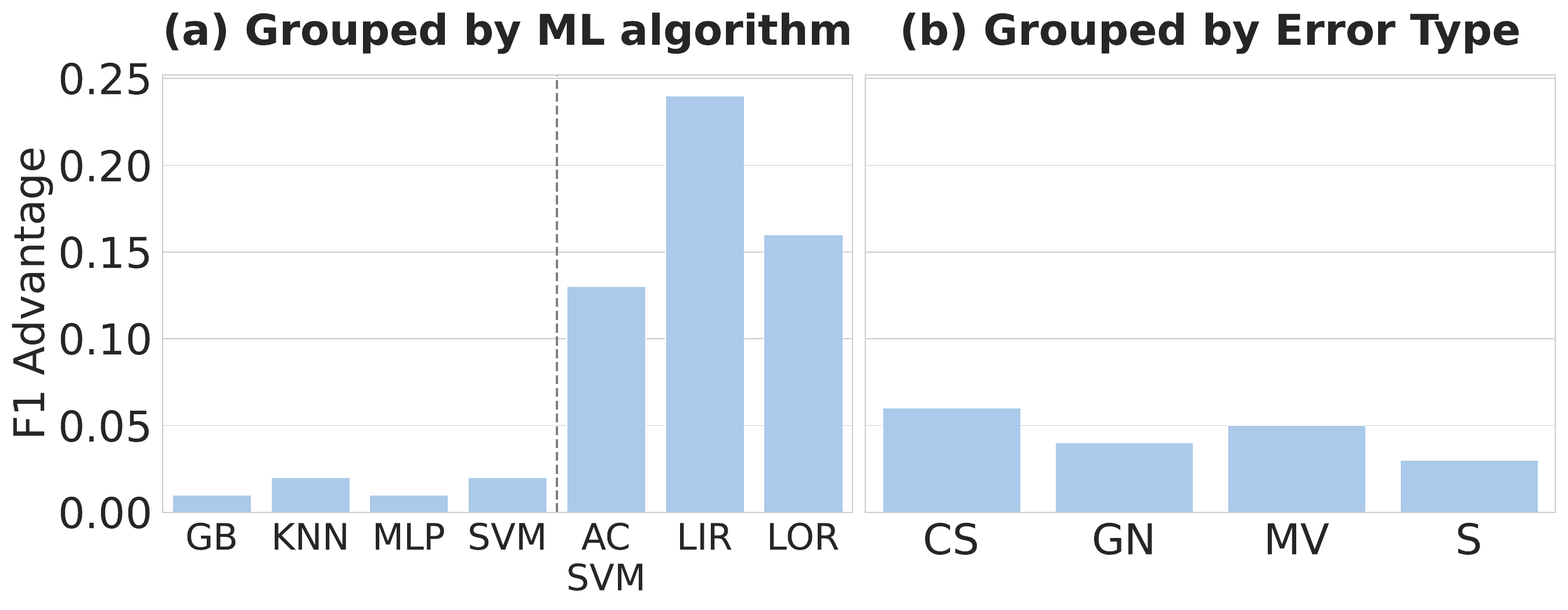}
\caption{\revision{Overall performance of \systemname. (CS -- Categorical Shift, GN -- Gaussian Noise, MV -- Missing Values, S -- Scaling).}}
\label{fig:overall_performance}
\end{figure}

For our fourth research question~(RQ~\ref{rq:3}), we examine the performance of \systemname across our considered error types respectively ML algorithms~(see Figure~\ref{fig:overall_performance}a and~b).
The experiments help to understand the adaptability and robustness of our approach regardless of a specific error type and invested budget.

\smallsection{Performance \revision{grouped by ML algorithm}.}
Figure~\ref{fig:overall_performance}a summarizes the F1 score differences \revision{grouped by ML algorithm, respectively} aggregated across all error types.
\revision{Here, we include our experiment results from the multi-error and single-error scenario.}
Each bar shows the mean advantage between FIR, RR, and AC compared to \systemname, highlighting its good performance independent of error type and invested budget.

It should be noted again that we tested only the ML algorithms LIR, LOR and AC-SVM in connection with AC, while we tested only GB, KNN, MLP and SVM in connection with FIR, RR, and CL.
The results align with trends from the previous sections, showing that \systemname's advantage over AC is significantly greater than over FIR, RR, and CL~(see Figure~\ref{fig:overall_performance}a).
In particular, LIR achieves the highest average F1 score advantage of 24\pt.
However, the results for LOR and AC-SVM also emphasize the superiority of \systemname with an average F1 score advantage between 12\pt and 15pt.
The differences between \systemname compared to FIR, RR, and CL are smaller, with SVM showing the highest average difference of 3\pt and the other ML algorithms (MLP, GB, KNN) ranging between 1\pt and 2\pt.

\smallsection{Performance \revision{grouped by error type}.}
For our third research question~(RQ~\ref{rq:3}), we examine \systemname's performance compared to the baselines, considering the mean F1 scores differences, but now grouped \revision{by error type, respectively aggregated} across the ML algorithms (Figure~\ref{fig:overall_performance}b).
This analysis shows how \systemname performs \revision{per error type}, regardless of the ML algorithm and budget.
\revision{Consequently, this perspective only includes the results from the single-error scenario.}

\systemname consistently outperforms the baselines across all error types, with minor variations.
It achieves the highest advantage of 6\pt in categorical shift errors, followed by 5\pt in missing value errors.
For Gaussian noise and scaling errors, the advantage is slightly lower, between 3\pt and 4\pt.
Again, the results are consistent with the trends from the previous sections, which show that \systemname performs slightly better compared to the baselines in the presence of categorical shifts or missing values than in the presence of Gaussian noise or scaling errors.

\smallsection{\revision{Conclusion}.}
\revision{The experiments conducted to address RQ1-RQ4 demonstrate that, with few exceptions, \systemname outperforms the baselines FIR, RR, and CL in the cleaning process, achieving up to a 52\pt and, on average, 5\pt advantage.
\systemname optimizes the cleaning process  compared to these baselines by giving the \textit{Cleaner} feature-wise cleaning recommendations.}

\revision{\systemname's superior performance in enhancing prediction accuracy stems from two key factors.
First, \systemname optimizes directly for a specific prediction metric, the F1 score, ensuring alignment to improve model performance.
This targeted focus allows for more precise prediction quality improvements.
Second, the prediction model of \systemname is adaptive.
It assesses the impact of cleaning per iteration by incrementally polluting the data to derive the effect of cleaning a particular feature.
\systemname adapts its predictive model individually to the circumstances per iteration and feature.
This method contrasts with FIR and CL, which rely on static or outdated information from previous dataset states, potentially leading to misguided cleaning efforts.
Likewise, AC, although it adapts using gradients of dirty records and prior average feature changes, also relies on outdated information, leading to suboptimal cleaning decisions.
The same applies trivially to RR, which randomly selects features for the next cleaning step.}

\subsection{Prediction accuracy}
\label{sec:prediction_accuracy}
\systemname predicts the feature-wise impact of data cleaning on ML prediction accuracy.
This section focuses on how closely the estimated F1 scores per cleaning step aligns with the actual outcomes for the features that \systemname recommended for cleaning~(RQ~\ref{rq:4}).

\begin{figure}[tb]
    \centering
    \includegraphics[width=1.0\linewidth]{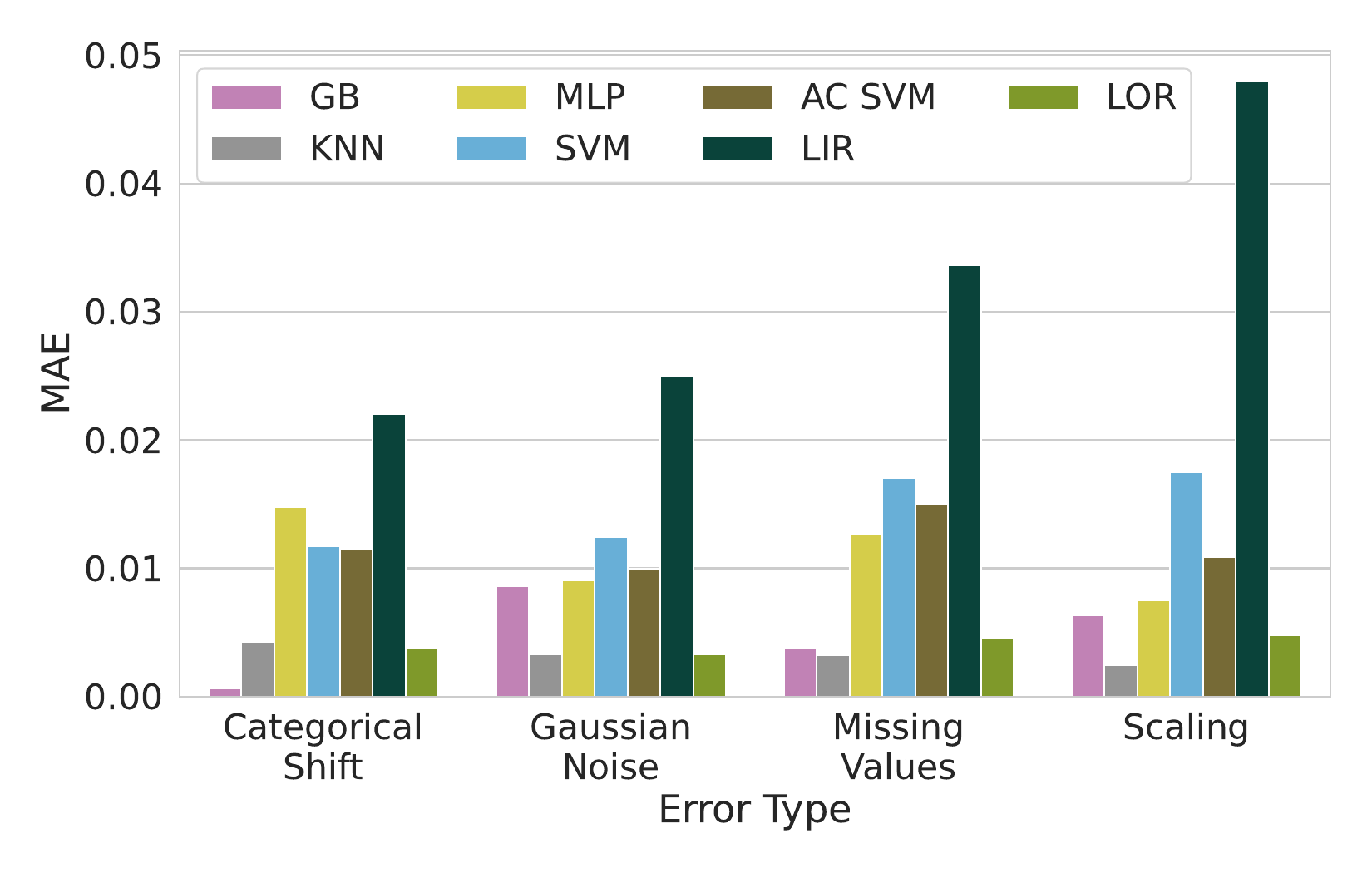}
    \caption{MAE of~\systemname's predictions.}
    \label{fig:prediction_accuracy}
\end{figure}

We assess the predictive accuracy using the Mean Absolute Error~(MAE) between predicted and the actual F1 scores after cleaning, focusing on the predictions that the~\emph{Recommender} finally used for cleaning.
Figure~\ref{fig:prediction_accuracy} presents the~MAE of \systemname grouped by error type and ML Algorithm across all datasets.

The MAE ranges from 0.0007 to 0.05, demonstrating \systemname's robust predictive model and ability to guide informed cleaning decisions.
However, there are variations across ML algorithms and error types.
For instance, the MAE for MLP ranges from 0.007 for scaling errors to 0.015 for categorical shifts. 
Compared to the overall minimum MAE of 0.0007 (by GB and categorical shift), \systemname’s predictions are less accurate when MLP is used.
\systemname behaves similarly with AC-SVM, where the MAE is between~0.009~(Gaussian noise) and~0.015~(missing values).

Figure~\ref{fig:bl_south_cat_pre2} shows that sudden jumps in F1 scores, especially for categorical shifts, lead to deviations between predicted and actual F1 scores.
In Figure~\ref{fig:bl_south_cat_pre2}, initial predictions match actual F1 scores, but sudden jumps at budgets 4 and 24 cause significant deviations.
The \emph{Recommender} compensates occasional inaccuracies by temporarily restoring a feature's cleaning step, optimizing the process so that \systemname outperforms the baselines.

Figure~\ref{fig:prediction_accuracy} also shows that KNN achieves the lowest MAE for all error types except categorical shift, with MAE ranging from 0.002 (missing values) and to 0.003 (scaling): \systemname makes particularly accurate predictions when KNN is used, aligning with its superior performance over FIR and RR in most cases.

Conversely, LIR exhibits the highest MAE, varying between 0.02 and~0.05.
Despite this, \systemname still significantly outperforms AC, with F1 score advantages of up to 50\pt, due to AC's unpredictable behavior and the restoring strategy of the \emph{Recommender}.
While \systemname shows variable predictive accuracy across different algorithms, it consistently guides experts, respectively the \textit{Cleaner}, towards more effective cleaning decisions.


\subsection{Runtime to produce a recommendation}
\systemname recommends to \revision{the} \emph{\revision{\textit{Cleaner}}} the next feature to clean until they have completely cleaned the dataset or the cleaning budget is spent.
After each recommended cleaning, they wait for the next suggestion before proceeding.
In our sixth research question~(RQ~\ref{rq:5}), we examine this waiting time, i.e., the runtime of \systemname per iteration.
We measured runtimes on a Slurm-managed compute cluster with two AMD EPYC 7742 processors, 64~cores, and 512~GB RAM, allocating up to 35~GB RAM and 40~CPU cores per job.
We use Scikit-learn~1.1.3 for ML model training and testing.

\begin{figure}[tb]
    \centering
    \includegraphics[width=1.0\linewidth]{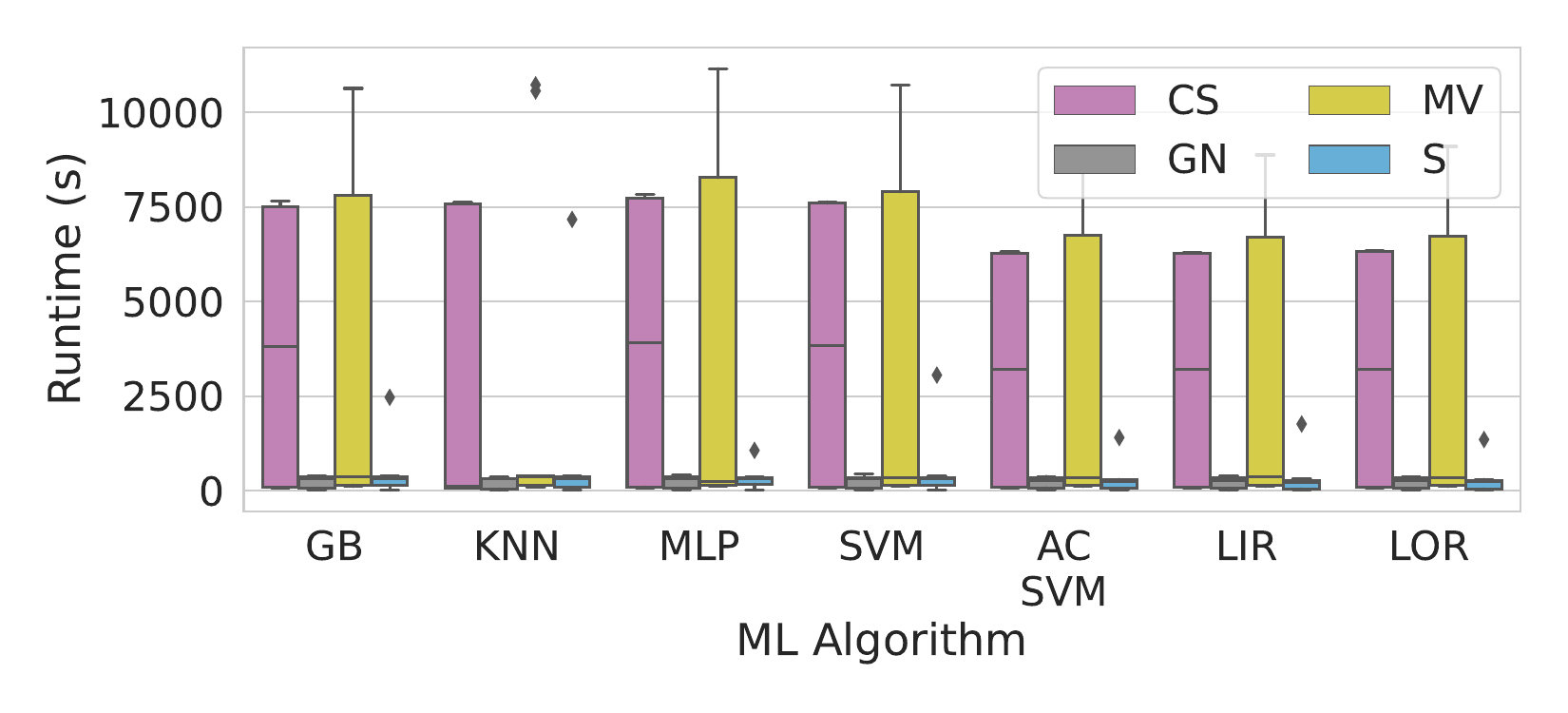}
    \caption{\systemname's runtimes. (CS -- Categorical Shift, GN -- Gaussian Noise, MV -- Missing Values, S -- Scaling).}
    \label{fig:runtimes}
\end{figure}

Figure~\ref{fig:runtimes} shows average runtimes per ML algorithm and error type across all datasets and pre-pollution settings, measured during the first iteration when the full extent of polluted features are considered, thus leading to the highest runtime.
%
\systemname shows the shortest runtimes for Gaussian noise (GN) and scaling errors (S) across all ML algorithms, with median runtimes between 230 and 330 seconds~(s) and averages between 220 and 500s.
The Airbnb dataset is an outlier~(in the context of~GN and~S) due to its high number of numerical features and its number of rows~(see Table~\ref{tbl:datasets}).
Since \systemname has to analyze the effect of cleaning each feature, the runtime scales with the number of features.

Categorical shift~(CS) and missing values~(MV) errors exhibit significantly higher runtimes across all ML algorithms, particularly in the upper quartiles, with medians between 125 and 3919s. 
The churn dataset largely contributes to these high runtimes for CS and MV. 
Excluding the churn dataset, CS and MV runtimes are typically under 400s. 
These longer runtimes are likely due to one-hot encoding, which increases the number of features and extends ML model training time.

\section{Conclusion} \label{sec:conclusion}
Data errors can significantly impact the prediction accuracy of \textit{machine learning}~(ML) models.
However, data cleaning in real-world scenarios is often associated with costs and expertise, usually requiring domain experts to configure and execute the cleaning process.

In this work, we introduced \systemname, a system that optimizes data cleaning efforts by aligning them with the underlying ML tasks.
\systemname recommends which feature to clean next, balancing cleaning costs against the potential for enhanced ML prediction accuracy.
We evaluated \systemname across seven diverse datasets, four ML algorithms, and four types of data errors, benchmarking it against a well-known method and other baselines.
Our results show \systemname achieves up to 52 and, on average, 5~percentage points higher F1 scores, indicating more efficient data cleaning.

Looking ahead, several directions for future research emerge.
One direction involves \revision{extending \systemname to capture and recommend multiple features to clean within one iteration.}
Another direction would be to extend \systemname for other ML tasks, such as regression and clustering.
It will also be valuable to add further error types, such as inconsistent representations or duplicates~\cite{budach2022effects}.

\subsection*{Acknowledgements}
This research was performed partially in the context of the \href{https://www.kitqar.de/de}{KITQAR} project, supported by Denkfabrik Digitale Arbeitsgemeinschaft im Bundesministerium für Arbeit und Soziales (BMAS). 
\balance

\bibliographystyle{ACM-Reference-Format}
\bibliography{references}


\begin{thebibliography}{41}


\ifx \showCODEN    \undefined \def \showCODEN     #1{\unskip}     \fi
\ifx \showDOI      \undefined \def \showDOI       #1{#1}\fi
\ifx \showISBNx    \undefined \def \showISBNx     #1{\unskip}     \fi
\ifx \showISBNxiii \undefined \def \showISBNxiii  #1{\unskip}     \fi
\ifx \showISSN     \undefined \def \showISSN      #1{\unskip}     \fi
\ifx \showLCCN     \undefined \def \showLCCN      #1{\unskip}     \fi
\ifx \shownote     \undefined \def \shownote      #1{#1}          \fi
\ifx \showarticletitle \undefined \def \showarticletitle #1{#1}   \fi
\ifx \showURL      \undefined \def \showURL       {\relax}        \fi
\providecommand\bibfield[2]{#2}
\providecommand\bibinfo[2]{#2}
\providecommand\natexlab[1]{#1}
\providecommand\showeprint[2][]{arXiv:#2}

\bibitem[\protect\citeauthoryear{Airbnb}{Airbnb}{2023}]%
        {airbnb_dataset}
Airbnb \bibinfo{year}{2023}\natexlab{}.
\newblock \bibinfo{booktitle}{\emph{Airbnb dataset (Last accessed: 2023-11-19)}}.
\newblock
\urldef\tempurl%
\url{https://www.dropbox.com/s/nerfrhbrseev928/CleanML-datasets-2020.zip?dl=0&file_subpath=%2Fdata%2FTitanic%2Fmissing_values}
\showURL{%
\tempurl}


\bibitem[\protect\citeauthoryear{Altman}{Altman}{1992}]%
        {altman1992introduction}
\bibfield{author}{\bibinfo{person}{Naomi~S Altman}.} \bibinfo{year}{1992}\natexlab{}.
\newblock \showarticletitle{An introduction to kernel and nearest-neighbor nonparametric regression}.
\newblock \bibinfo{journal}{\emph{The American Statistician}} \bibinfo{volume}{46}, \bibinfo{number}{3} (\bibinfo{year}{1992}), \bibinfo{pages}{175--185}.
\newblock


\bibitem[\protect\citeauthoryear{Chang}{Chang}{2023}]%
        {chang_quantum_2023}
\bibfield{author}{\bibinfo{person}{Kenneth Chang}.} \bibinfo{year}{2023}\natexlab{}.
\newblock \showarticletitle{Quantum Computing Advance Begins New Era, {IBM} Says}.
\newblock \bibinfo{journal}{\emph{The New York Times}} (\bibinfo{year}{2023}).
\newblock
\showISSN{0362-4331}
\urldef\tempurl%
\url{https://www.nytimes.com/2023/06/14/science/ibm-quantum-computing.html}
\showURL{%
\tempurl}


\bibitem[\protect\citeauthoryear{Chu, Ouzzani, Morcos, Ilyas, Papotti, Tang, and Ye}{Chu et~al\mbox{.}}{2015}]%
        {DBLP:journals/pvldb/ChuOMIP0Y15}
\bibfield{author}{\bibinfo{person}{Xu Chu}, \bibinfo{person}{Mourad Ouzzani}, \bibinfo{person}{John Morcos}, \bibinfo{person}{Ihab~F. Ilyas}, \bibinfo{person}{Paolo Papotti}, \bibinfo{person}{Nan Tang}, {and} \bibinfo{person}{Yin Ye}.} \bibinfo{year}{2015}\natexlab{}.
\newblock \showarticletitle{{KATARA:} Reliable Data Cleaning with Knowledge Bases and Crowdsourcing}.
\newblock \bibinfo{journal}{\emph{PVLDB}} \bibinfo{volume}{8}, \bibinfo{number}{12} (\bibinfo{year}{2015}), \bibinfo{pages}{1952--1955}.
\newblock
\urldef\tempurl%
\url{https://doi.org/10.14778/2824032.2824109}
\showDOI{\tempurl}


\bibitem[\protect\citeauthoryear{CleanML}{CleanML}{2023}]%
        {cleanmldatasetdescriptionspdf}
CleanML \bibinfo{year}{2023}\natexlab{}.
\newblock \bibinfo{title}{{CleanML}/{DatasetDescriptions}.pdf at master · chu-data-lab/{CleanML} (Last accessed: 2023-11-19)}.
\newblock
\newblock
\urldef\tempurl%
\url{https://github.com/chu-data-lab/CleanML/blob/master/DatasetDescriptions.pdf}
\showURL{%
\tempurl}


\bibitem[\protect\citeauthoryear{Cortes and Vapnik}{Cortes and Vapnik}{1995}]%
        {cortes1995support}
\bibfield{author}{\bibinfo{person}{Corinna Cortes} {and} \bibinfo{person}{Vladimir Vapnik}.} \bibinfo{year}{1995}\natexlab{}.
\newblock \showarticletitle{Support-vector networks}.
\newblock \bibinfo{journal}{\emph{Machine learning}}  \bibinfo{volume}{20} (\bibinfo{year}{1995}), \bibinfo{pages}{273--297}.
\newblock


\bibitem[\protect\citeauthoryear{Credit}{Credit}{2023}]%
        {credit_dataset}
Credit \bibinfo{year}{2023}\natexlab{}.
\newblock \bibinfo{booktitle}{\emph{Credit dataset (Last accessed: 2023-11-19)}}.
\newblock
\urldef\tempurl%
\url{https://www.dropbox.com/s/nerfrhbrseev928/CleanML-datasets-2020.zip?dl=0&file_subpath=%2Fdata%2FTitanic%2Fmissing_values}
\showURL{%
\tempurl}


\bibitem[\protect\citeauthoryear{Dallachiesa, Ebaid, Eldawy, Elmagarmid, Ilyas, Ouzzani, and Tang}{Dallachiesa et~al\mbox{.}}{2013}]%
        {DBLP:conf/sigmod/DallachiesaEEEIOT13}
\bibfield{author}{\bibinfo{person}{Michele Dallachiesa}, \bibinfo{person}{Amr Ebaid}, \bibinfo{person}{Ahmed Eldawy}, \bibinfo{person}{Ahmed~K. Elmagarmid}, \bibinfo{person}{Ihab~F. Ilyas}, \bibinfo{person}{Mourad Ouzzani}, {and} \bibinfo{person}{Nan Tang}.} \bibinfo{year}{2013}\natexlab{}.
\newblock \showarticletitle{{NADEEF:} a commodity data cleaning system}. In \bibinfo{booktitle}{\emph{Proceedings of the International Conference on Management of Data (SIGMOD)}}. \bibinfo{publisher}{{ACM}}, \bibinfo{pages}{541--552}.
\newblock
\urldef\tempurl%
\url{https://doi.org/10.1145/2463676.2465327}
\showDOI{\tempurl}


\bibitem[\protect\citeauthoryear{Emmanuel, Maupong, Mpoeleng, Semong, Mphago, and Tabona}{Emmanuel et~al\mbox{.}}{2021}]%
        {emmanuel2021survey}
\bibfield{author}{\bibinfo{person}{Tlamelo Emmanuel}, \bibinfo{person}{Thabiso~M. Maupong}, \bibinfo{person}{Dimane Mpoeleng}, \bibinfo{person}{Thabo Semong}, \bibinfo{person}{Banyatsang Mphago}, {and} \bibinfo{person}{Oteng Tabona}.} \bibinfo{year}{2021}\natexlab{}.
\newblock \showarticletitle{A survey on missing data in machine learning}.
\newblock \bibinfo{journal}{\emph{J. Big Data}} \bibinfo{volume}{8}, \bibinfo{number}{1} (\bibinfo{year}{2021}), \bibinfo{pages}{140}.
\newblock
\urldef\tempurl%
\url{https://doi.org/10.1186/s40537-021-00516-9}
\showDOI{\tempurl}


\bibitem[\protect\citeauthoryear{Feurer, Klein, Eggensperger, Springenberg, Blum, and Hutter}{Feurer et~al\mbox{.}}{2015}]%
        {feurer2015efficient}
\bibfield{author}{\bibinfo{person}{Matthias Feurer}, \bibinfo{person}{Aaron Klein}, \bibinfo{person}{Katharina Eggensperger}, \bibinfo{person}{Jost~Tobias Springenberg}, \bibinfo{person}{Manuel Blum}, {and} \bibinfo{person}{Frank Hutter}.} \bibinfo{year}{2015}\natexlab{}.
\newblock \showarticletitle{Efficient and Robust Automated Machine Learning}. In \bibinfo{booktitle}{\emph{Proceedings of the International Conference on Neural Information Processing Systems (NeurIPS)}}. \bibinfo{pages}{2962--2970}.
\newblock
\urldef\tempurl%
\url{https://proceedings.neurips.cc/paper/2015/hash/11d0e6287202fced83f79975ec59a3a6-Abstract.html}
\showURL{%
\tempurl}


\bibitem[\protect\citeauthoryear{Foroni, Lissandrini, and Velegrakis}{Foroni et~al\mbox{.}}{2021}]%
        {DBLP:conf/icde/ForoniLV21}
\bibfield{author}{\bibinfo{person}{Daniele Foroni}, \bibinfo{person}{Matteo Lissandrini}, {and} \bibinfo{person}{Yannis Velegrakis}.} \bibinfo{year}{2021}\natexlab{}.
\newblock \showarticletitle{Estimating the extent of the effects of Data Quality through Observations}. In \bibinfo{booktitle}{\emph{Proceedings of the International Conference on Data Engineering (ICDE)}}. \bibinfo{publisher}{{IEEE}}, \bibinfo{pages}{1913--1918}.
\newblock
\urldef\tempurl%
\url{https://doi.org/10.1109/ICDE51399.2021.00176}
\showDOI{\tempurl}


\bibitem[\protect\citeauthoryear{Friedman}{Friedman}{2001}]%
        {10.1214/aos/1013203451}
\bibfield{author}{\bibinfo{person}{Jerome~H. Friedman}.} \bibinfo{year}{2001}\natexlab{}.
\newblock \showarticletitle{{Greedy function approximation: A gradient boosting machine.}}
\newblock \bibinfo{journal}{\emph{The Annals of Statistics}} \bibinfo{volume}{29}, \bibinfo{number}{5} (\bibinfo{year}{2001}), \bibinfo{pages}{1189 -- 1232}.
\newblock
\urldef\tempurl%
\url{https://doi.org/10.1214/aos/1013203451}
\showDOI{\tempurl}


\bibitem[\protect\citeauthoryear{Groemping}{Groemping}{2019}]%
        {groemping2019south}
\bibfield{author}{\bibinfo{person}{Ulrike Groemping}.} \bibinfo{year}{2019}\natexlab{}.
\newblock \showarticletitle{South German credit data: Correcting a widely used data set}.
\newblock \bibinfo{journal}{\emph{Rep. Math., Phys. Chem., Berlin, Germany, Tech. Rep}}  \bibinfo{volume}{4} (\bibinfo{year}{2019}), \bibinfo{pages}{2019}.
\newblock


\bibitem[\protect\citeauthoryear{He, Zhao, and Chu}{He et~al\mbox{.}}{2021}]%
        {DBLP:journals/kbs/HeZC21}
\bibfield{author}{\bibinfo{person}{Xin He}, \bibinfo{person}{Kaiyong Zhao}, {and} \bibinfo{person}{Xiaowen Chu}.} \bibinfo{year}{2021}\natexlab{}.
\newblock \showarticletitle{AutoML: {A} survey of the state-of-the-art}.
\newblock \bibinfo{journal}{\emph{Knowl. Based Syst.}}  \bibinfo{volume}{212} (\bibinfo{year}{2021}), \bibinfo{pages}{106622}.
\newblock
\urldef\tempurl%
\url{https://doi.org/10.1016/j.knosys.2020.106622}
\showDOI{\tempurl}


\bibitem[\protect\citeauthoryear{Heidari, McGrath, Ilyas, and Rekatsinas}{Heidari et~al\mbox{.}}{2019}]%
        {DBLP:conf/sigmod/HeidariMIR19}
\bibfield{author}{\bibinfo{person}{Alireza Heidari}, \bibinfo{person}{Joshua McGrath}, \bibinfo{person}{Ihab~F. Ilyas}, {and} \bibinfo{person}{Theodoros Rekatsinas}.} \bibinfo{year}{2019}\natexlab{}.
\newblock \showarticletitle{HoloDetect: Few-Shot Learning for Error Detection}. In \bibinfo{booktitle}{\emph{Proceedings of the International Conference on Management of Data (SIGMOD)}}. \bibinfo{publisher}{{ACM}}, \bibinfo{pages}{829--846}.
\newblock
\urldef\tempurl%
\url{https://doi.org/10.1145/3299869.3319888}
\showDOI{\tempurl}


\bibitem[\protect\citeauthoryear{Hofmann}{Hofmann}{1994}]%
        {statlog_dataset}
\bibfield{author}{\bibinfo{person}{Hans Hofmann}.} \bibinfo{year}{1994}\natexlab{}.
\newblock \bibinfo{title}{{Statlog (German Credit Data)}}.
\newblock \bibinfo{howpublished}{UCI Machine Learning Repository}.
\newblock
\newblock
\shownote{{DOI}: https://doi.org/10.24432/C5NC77.}


\bibitem[\protect\citeauthoryear{Ilyas and Rekatsinas}{Ilyas and Rekatsinas}{2022}]%
        {DBLP:journals/jdiq/IlyasR22}
\bibfield{author}{\bibinfo{person}{Ihab~F. Ilyas} {and} \bibinfo{person}{Theodoros Rekatsinas}.} \bibinfo{year}{2022}\natexlab{}.
\newblock \showarticletitle{Machine Learning and Data Cleaning: Which Serves the Other?}
\newblock \bibinfo{journal}{\emph{Journal on Data and Information Quality (JDIQ)}} \bibinfo{volume}{14}, \bibinfo{number}{3} (\bibinfo{year}{2022}), \bibinfo{pages}{13:1--13:11}.
\newblock
\urldef\tempurl%
\url{https://doi.org/10.1145/3506712}
\showDOI{\tempurl}


\bibitem[\protect\citeauthoryear{Karlas, Li, Wu, G{\"{u}}rel, Chu, Wu, and Zhang}{Karlas et~al\mbox{.}}{2020}]%
        {karlavs2020nearest}
\bibfield{author}{\bibinfo{person}{Bojan Karlas}, \bibinfo{person}{Peng Li}, \bibinfo{person}{Renzhi Wu}, \bibinfo{person}{Nezihe~Merve G{\"{u}}rel}, \bibinfo{person}{Xu Chu}, \bibinfo{person}{Wentao Wu}, {and} \bibinfo{person}{Ce Zhang}.} \bibinfo{year}{2020}\natexlab{}.
\newblock \showarticletitle{Nearest Neighbor Classifiers over Incomplete Information: From Certain Answers to Certain Predictions}.
\newblock \bibinfo{journal}{\emph{PVLDB}} \bibinfo{volume}{14}, \bibinfo{number}{3} (\bibinfo{year}{2020}), \bibinfo{pages}{255--267}.
\newblock
\urldef\tempurl%
\url{https://doi.org/10.5555/3430915.3442426}
\showDOI{\tempurl}


\bibitem[\protect\citeauthoryear{Krishnan, Franklin, Goldberg, and Wu}{Krishnan et~al\mbox{.}}{2017}]%
        {boostclean}
\bibfield{author}{\bibinfo{person}{Sanjay Krishnan}, \bibinfo{person}{Michael~J. Franklin}, \bibinfo{person}{Ken Goldberg}, {and} \bibinfo{person}{Eugene Wu}.} \bibinfo{year}{2017}\natexlab{}.
\newblock \showarticletitle{BoostClean: Automated Error Detection and Repair for Machine Learning}.
\newblock \bibinfo{journal}{\emph{CoRR}}  \bibinfo{volume}{abs/1711.01299} (\bibinfo{year}{2017}).
\newblock
\showeprint[arXiv]{1711.01299}
\urldef\tempurl%
\url{http://arxiv.org/abs/1711.01299}
\showURL{%
\tempurl}


\bibitem[\protect\citeauthoryear{Krishnan, Wang, Wu, Franklin, and Goldberg}{Krishnan et~al\mbox{.}}{2016}]%
        {krishnan2016activeclean}
\bibfield{author}{\bibinfo{person}{Sanjay Krishnan}, \bibinfo{person}{Jiannan Wang}, \bibinfo{person}{Eugene Wu}, \bibinfo{person}{Michael~J. Franklin}, {and} \bibinfo{person}{Ken Goldberg}.} \bibinfo{year}{2016}\natexlab{}.
\newblock \showarticletitle{ActiveClean: Interactive Data Cleaning For Statistical Modeling}.
\newblock \bibinfo{journal}{\emph{PVLDB}} \bibinfo{volume}{9}, \bibinfo{number}{12} (\bibinfo{year}{2016}), \bibinfo{pages}{948--959}.
\newblock
\urldef\tempurl%
\url{https://doi.org/10.14778/2994509.2994514}
\showDOI{\tempurl}


\bibitem[\protect\citeauthoryear{Li, Chen, Chu, and Rong}{Li et~al\mbox{.}}{2023}]%
        {diffprep}
\bibfield{author}{\bibinfo{person}{Peng Li}, \bibinfo{person}{Zhiyi Chen}, \bibinfo{person}{Xu Chu}, {and} \bibinfo{person}{Kexin Rong}.} \bibinfo{year}{2023}\natexlab{}.
\newblock \showarticletitle{DiffPrep: Differentiable Data Preprocessing Pipeline Search for Learning over Tabular Data}.
\newblock \bibinfo{journal}{\emph{Proceedings of the International Conference on Management of Data (SIGMOD)}} \bibinfo{volume}{1}, \bibinfo{number}{2} (\bibinfo{year}{2023}), \bibinfo{pages}{183:1--183:26}.
\newblock
\urldef\tempurl%
\url{https://doi.org/10.1145/3589328}
\showDOI{\tempurl}


\bibitem[\protect\citeauthoryear{Li, Rao, Blase, Zhang, Chu, and Zhang}{Li et~al\mbox{.}}{2021}]%
        {DBLP:conf/icde/LiRBZCZ21}
\bibfield{author}{\bibinfo{person}{Peng Li}, \bibinfo{person}{Xi Rao}, \bibinfo{person}{Jennifer Blase}, \bibinfo{person}{Yue Zhang}, \bibinfo{person}{Xu Chu}, {and} \bibinfo{person}{Ce Zhang}.} \bibinfo{year}{2021}\natexlab{}.
\newblock \showarticletitle{CleanML: {A} Study for Evaluating the Impact of Data Cleaning on {ML} Classification Tasks}. In \bibinfo{booktitle}{\emph{Proceedings of the International Conference on Data Engineering (ICDE)}}. \bibinfo{publisher}{{IEEE}}, \bibinfo{pages}{13--24}.
\newblock
\urldef\tempurl%
\url{https://doi.org/10.1109/ICDE51399.2021.00009}
\showDOI{\tempurl}


\bibitem[\protect\citeauthoryear{Lim}{Lim}{1997}]%
        {misc_contraceptive_method_choice_30}
\bibfield{author}{\bibinfo{person}{Tjen-Sien Lim}.} \bibinfo{year}{1997}\natexlab{}.
\newblock \bibinfo{title}{{Contraceptive Method Choice}}.
\newblock \bibinfo{howpublished}{UCI Machine Learning Repository}.
\newblock
\newblock
\shownote{{DOI}: https://doi.org/10.24432/C59W2D.}


\bibitem[\protect\citeauthoryear{Lundberg and Lee}{Lundberg and Lee}{2017}]%
        {lundberg2017unified}
\bibfield{author}{\bibinfo{person}{Scott~M. Lundberg} {and} \bibinfo{person}{Su{-}In Lee}.} \bibinfo{year}{2017}\natexlab{}.
\newblock \showarticletitle{A Unified Approach to Interpreting Model Predictions}. In \bibinfo{booktitle}{\emph{Proceedings of the International Conference on Neural Information Processing Systems (NeurIPS)}}. \bibinfo{pages}{4765--4774}.
\newblock
\urldef\tempurl%
\url{http://papers.nips.cc/paper/7062-a-unified-approach-to-interpreting-model-predictions.pdf}
\showURL{%
\tempurl}


\bibitem[\protect\citeauthoryear{Mazumder, Banbury, Yao, Karlas, Rojas, Diamos, Diamos, He, Parrish, Kirk, Quaye, Rastogi, Kiela, Jurado, Kanter, Mosquera, Cukierski, Ciro, Aroyo, Acun, Chen, Raje, Bartolo, Eyuboglu, Ghorbani, Goodman, Howard, Inel, Kane, Kirkpatrick, Sculley, Kuo, Mueller, Thrush, Vanschoren, Warren, Williams, Yeung, Ardalani, Paritosh, Zhang, Zou, Wu, Coleman, Ng, Mattson, and Reddi}{Mazumder et~al\mbox{.}}{2023}]%
        {mazumder2022dataperf}
\bibfield{author}{\bibinfo{person}{Mark Mazumder}, \bibinfo{person}{Colby~R. Banbury}, \bibinfo{person}{Xiaozhe Yao}, \bibinfo{person}{Bojan Karlas}, \bibinfo{person}{William~Gaviria Rojas}, \bibinfo{person}{Sudnya~Frederick Diamos}, \bibinfo{person}{Greg Diamos}, \bibinfo{person}{Lynn He}, \bibinfo{person}{Alicia Parrish}, \bibinfo{person}{Hannah~Rose Kirk}, \bibinfo{person}{Jessica Quaye}, \bibinfo{person}{Charvi Rastogi}, \bibinfo{person}{Douwe Kiela}, \bibinfo{person}{David Jurado}, \bibinfo{person}{David Kanter}, \bibinfo{person}{Rafael Mosquera}, \bibinfo{person}{Will Cukierski}, \bibinfo{person}{Juan Ciro}, \bibinfo{person}{Lora Aroyo}, \bibinfo{person}{Bilge Acun}, \bibinfo{person}{Lingjiao Chen}, \bibinfo{person}{Mehul Raje}, \bibinfo{person}{Max Bartolo}, \bibinfo{person}{Evan~Sabri Eyuboglu}, \bibinfo{person}{Amirata Ghorbani}, \bibinfo{person}{Emmett~D. Goodman}, \bibinfo{person}{Addison Howard}, \bibinfo{person}{Oana Inel}, \bibinfo{person}{Tariq Kane}, \bibinfo{person}{Christine~R. Kirkpatrick},
  \bibinfo{person}{D. Sculley}, \bibinfo{person}{Tzu{-}Sheng Kuo}, \bibinfo{person}{Jonas~W. Mueller}, \bibinfo{person}{Tristan Thrush}, \bibinfo{person}{Joaquin Vanschoren}, \bibinfo{person}{Margaret Warren}, \bibinfo{person}{Adina Williams}, \bibinfo{person}{Serena Yeung}, \bibinfo{person}{Newsha Ardalani}, \bibinfo{person}{Praveen~K. Paritosh}, \bibinfo{person}{Ce Zhang}, \bibinfo{person}{James~Y. Zou}, \bibinfo{person}{Carole{-}Jean Wu}, \bibinfo{person}{Cody Coleman}, \bibinfo{person}{Andrew~Y. Ng}, \bibinfo{person}{Peter Mattson}, {and} \bibinfo{person}{Vijay~Janapa Reddi}.} \bibinfo{year}{2023}\natexlab{}.
\newblock \showarticletitle{DataPerf: Benchmarks for Data-Centric {AI} Development}. In \bibinfo{booktitle}{\emph{Proceedings of the International Conference on Neural Information Processing Systems (NeurIPS)}}.
\newblock


\bibitem[\protect\citeauthoryear{Mohammed, Budach, Feuerpfeil, Ihde, Nathansen, Noack, Patzlaff, Naumann, and Harmouch}{Mohammed et~al\mbox{.}}{2024}]%
        {budach2022effects}
\bibfield{author}{\bibinfo{person}{Sedir Mohammed}, \bibinfo{person}{Lukas Budach}, \bibinfo{person}{Moritz Feuerpfeil}, \bibinfo{person}{Nina Ihde}, \bibinfo{person}{Andrea Nathansen}, \bibinfo{person}{Nele Noack}, \bibinfo{person}{Hendrik Patzlaff}, \bibinfo{person}{Felix Naumann}, {and} \bibinfo{person}{Hazar Harmouch}.} \bibinfo{year}{2024}\natexlab{}.
\newblock \bibinfo{title}{The Effects of Data Quality on Machine Learning Performance}.
\newblock
\newblock
\urldef\tempurl%
\url{https://arxiv.org/abs/2207.14529}
\showURL{%
\tempurl}


\bibitem[\protect\citeauthoryear{Neutatz, Chen, Abedjan, and Wu}{Neutatz et~al\mbox{.}}{2021}]%
        {neutatz2021cleaning}
\bibfield{author}{\bibinfo{person}{Felix Neutatz}, \bibinfo{person}{Binger Chen}, \bibinfo{person}{Ziawasch Abedjan}, {and} \bibinfo{person}{Eugene Wu}.} \bibinfo{year}{2021}\natexlab{}.
\newblock \showarticletitle{From Cleaning before {ML} to Cleaning for {ML}}.
\newblock \bibinfo{journal}{\emph{IEEE Data Engineering Bulletin}} \bibinfo{volume}{44}, \bibinfo{number}{1} (\bibinfo{year}{2021}), \bibinfo{pages}{24--41}.
\newblock
\urldef\tempurl%
\url{http://sites.computer.org/debull/A21mar/p24.pdf}
\showURL{%
\tempurl}


\bibitem[\protect\citeauthoryear{Neutatz, Chen, Alkhatib, Ye, and Abedjan}{Neutatz et~al\mbox{.}}{2022}]%
        {neutatz2022data}
\bibfield{author}{\bibinfo{person}{Felix Neutatz}, \bibinfo{person}{Binger Chen}, \bibinfo{person}{Yazan Alkhatib}, \bibinfo{person}{Jingwen Ye}, {and} \bibinfo{person}{Ziawasch Abedjan}.} \bibinfo{year}{2022}\natexlab{}.
\newblock \showarticletitle{Data Cleaning and AutoML: Would an Optimizer Choose to Clean?}
\newblock \bibinfo{journal}{\emph{Datenbank-Spektrum}} \bibinfo{volume}{22}, \bibinfo{number}{2} (\bibinfo{year}{2022}), \bibinfo{pages}{121--130}.
\newblock
\urldef\tempurl%
\url{https://doi.org/10.1007/s13222-022-00413-2}
\showDOI{\tempurl}


\bibitem[\protect\citeauthoryear{Neutatz, Mahdavi, and Abedjan}{Neutatz et~al\mbox{.}}{2019}]%
        {DBLP:conf/cikm/NeutatzMA19}
\bibfield{author}{\bibinfo{person}{Felix Neutatz}, \bibinfo{person}{Mohammad Mahdavi}, {and} \bibinfo{person}{Ziawasch Abedjan}.} \bibinfo{year}{2019}\natexlab{}.
\newblock \showarticletitle{{ED2:} {A} Case for Active Learning in Error Detection}. In \bibinfo{booktitle}{\emph{Proceedings of the International Conference on Information and Knowledge Management (CIKM)}}. \bibinfo{publisher}{{ACM}}, \bibinfo{pages}{2249--2252}.
\newblock
\urldef\tempurl%
\url{https://doi.org/10.1145/3357384.3358129}
\showDOI{\tempurl}


\bibitem[\protect\citeauthoryear{Polyzotis and Zaharia}{Polyzotis and Zaharia}{2021}]%
        {polyzotis2021can}
\bibfield{author}{\bibinfo{person}{Neoklis Polyzotis} {and} \bibinfo{person}{Matei Zaharia}.} \bibinfo{year}{2021}\natexlab{}.
\newblock \showarticletitle{What can Data-Centric {AI} Learn from Data and {ML} Engineering?}
\newblock \bibinfo{journal}{\emph{CoRR}}  \bibinfo{volume}{abs/2112.06439} (\bibinfo{year}{2021}).
\newblock
\showeprint[arXiv]{2112.06439}
\urldef\tempurl%
\url{https://arxiv.org/abs/2112.06439}
\showURL{%
\tempurl}


\bibitem[\protect\citeauthoryear{Qahtan, Elmagarmid, Castro~Fernandez, Ouzzani, and Tang}{Qahtan et~al\mbox{.}}{2018}]%
        {FAHES2018}
\bibfield{author}{\bibinfo{person}{Abdulhakim~A. Qahtan}, \bibinfo{person}{Ahmed Elmagarmid}, \bibinfo{person}{Raul Castro~Fernandez}, \bibinfo{person}{Mourad Ouzzani}, {and} \bibinfo{person}{Nan Tang}.} \bibinfo{year}{2018}\natexlab{}.
\newblock \showarticletitle{FAHES: A Robust Disguised Missing Values Detector}. In \bibinfo{booktitle}{\emph{Proceedings of the International Conference on Knowledge discovery and data mining (SIGKDD)}}. \bibinfo{publisher}{ACM}, \bibinfo{address}{New York, NY, USA}, \bibinfo{pages}{2100–2109}.
\newblock
\showISBNx{9781450355520}
\urldef\tempurl%
\url{https://doi.org/10.1145/3219819.3220109}
\showDOI{\tempurl}


\bibitem[\protect\citeauthoryear{Rekatsinas, Chu, Ilyas, and R{\'{e}}}{Rekatsinas et~al\mbox{.}}{2017}]%
        {DBLP:journals/corr/RekatsinasCIR17}
\bibfield{author}{\bibinfo{person}{Theodoros Rekatsinas}, \bibinfo{person}{Xu Chu}, \bibinfo{person}{Ihab~F. Ilyas}, {and} \bibinfo{person}{Christopher R{\'{e}}}.} \bibinfo{year}{2017}\natexlab{}.
\newblock \showarticletitle{HoloClean: Holistic Data Repairs with Probabilistic Inference}.
\newblock \bibinfo{journal}{\emph{PVLDB}} \bibinfo{volume}{10}, \bibinfo{number}{11} (\bibinfo{year}{2017}), \bibinfo{pages}{1190--1201}.
\newblock
\urldef\tempurl%
\url{https://doi.org/10.14778/3137628.3137631}
\showDOI{\tempurl}


\bibitem[\protect\citeauthoryear{Roesler}{Roesler}{2013}]%
        {misc_eeg_eye_state_264}
\bibfield{author}{\bibinfo{person}{Oliver Roesler}.} \bibinfo{year}{2013}\natexlab{}.
\newblock \bibinfo{title}{{EEG Eye State}}.
\newblock \bibinfo{howpublished}{UCI Machine Learning Repository}.
\newblock
\newblock
\shownote{{DOI}: https://doi.org/10.24432/C57G7J.}


\bibitem[\protect\citeauthoryear{Rubin}{Rubin}{1976}]%
        {rubin1976inference}
\bibfield{author}{\bibinfo{person}{Donald~B Rubin}.} \bibinfo{year}{1976}\natexlab{}.
\newblock \showarticletitle{Inference and missing data}.
\newblock \bibinfo{journal}{\emph{Biometrika}} \bibinfo{volume}{63}, \bibinfo{number}{3} (\bibinfo{year}{1976}), \bibinfo{pages}{581--592}.
\newblock


\bibitem[\protect\citeauthoryear{Schelter, Rukat, and Biessmann}{Schelter et~al\mbox{.}}{2021}]%
        {Jenga}
\bibfield{author}{\bibinfo{person}{S. Schelter}, \bibinfo{person}{T. Rukat}, {and} \bibinfo{person}{F. Biessmann}.} \bibinfo{year}{2021}\natexlab{}.
\newblock \showarticletitle{{JENGA} - {A} Framework to Study the Impact of Data Errors on the Predictions of Machine Learning Models}. In \bibinfo{booktitle}{\emph{Proceedings of the International Conference on Extending Database Technology (EDBT)}}. \bibinfo{publisher}{OpenProceedings.org}, \bibinfo{pages}{529--534}.
\newblock
\urldef\tempurl%
\url{https://doi.org/10.5441/002/edbt.2021.63}
\showURL{%
\tempurl}


\bibitem[\protect\citeauthoryear{Specht}{Specht}{1991}]%
        {specht1991general}
\bibfield{author}{\bibinfo{person}{Donald~F Specht}.} \bibinfo{year}{1991}\natexlab{}.
\newblock \showarticletitle{A general regression neural network}.
\newblock \bibinfo{journal}{\emph{IEEE Transactions on Neural Networks}} \bibinfo{volume}{2}, \bibinfo{number}{6} (\bibinfo{year}{1991}), \bibinfo{pages}{568--576}.
\newblock


\bibitem[\protect\citeauthoryear{Telco}{Telco}{2023}]%
        {noauthor_telco_nodate}
Telco \bibinfo{year}{2023}\natexlab{}.
\newblock \bibinfo{booktitle}{\emph{Telco Customer Churn (Last accessed: 2023-11-22)}}.
\newblock
\urldef\tempurl%
\url{https://www.kaggle.com/datasets/blastchar/telco-customer-churn}
\showURL{%
\tempurl}


\bibitem[\protect\citeauthoryear{Titanic}{Titanic}{2023}]%
        {titanic_dataset}
Titanic \bibinfo{year}{2023}\natexlab{}.
\newblock \bibinfo{title}{Titanic dataset (Last accessed: 2023-11-19)}.
\newblock
\newblock
\urldef\tempurl%
\url{https://www.dropbox.com/s/nerfrhbrseev928/CleanML-datasets-2020.zip?dl=0&file_subpath=%2Fdata%2FTitanic%2Fmissing_values}
\showURL{%
\tempurl}


\bibitem[\protect\citeauthoryear{Whang, Roh, Song, and Lee}{Whang et~al\mbox{.}}{2023}]%
        {whang2023data}
\bibfield{author}{\bibinfo{person}{Steven~Euijong Whang}, \bibinfo{person}{Yuji Roh}, \bibinfo{person}{Hwanjun Song}, {and} \bibinfo{person}{Jae{-}Gil Lee}.} \bibinfo{year}{2023}\natexlab{}.
\newblock \showarticletitle{Data collection and quality challenges in deep learning: a data-centric {AI} perspective}.
\newblock \bibinfo{journal}{\emph{VLDB Journal}} \bibinfo{volume}{32}, \bibinfo{number}{4} (\bibinfo{year}{2023}), \bibinfo{pages}{791--813}.
\newblock
\urldef\tempurl%
\url{https://doi.org/10.1007/s00778-022-00775-9}
\showDOI{\tempurl}


\bibitem[\protect\citeauthoryear{Yakout, Berti{-}{\'{E}}quille, and Elmagarmid}{Yakout et~al\mbox{.}}{2013}]%
        {DBLP:conf/sigmod/YakoutBE13}
\bibfield{author}{\bibinfo{person}{Mohamed Yakout}, \bibinfo{person}{Laure Berti{-}{\'{E}}quille}, {and} \bibinfo{person}{Ahmed~K. Elmagarmid}.} \bibinfo{year}{2013}\natexlab{}.
\newblock \showarticletitle{Don't be SCAREd: use SCalable Automatic REpairing with maximal likelihood and bounded changes}. In \bibinfo{booktitle}{\emph{Proceedings of the International Conference on Management of Data (SIGMOD)}}. \bibinfo{publisher}{{ACM}}, \bibinfo{pages}{553--564}.
\newblock
\urldef\tempurl%
\url{https://doi.org/10.1145/2463676.2463706}
\showDOI{\tempurl}


\bibitem[\protect\citeauthoryear{Zha, Bhat, Lai, Yang, Jiang, Zhong, and Hu}{Zha et~al\mbox{.}}{2023}]%
        {zha2023data}
\bibfield{author}{\bibinfo{person}{Daochen Zha}, \bibinfo{person}{Zaid~Pervaiz Bhat}, \bibinfo{person}{Kwei{-}Herng Lai}, \bibinfo{person}{Fan Yang}, \bibinfo{person}{Zhimeng Jiang}, \bibinfo{person}{Shaochen Zhong}, {and} \bibinfo{person}{Xia Hu}.} \bibinfo{year}{2023}\natexlab{}.
\newblock \showarticletitle{Data-centric Artificial Intelligence: {A} Survey}.
\newblock \bibinfo{journal}{\emph{CoRR}}  \bibinfo{volume}{abs/2303.10158} (\bibinfo{year}{2023}).
\newblock
\urldef\tempurl%
\url{https://doi.org/10.48550/arXiv.2303.10158}
\showDOI{\tempurl}
\showeprint[arXiv]{2303.10158}


\end{thebibliography}
\newpage
\appendix
\onecolumn

\section{Comparison to baselines for multiple error types and diverse cost functions}

\begin{figure*}[h!]
    \centering
    \begin{subfigure}{0.24\textwidth}
        \includegraphics[width=\linewidth]{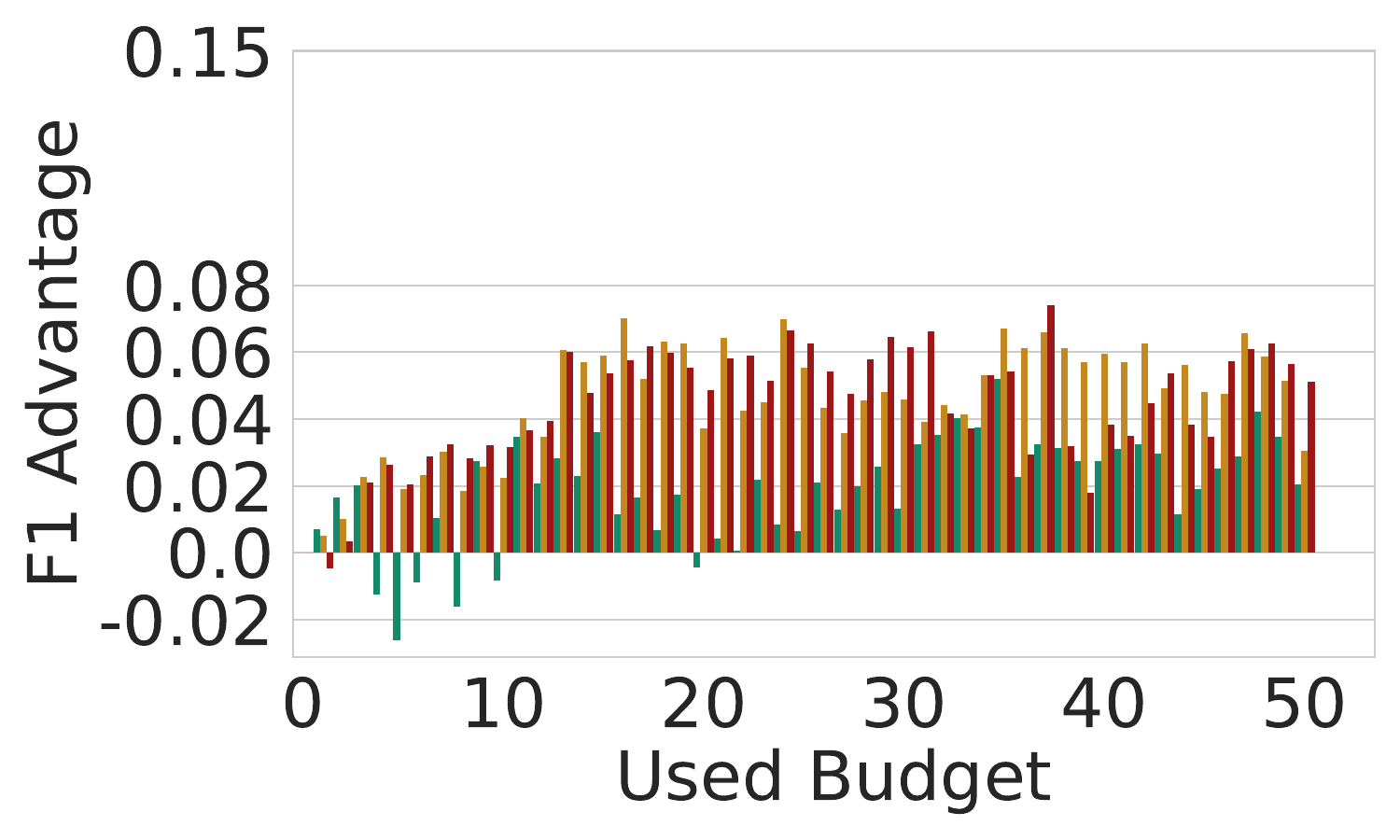}
        \caption{CMC}
    \end{subfigure}
    \begin{subfigure}{0.24\textwidth}
        \includegraphics[width=\linewidth]{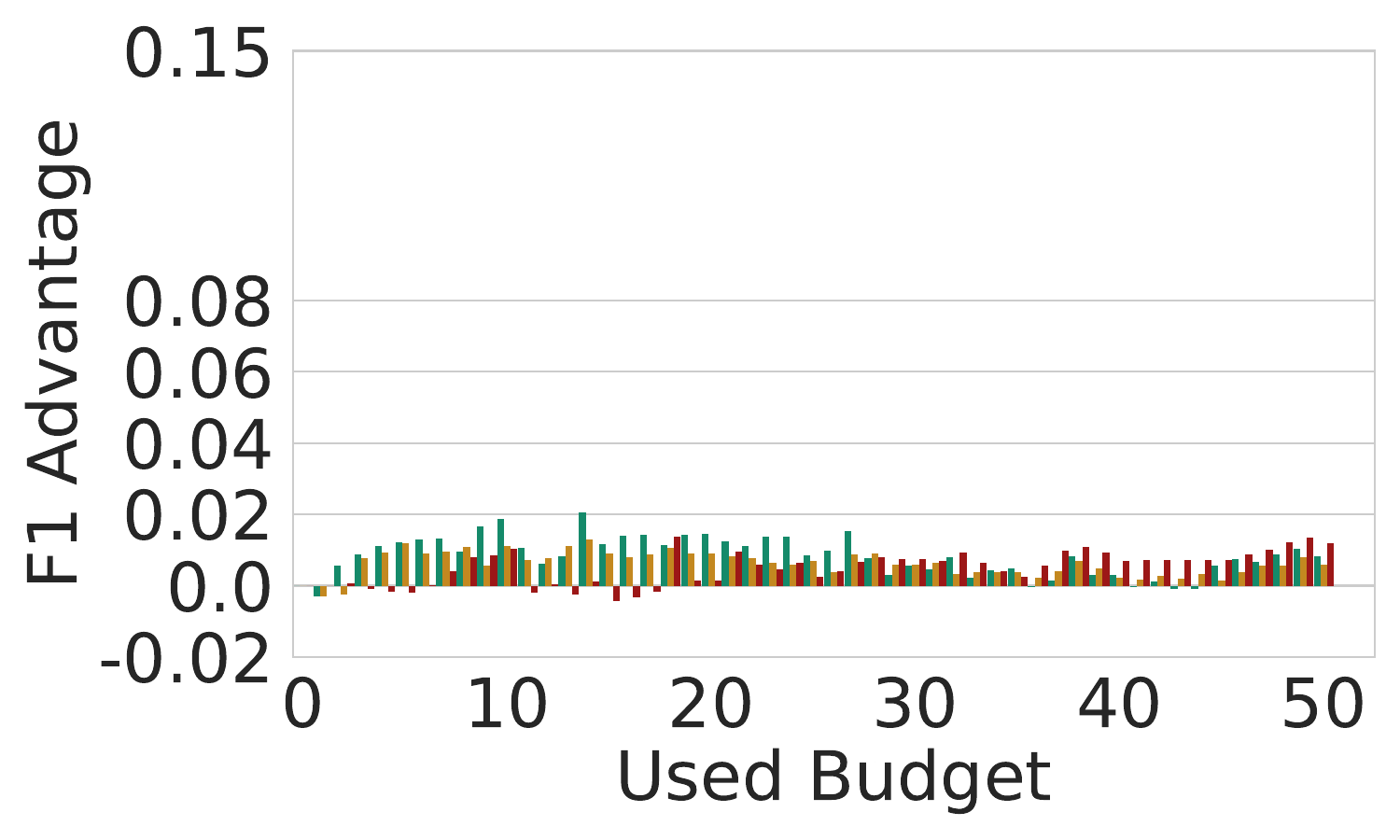}
        \caption{Churn}
    \end{subfigure}
    \begin{subfigure}{0.24\textwidth}
        \includegraphics[width=\linewidth]{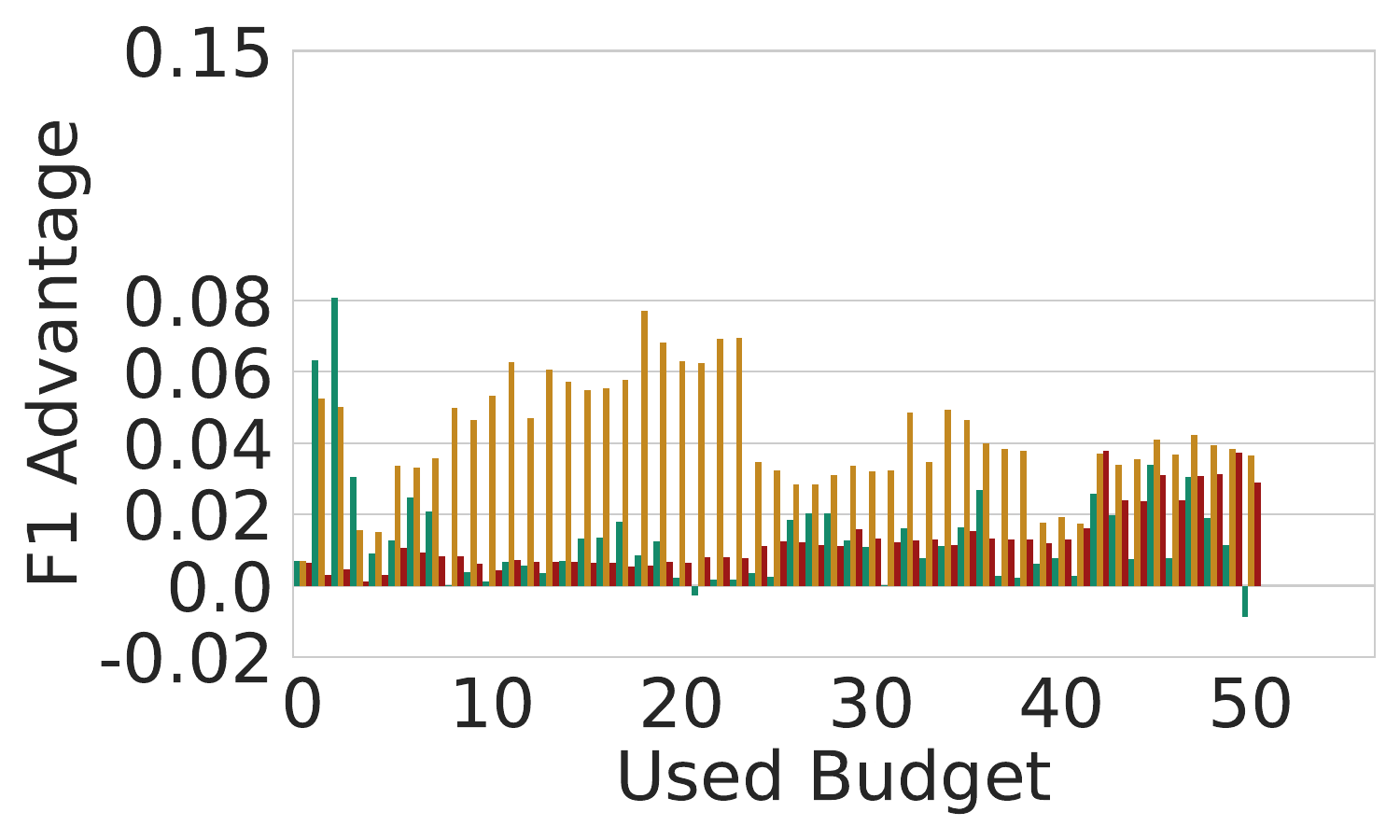}
        \caption{EEG}
    \end{subfigure}
    \begin{subfigure}{0.24\textwidth}
        \includegraphics[width=\linewidth]{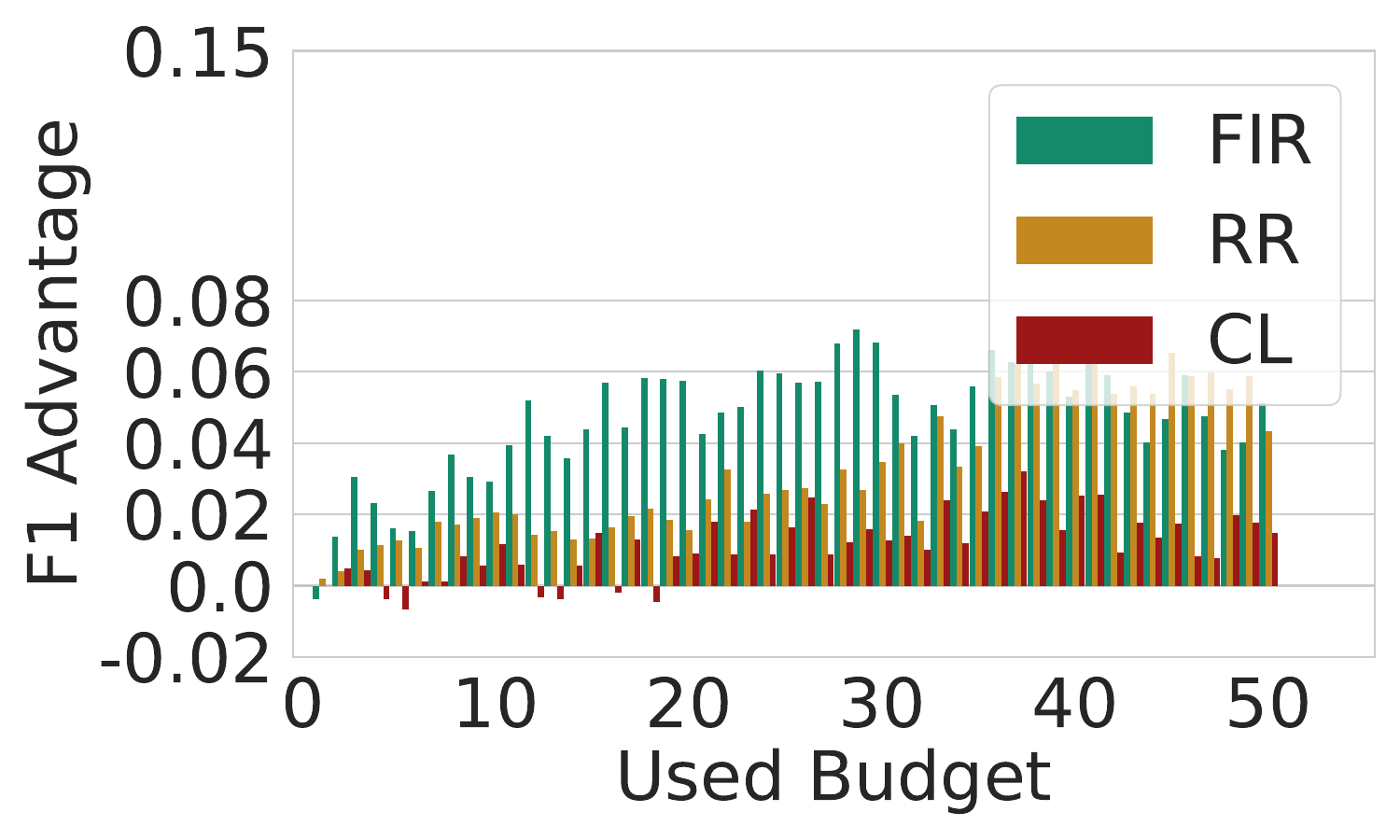}
        \caption{S-Credit}
    \end{subfigure}
    \caption{Comparison of~\systemname with the baselines for MLP across multiple error types and cost functions.}
    \label{fig:agg_bl_multi_error_results_mlp}
\end{figure*}

\begin{figure*}[h!]
    \centering
    \begin{subfigure}{0.24\textwidth}
        \includegraphics[width=\linewidth]{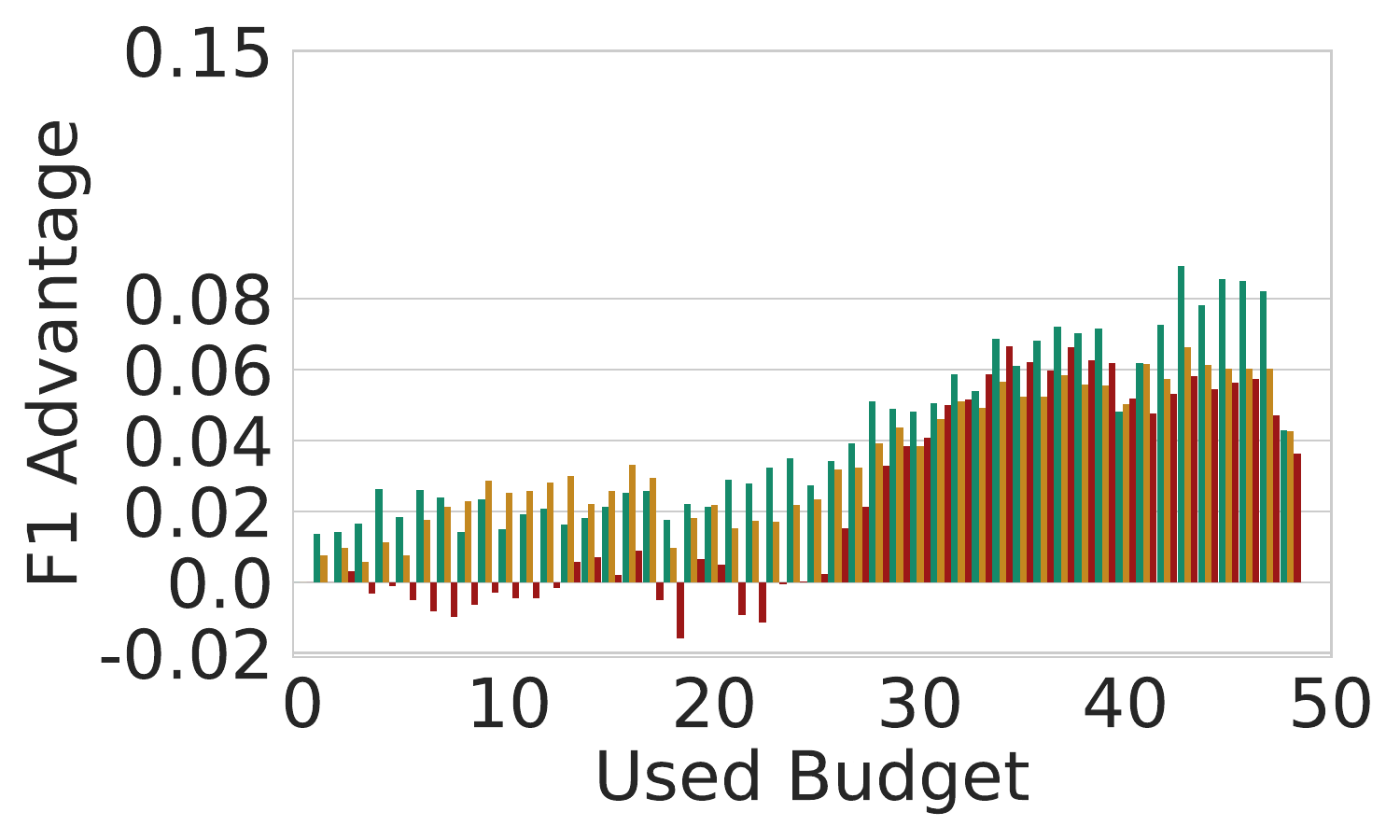}
        \caption{CMC}
    \end{subfigure}
    \begin{subfigure}{0.24\textwidth}
        \includegraphics[width=\linewidth]{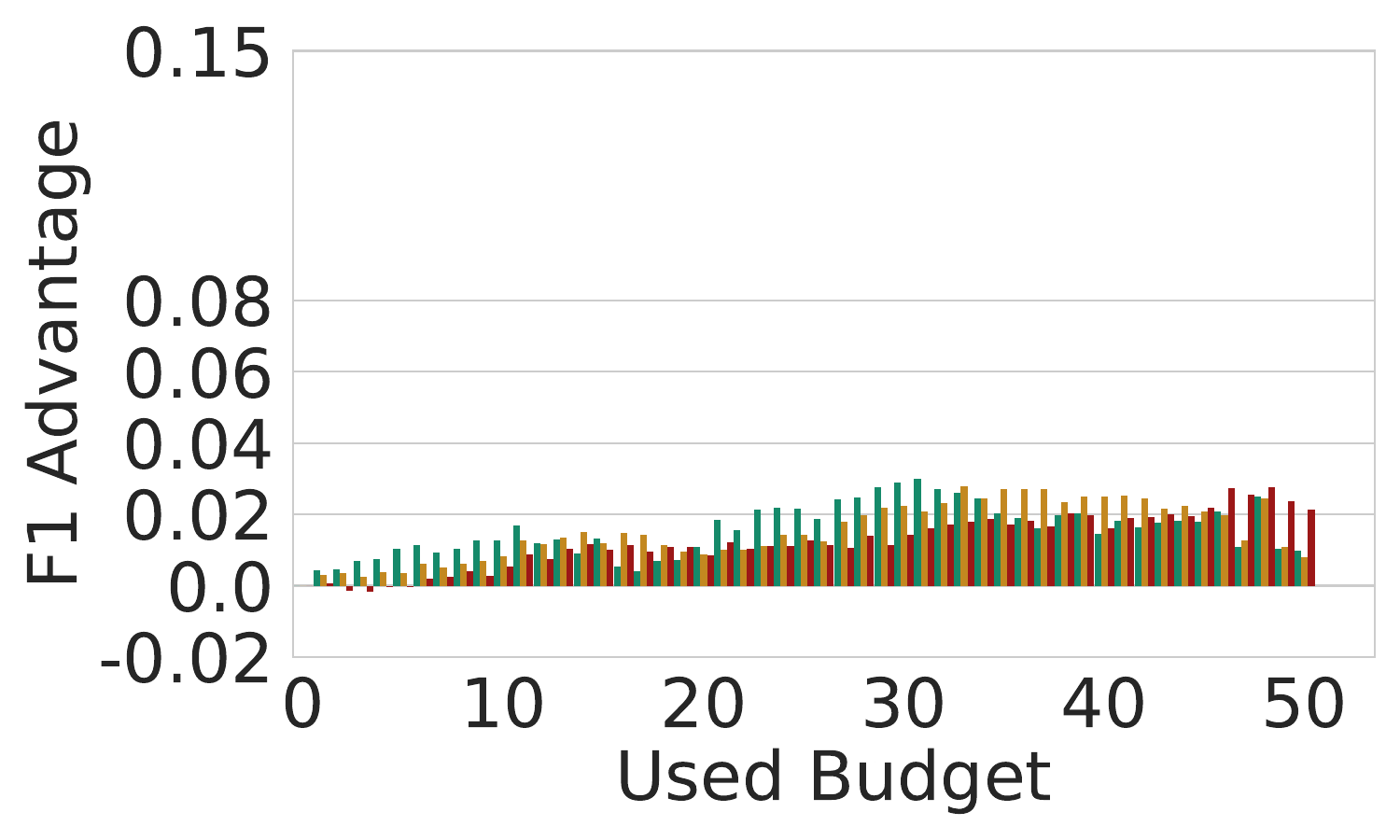}
        \caption{Churn}
    \end{subfigure}
    \begin{subfigure}{0.24\textwidth}
        \includegraphics[width=\linewidth]{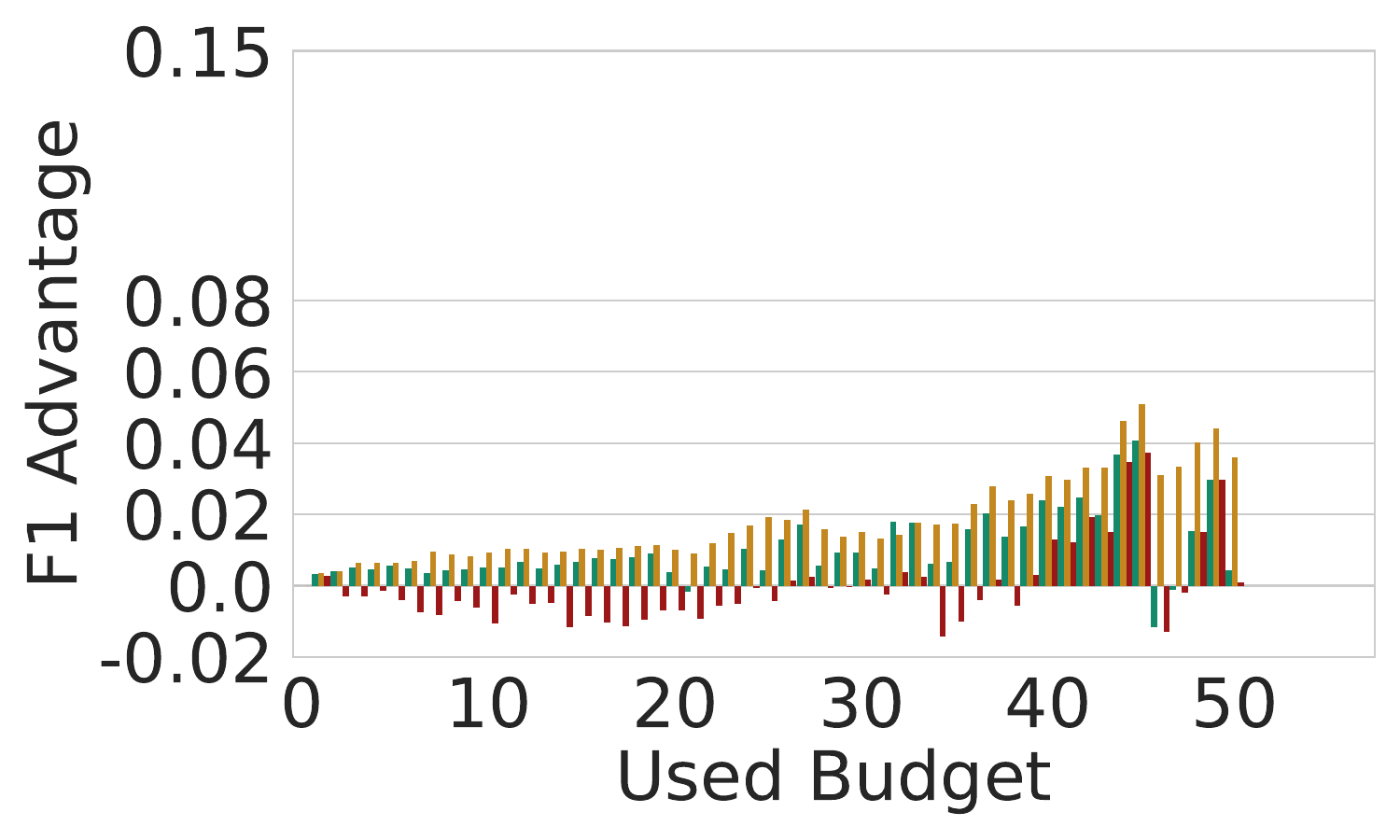}
        \caption{EEG}
    \end{subfigure}
    \begin{subfigure}{0.24\textwidth}
        \includegraphics[width=\linewidth]{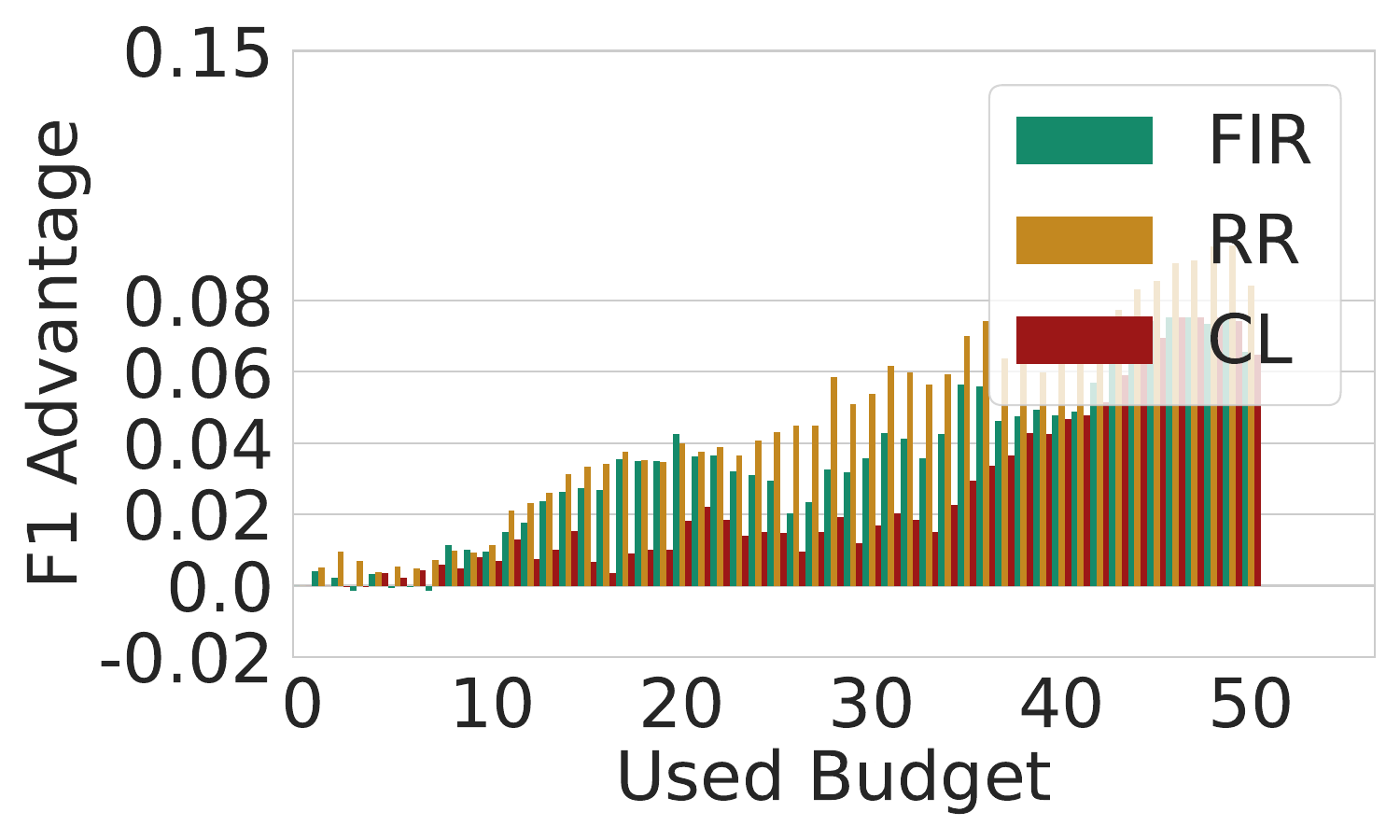}
        \caption{S-Credit}
    \end{subfigure}
    \caption{Comparison of~\systemname with the baselines for KNN across multiple error types and cost functions.}
    \label{fig:agg_bl_multi_error_results_knn}
\end{figure*}

\begin{figure*}[h!]
    \centering
    \begin{subfigure}{0.24\textwidth}
        \includegraphics[width=\linewidth]{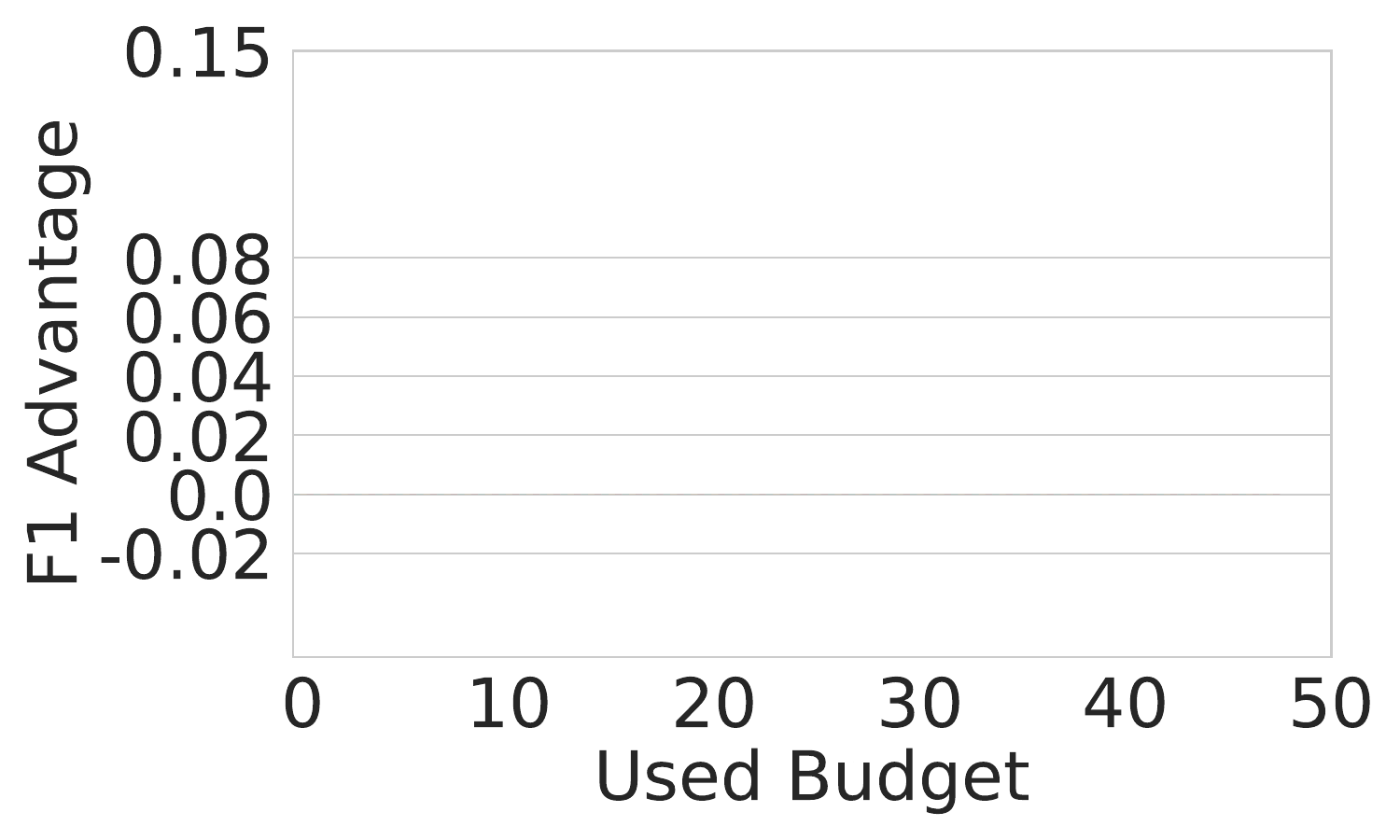}
        \caption{CMC}
    \end{subfigure}
    \begin{subfigure}{0.24\textwidth}
        \includegraphics[width=\linewidth]{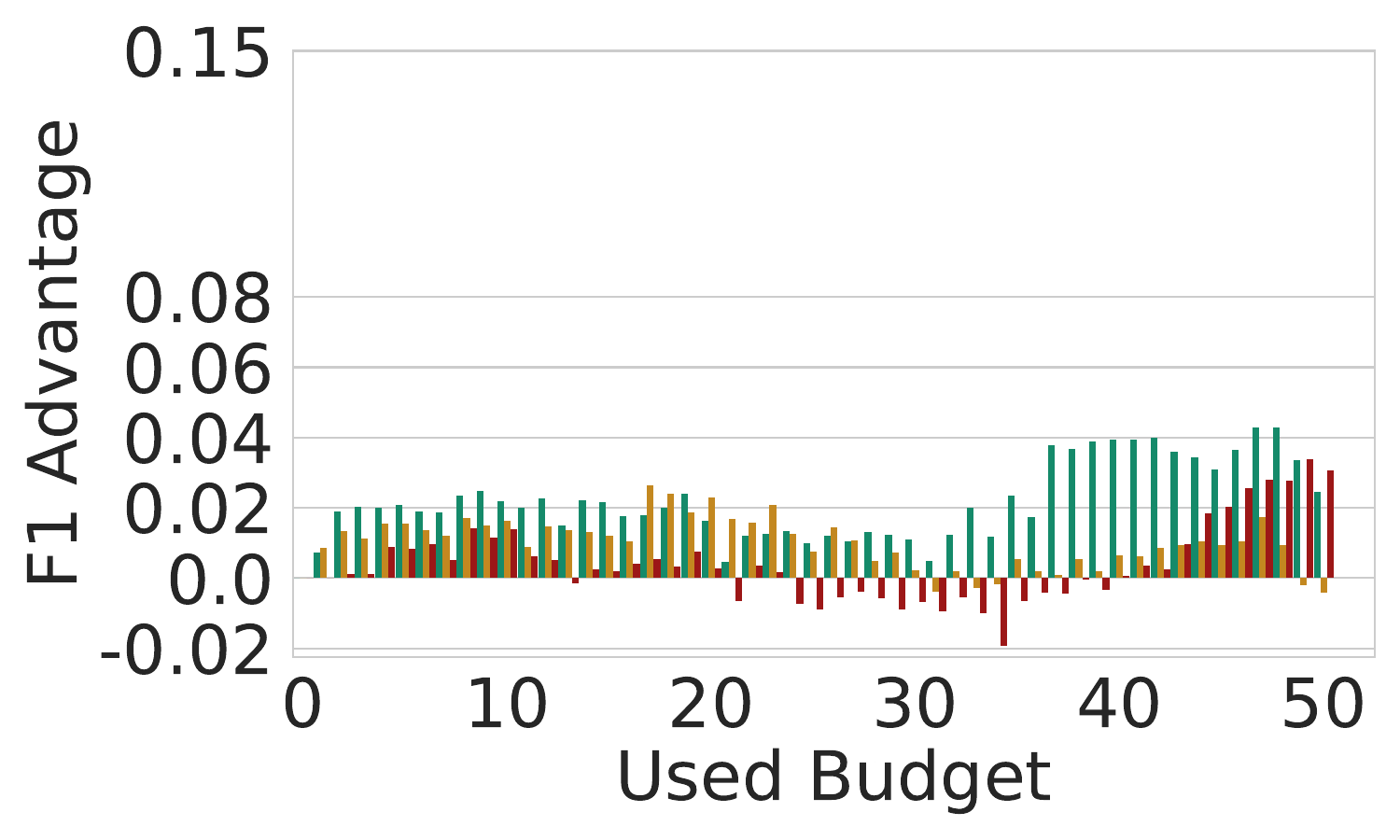}
        \caption{Churn}
    \end{subfigure}
    \begin{subfigure}{0.24\textwidth}
        \includegraphics[width=\linewidth]{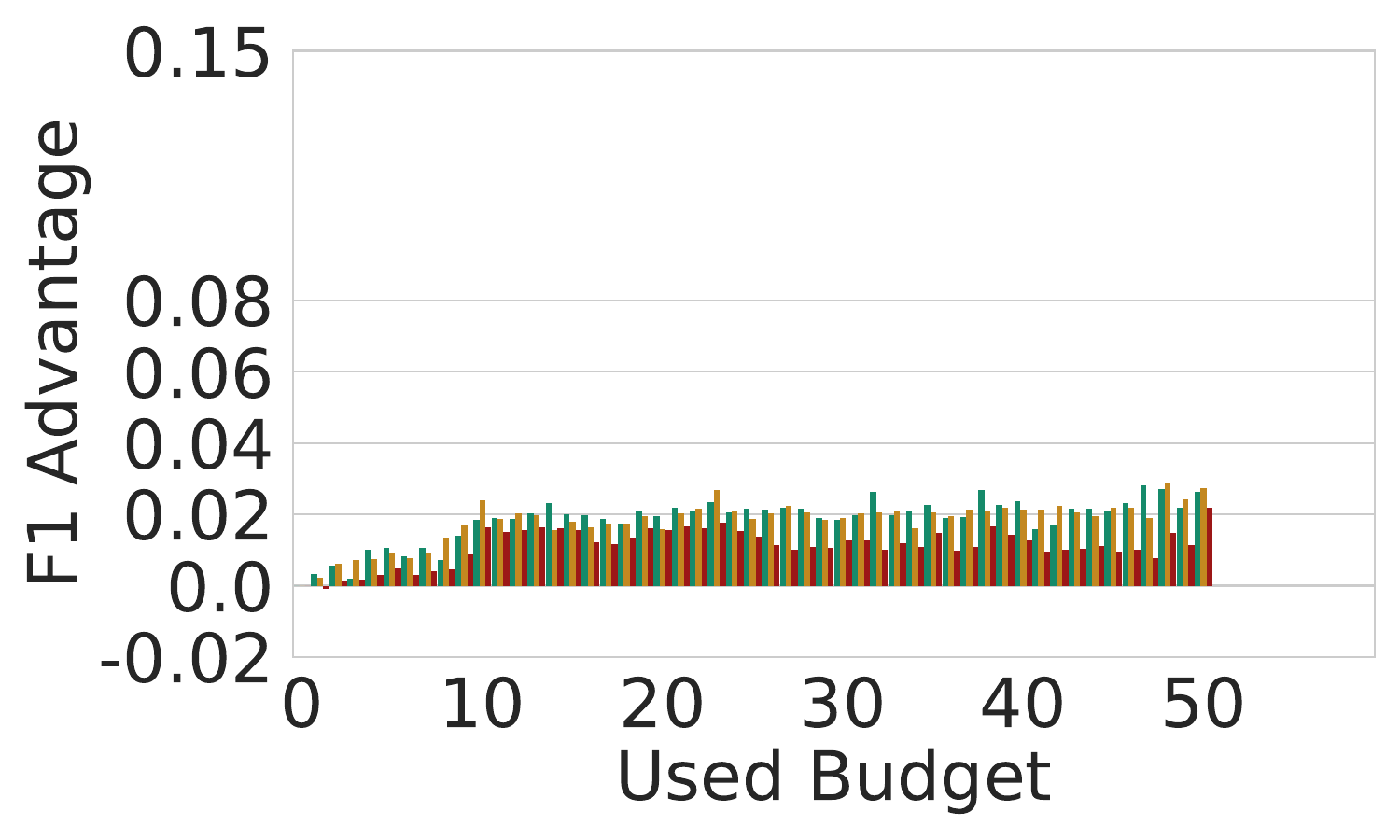}
        \caption{EEG}
    \end{subfigure}
    \begin{subfigure}{0.24\textwidth}
        \includegraphics[width=\linewidth]{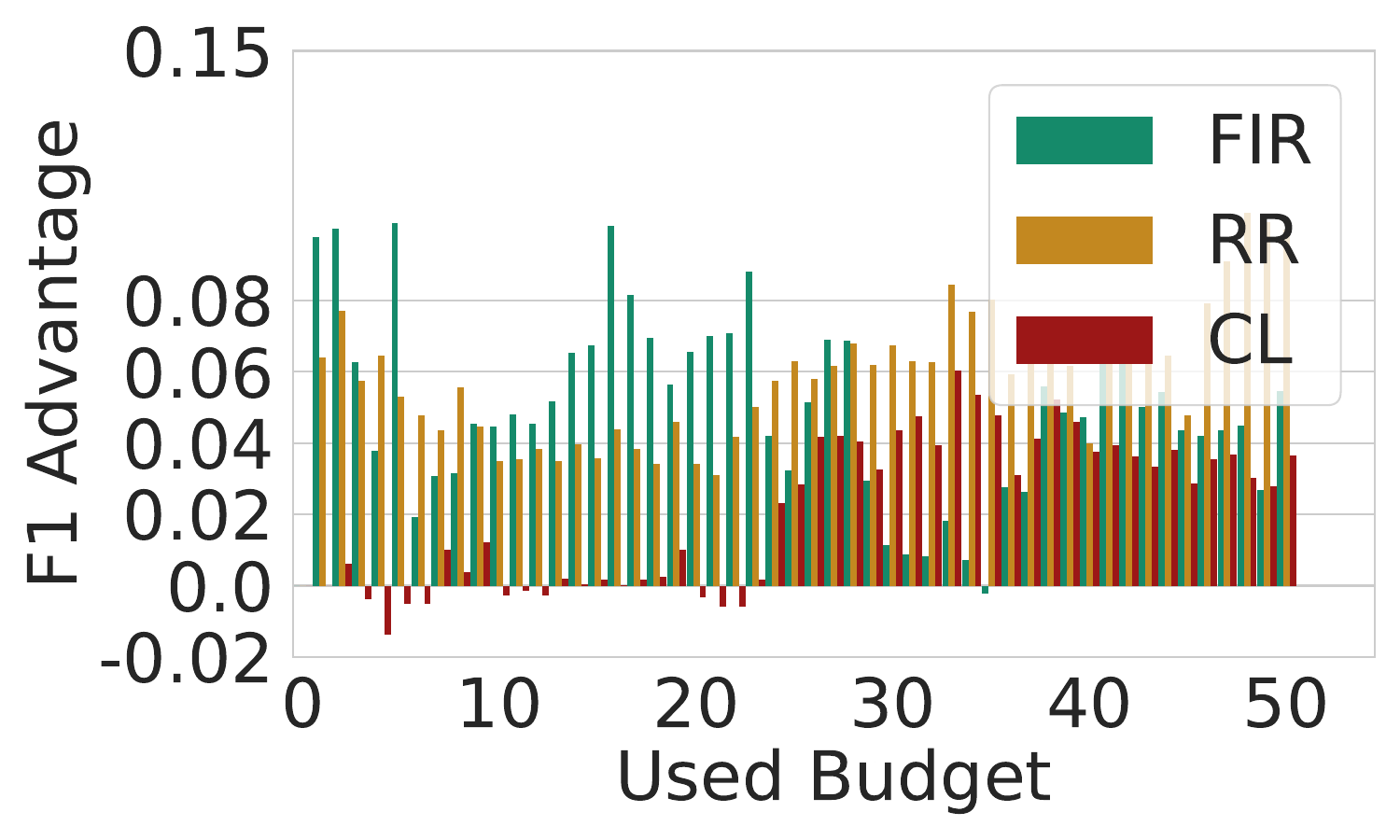}
        \caption{S-Credit}
    \end{subfigure}
    \caption{Comparison of~\systemname with the baselines for GB across multiple error types and cost functions. (For CMC, there is no difference in the F1 scores between the dirty and cleaned states).}
    \label{fig:agg_bl_multi_error_results_gb}
\end{figure*}

\begin{figure*}[h!]
    \centering
    \begin{subfigure}{0.24\textwidth}
        \includegraphics[width=\linewidth]{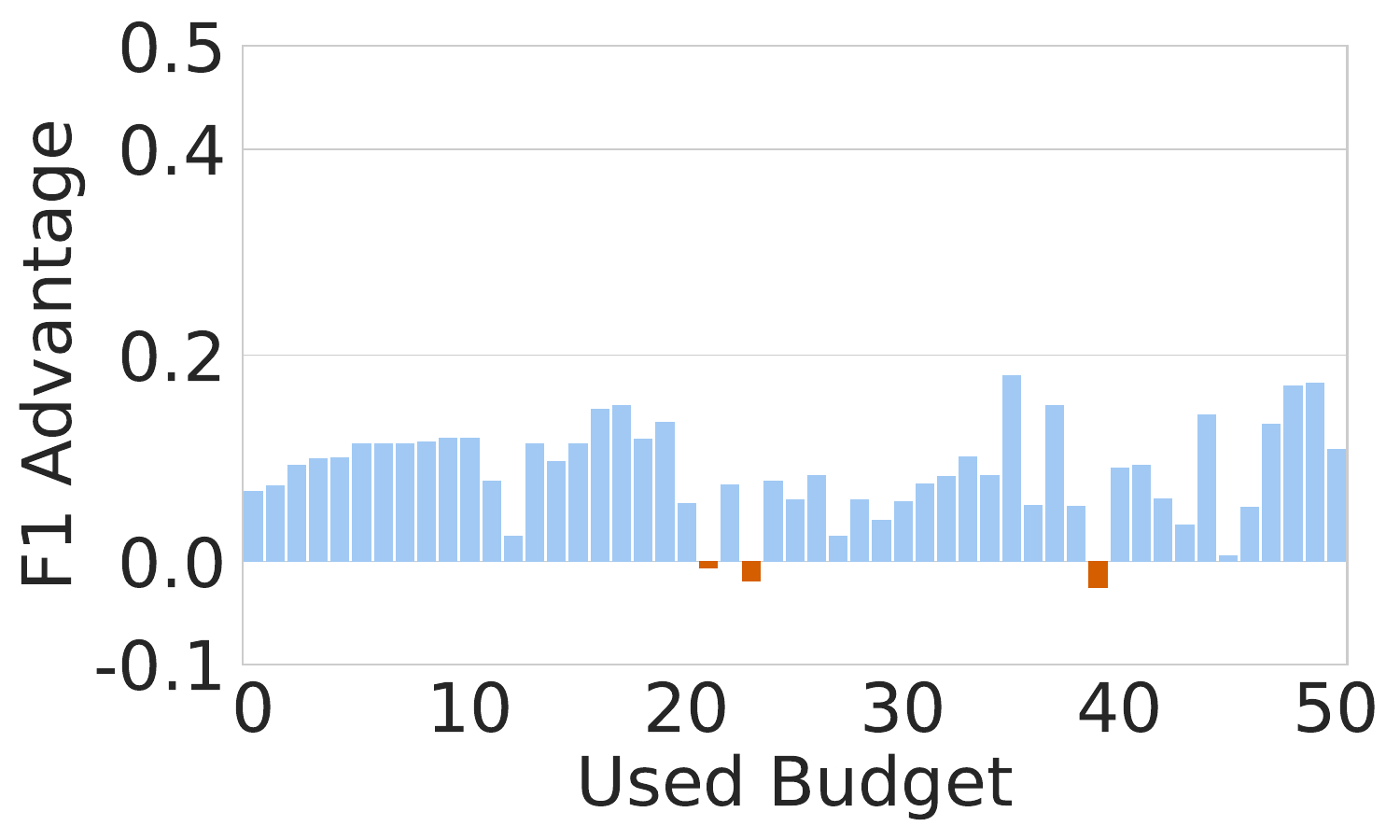}
        \caption{CMC}
    \end{subfigure}
    \begin{subfigure}{0.24\textwidth}
        \includegraphics[width=\linewidth]{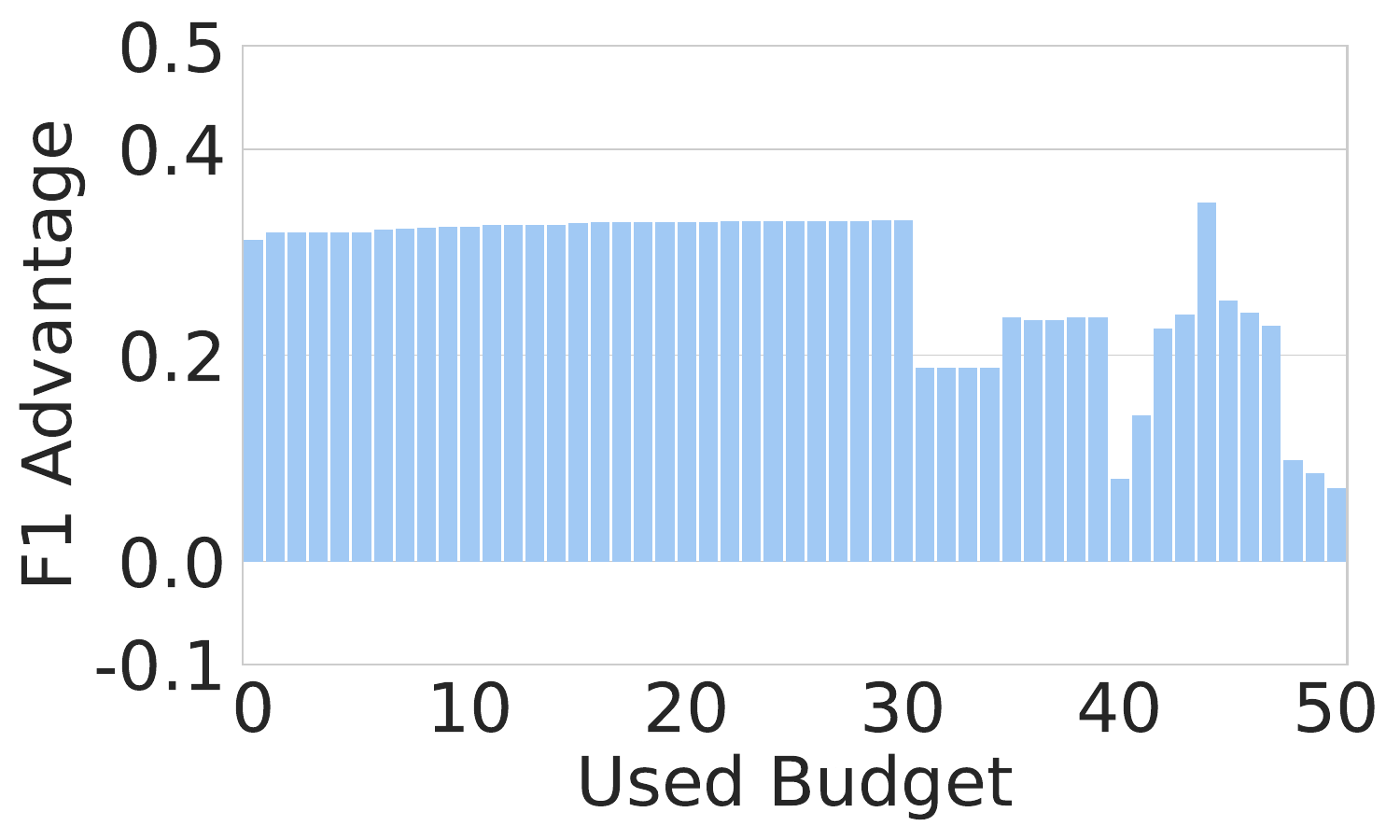}
        \caption{Churn}
    \end{subfigure}
    \begin{subfigure}{0.24\textwidth}
        \includegraphics[width=\linewidth]{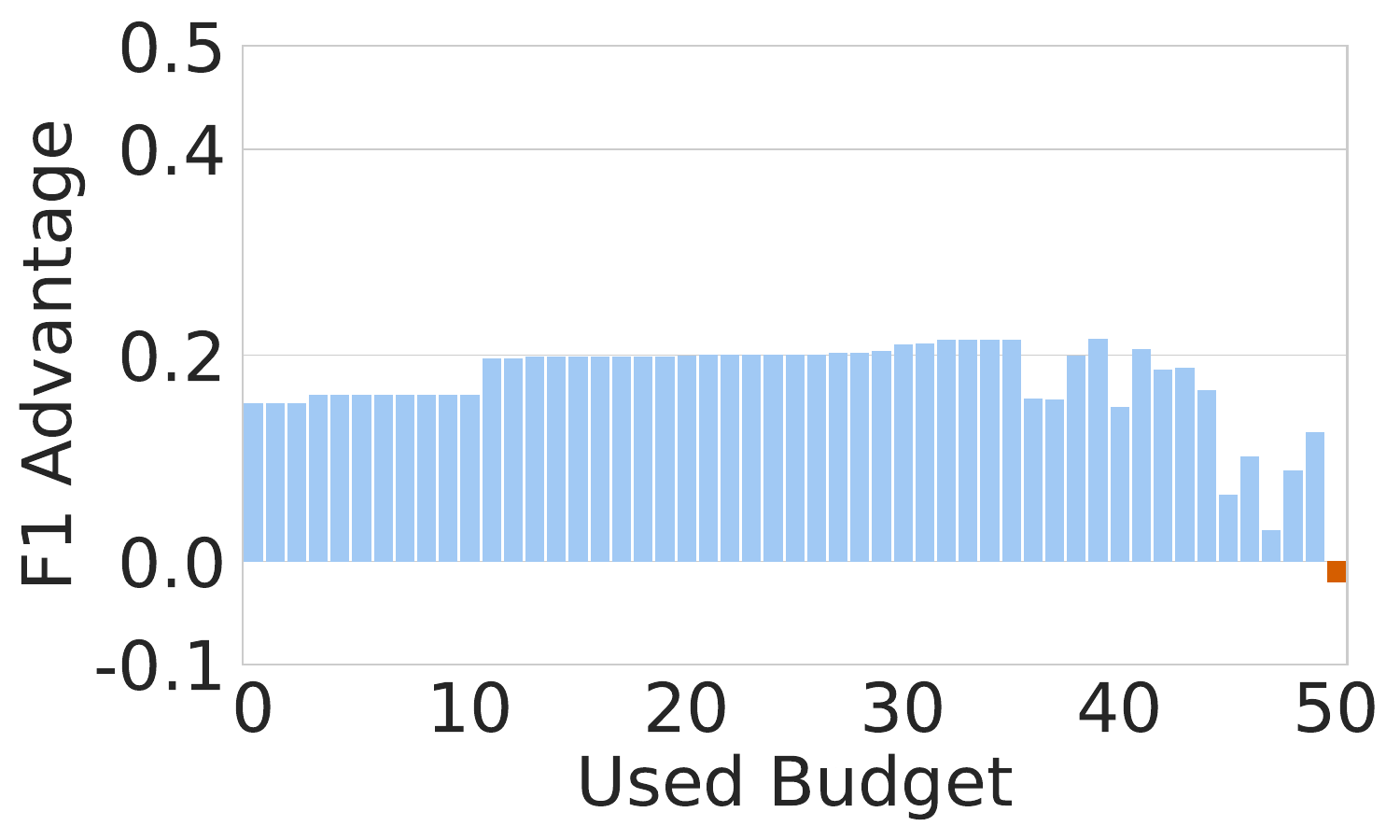}
        \caption{EEG}
    \end{subfigure}
    \begin{subfigure}{0.24\textwidth}
        \includegraphics[width=\linewidth]{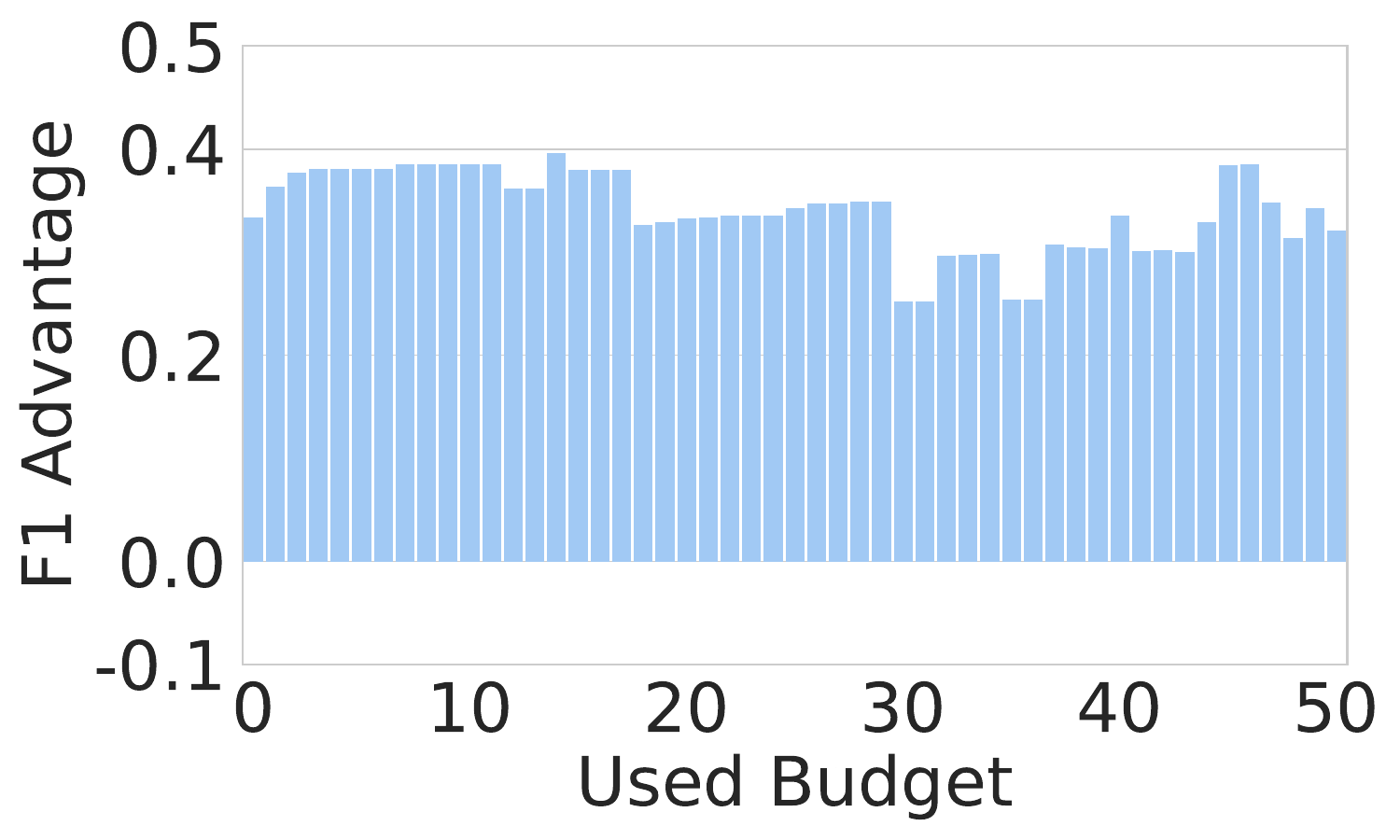}
        \caption{S-Credit}
    \end{subfigure}
    \caption{Comparison of~\systemname with AC for LOG across multiple error types and cost functions.}
    \label{fig:agg_ac_multi_error_results_log}
\end{figure*}

\begin{figure*}[h!]
    \centering
    \begin{subfigure}{0.24\textwidth}
        \includegraphics[width=\linewidth]{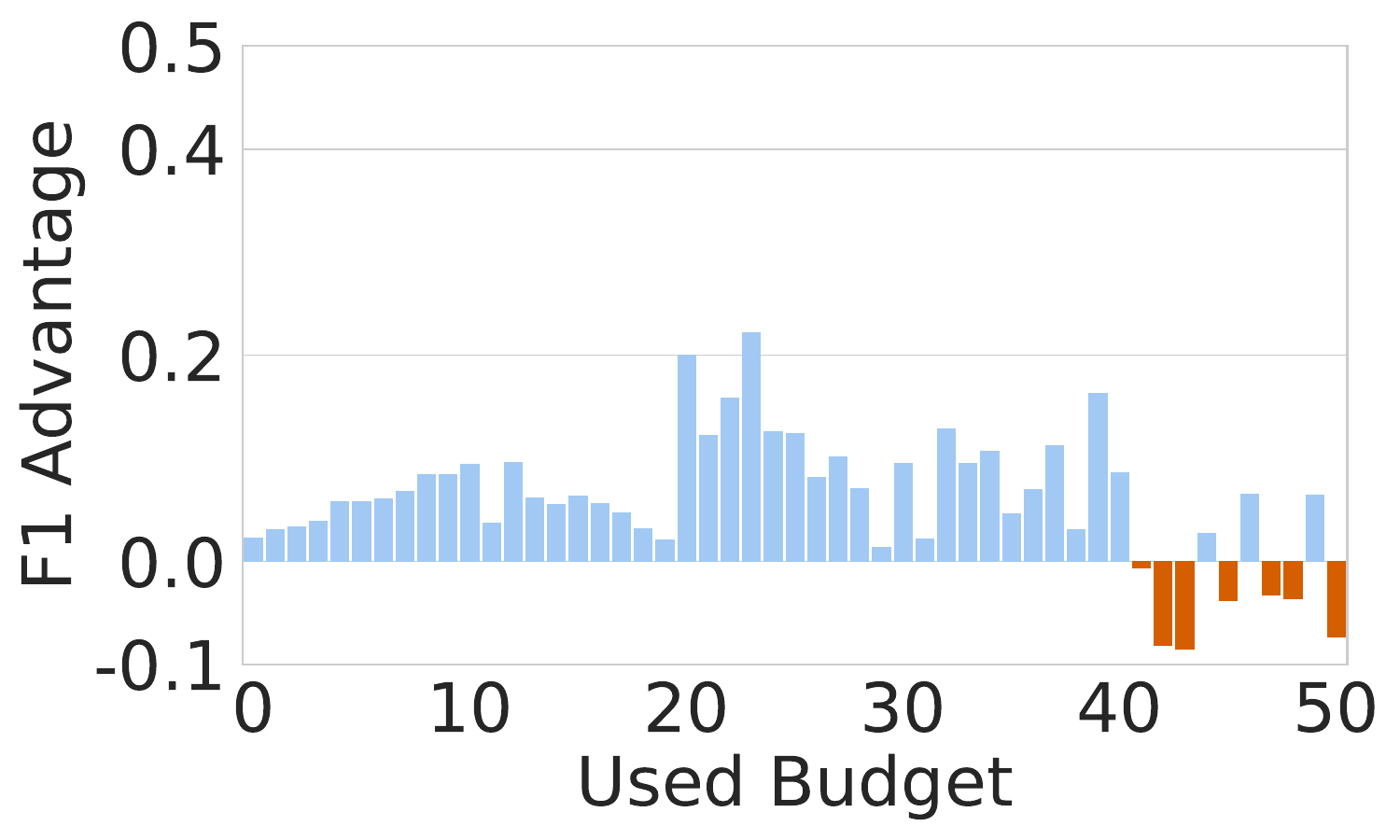}
        \caption{CMC}
    \end{subfigure}
    \begin{subfigure}{0.24\textwidth}
        \includegraphics[width=\linewidth]{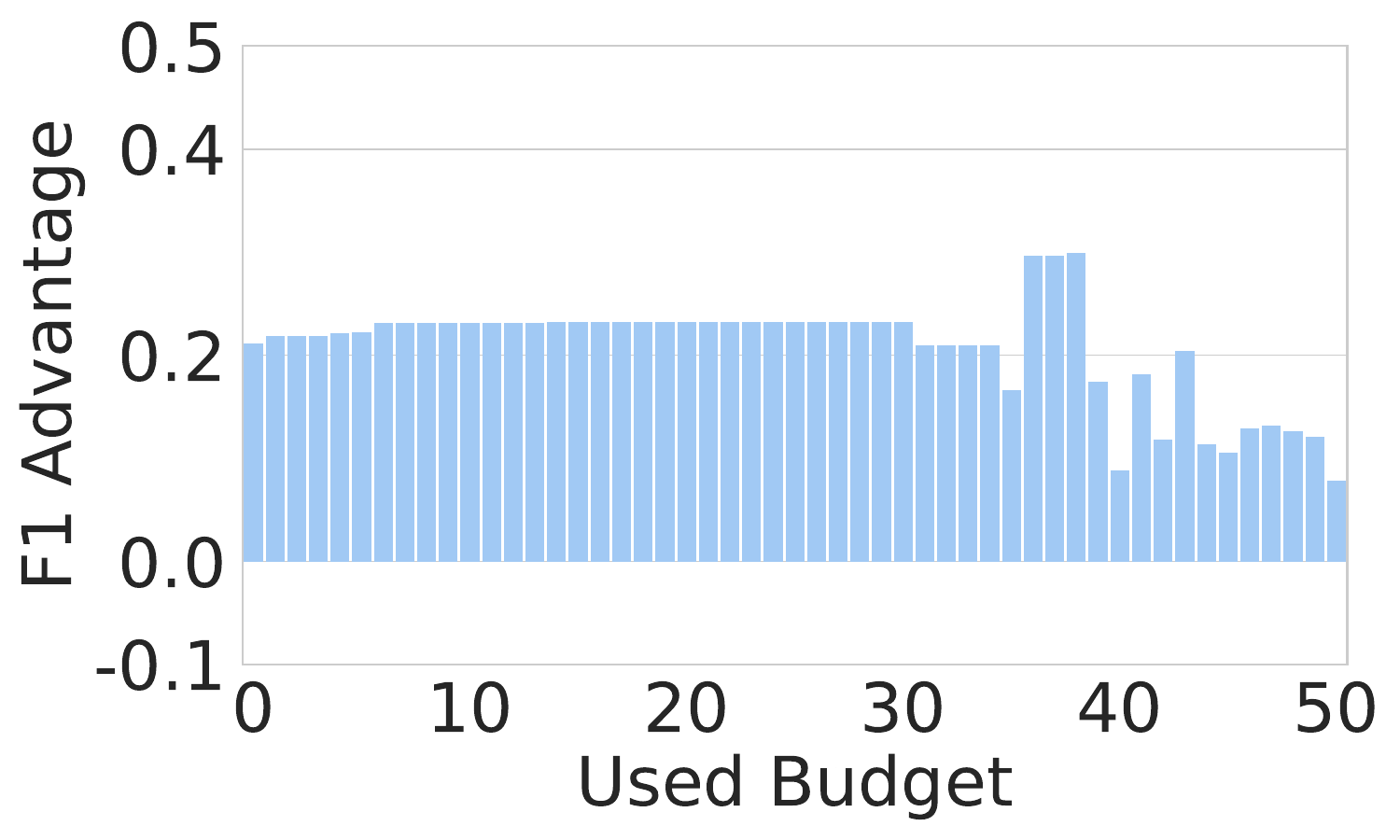}
        \caption{Churn}
    \end{subfigure}
    \begin{subfigure}{0.24\textwidth}
        \includegraphics[width=\linewidth]{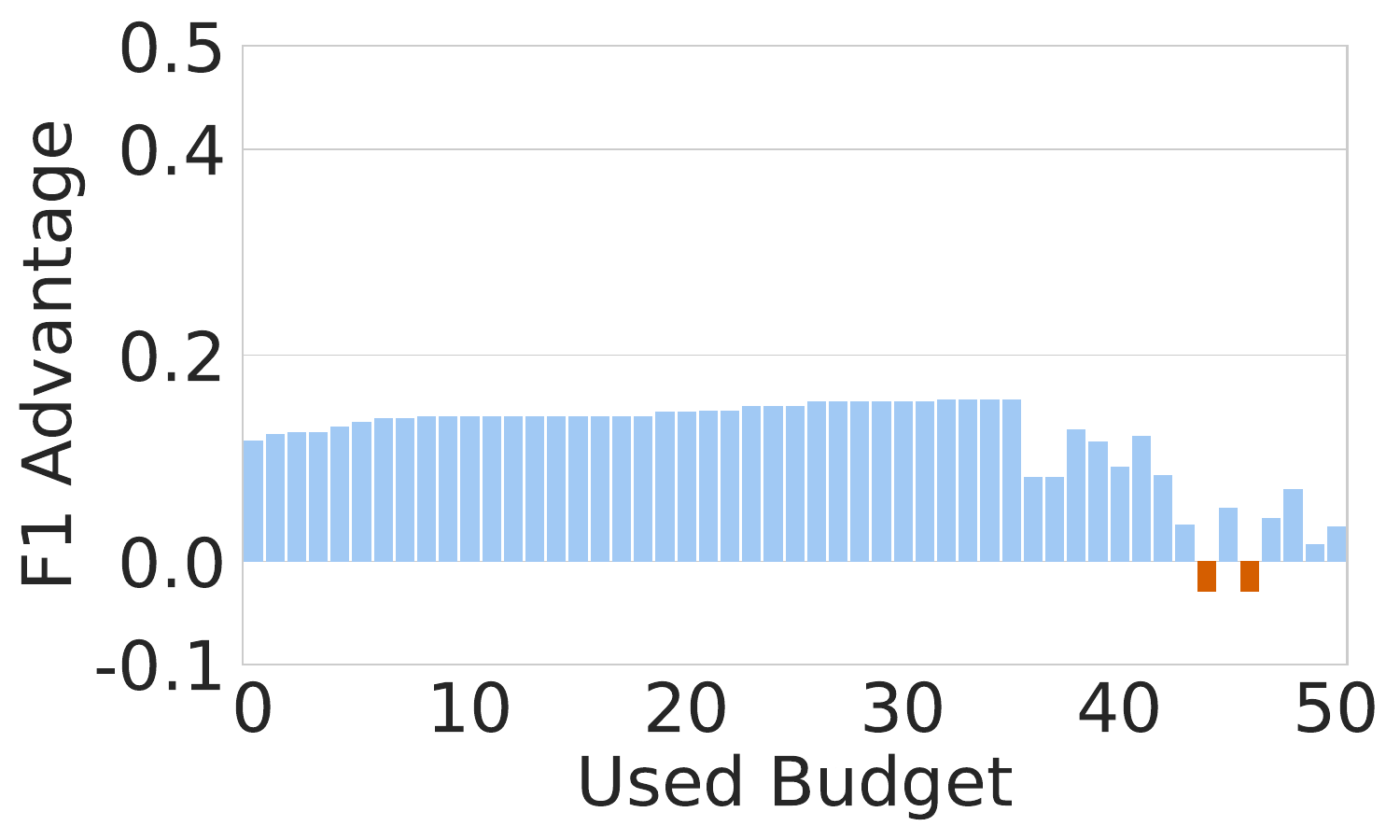}
        \caption{EEG}
    \end{subfigure}
    \begin{subfigure}{0.24\textwidth}
        \includegraphics[width=\linewidth]{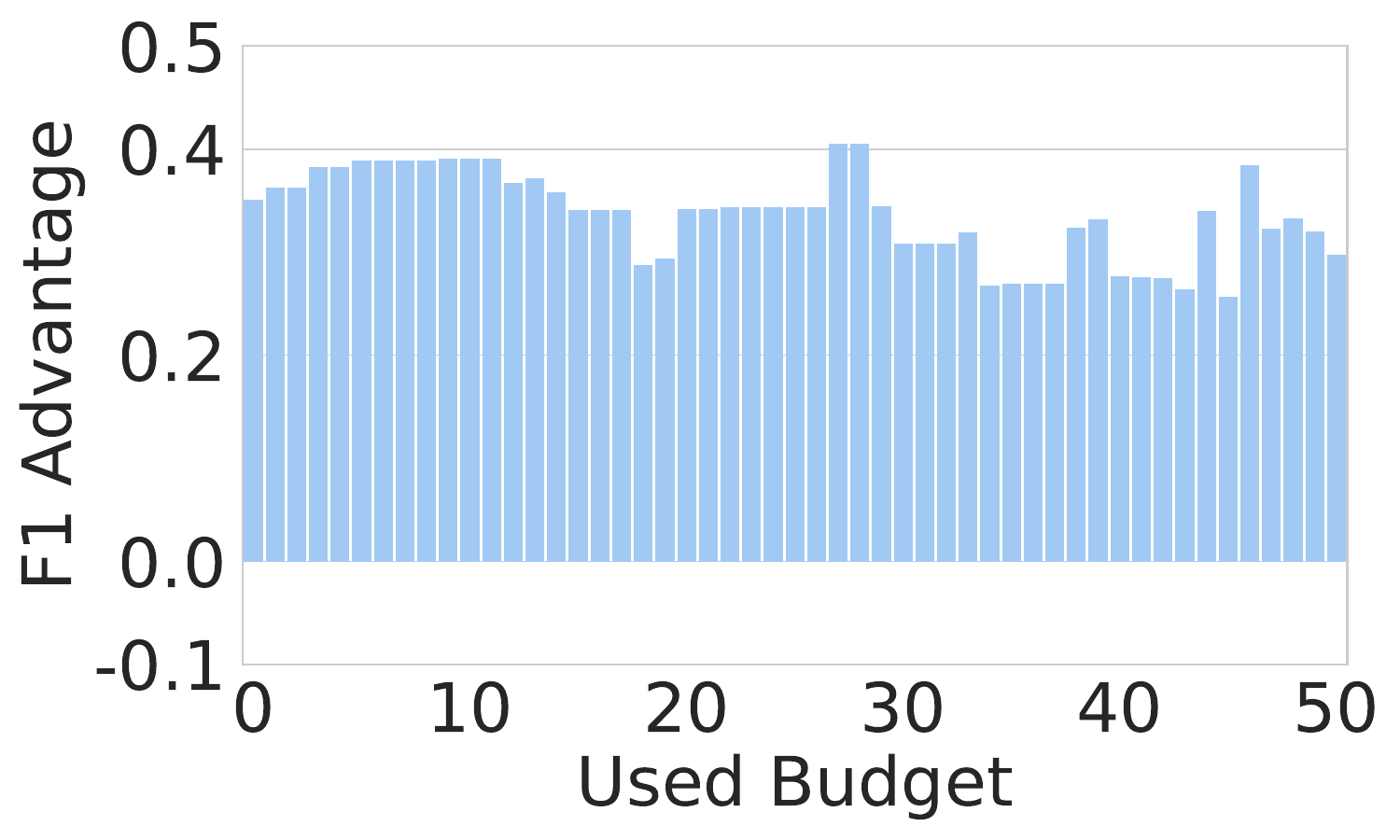}
        \caption{S-Credit}
    \end{subfigure}
    \caption{Comparison of \systemname with AC for AC-SVM across multiple error types and cost functions.}
    \label{fig:agg_ac_multi_error_results_svm}
\end{figure*}

\clearpage
\section{Comparison to FIR, RR, and CL for a single error type}

\begin{figure*}[h!]
    \centering
    \raisebox{1.4\height}{\rotatebox{90}{\textbf{CMC}}}\hspace{0.3em}%
    \begin{subfigure}{0.24\textwidth}
        \includegraphics[width=\linewidth]{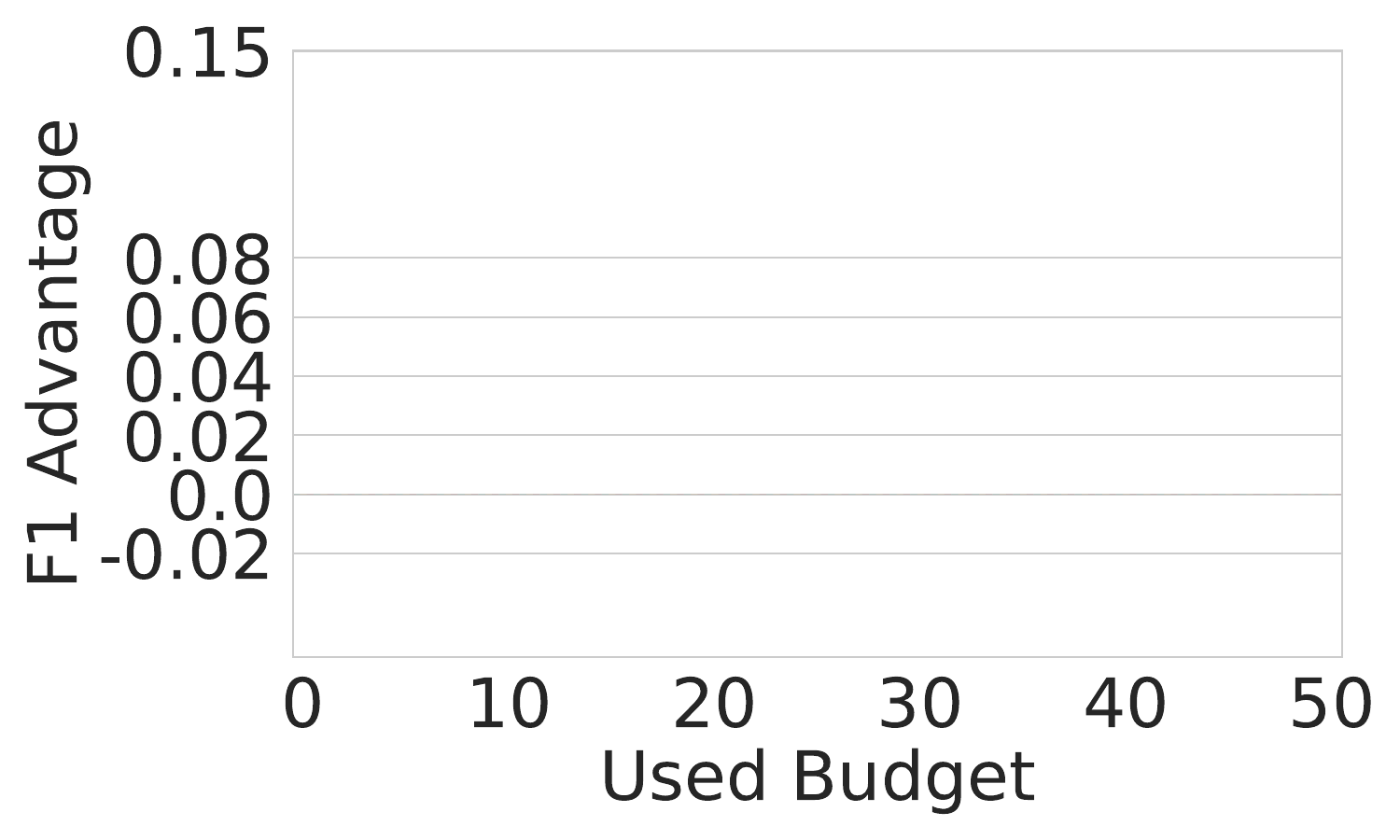}
    \end{subfigure}\hfill
    \begin{subfigure}{0.24\textwidth}
        \includegraphics[width=\linewidth]{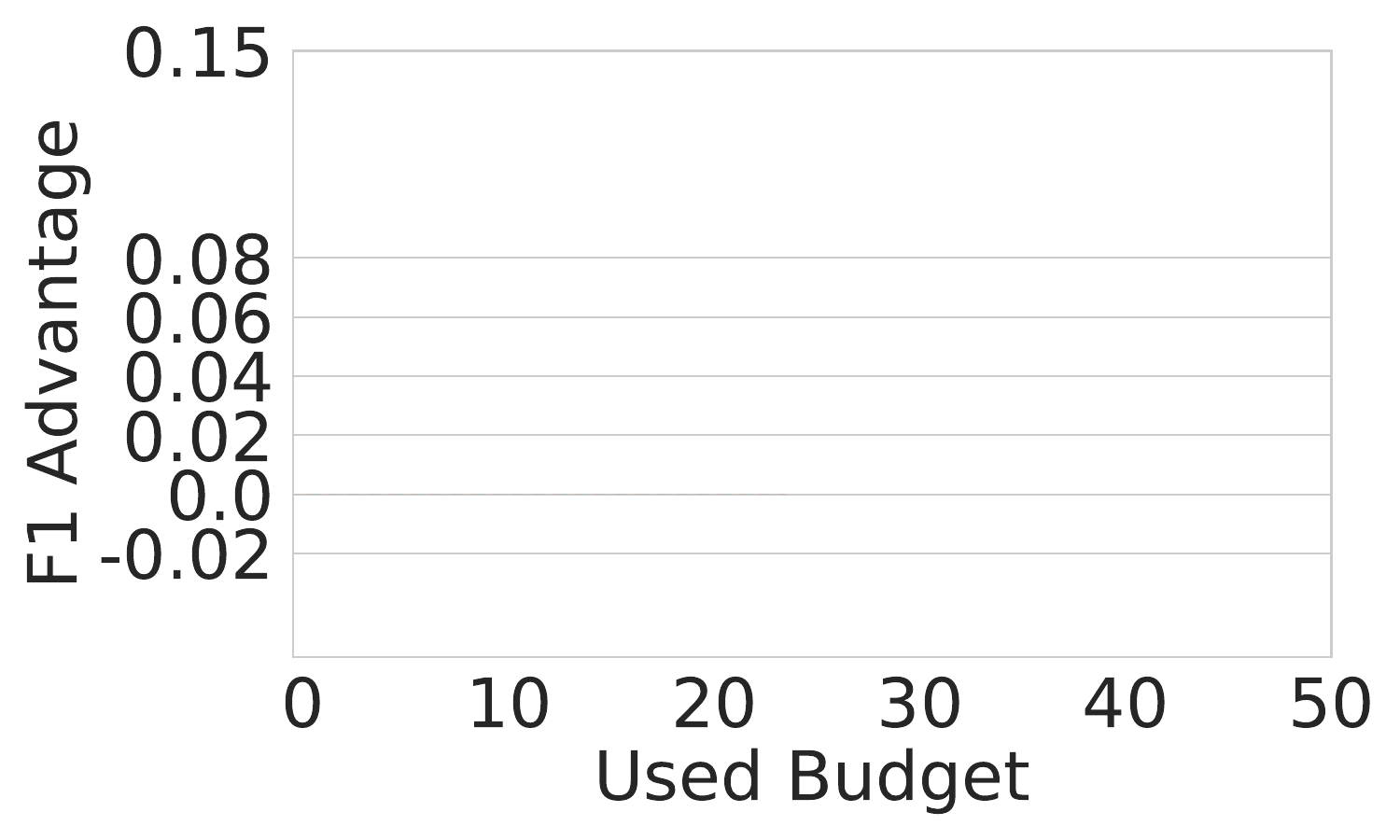}
    \end{subfigure}\hfill
    \begin{subfigure}{0.24\textwidth}
        \includegraphics[width=\linewidth]{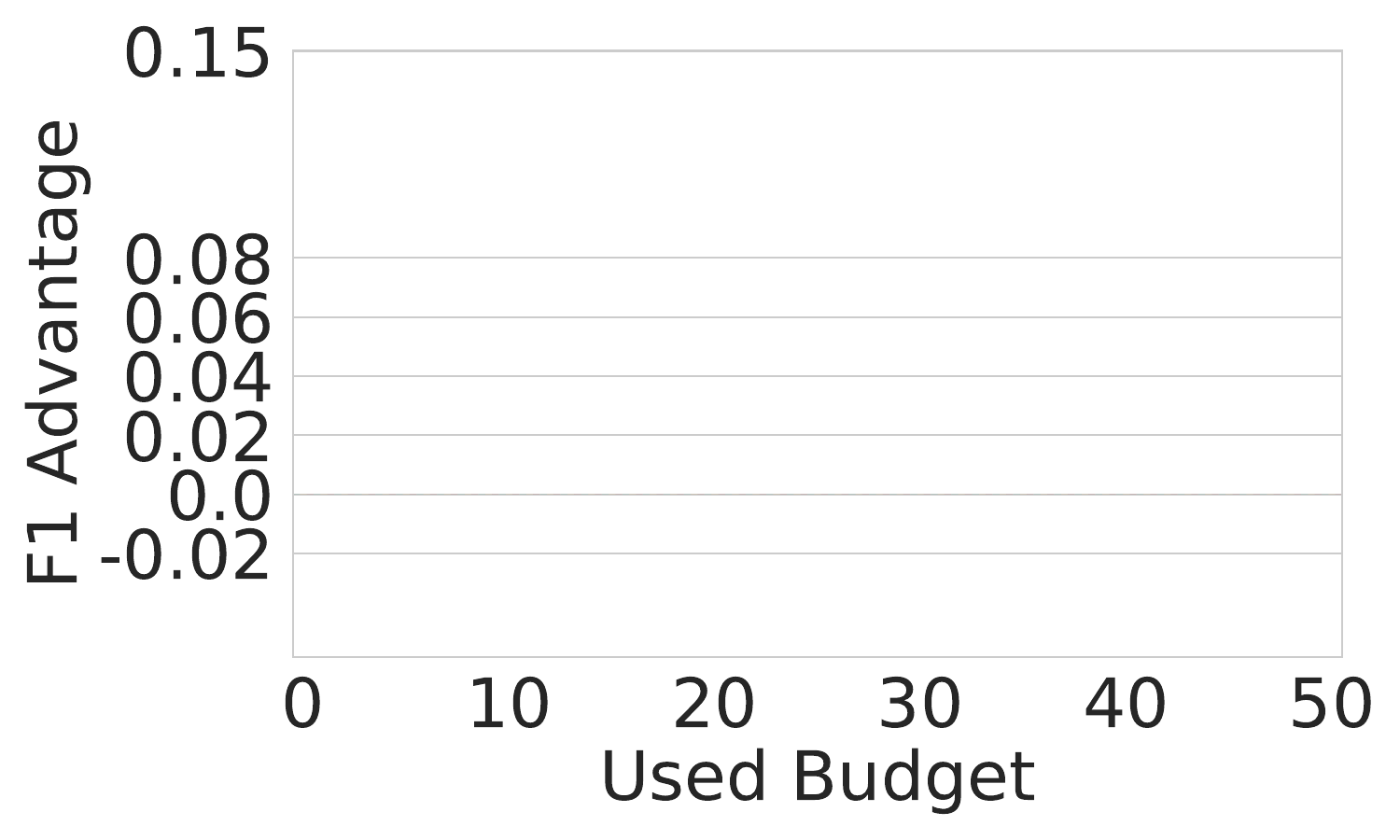}
    \end{subfigure}\hfill
    \begin{subfigure}{0.24\textwidth}
        \includegraphics[width=\linewidth]{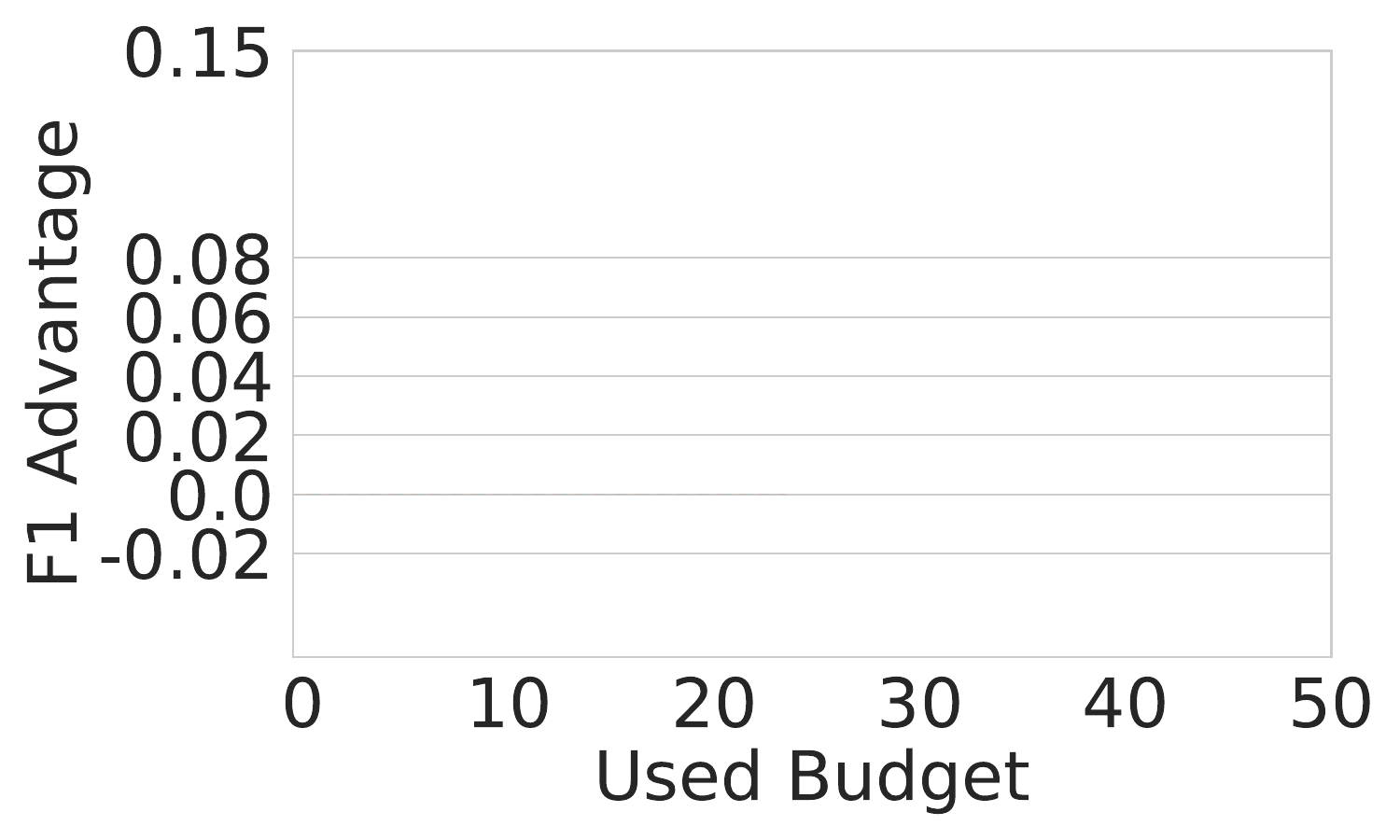}
    \end{subfigure}
    
    \vspace{-0.1em}

        \raisebox{1.2\height}{\rotatebox{90}{\textbf{Churn}}}\hspace{0.3em}%
    \begin{subfigure}{0.24\textwidth}
        \includegraphics[width=\linewidth]{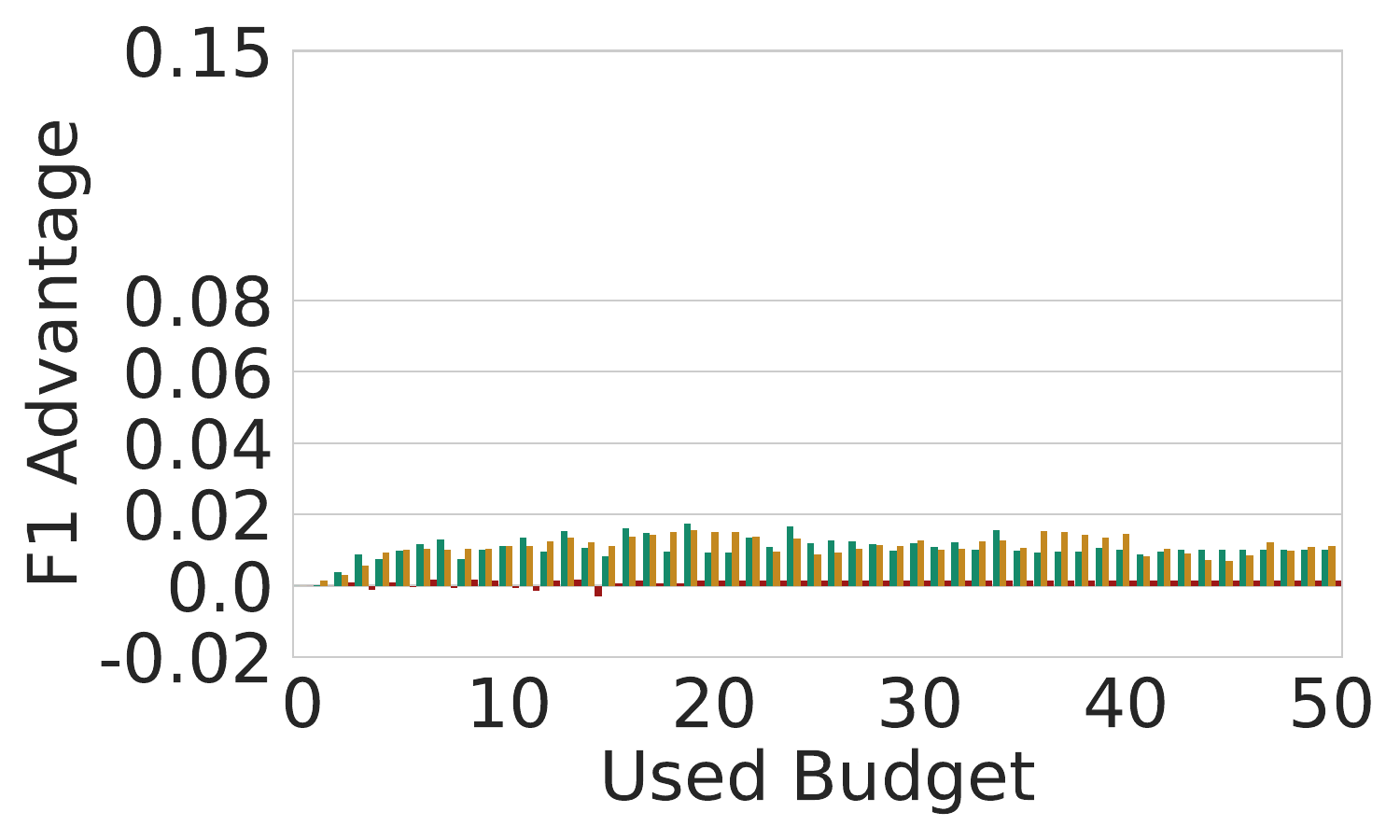}
    \end{subfigure}\hfill
    \begin{subfigure}{0.24\textwidth}
        \includegraphics[width=\linewidth]{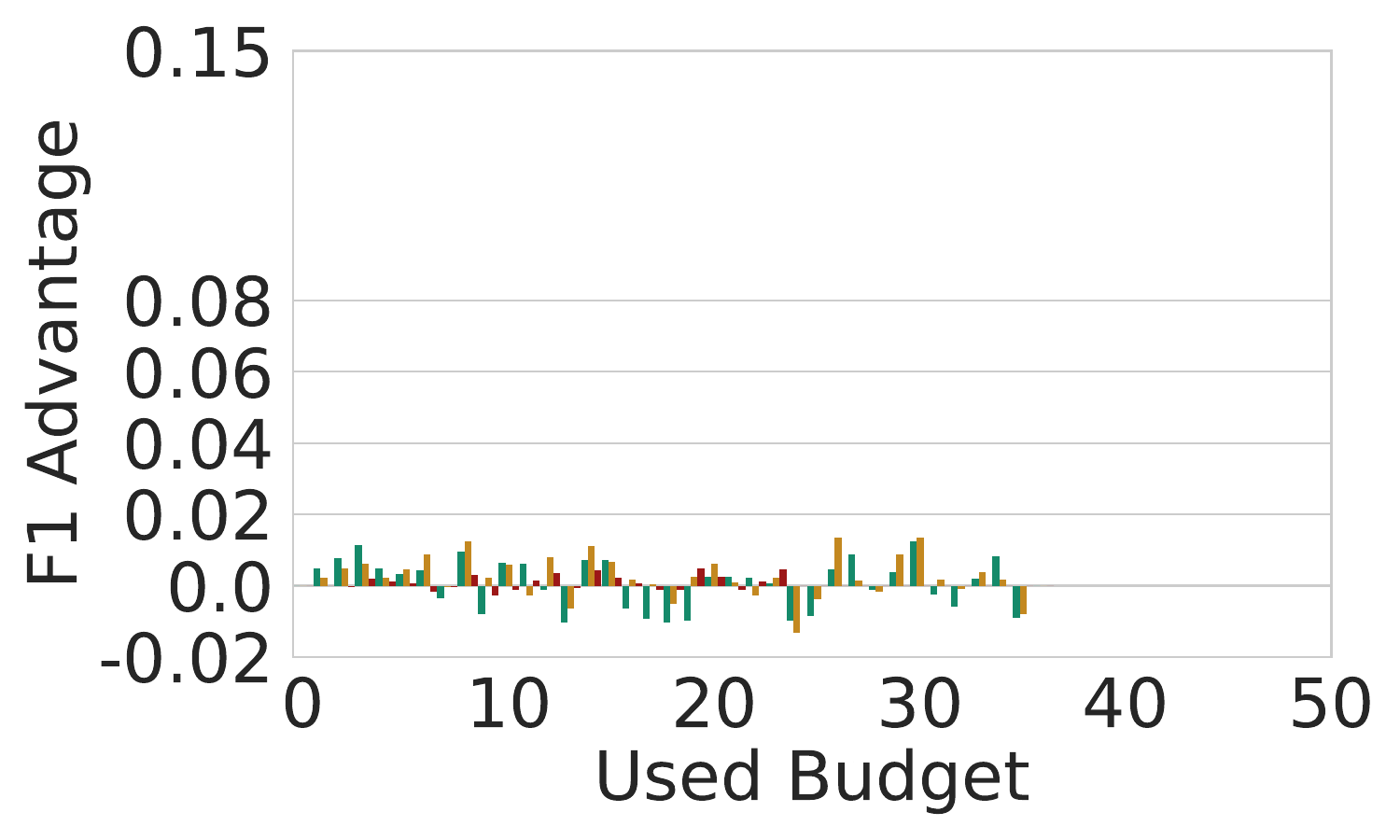}
    \end{subfigure}\hfill
    \begin{subfigure}{0.24\textwidth}
        \includegraphics[width=\linewidth]{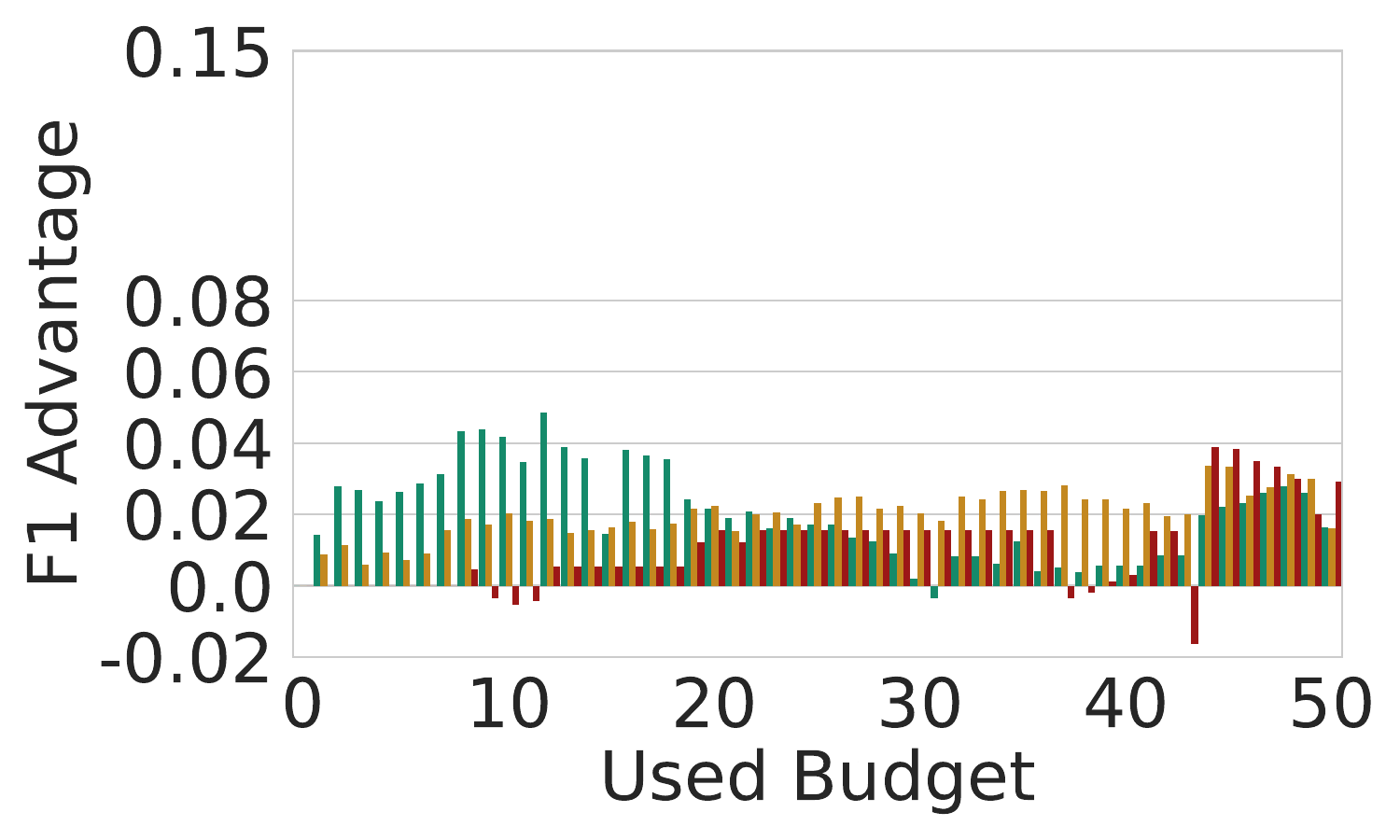}
    \end{subfigure}\hfill
    \begin{subfigure}{0.24\textwidth}
        \includegraphics[width=\linewidth]{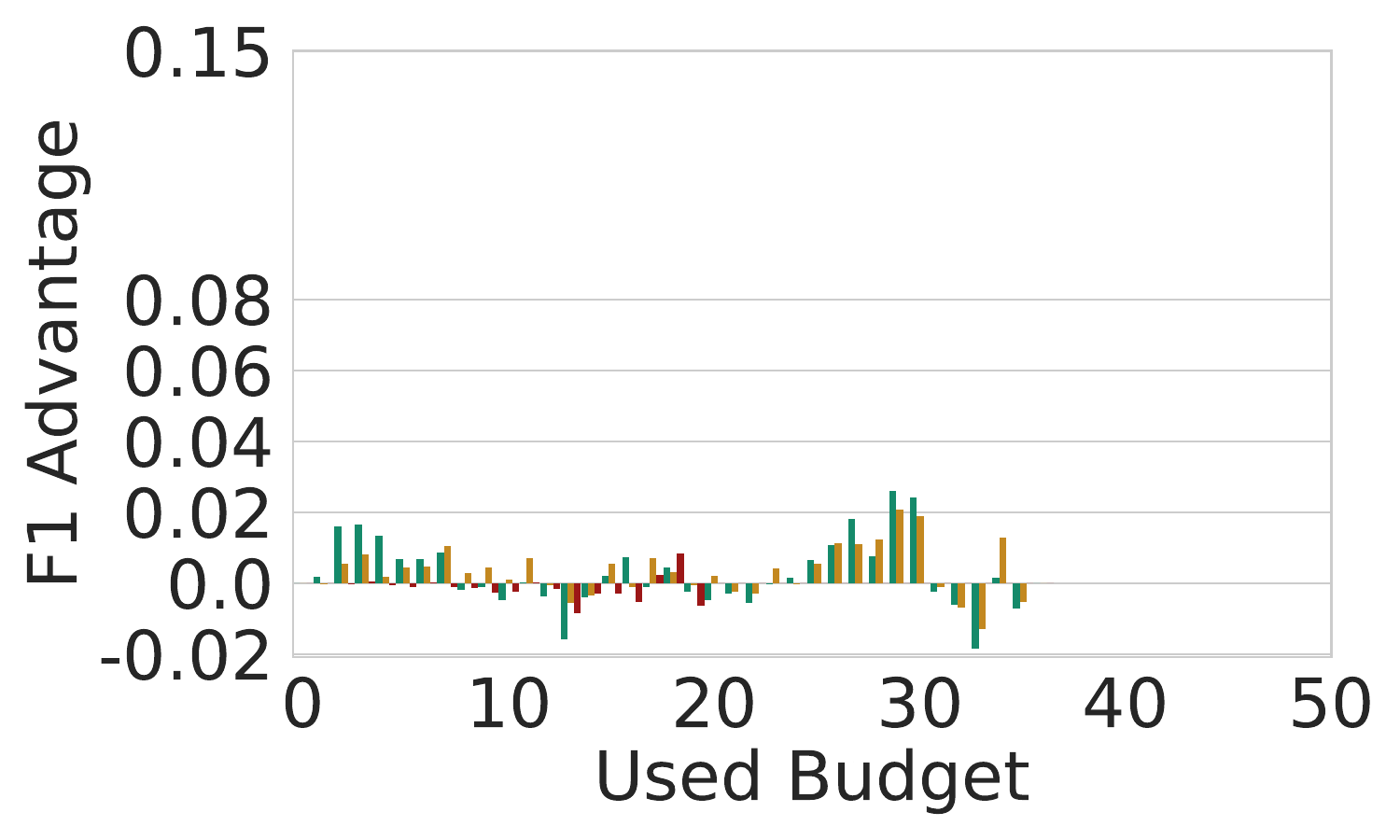}
    \end{subfigure}
    
    \vspace{-0.1em}

    \raisebox{2.\height}{\rotatebox{90}{\textbf{EEG}}}\hspace{0.3em}%
    \begin{subfigure}{0.24\textwidth}
        \centering\raisebox{3.85\height}{\parbox{0.75\linewidth}{\texttt{EEG only contains numerical features.}}}
    \end{subfigure}\hfill
    \begin{subfigure}{0.24\textwidth}
        \includegraphics[width=\linewidth]{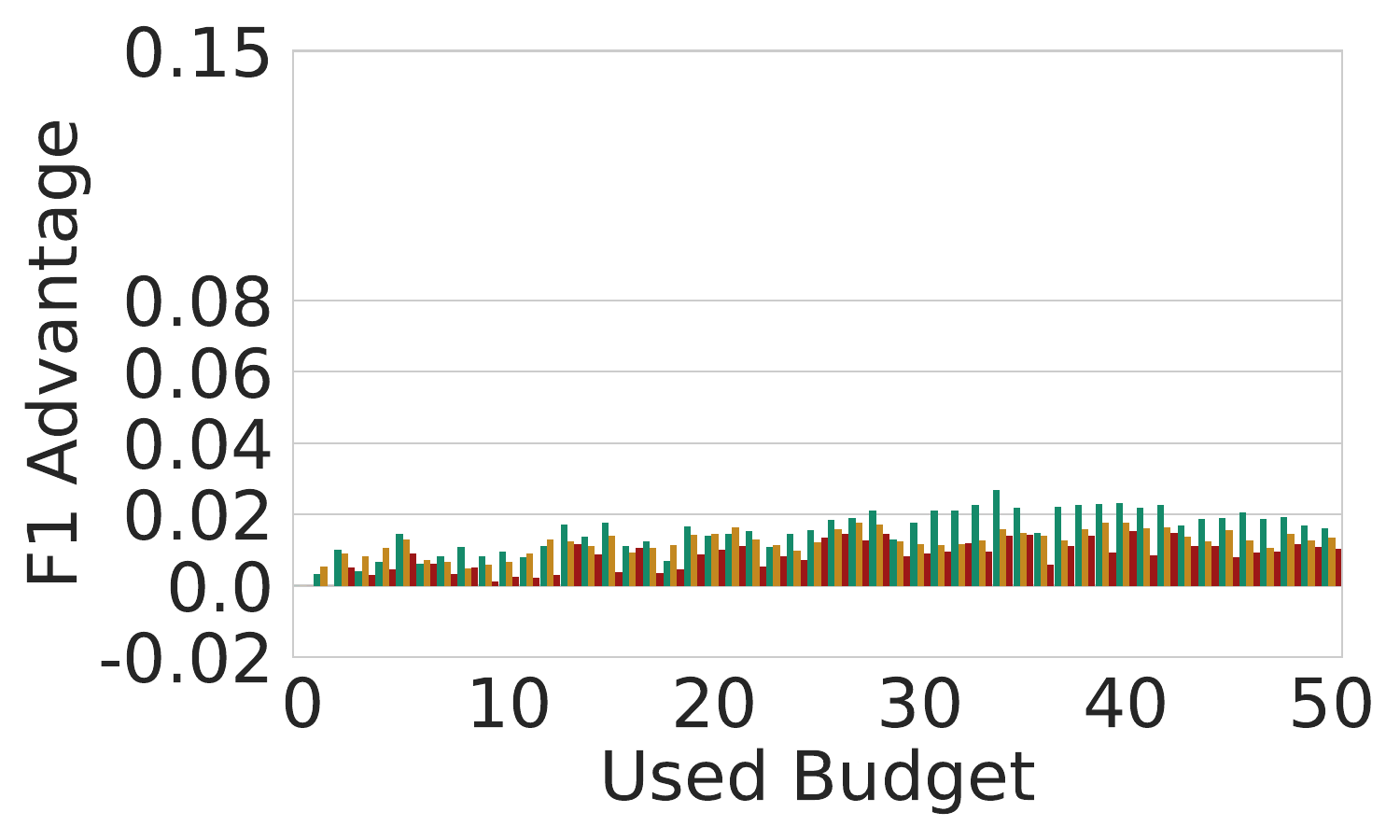}
    \end{subfigure}\hfill
    \begin{subfigure}{0.24\textwidth}
        \includegraphics[width=\linewidth]{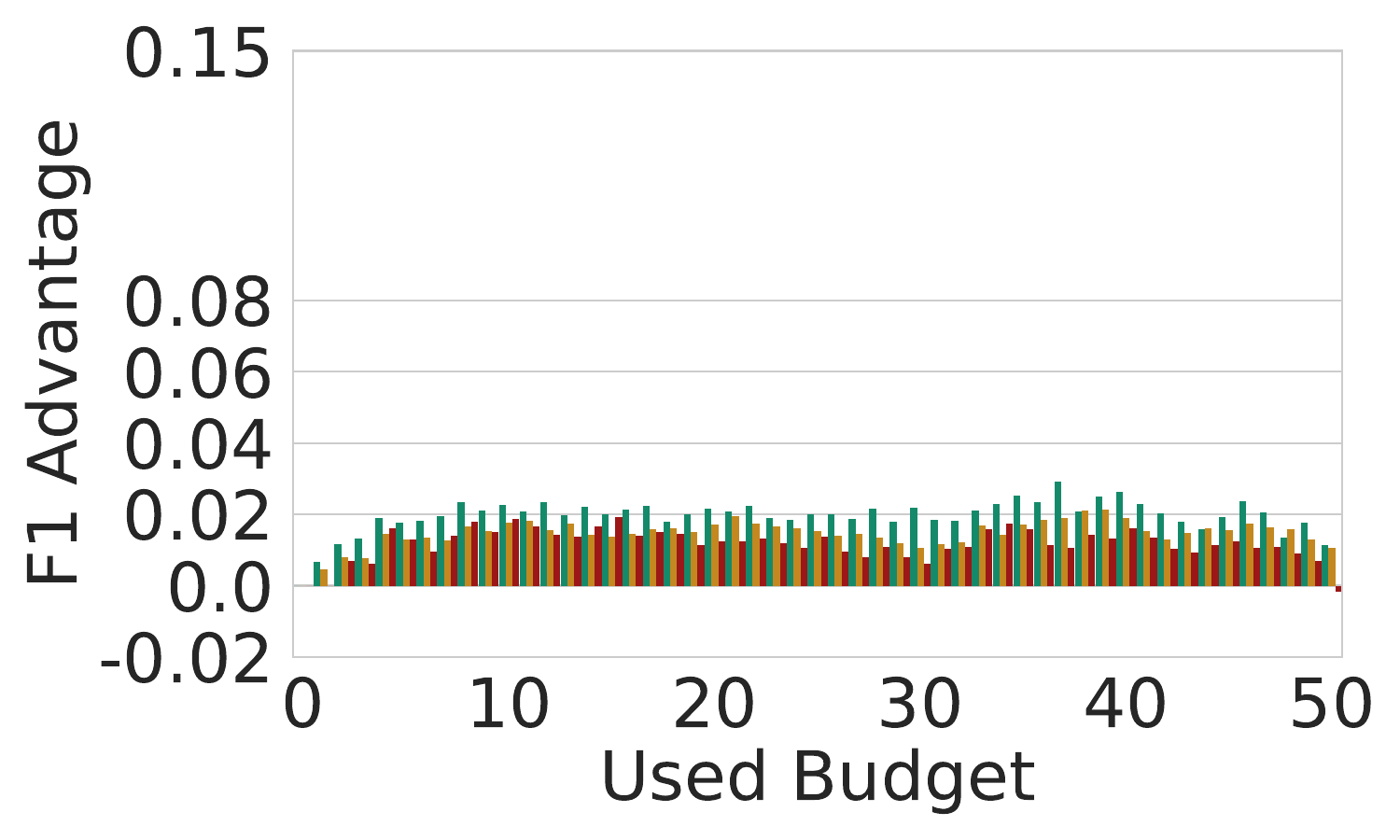}
    \end{subfigure}\hfill
    \begin{subfigure}{0.24\textwidth}
        \includegraphics[width=\linewidth]{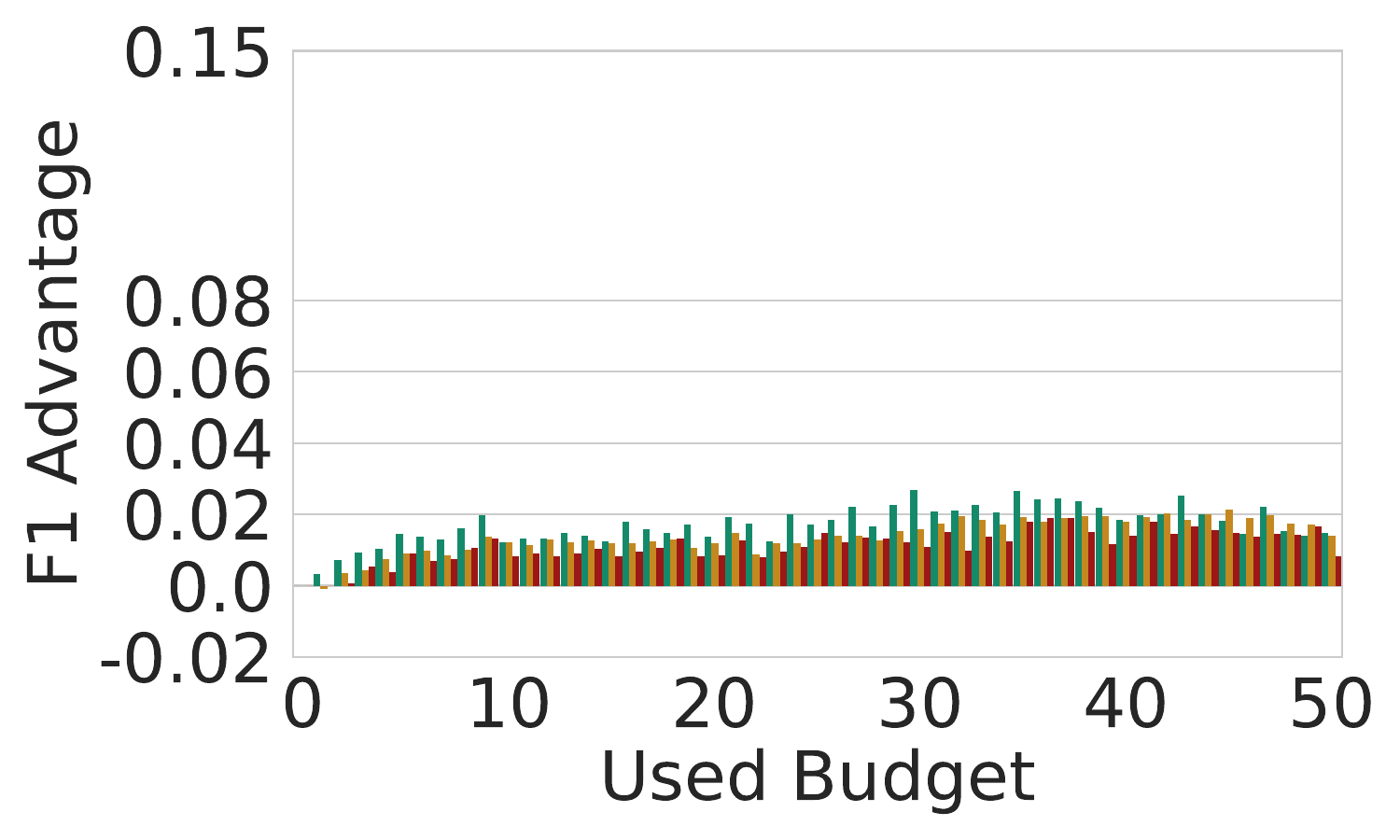}
    \end{subfigure}
    
    \vspace{-0.1em}
    
    \raisebox{1.2\height}{\rotatebox{90}{\textbf{S-Credit}}}\hspace{0.3em}%
    \begin{subfigure}{0.24\textwidth}
        \includegraphics[width=\linewidth]{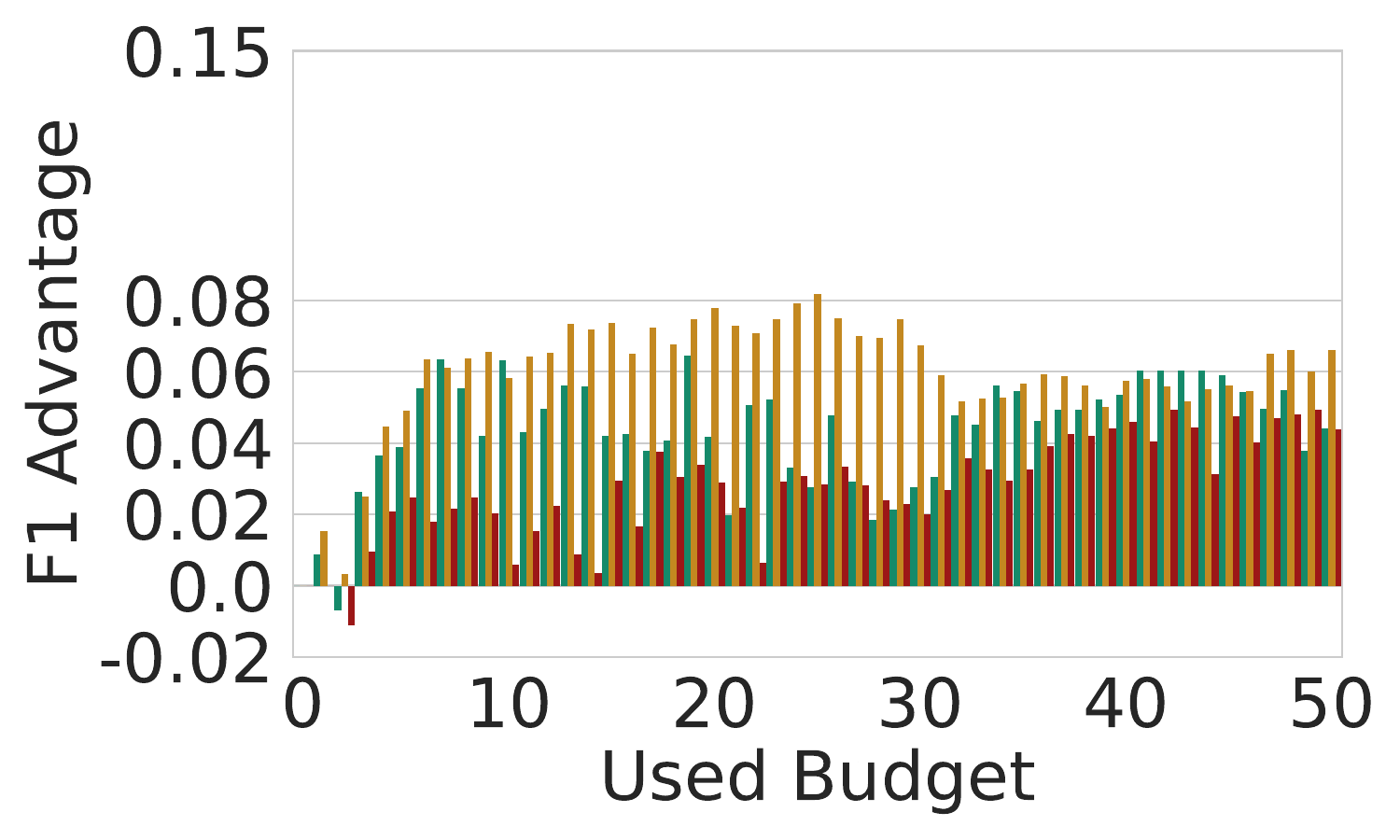}
        \caption{Categorical Shift}
    \end{subfigure}\hfill
    \begin{subfigure}{0.24\textwidth}
        \includegraphics[width=\linewidth]{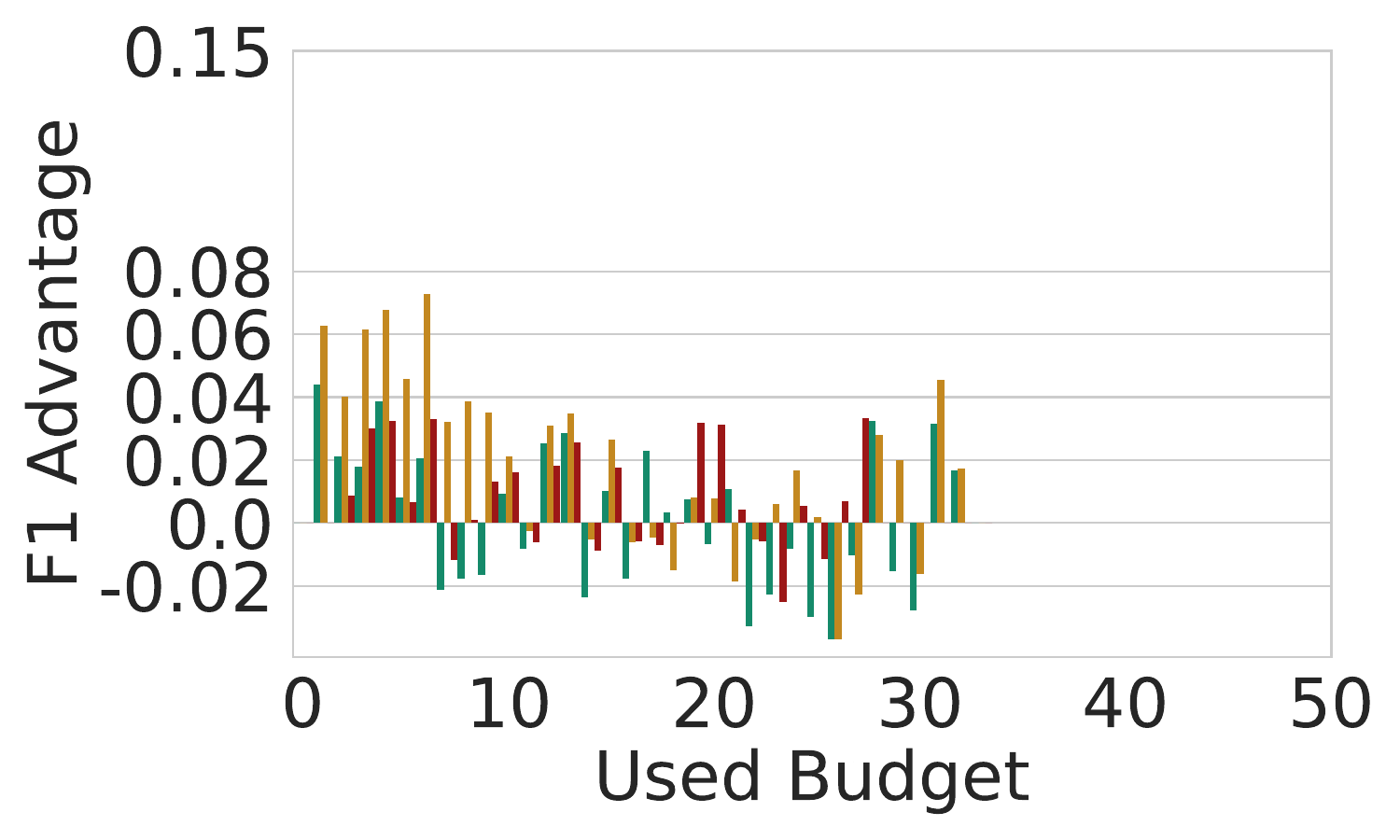}
        \caption{Gaussian Noise}
    \end{subfigure}\hfill
    \begin{subfigure}{0.24\textwidth}
        \includegraphics[width=\linewidth]{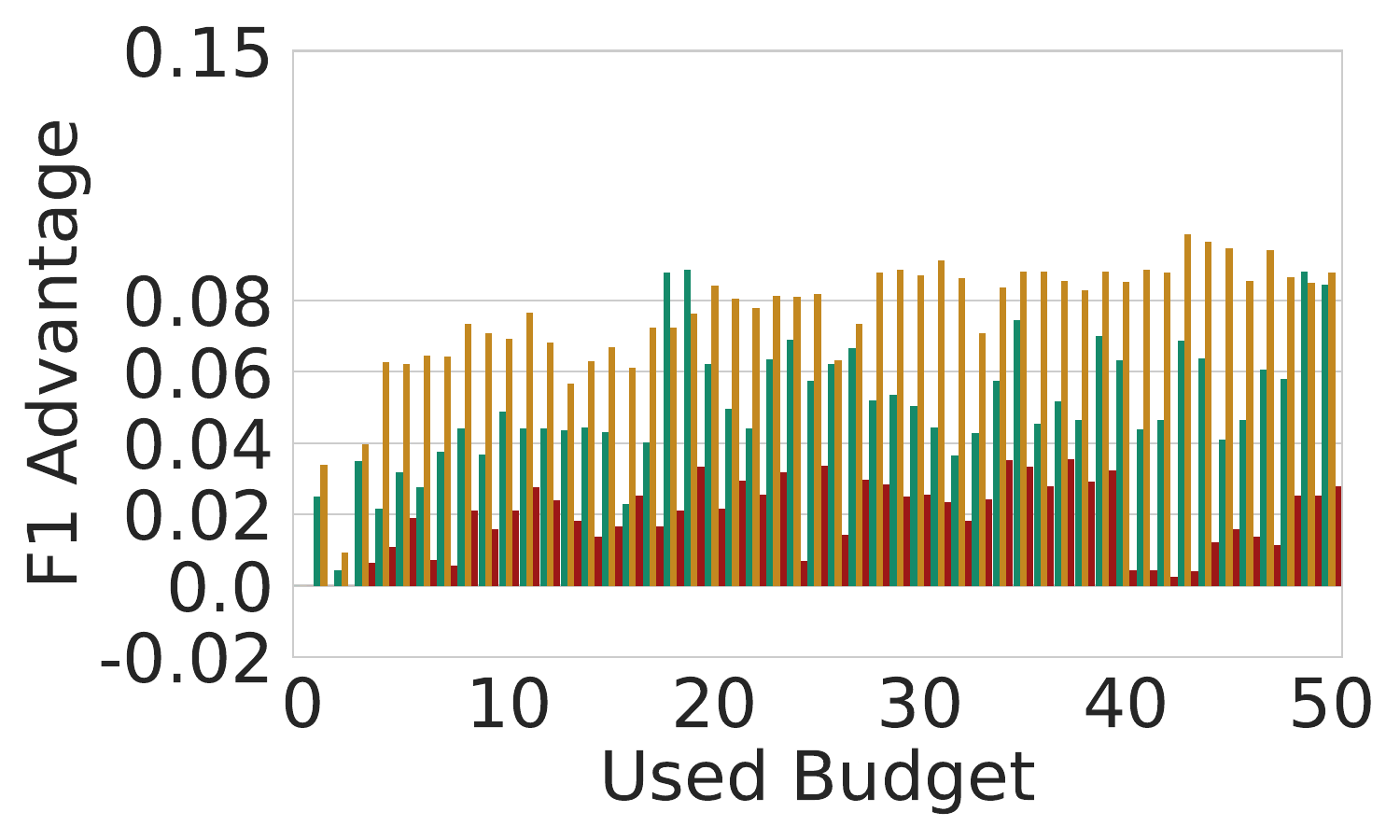}
        \caption{Missing Values}
    \end{subfigure}\hfill
    \begin{subfigure}{0.24\textwidth}
        \includegraphics[width=\linewidth]{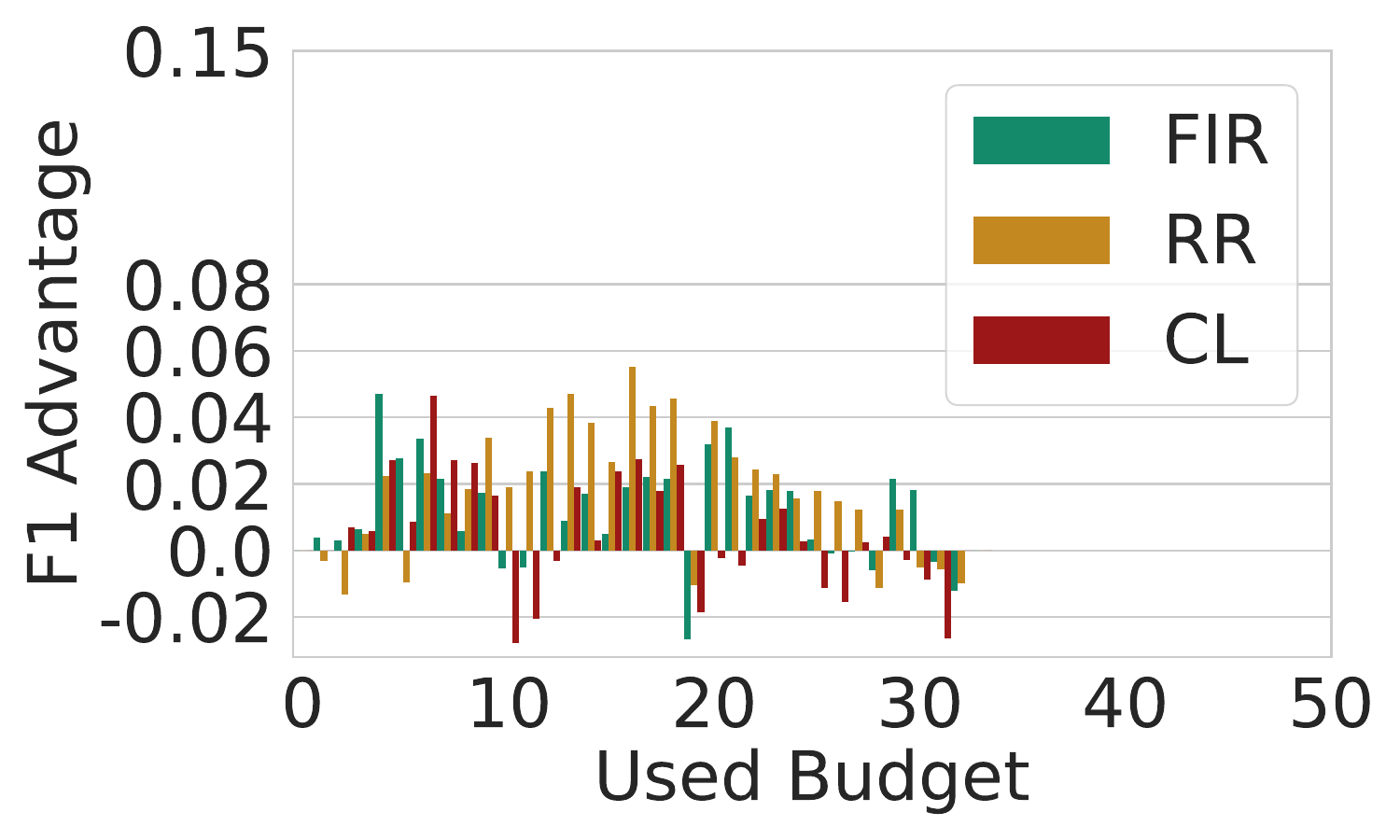}
        \caption{Scaling}
    \end{subfigure}
    \caption{Comparison of~\systemname with the baselines FIR, RR and CL for GB across error types.}
    \label{fig:agg_bl_results_gb}
\end{figure*}

\begin{figure*}[h!]
    \centering
    \begin{subfigure}{0.24\textwidth}
        \includegraphics[width=\linewidth]{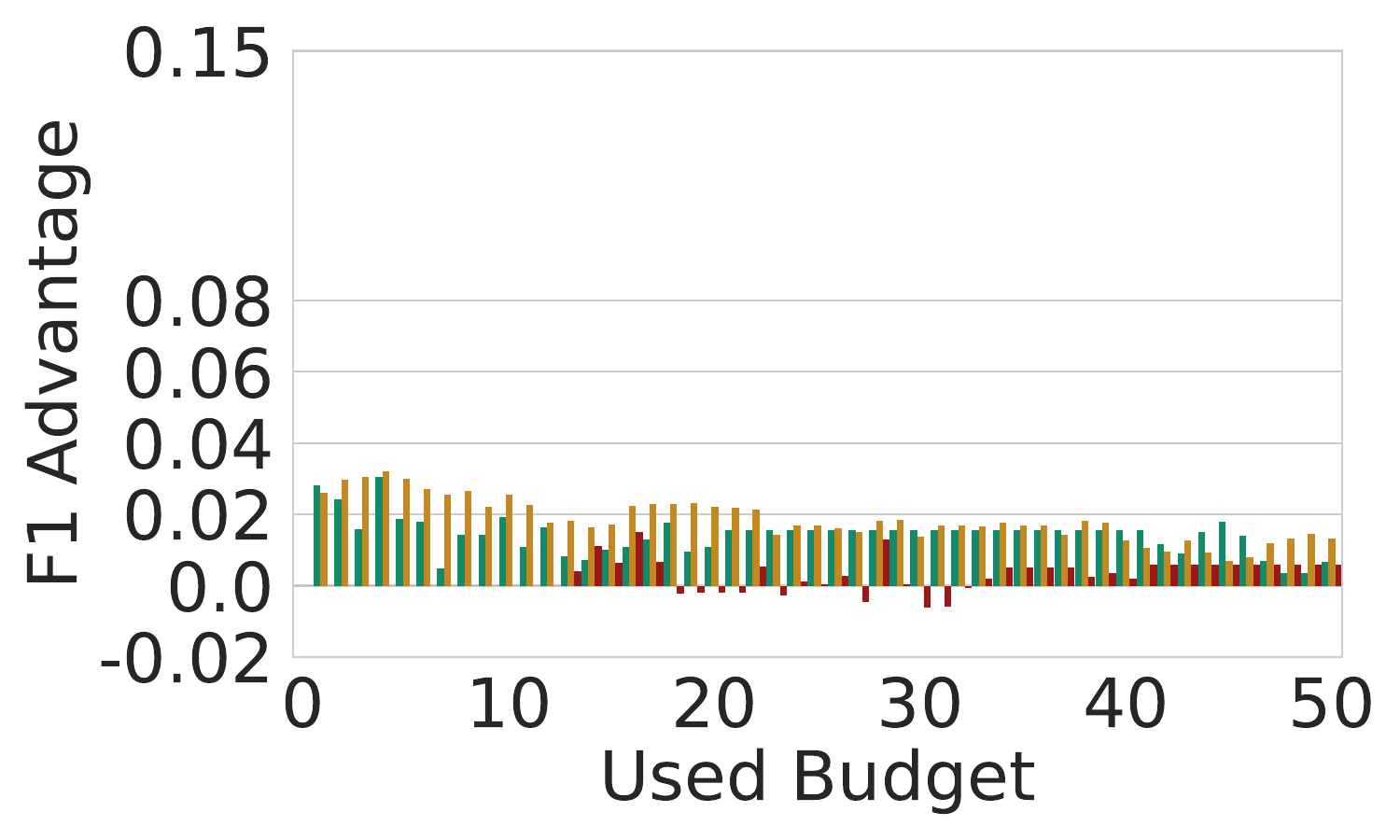}
        \caption{Airbnb - Scaling}
    \end{subfigure}
    \begin{subfigure}{0.24\textwidth}
        \includegraphics[width=\linewidth]{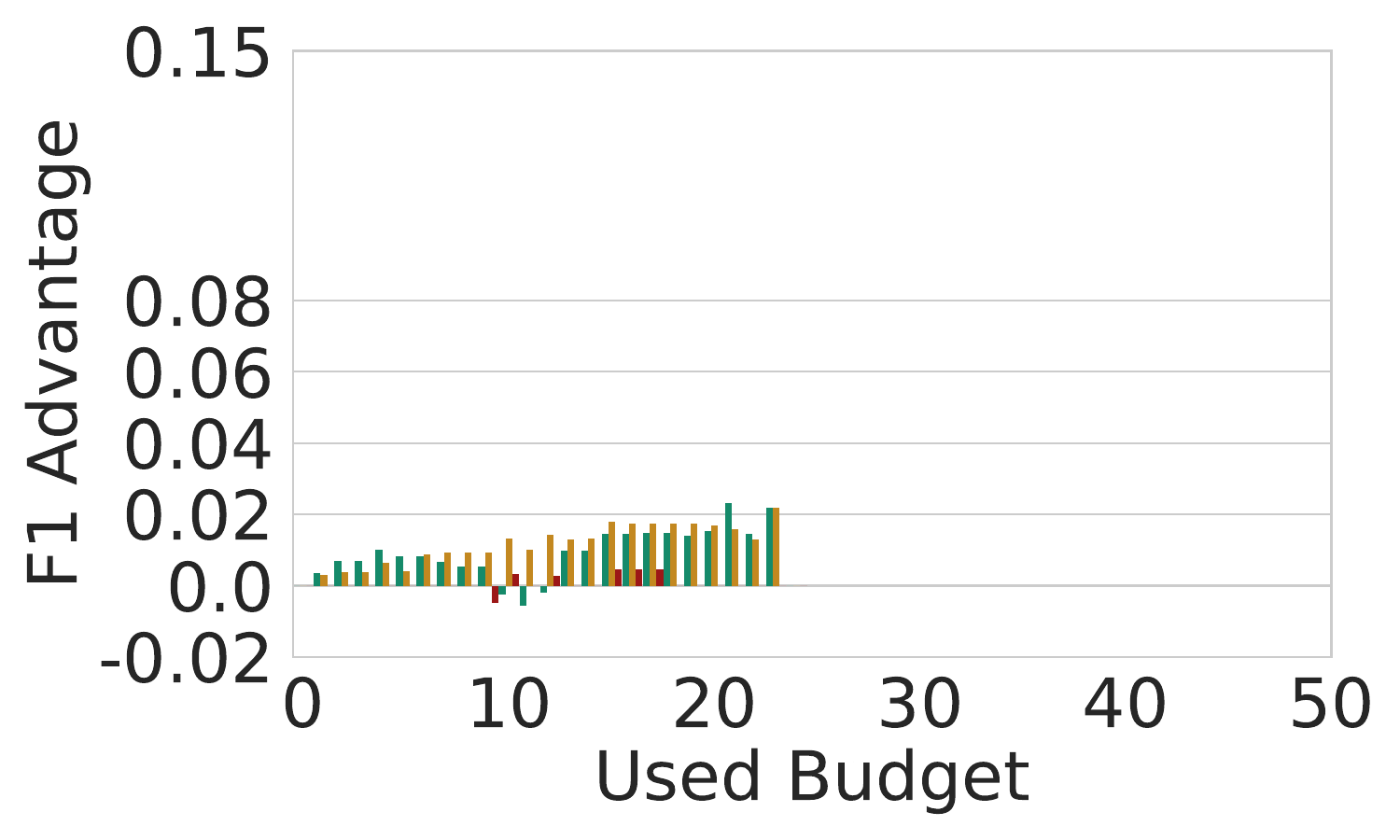}
        \caption{Credit - Scaling}
    \end{subfigure}
    \begin{subfigure}{0.24\textwidth}
        \includegraphics[width=\linewidth]{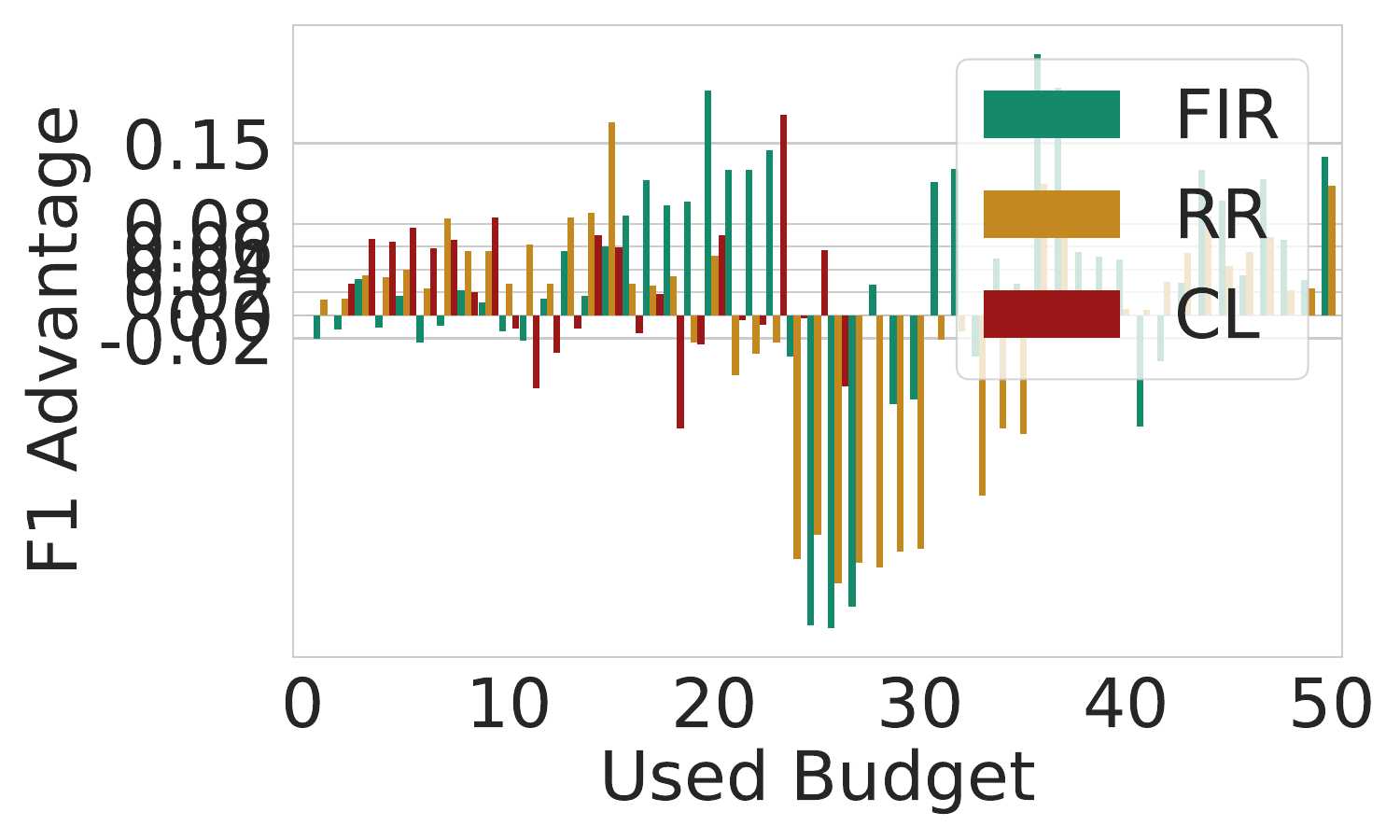}
        \caption{Titanic - Missing Values}
    \end{subfigure}
    \caption{Comparison of \systemname with FIR and RR for GB across error types, for dataset from CleanML. (For CMC, there is no difference in the F1 scores between the dirty and cleaned states).}
    \label{fig:agg_bl_results_gb2}
\end{figure*}

\begin{figure*}[h!]
    \centering
    \raisebox{1.4\height}{\rotatebox{90}{\textbf{CMC}}}\hspace{0.3em}%
    \begin{subfigure}{0.24\textwidth}
        \includegraphics[width=\linewidth]{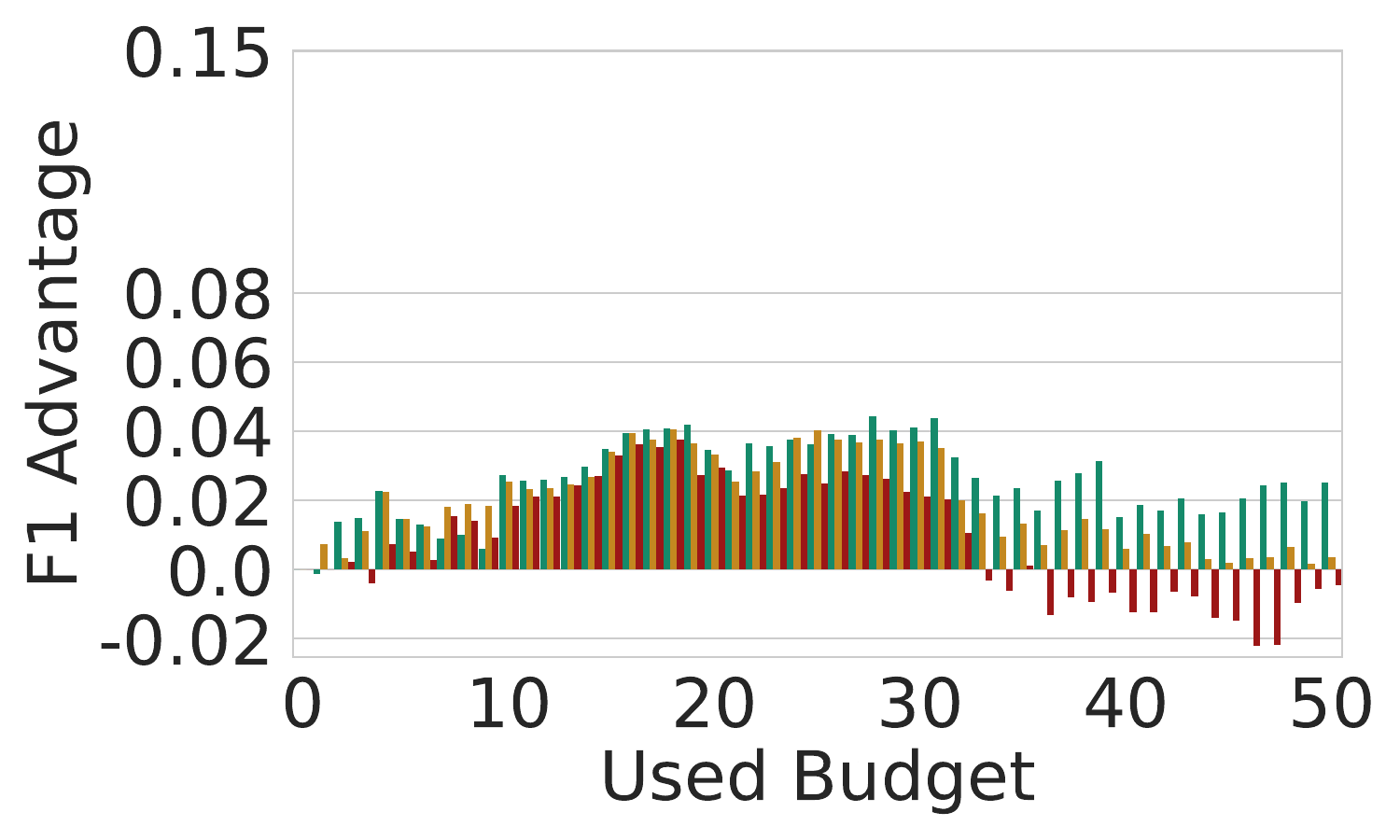}
    \end{subfigure}\hfill
    \begin{subfigure}{0.24\textwidth}
        \includegraphics[width=\linewidth]{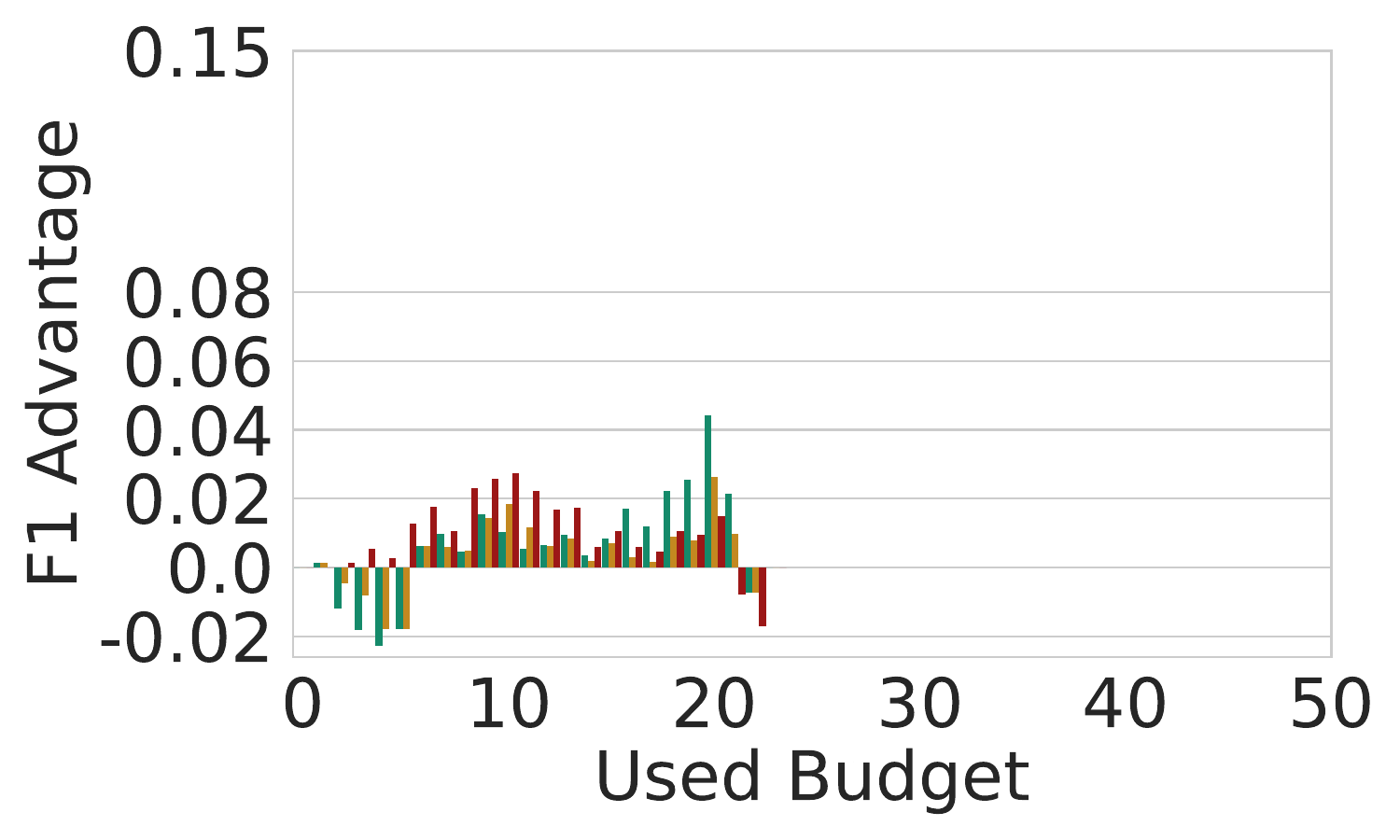}
    \end{subfigure}\hfill
    \begin{subfigure}{0.24\textwidth}
        \includegraphics[width=\linewidth]{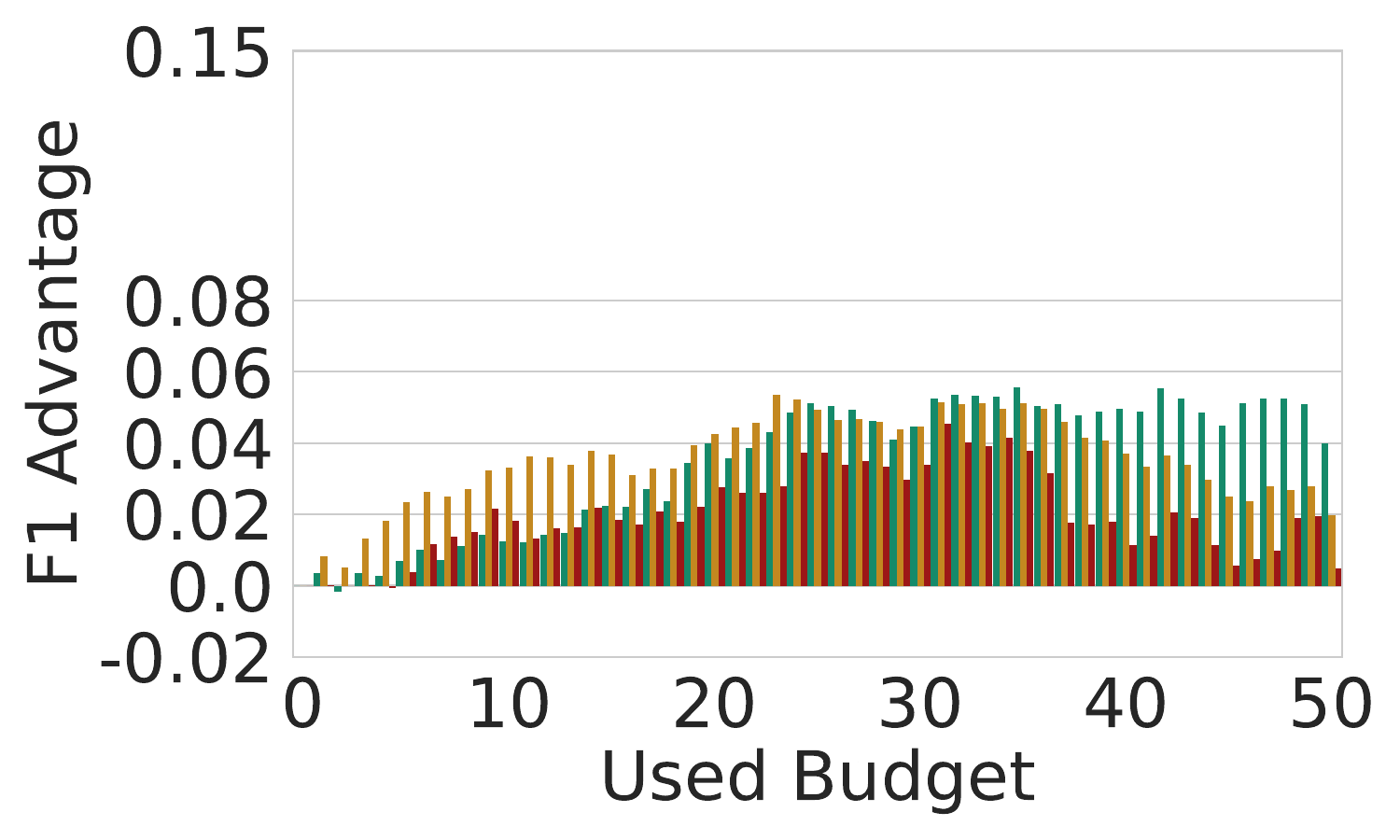}
    \end{subfigure}\hfill
    \begin{subfigure}{0.24\textwidth}
        \includegraphics[width=\linewidth]{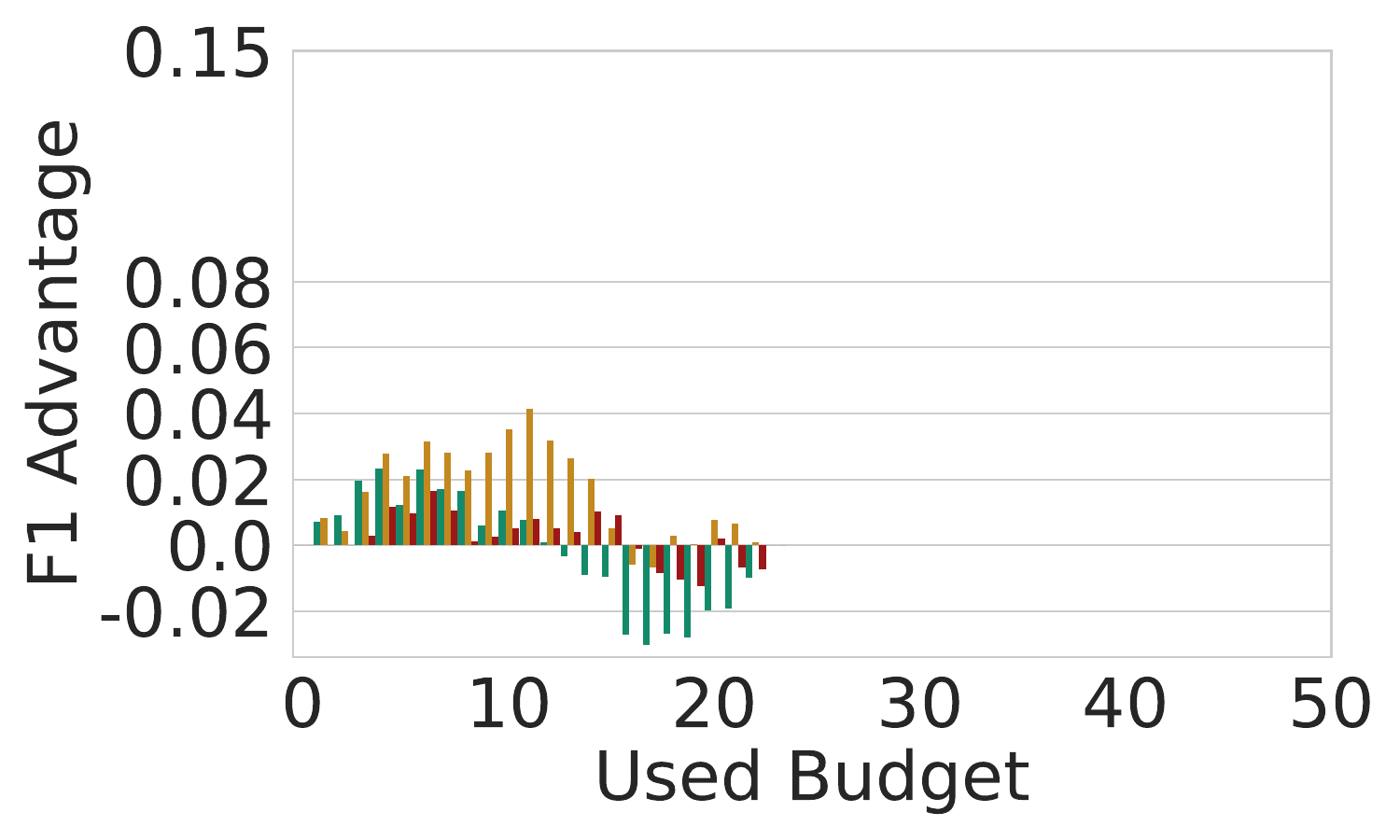}
    \end{subfigure}
    
    \vspace{-0.1em}

        \raisebox{1.2\height}{\rotatebox{90}{\textbf{Churn}}}\hspace{0.3em}%
    \begin{subfigure}{0.24\textwidth}
        \includegraphics[width=\linewidth]{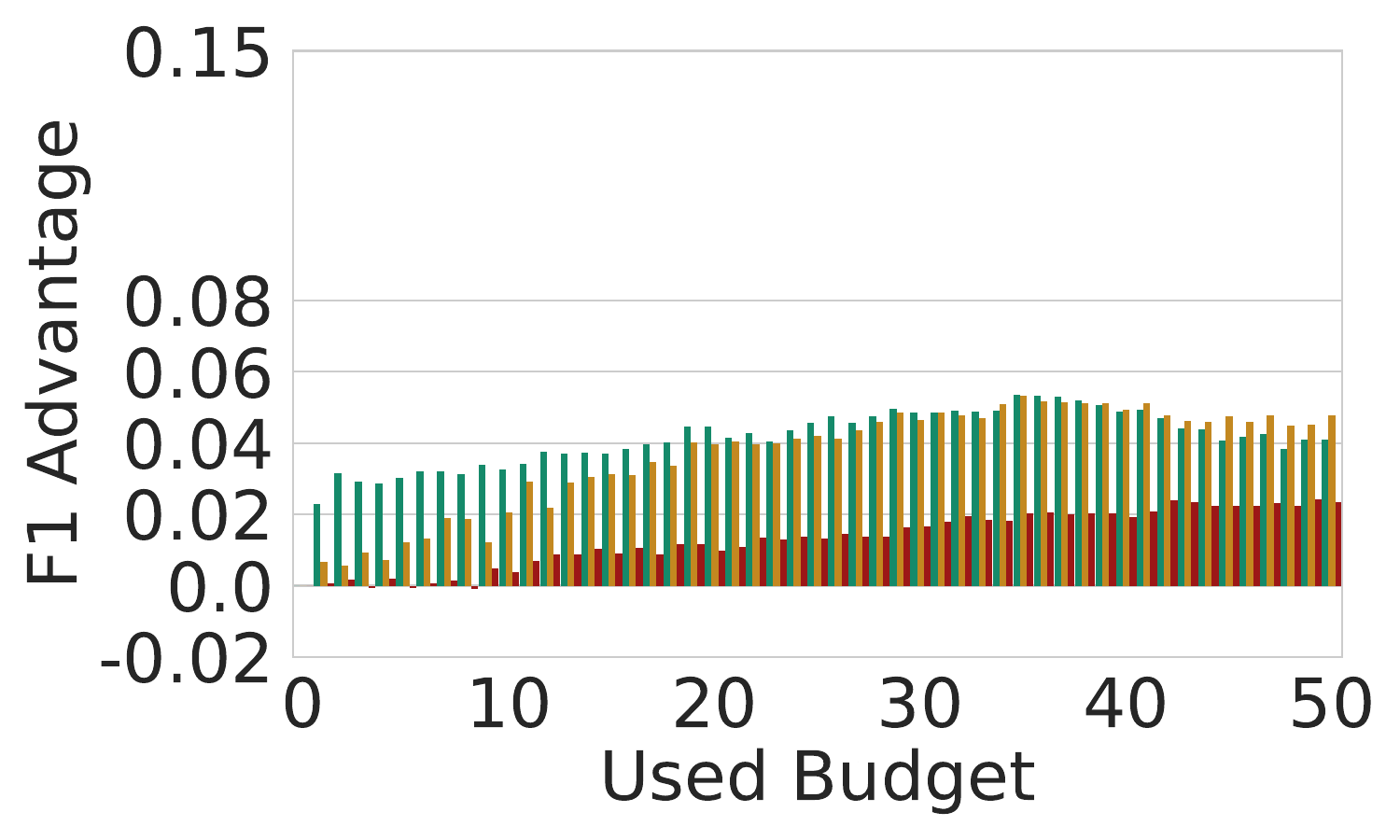}
    \end{subfigure}\hfill
    \begin{subfigure}{0.24\textwidth}
        \includegraphics[width=\linewidth]{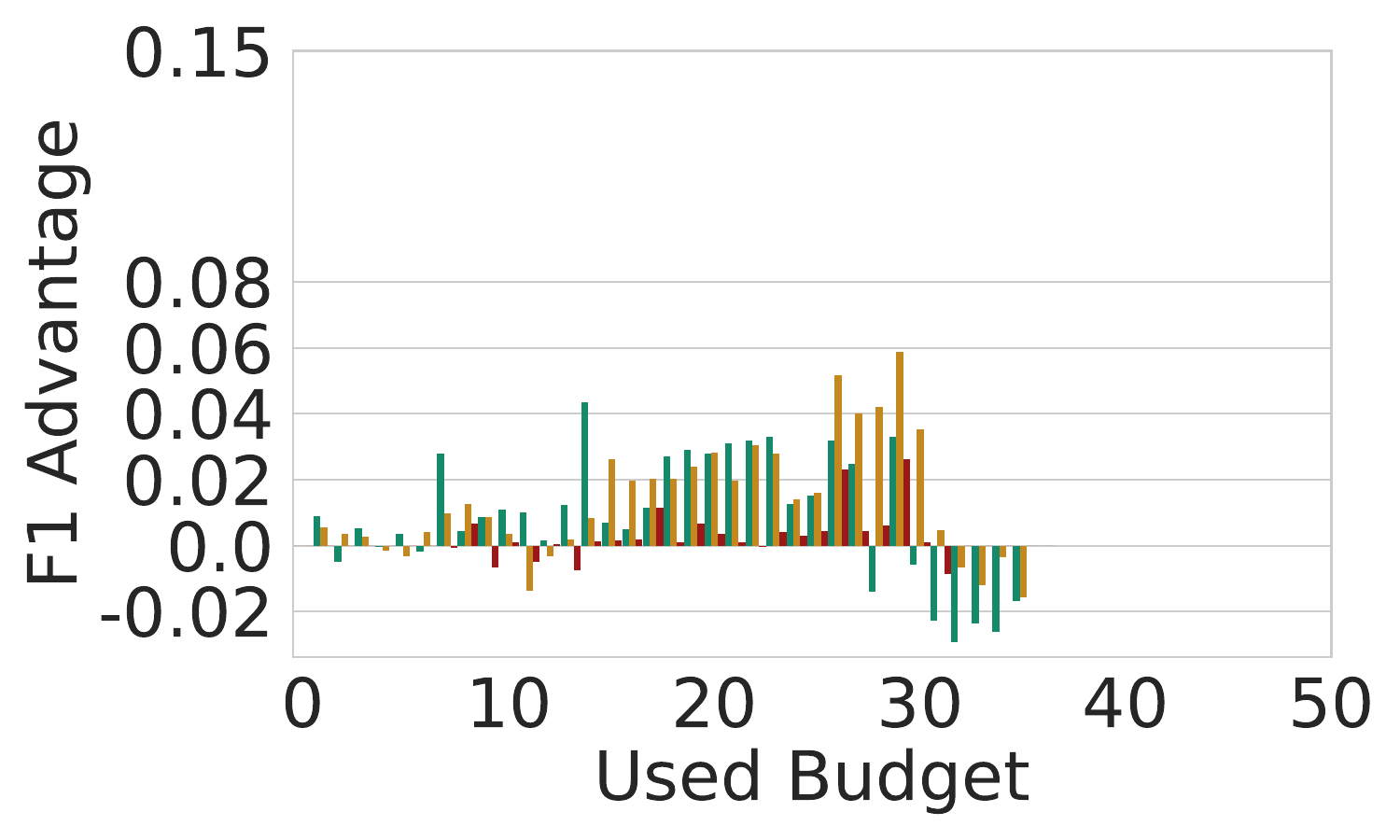}
    \end{subfigure}\hfill
    \begin{subfigure}{0.24\textwidth}
        \includegraphics[width=\linewidth]{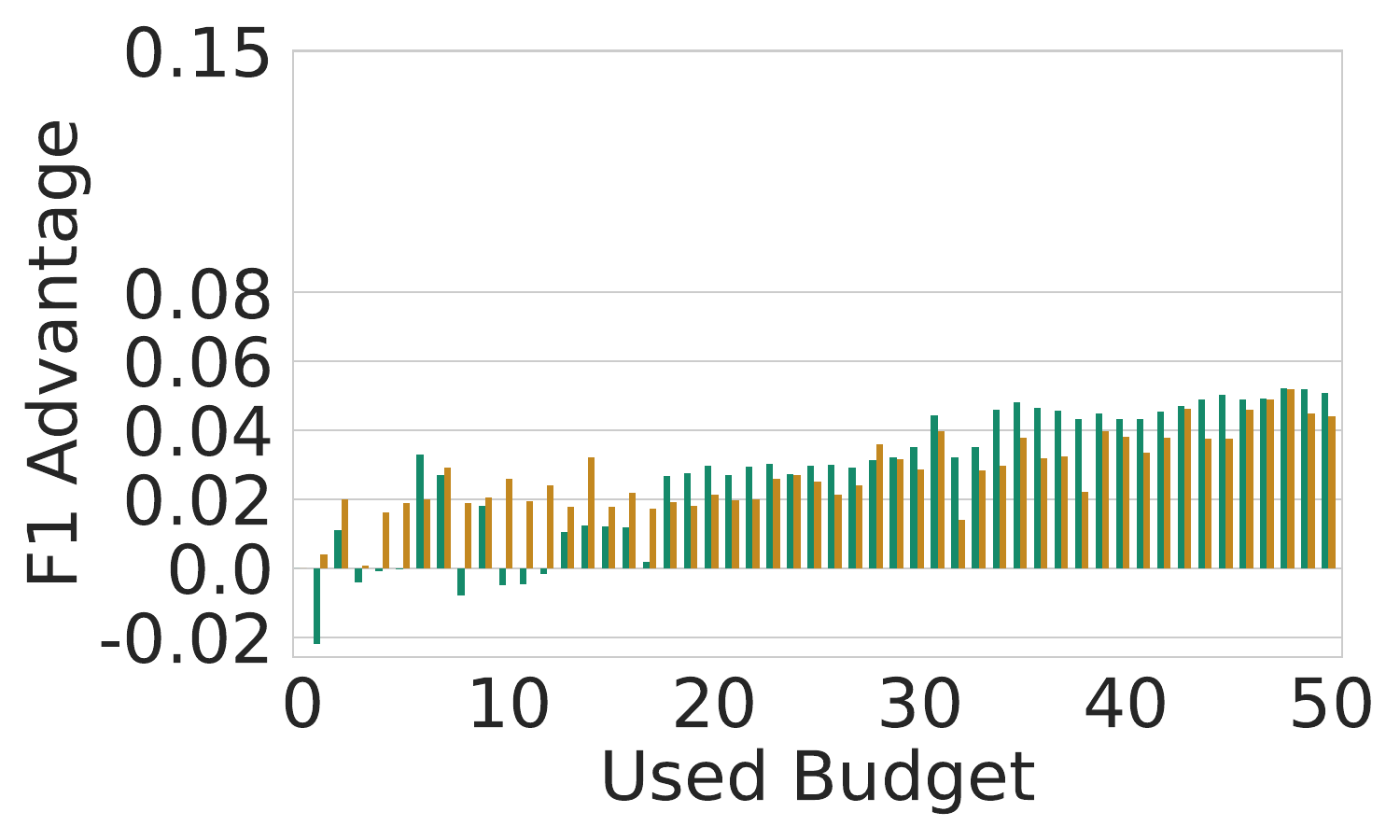}
    \end{subfigure}\hfill
    \begin{subfigure}{0.24\textwidth}
        \includegraphics[width=\linewidth]{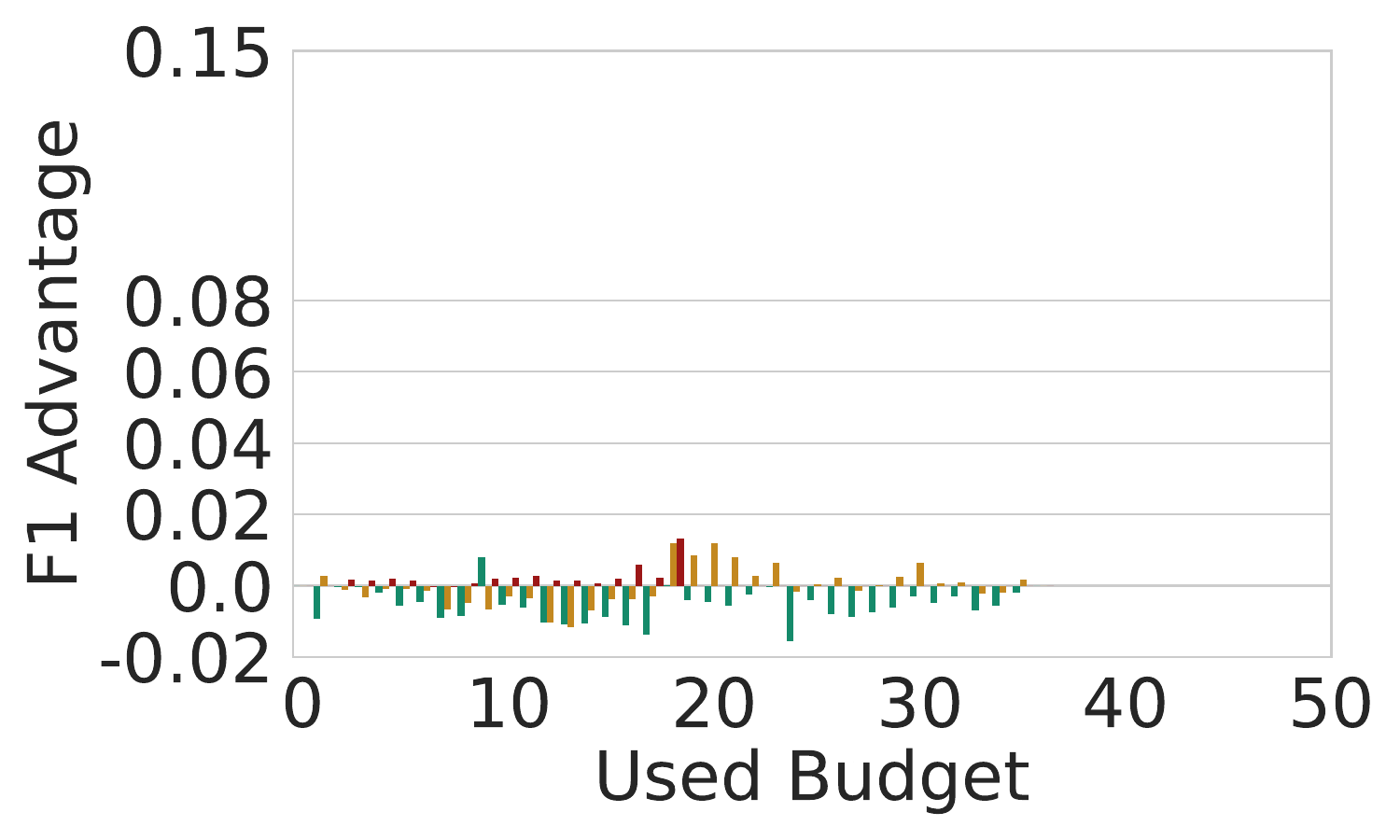}
    \end{subfigure}
    
    \vspace{-0.1em}

    \raisebox{2.\height}{\rotatebox{90}{\textbf{EEG}}}\hspace{0.3em}%
    \begin{subfigure}{0.24\textwidth}
        \centering\raisebox{3.85\height}{\parbox{0.75\linewidth}{\texttt{EEG only contains numerical features.}}}
    \end{subfigure}\hfill
    \begin{subfigure}{0.24\textwidth}
        \includegraphics[width=\linewidth]{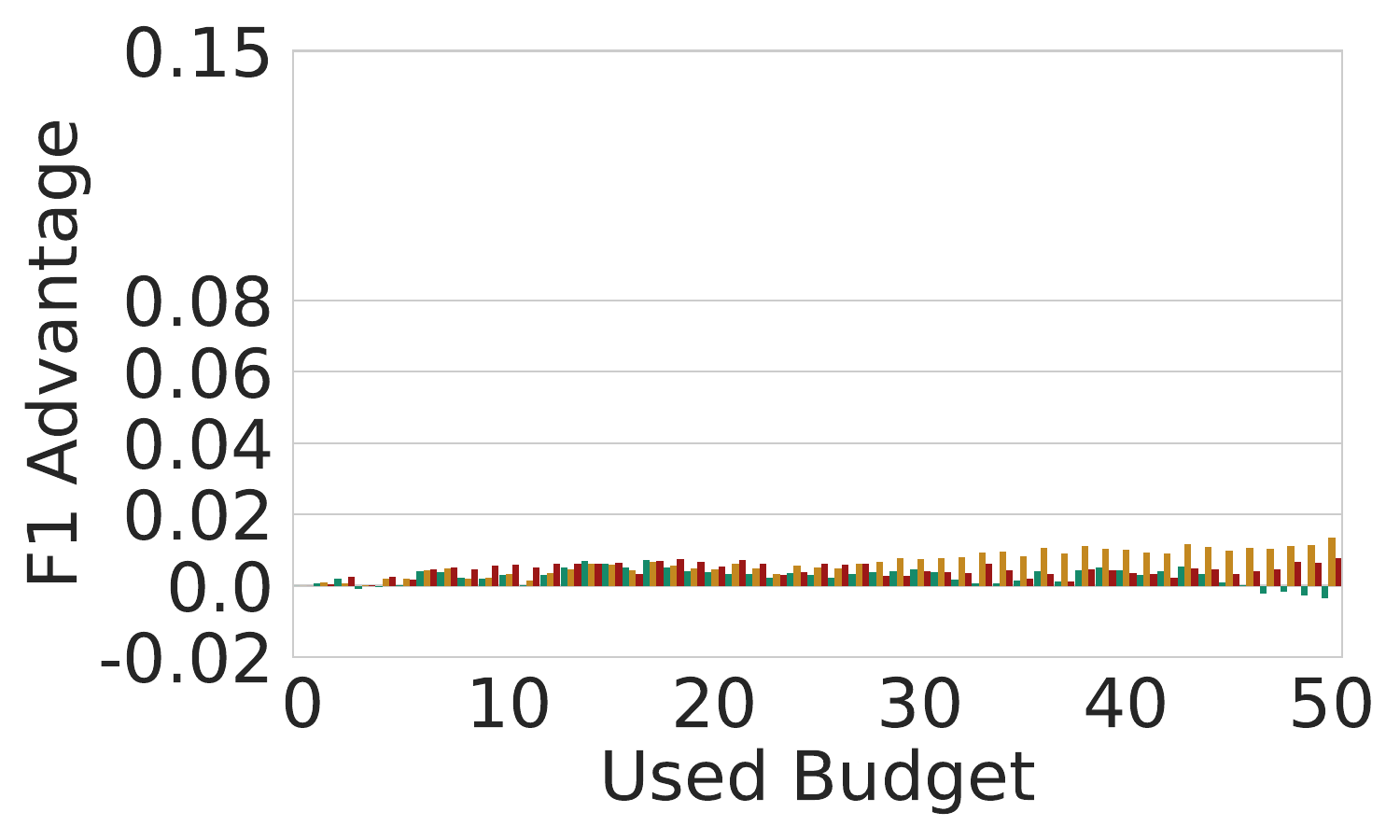}
    \end{subfigure}\hfill
    \begin{subfigure}{0.24\textwidth}
        \includegraphics[width=\linewidth]{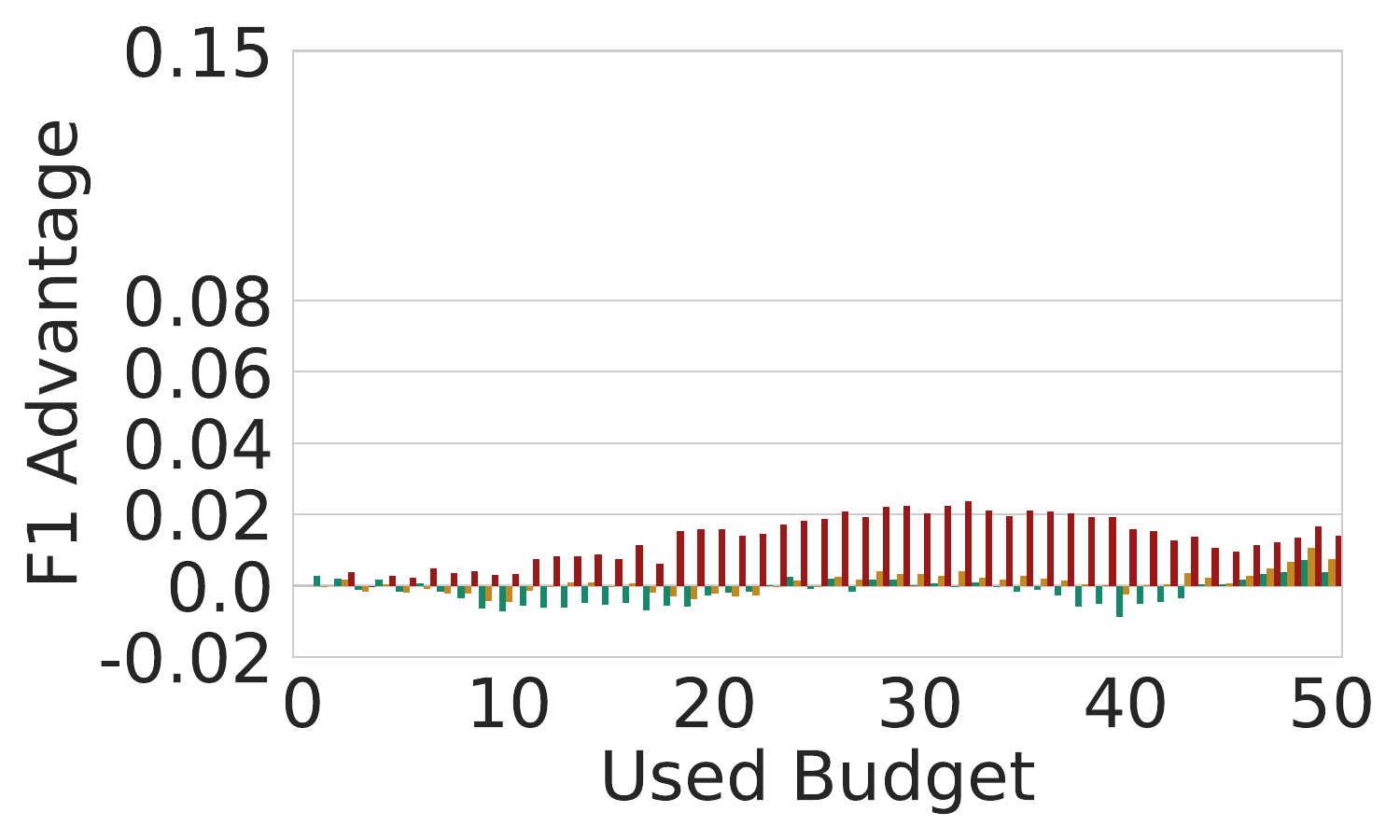}
    \end{subfigure}\hfill
    \begin{subfigure}{0.24\textwidth}
        \includegraphics[width=\linewidth]{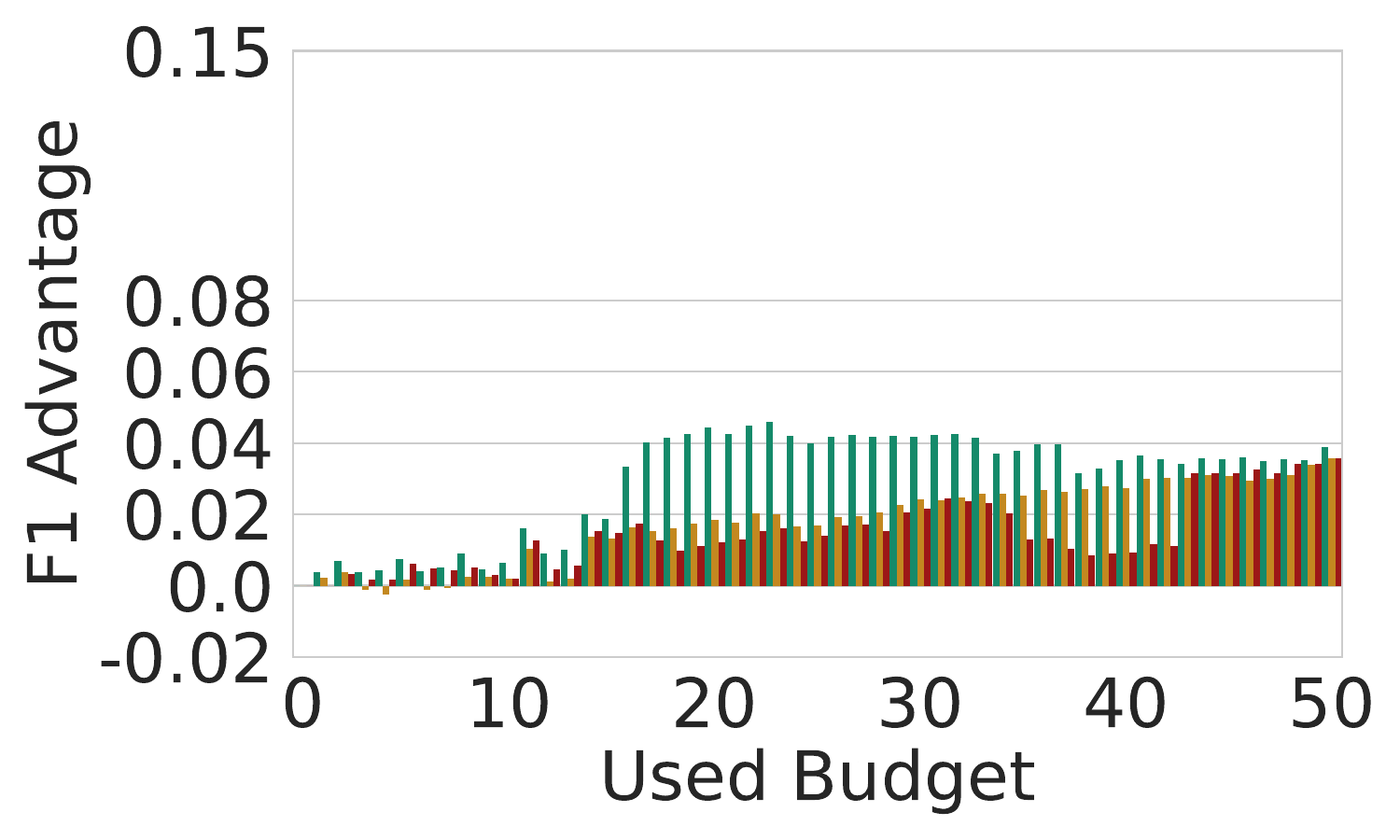}
    \end{subfigure}
    
    \vspace{-0.1em}
    
    \raisebox{1.2\height}{\rotatebox{90}{\textbf{S-Credit}}}\hspace{0.3em}%
    \begin{subfigure}{0.24\textwidth}
        \includegraphics[width=\linewidth]{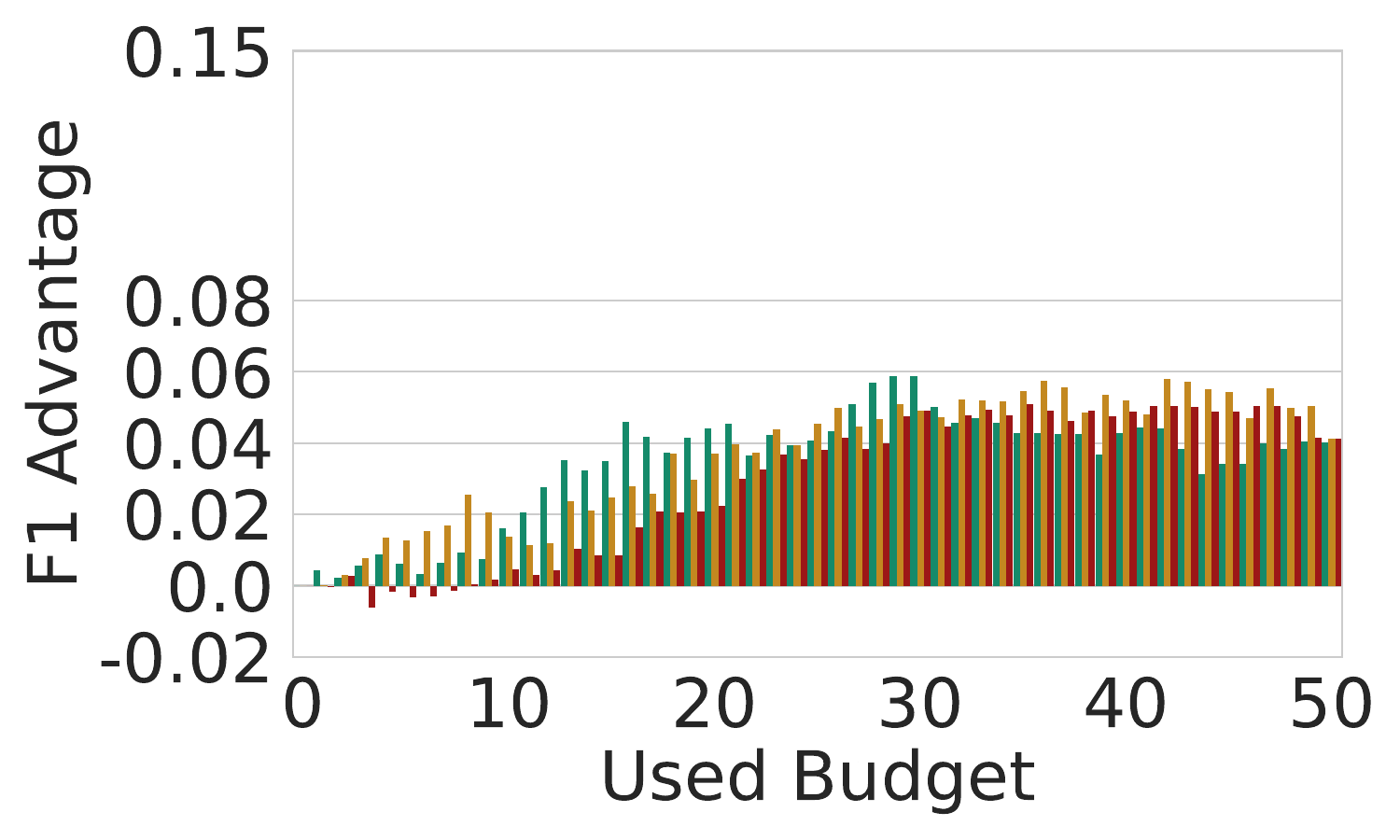}
        \caption{Categorical Shift}
    \end{subfigure}\hfill
    \begin{subfigure}{0.24\textwidth}
        \includegraphics[width=\linewidth]{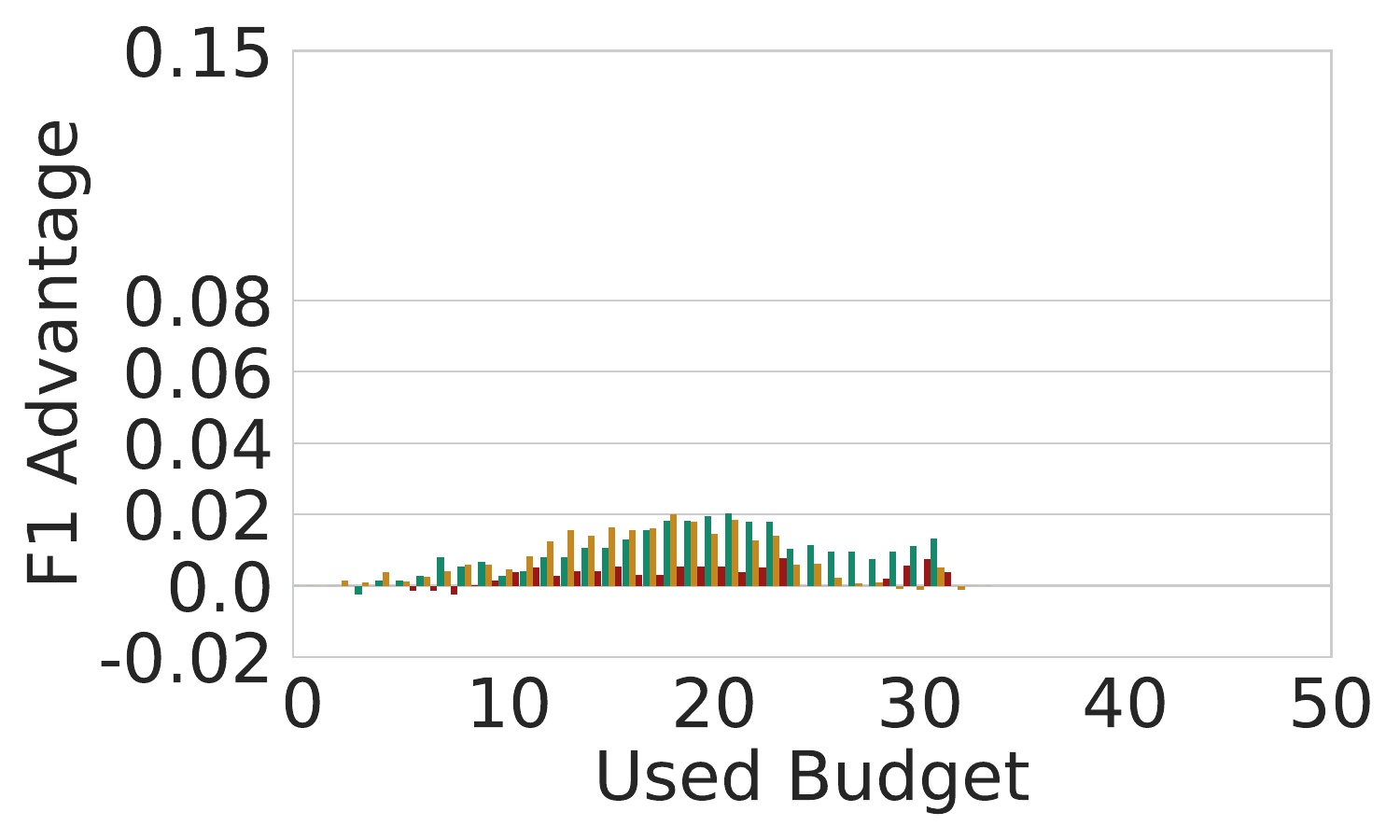}
        \caption{Gaussian Noise}
    \end{subfigure}\hfill
    \begin{subfigure}{0.24\textwidth}
        \includegraphics[width=\linewidth]{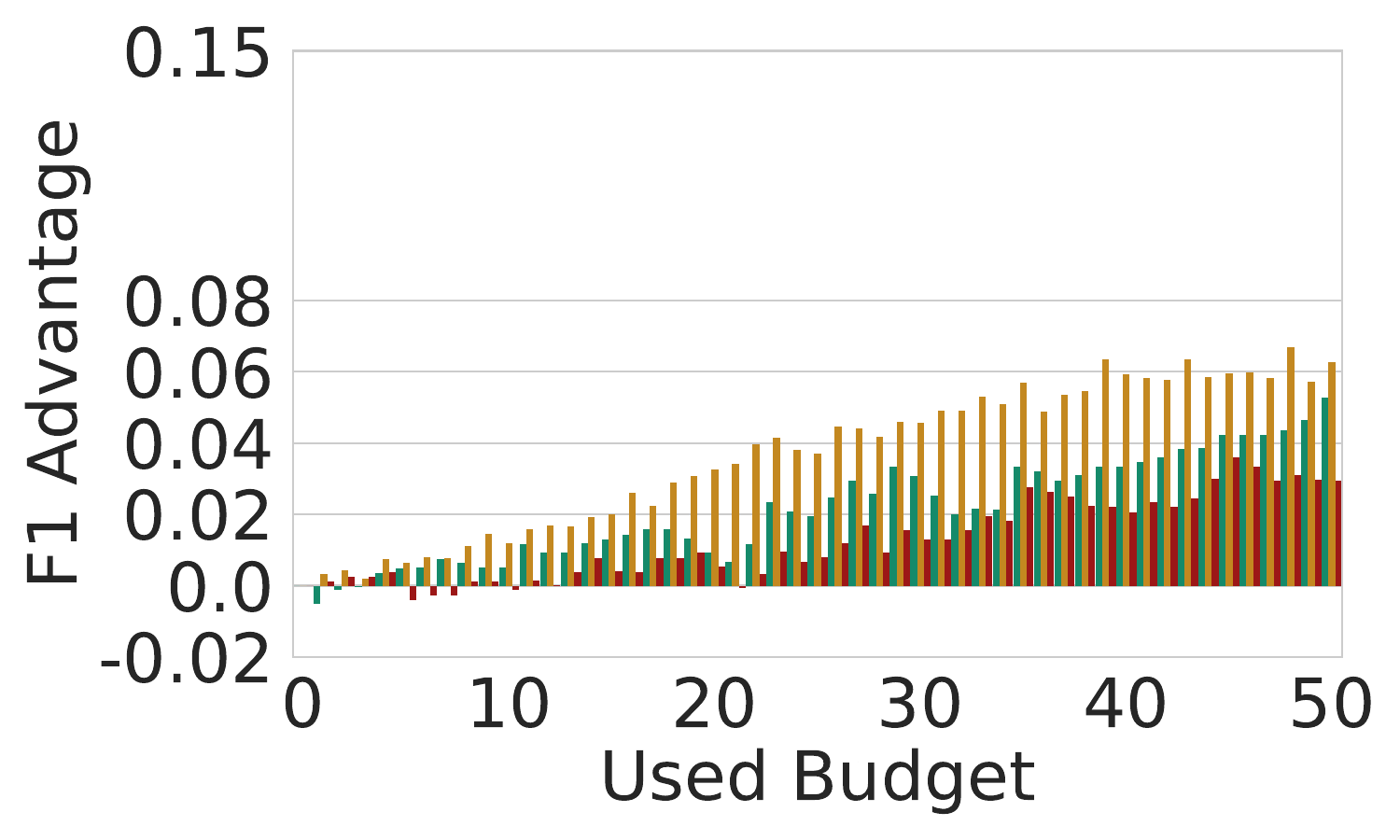}
        \caption{Missing Values}
    \end{subfigure}\hfill
    \begin{subfigure}{0.24\textwidth}
        \includegraphics[width=\linewidth]{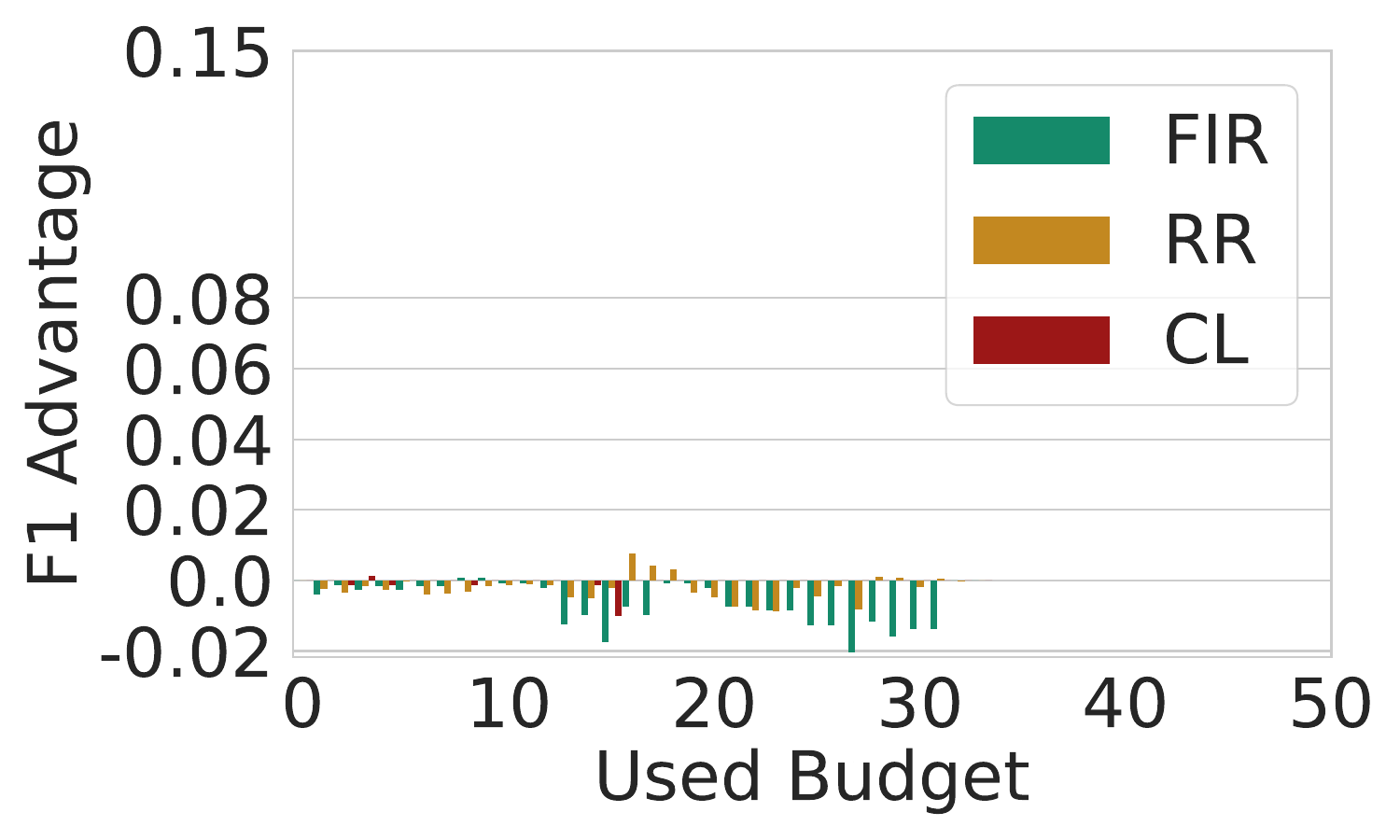}
        \caption{Scaling}
    \end{subfigure}
    \caption{Comparison of \systemname with FIR and RR for KNN across error types.}
    \label{fig:agg_bl_results_knn}
\end{figure*}

\begin{figure*}[h!]
    \centering
    \begin{subfigure}{0.24\textwidth}
        \includegraphics[width=\linewidth]{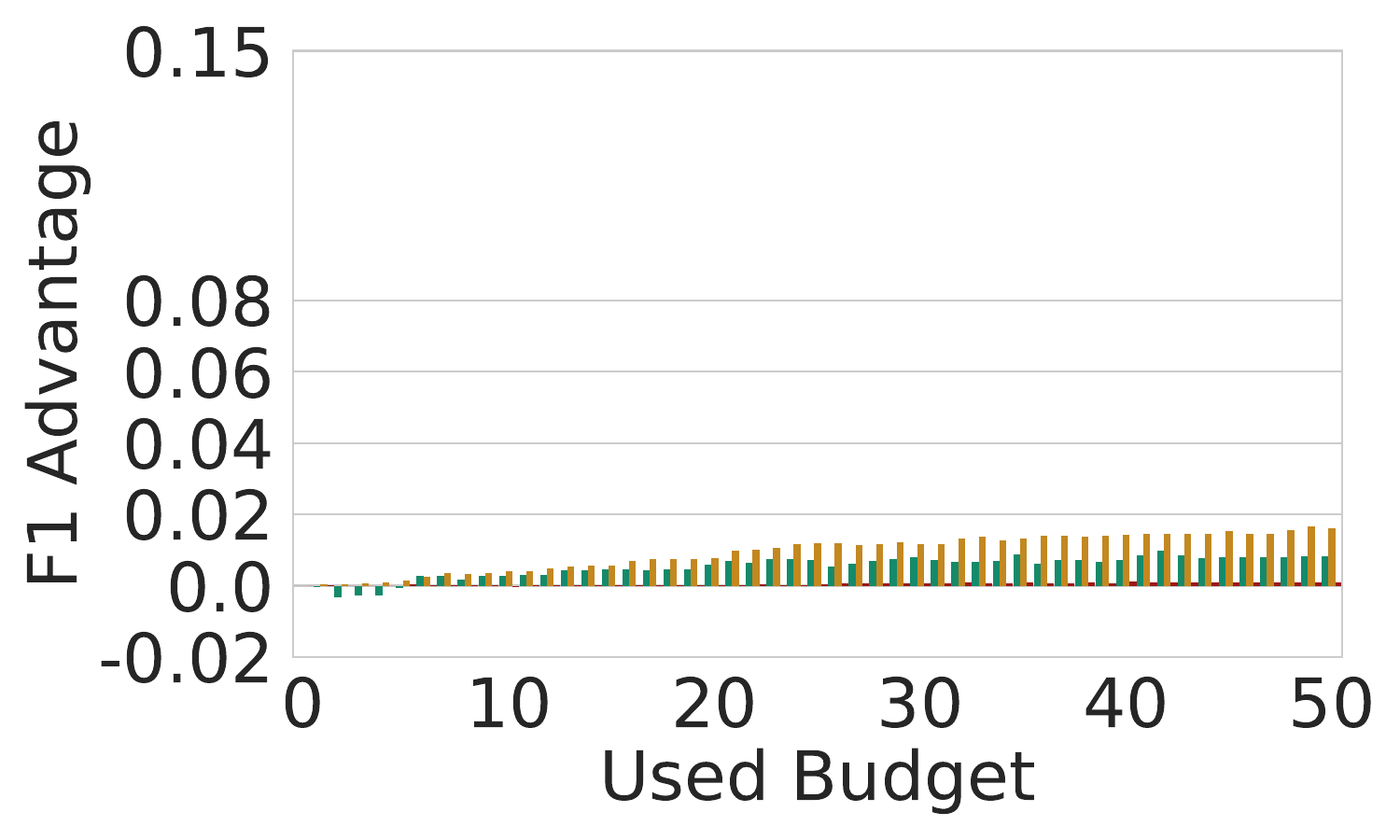}
        \caption{Airbnb - Scaling}
    \end{subfigure}
    \begin{subfigure}{0.24\textwidth}
        \includegraphics[width=\linewidth]{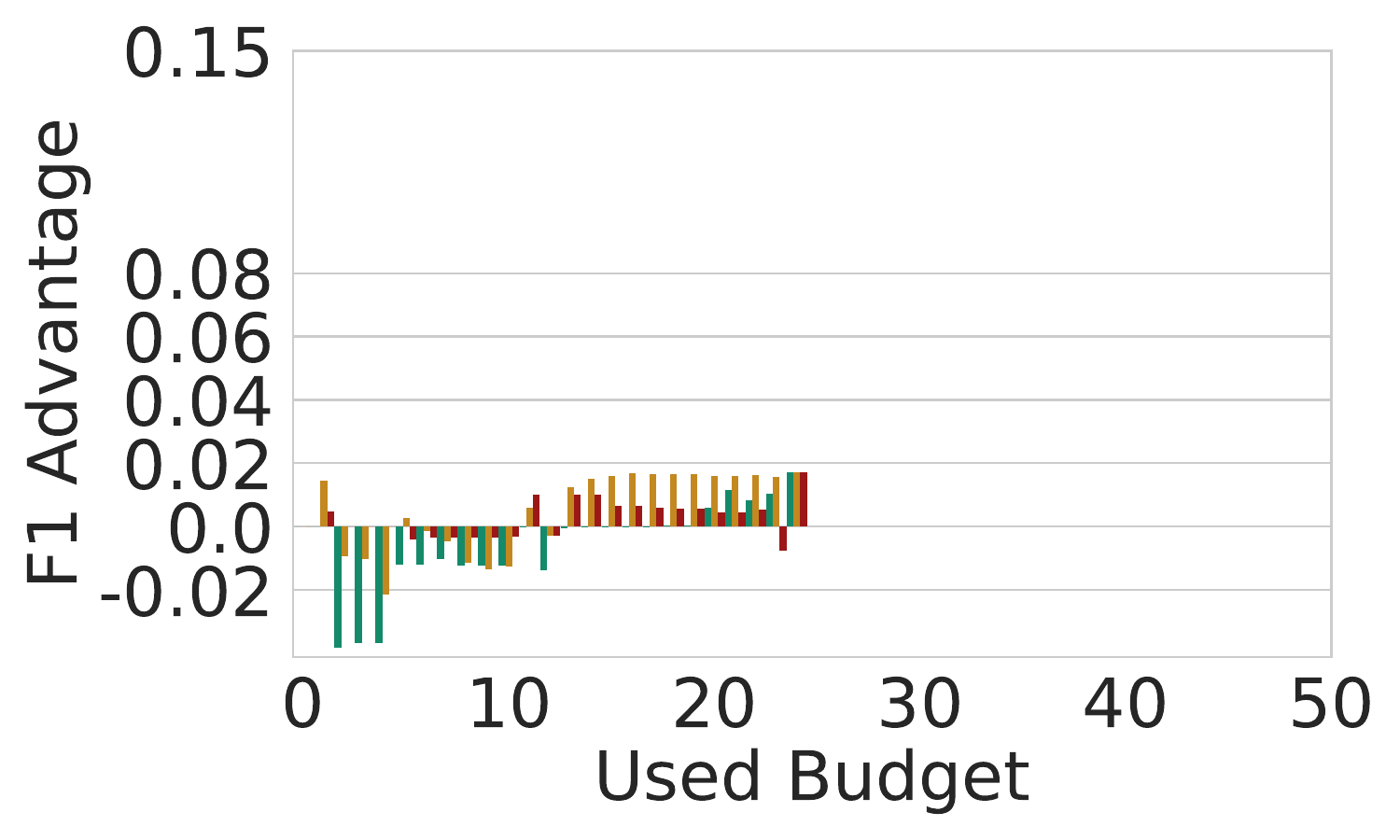}
        \caption{Credit - Scaling}
    \end{subfigure}
    \begin{subfigure}{0.24\textwidth}
        \includegraphics[width=\linewidth]{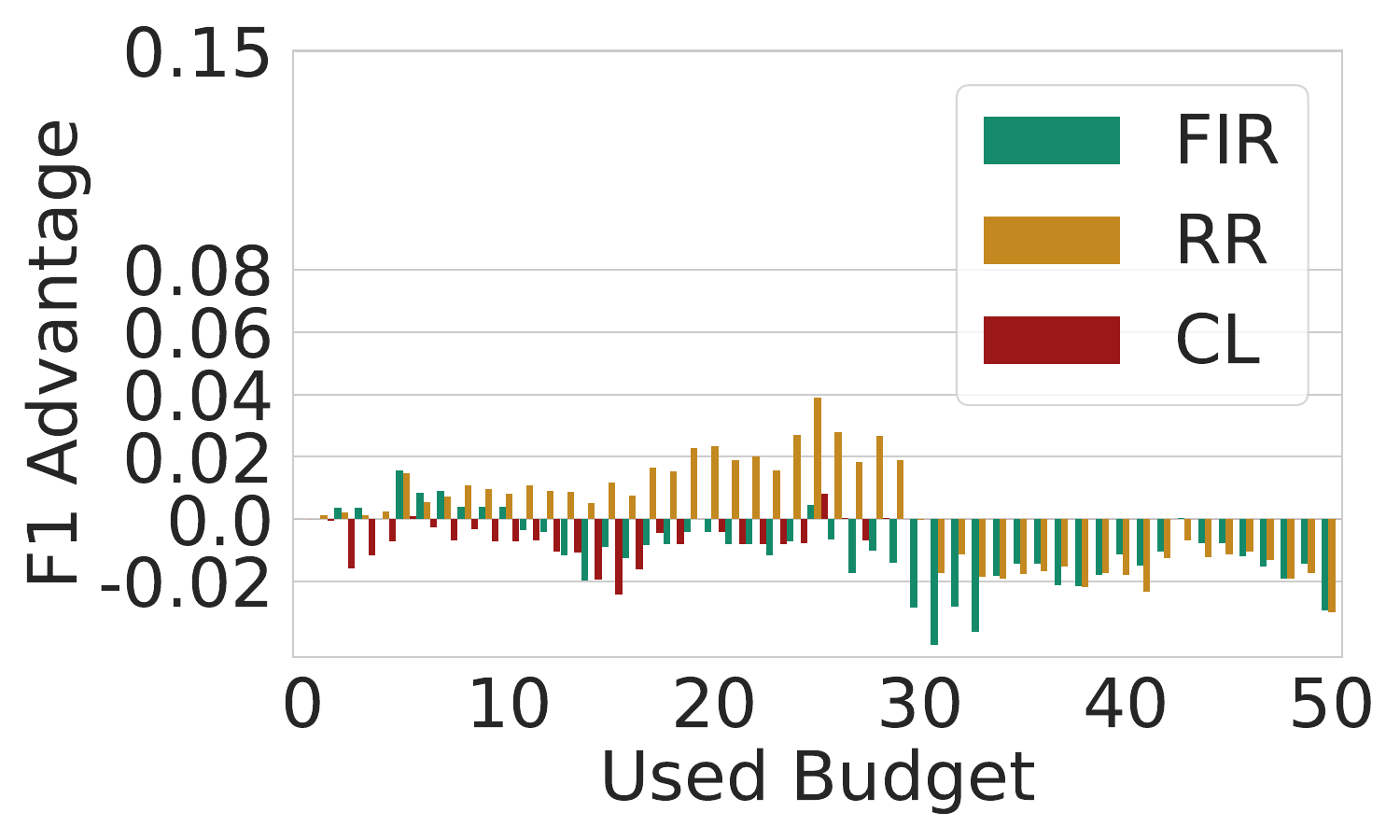}
        \caption{Titanic - Missing Values}
    \end{subfigure}
    \caption{Comparison of~\systemname with the baselines FIR, RR and CL for KNN across error types, for datasets from CleanML.}
    \label{fig:agg_bl_results_knn2}
\end{figure*}

\begin{figure*}[h!]
    \centering
    \raisebox{1.4\height}{\rotatebox{90}{\textbf{CMC}}}\hspace{0.3em}%
    \begin{subfigure}{0.24\textwidth}
        \includegraphics[width=\linewidth]{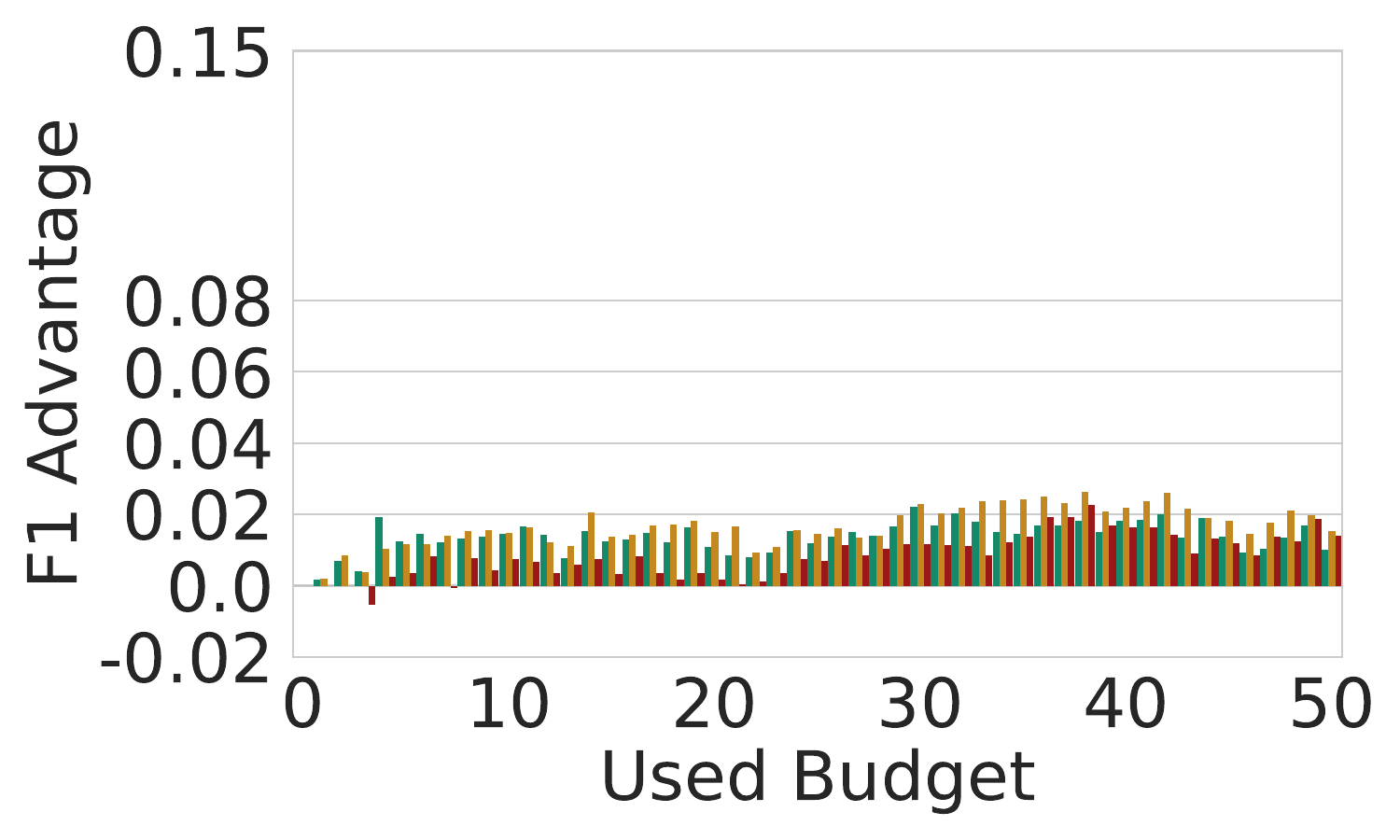}
    \end{subfigure}\hfill
    \begin{subfigure}{0.24\textwidth}
        \includegraphics[width=\linewidth]{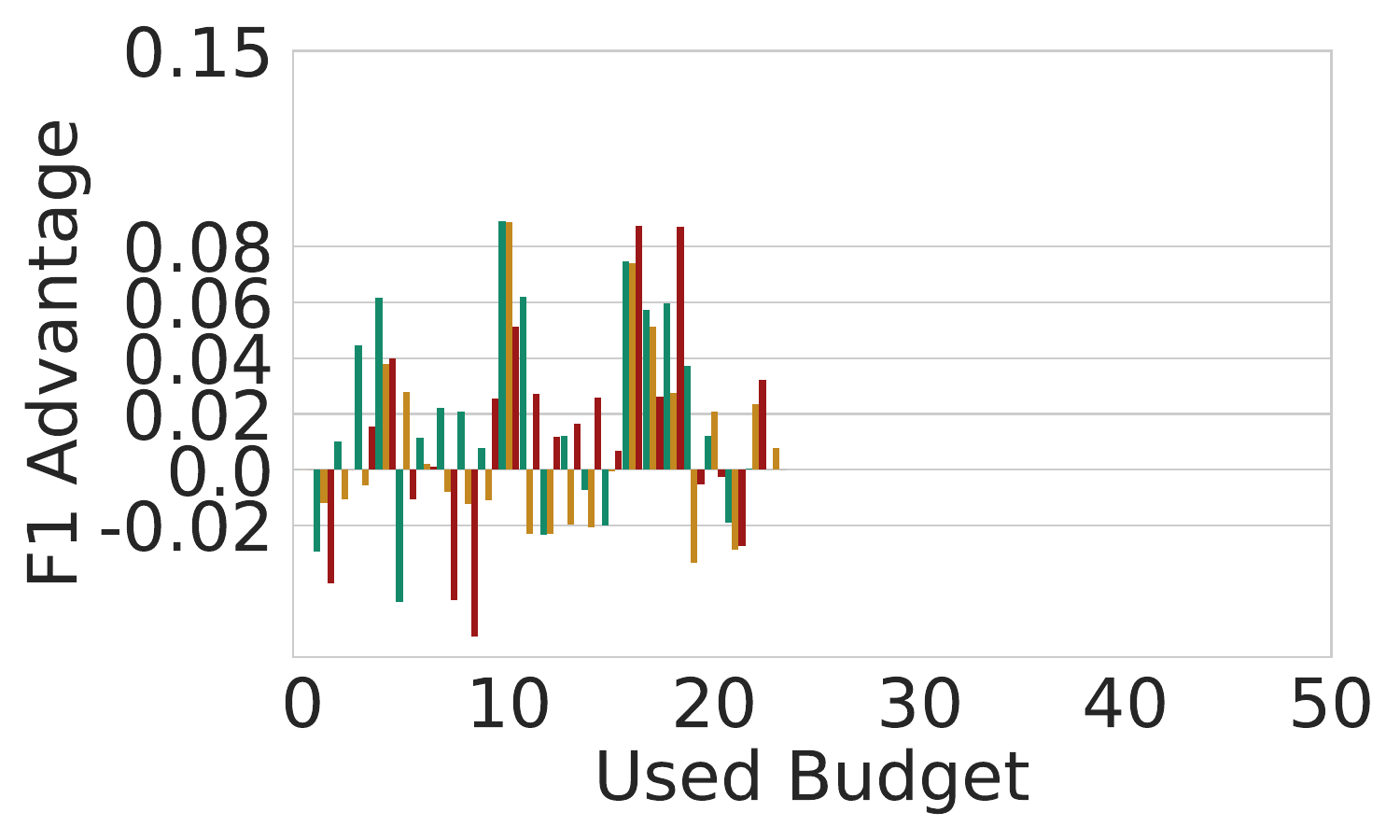}
    \end{subfigure}\hfill
    \begin{subfigure}{0.24\textwidth}
        \includegraphics[width=\linewidth]{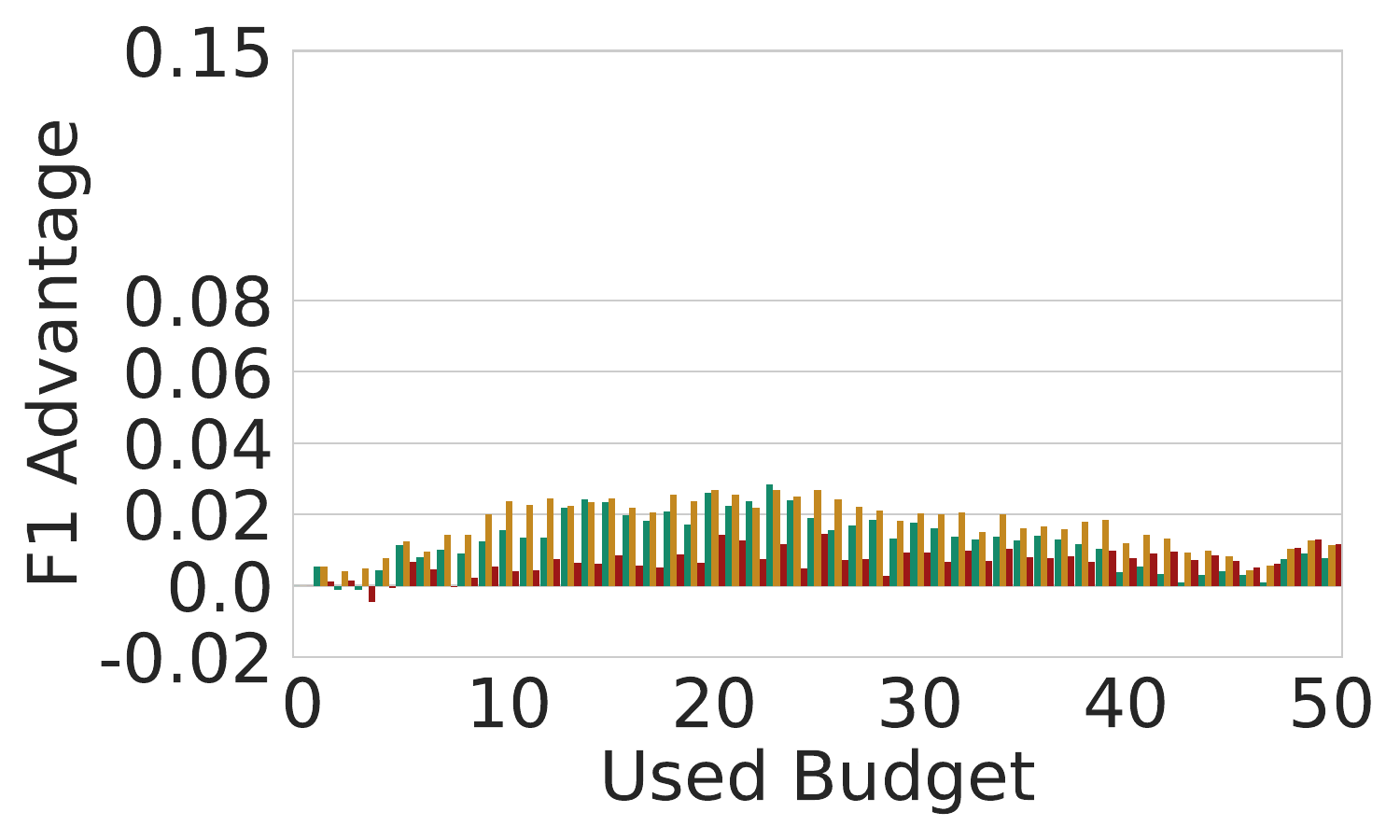}
    \end{subfigure}\hfill
    \begin{subfigure}{0.24\textwidth}
        \includegraphics[width=\linewidth]{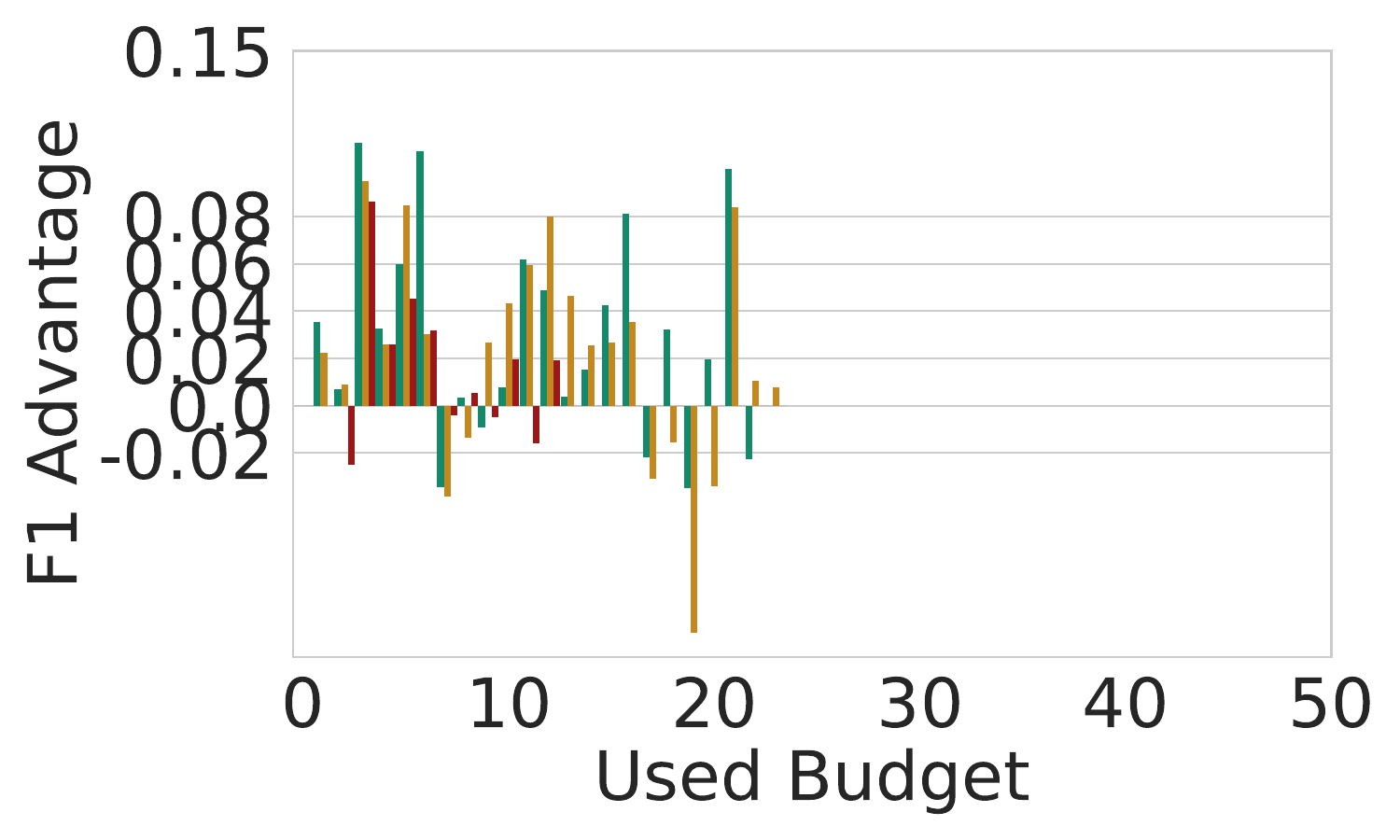}
    \end{subfigure}
    
    \vspace{-0.1em}

        \raisebox{1.2\height}{\rotatebox{90}{\textbf{Churn}}}\hspace{0.3em}%
    \begin{subfigure}{0.24\textwidth}
        \includegraphics[width=\linewidth]{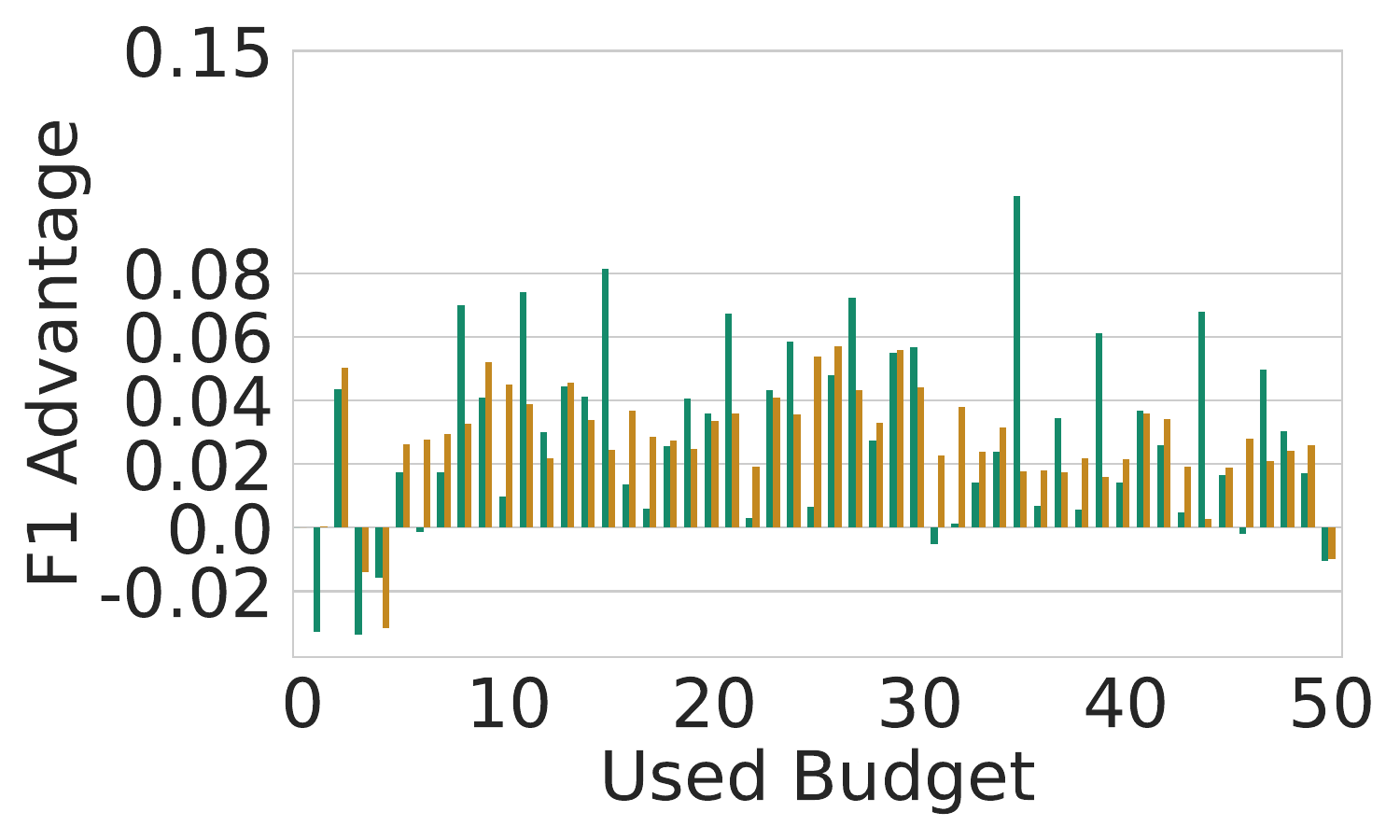}
    \end{subfigure}\hfill
    \begin{subfigure}{0.24\textwidth}
        \includegraphics[width=\linewidth]{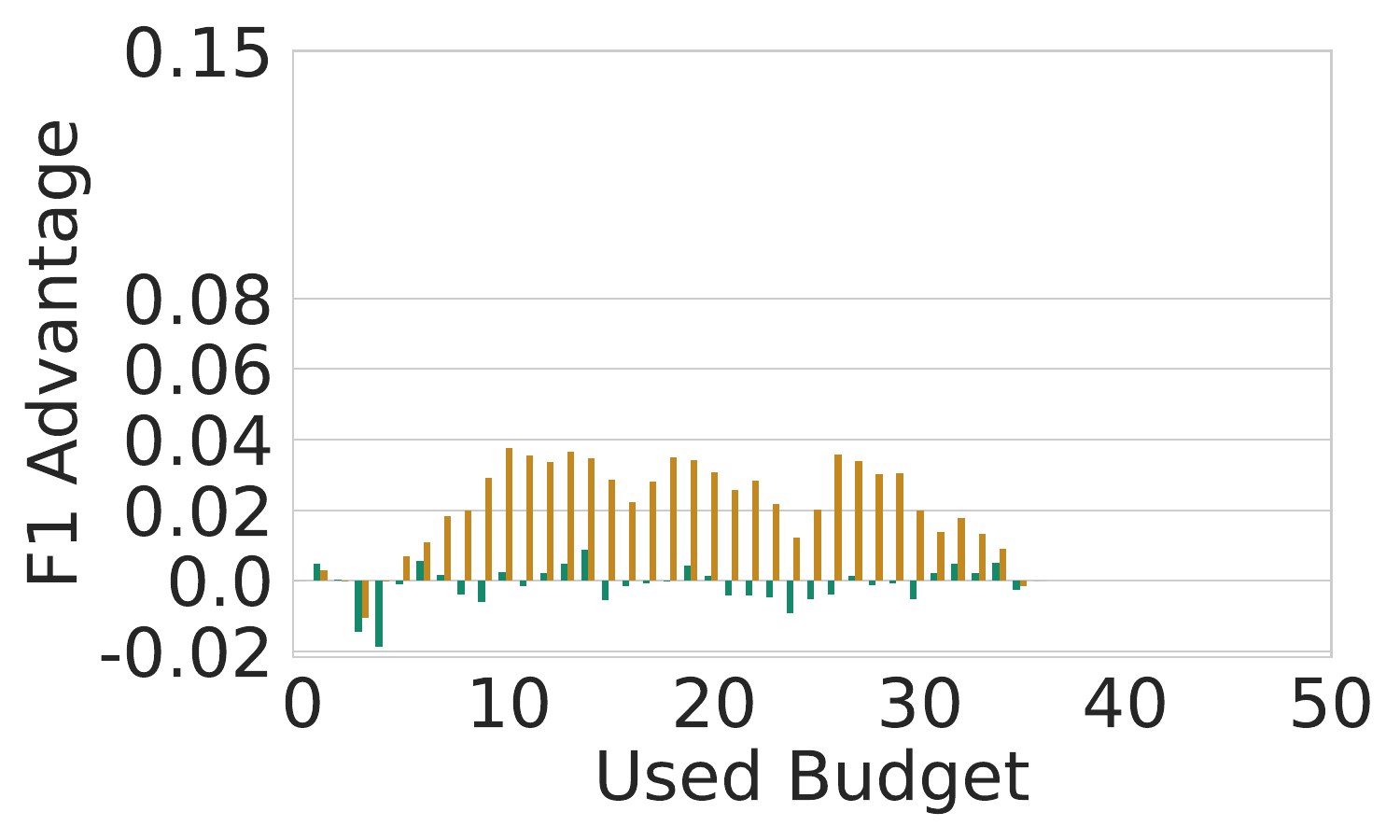}
    \end{subfigure}\hfill
    \begin{subfigure}{0.24\textwidth}
        \includegraphics[width=\linewidth]{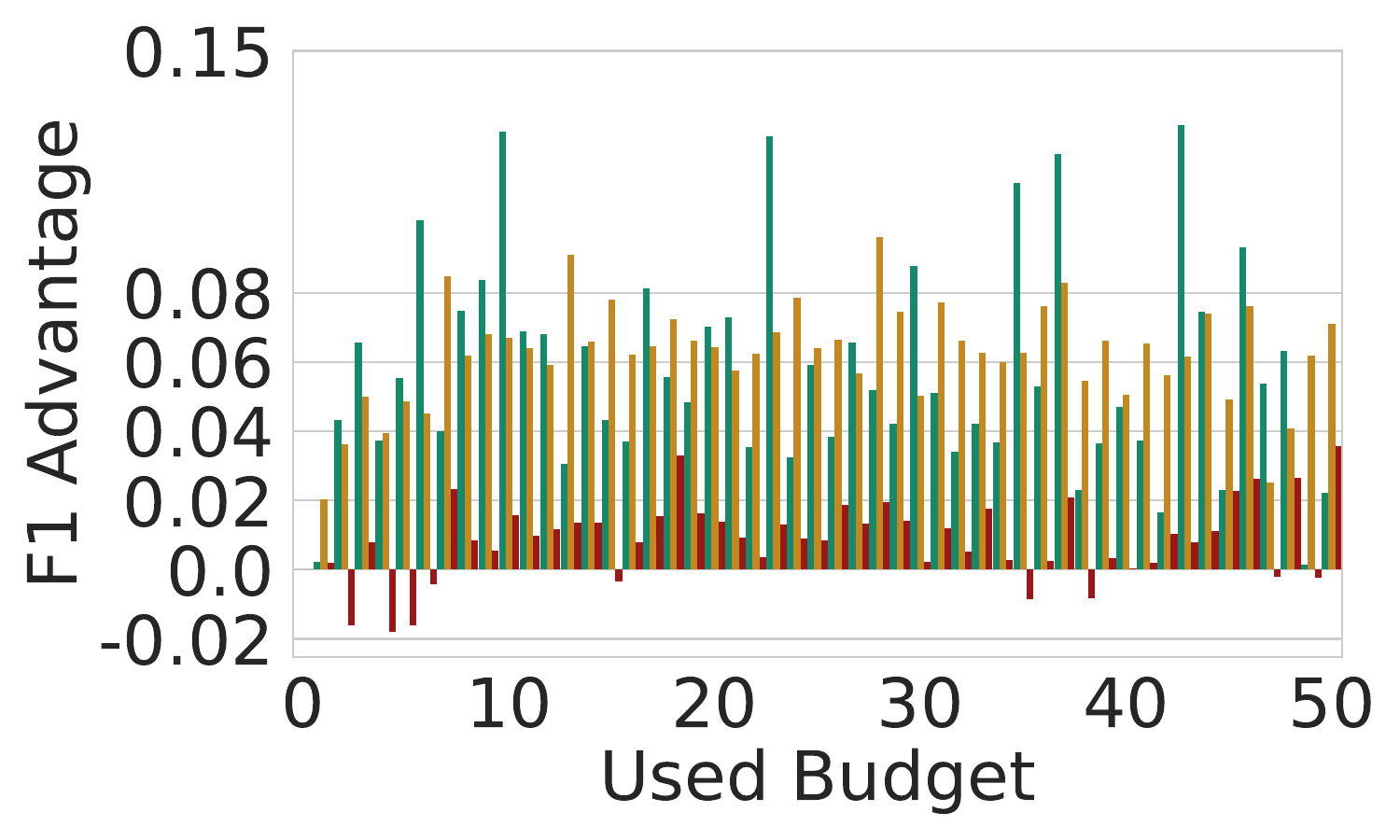}
    \end{subfigure}\hfill
    \begin{subfigure}{0.24\textwidth}
        \includegraphics[width=\linewidth]{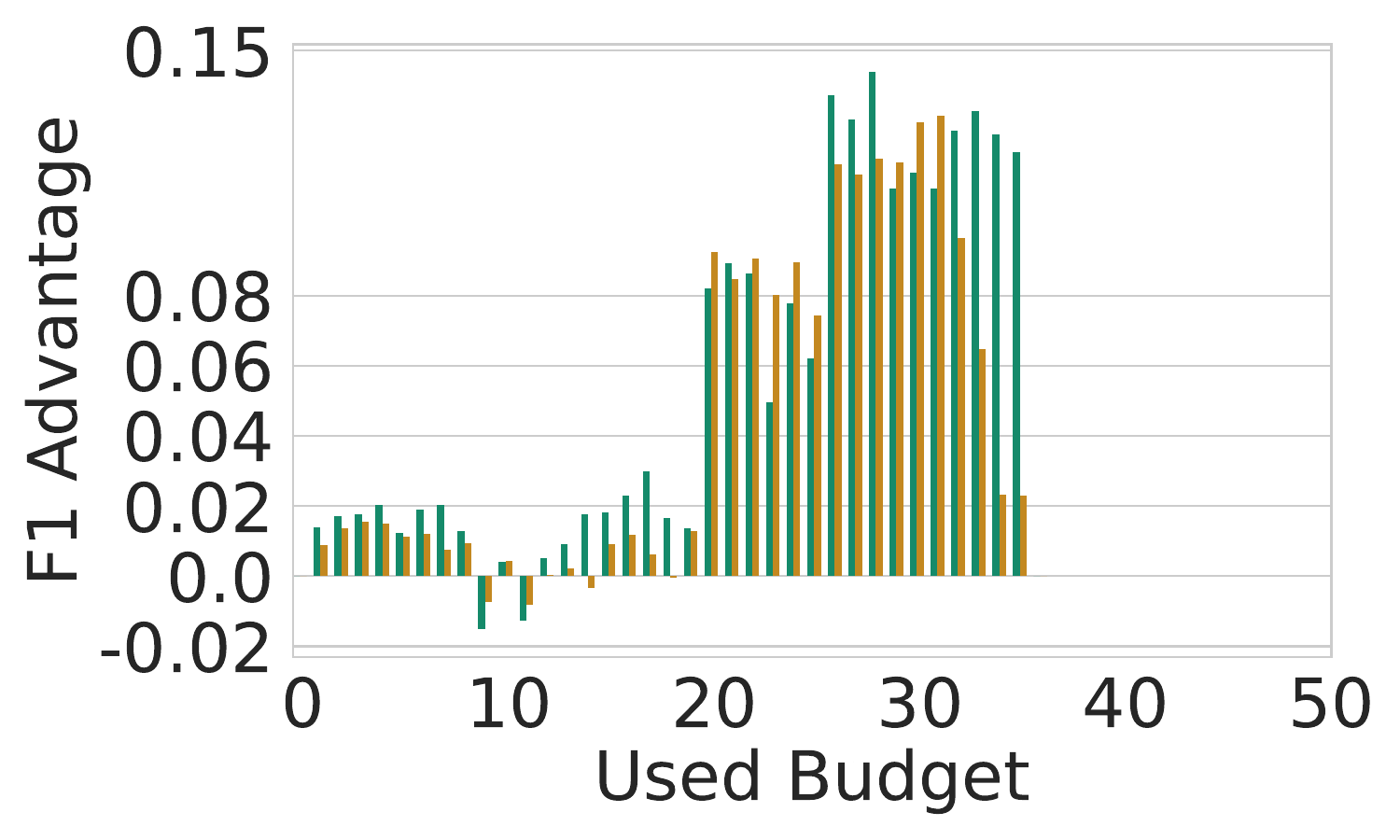}
    \end{subfigure}
    
    \vspace{-0.1em}

    \raisebox{2.\height}{\rotatebox{90}{\textbf{EEG}}}\hspace{0.3em}%
    \begin{subfigure}{0.24\textwidth}
        \centering\raisebox{3.85\height}{\parbox{0.75\linewidth}{\texttt{EEG only contains numerical features.}}}
    \end{subfigure}\hfill
    \begin{subfigure}{0.24\textwidth}
        \includegraphics[width=\linewidth]{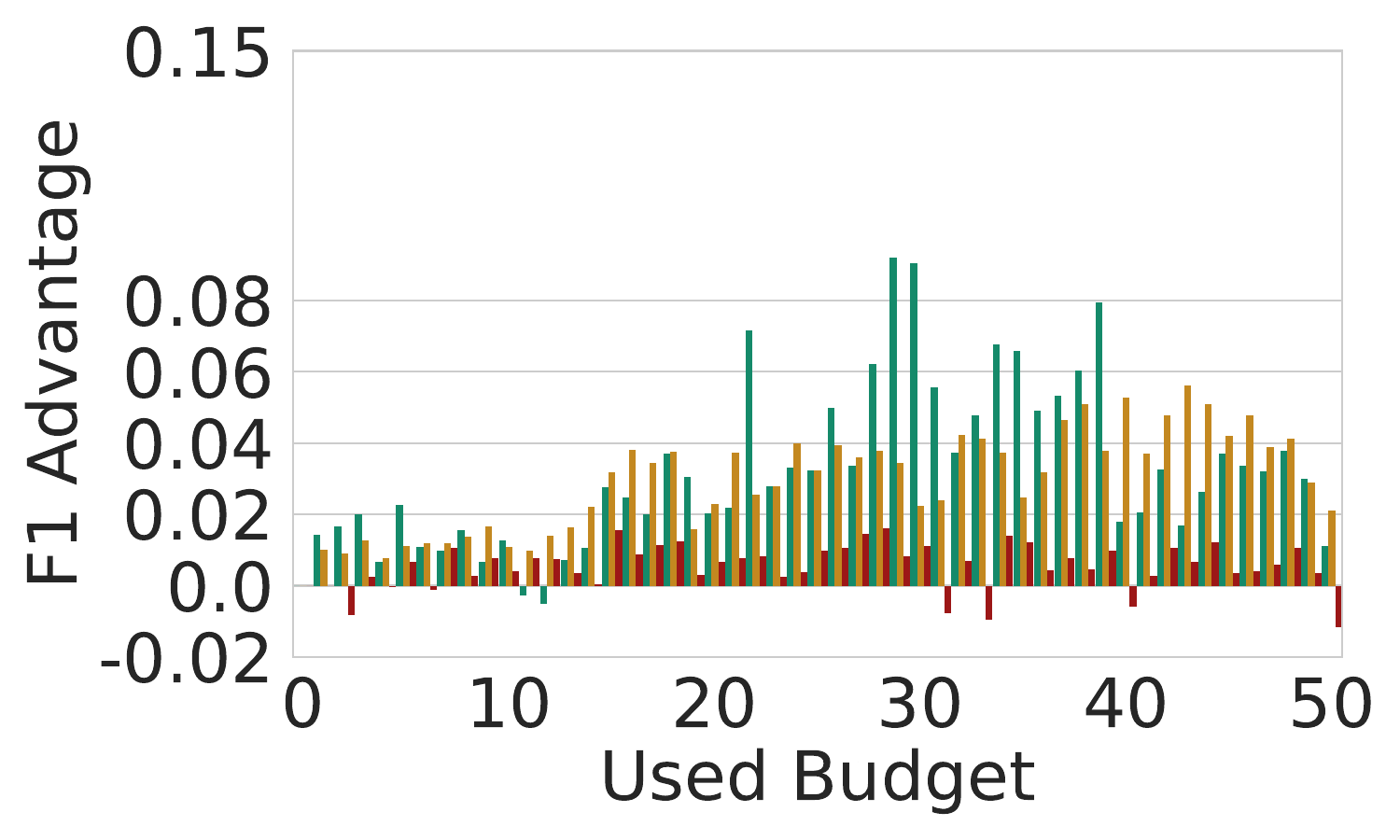}
    \end{subfigure}\hfill
    \begin{subfigure}{0.24\textwidth}
        \includegraphics[width=\linewidth]{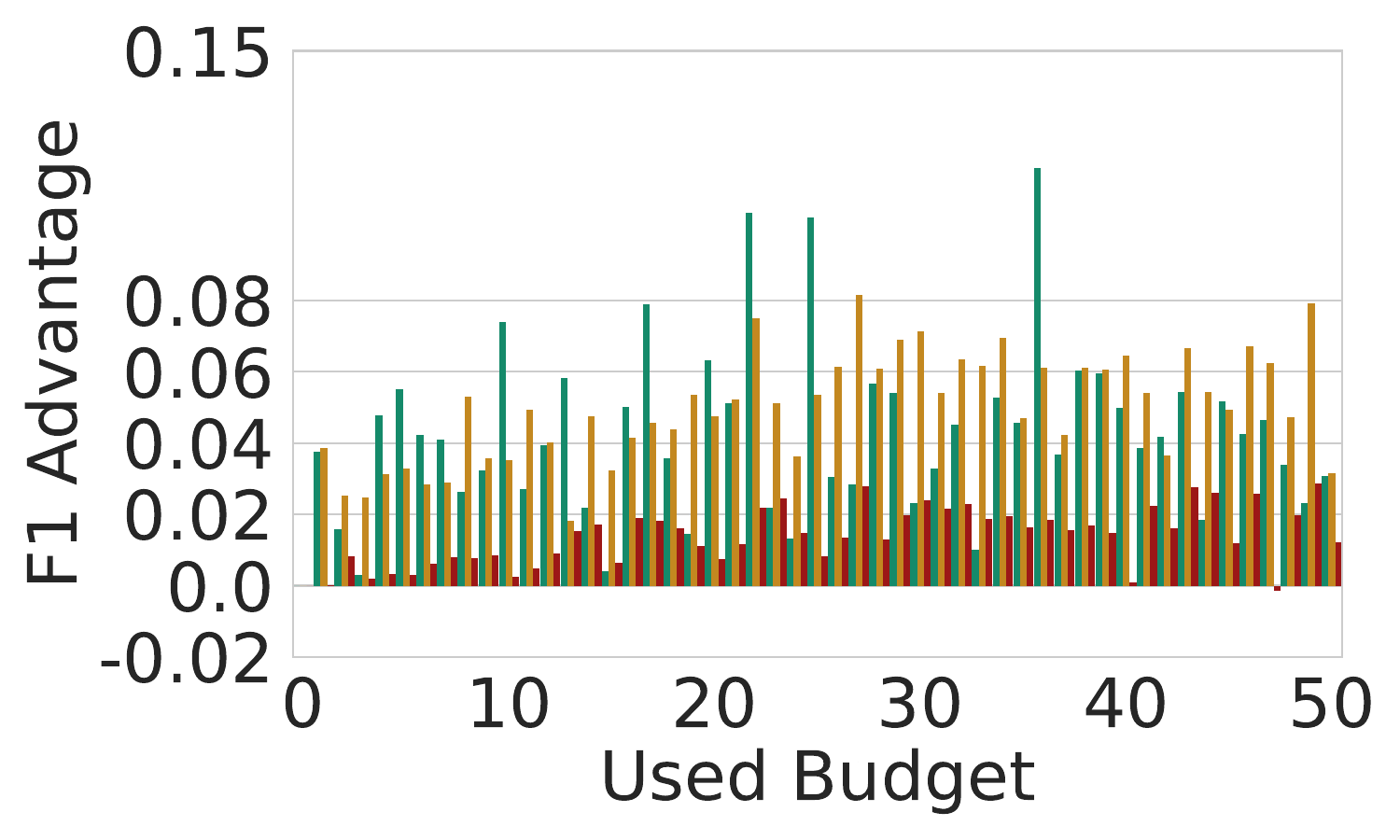}
    \end{subfigure}\hfill
    \begin{subfigure}{0.24\textwidth}
        \includegraphics[width=\linewidth]{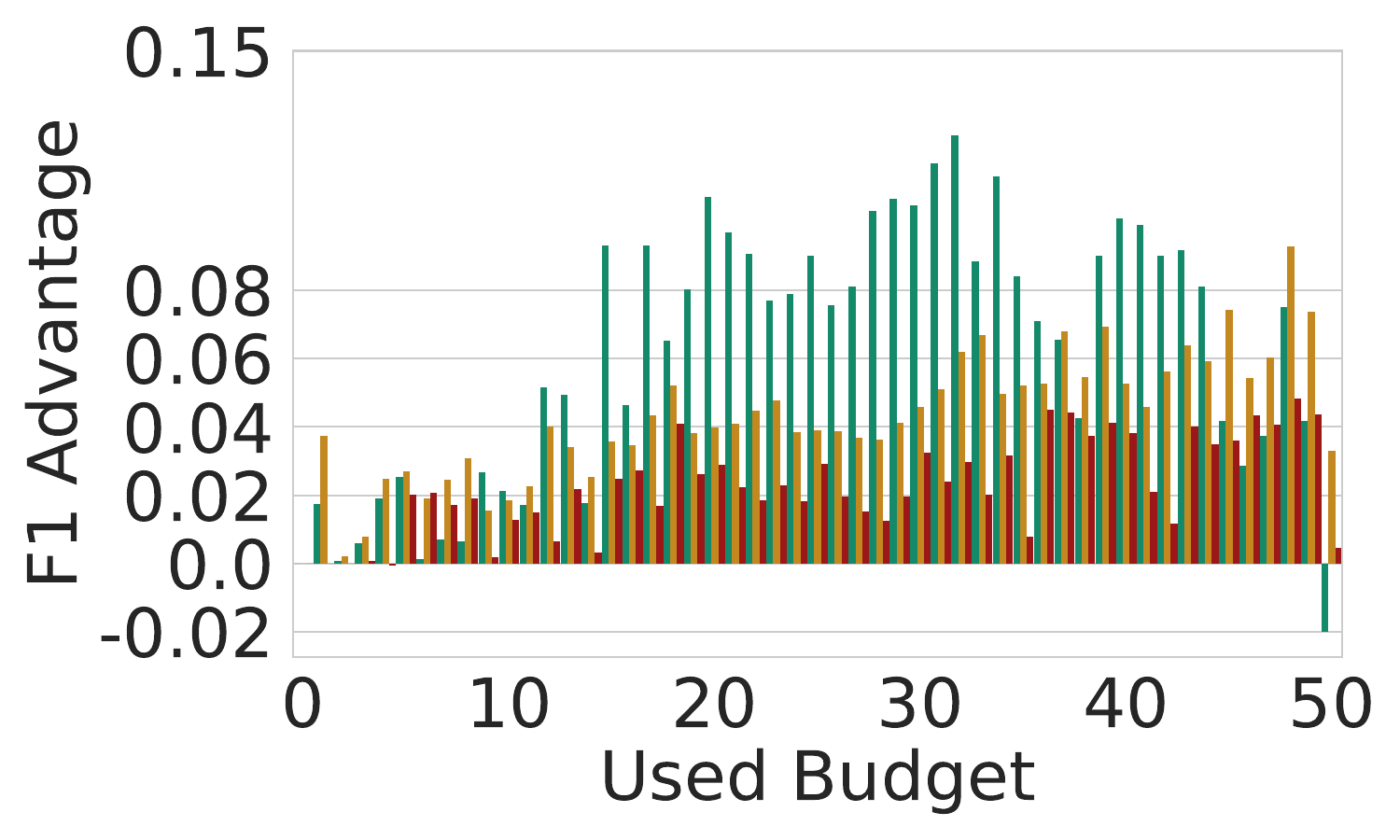}
    \end{subfigure}
    
    \vspace{-0.1em}
    
    \raisebox{1.2\height}{\rotatebox{90}{\textbf{S-Credit}}}\hspace{0.3em}%
    \begin{subfigure}{0.24\textwidth}
        \includegraphics[width=\linewidth]{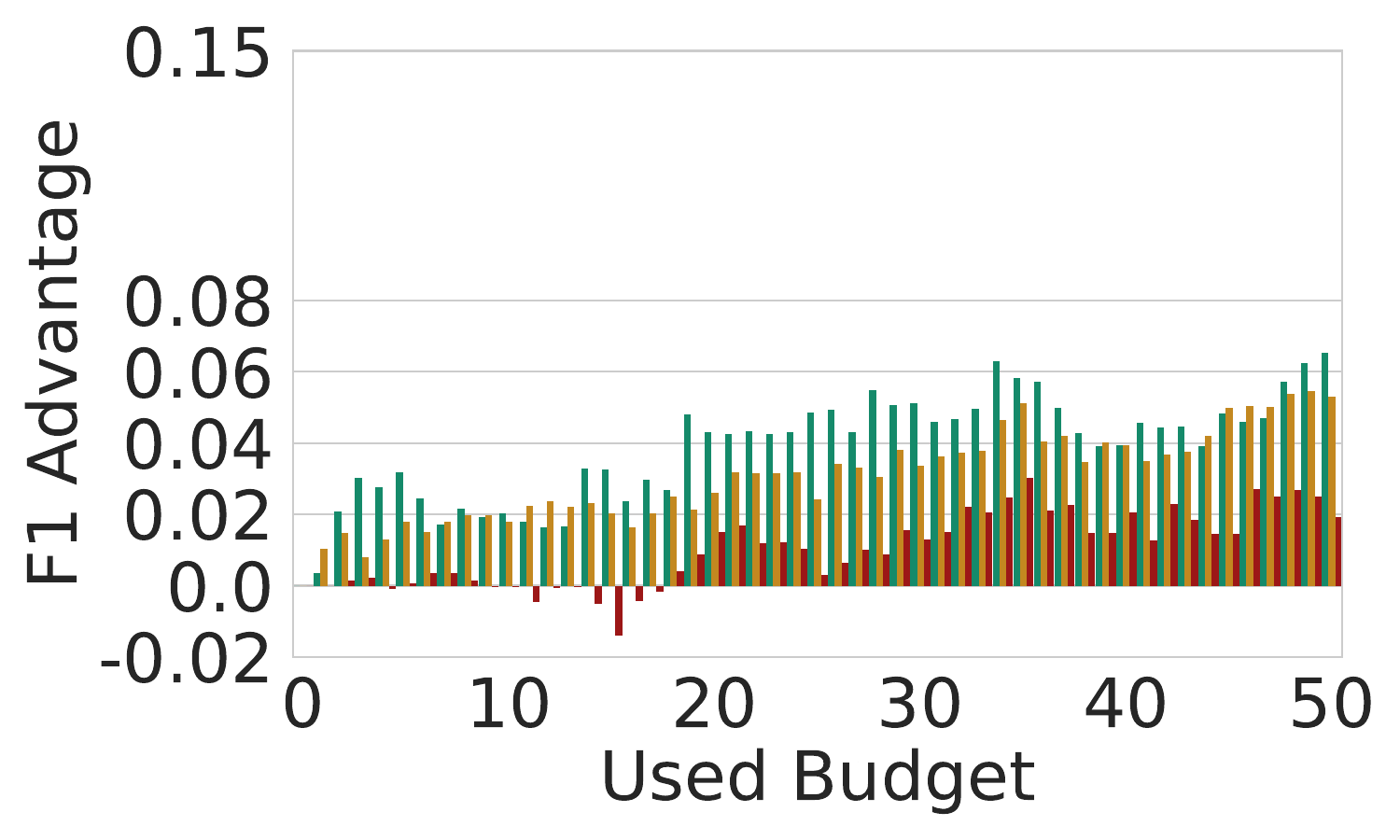}
        \caption{Categorical Shift}
    \end{subfigure}\hfill
    \begin{subfigure}{0.24\textwidth}
        \includegraphics[width=\linewidth]{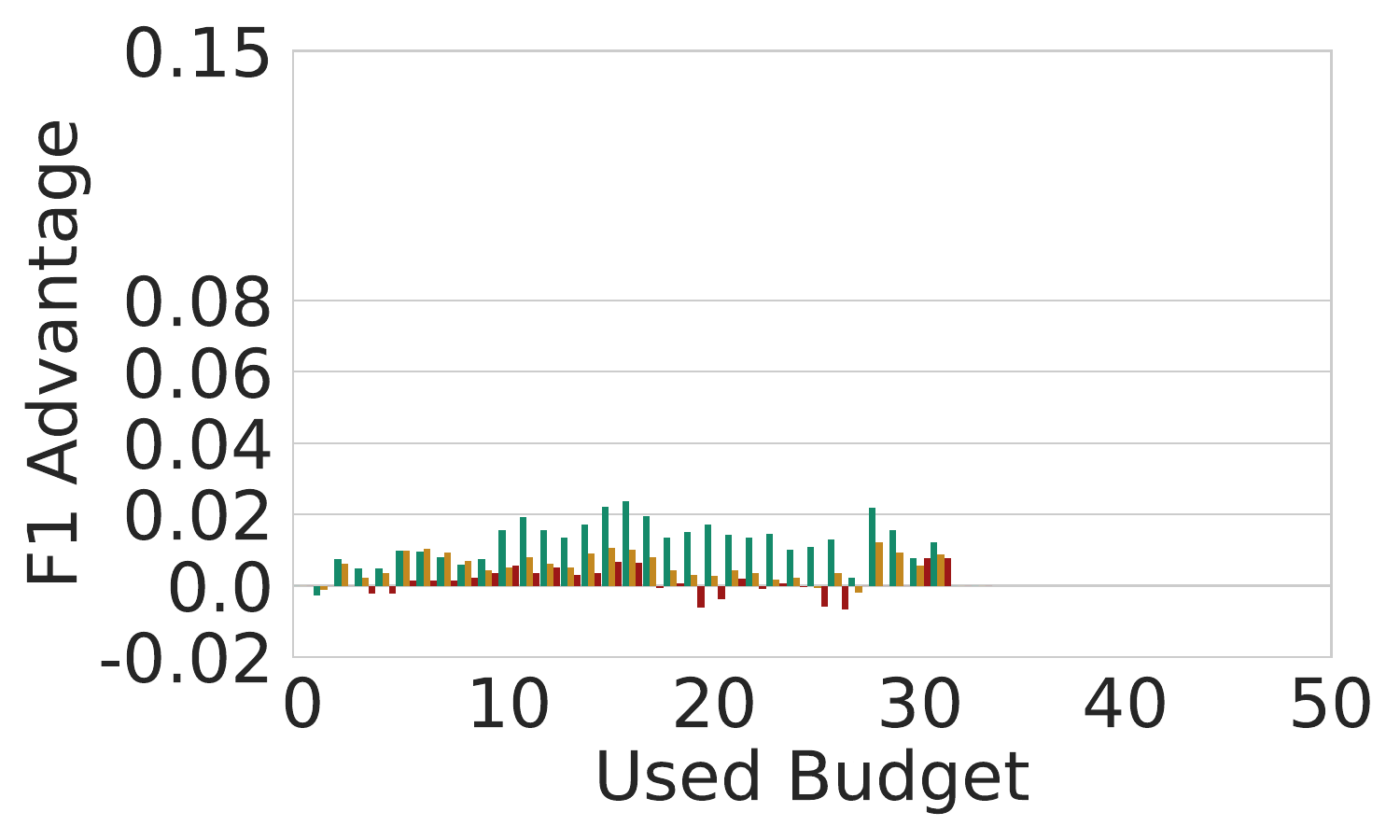}
        \caption{Gaussian Noise}
    \end{subfigure}\hfill
    \begin{subfigure}{0.24\textwidth}
        \includegraphics[width=\linewidth]{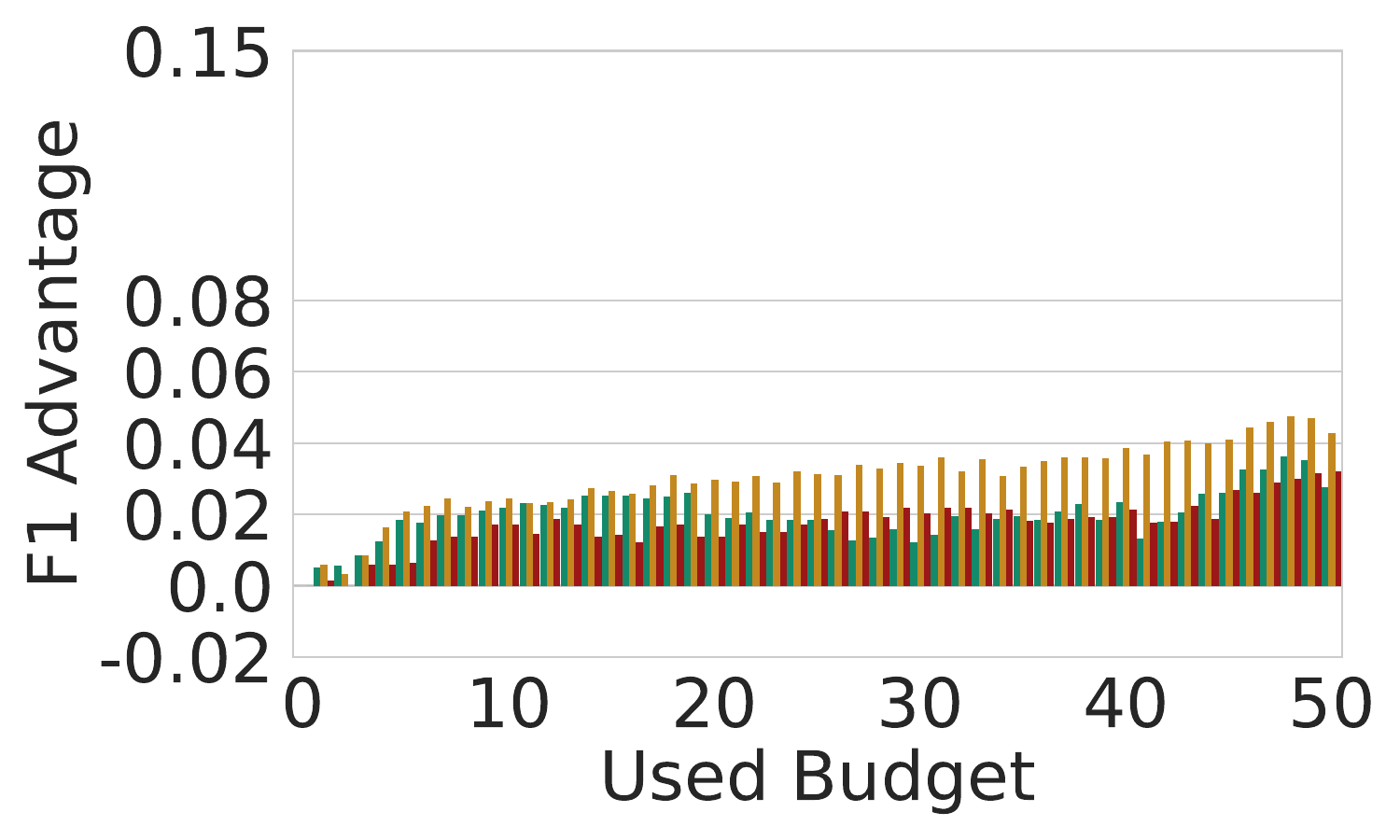}
        \caption{Missing Values}
    \end{subfigure}\hfill
    \begin{subfigure}{0.24\textwidth}
        \includegraphics[width=\linewidth]{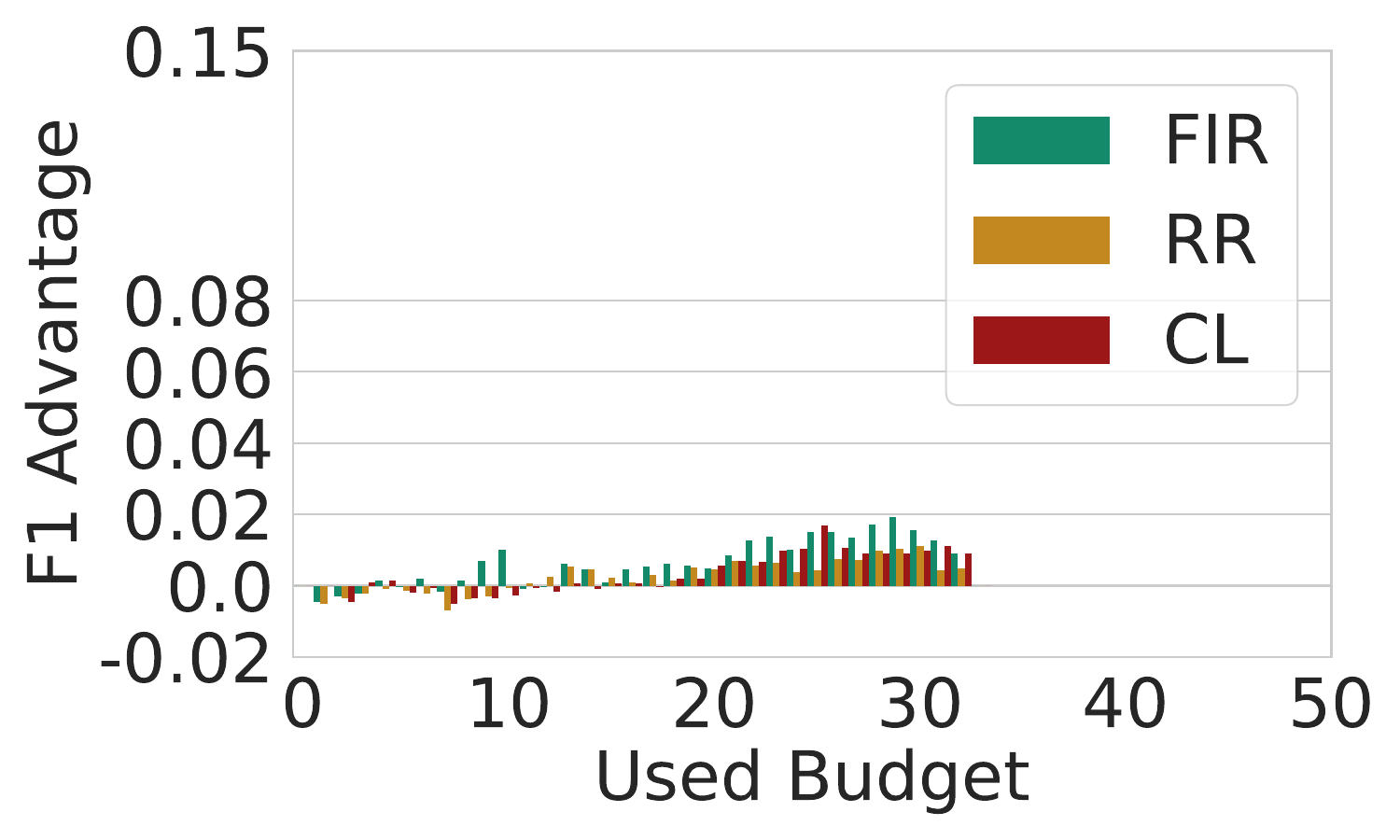}
        \caption{Scaling}
    \end{subfigure}
    \caption{Comparison of \systemname with FIR and RR for SVM across error types.}
    \label{fig:agg_bl_results_svm}
\end{figure*}

\begin{figure*}[h!]
    \centering
    \begin{subfigure}{0.24\textwidth}
        \includegraphics[width=\linewidth]{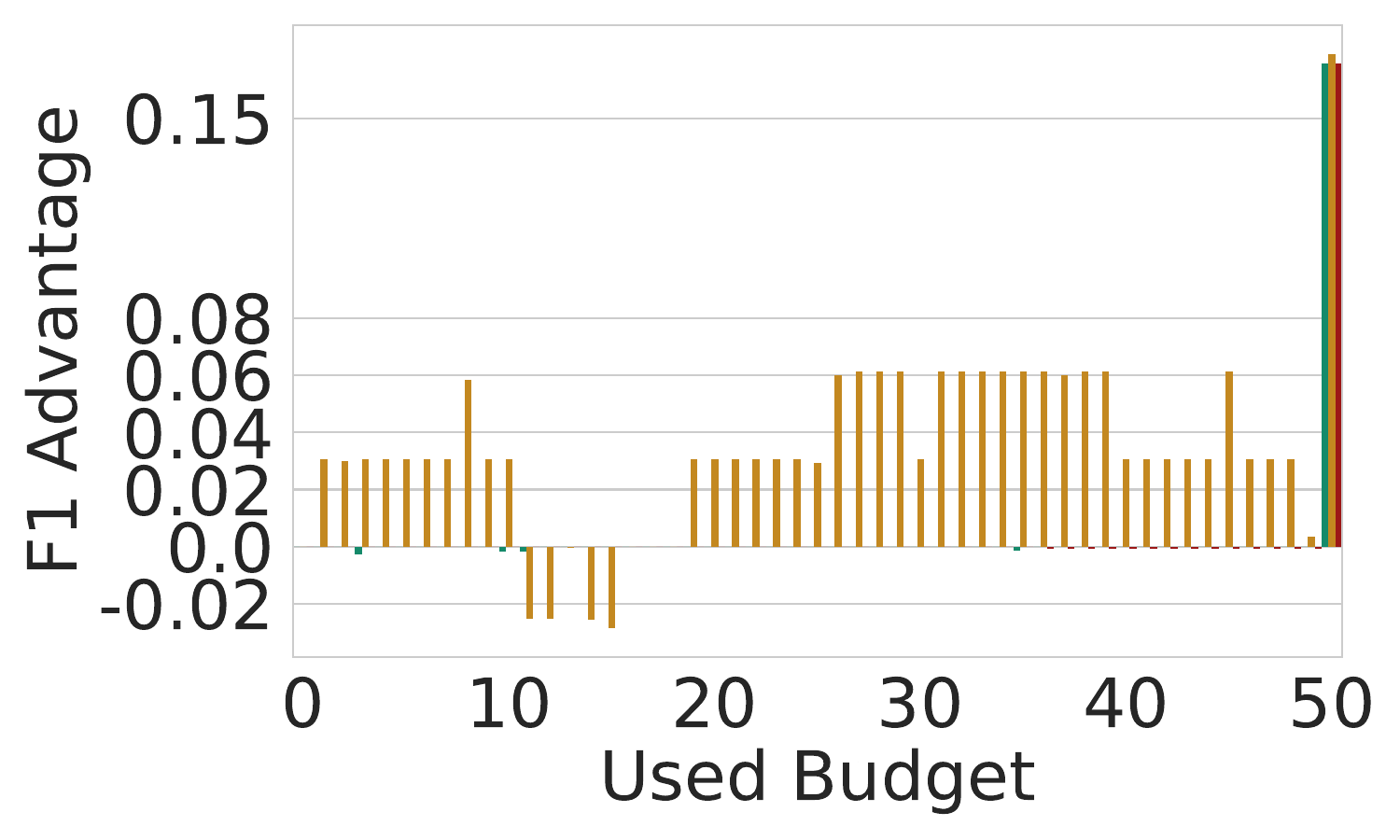}
        \caption{Airbnb - Scaling}
    \end{subfigure}
    \begin{subfigure}{0.24\textwidth}
        \includegraphics[width=\linewidth]{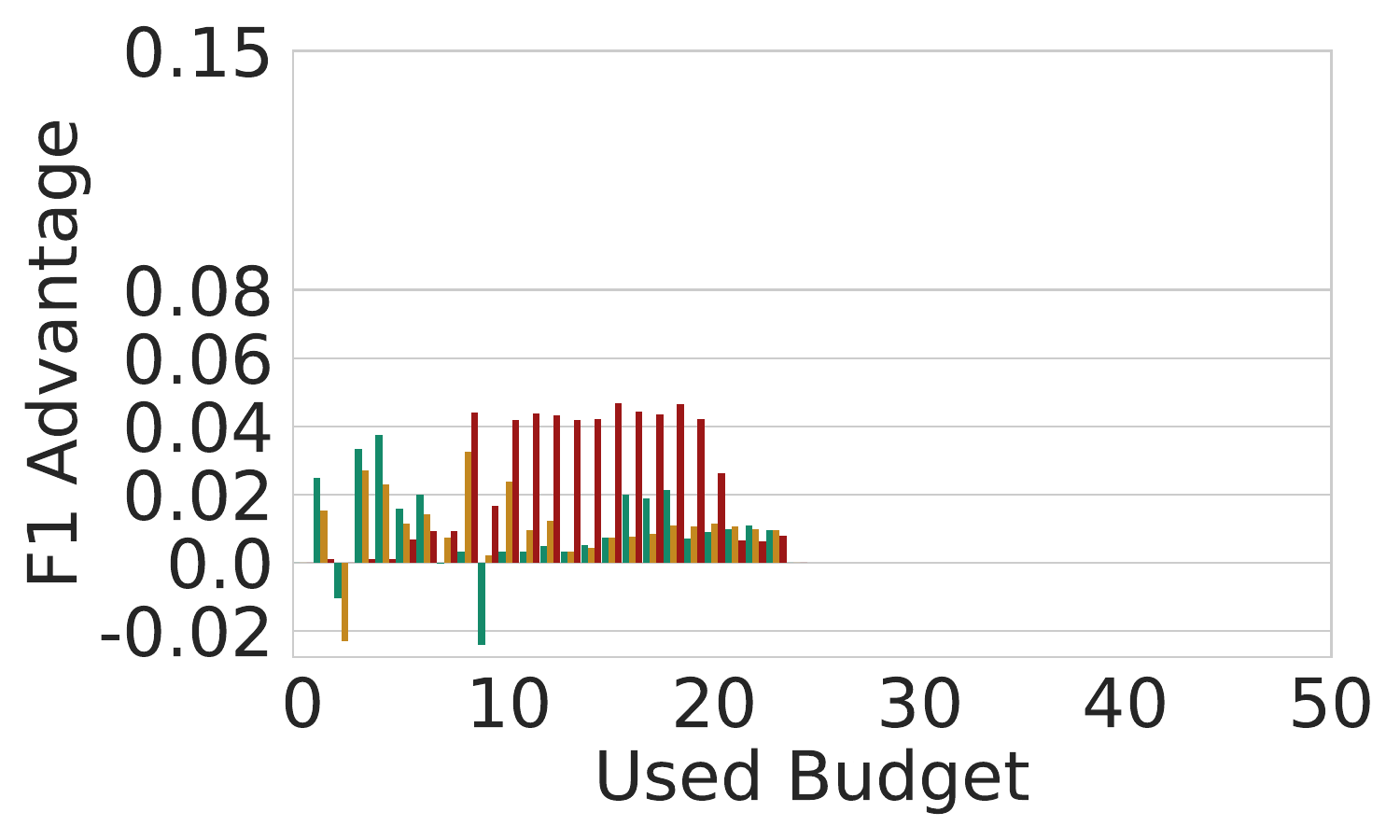}
        \caption{Credit - Scaling}
    \end{subfigure}
    \begin{subfigure}{0.24\textwidth}
        \includegraphics[width=\linewidth]{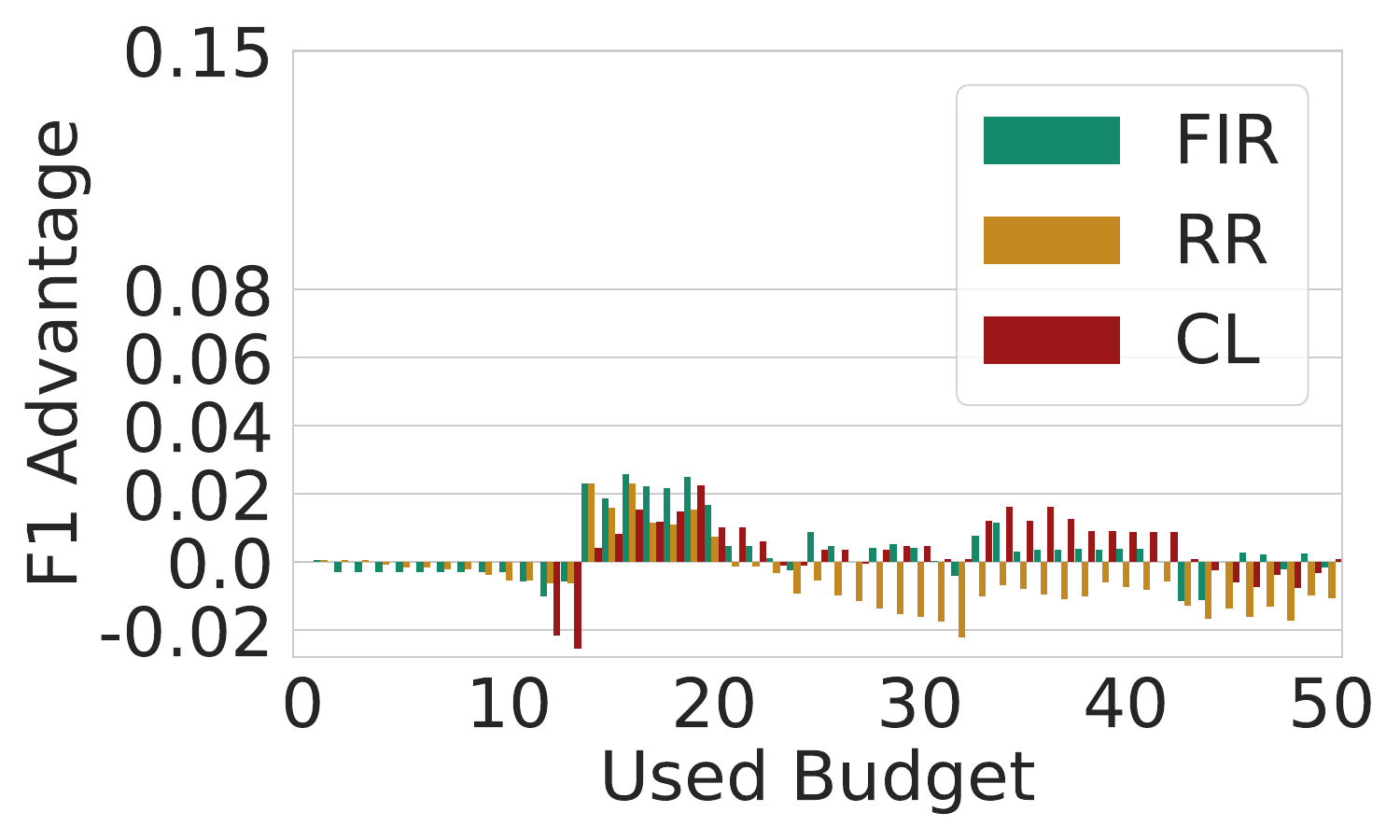}
        \caption{Titanic - Missing Values}
    \end{subfigure}
    \caption{Comparison of~\systemname with the baselines FIR, RR and CL for SVM across error types, for datasets from CleanML.}
    \label{fig:agg_bl_results_svm2}
\end{figure*}

\clearpage
\section{Comparison to AC for a single error type}

\begin{figure*}[h!]
    \centering
    \raisebox{1.4\height}{\rotatebox{90}{\textbf{CMC}}}\hspace{0.3em}%
    \begin{subfigure}{0.24\textwidth}
        \includegraphics[width=\linewidth]{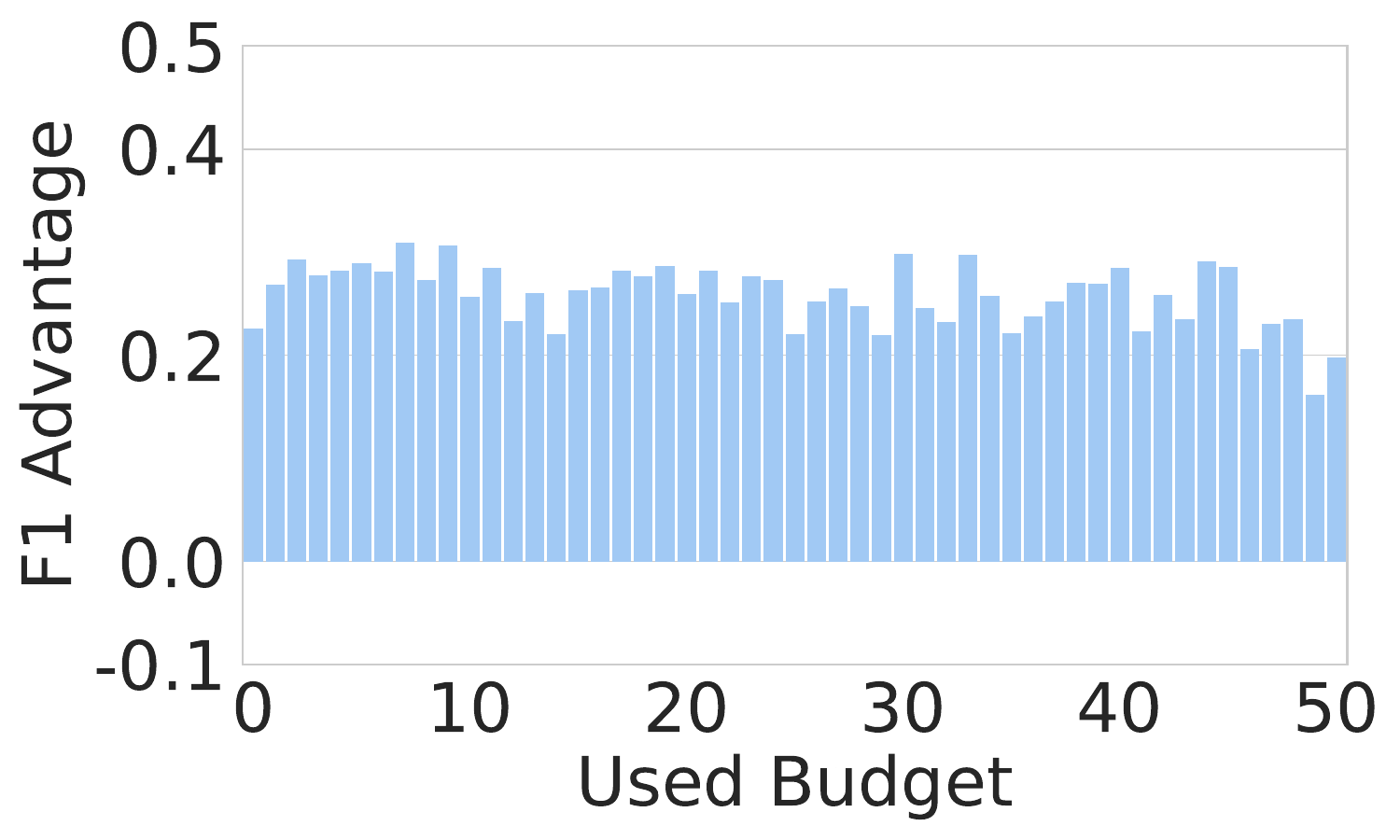}
    \end{subfigure}\hfill
    \begin{subfigure}{0.24\textwidth}
        \includegraphics[width=\linewidth]{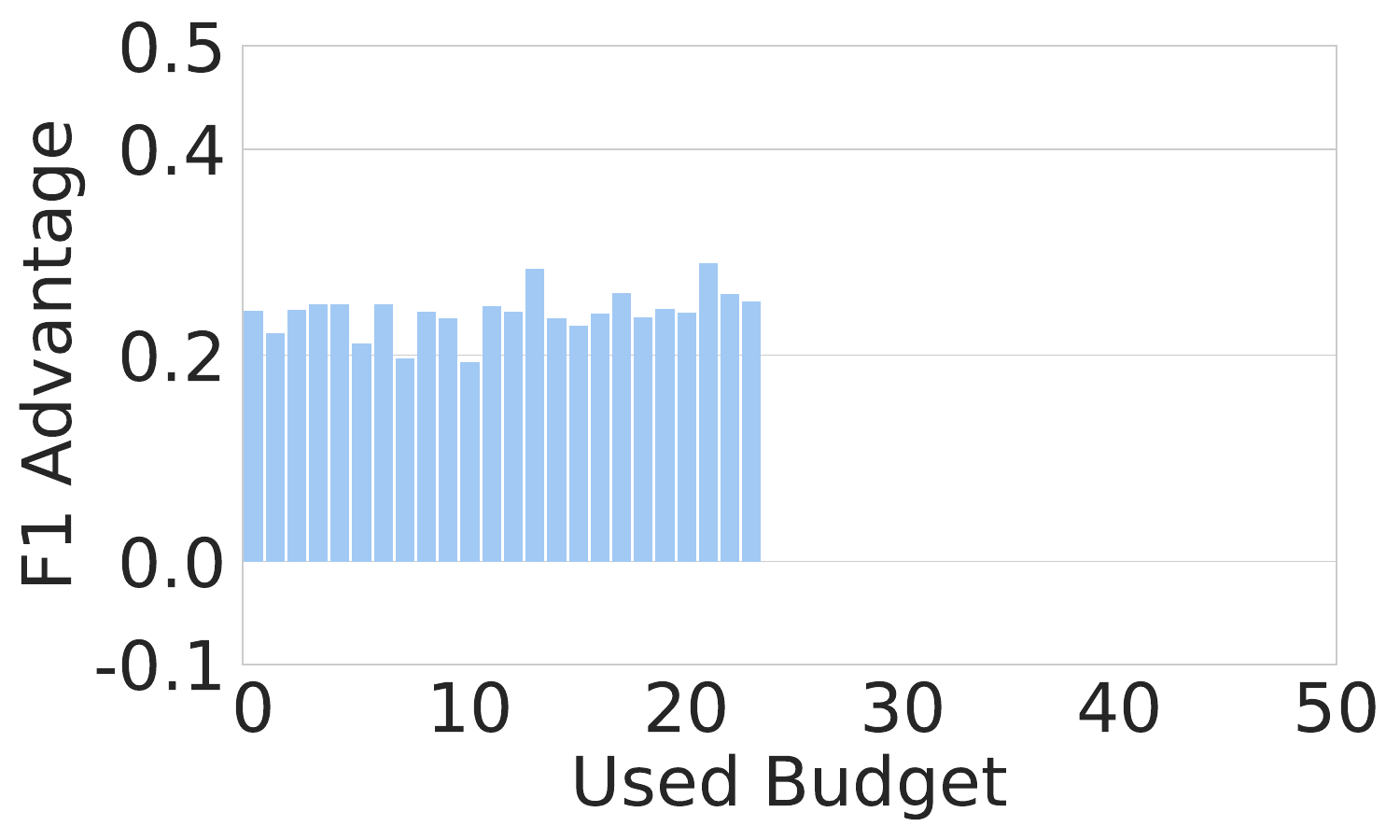}
    \end{subfigure}\hfill
    \begin{subfigure}{0.24\textwidth}
        \includegraphics[width=\linewidth]{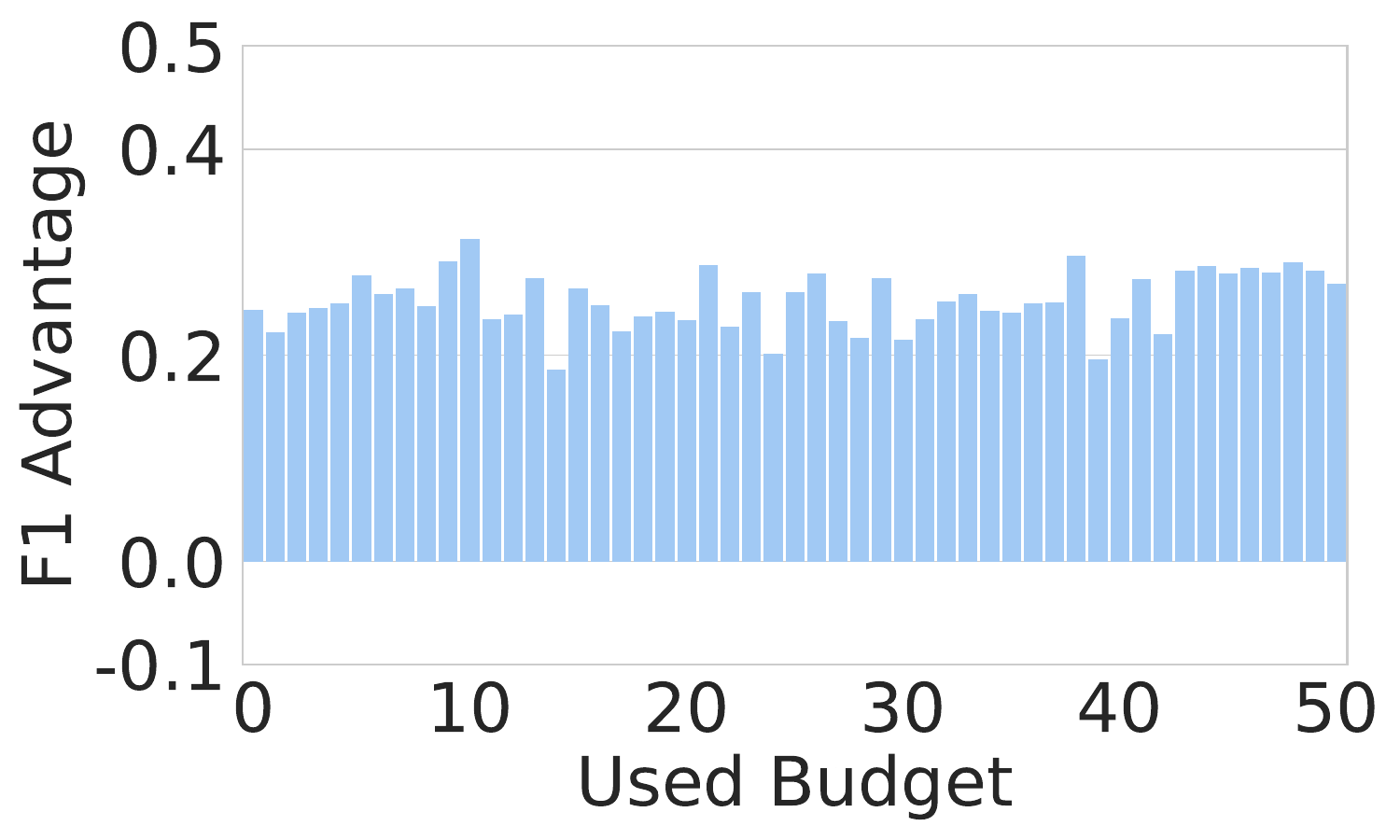}
    \end{subfigure}\hfill
    \begin{subfigure}{0.24\textwidth}
        \includegraphics[width=\linewidth]{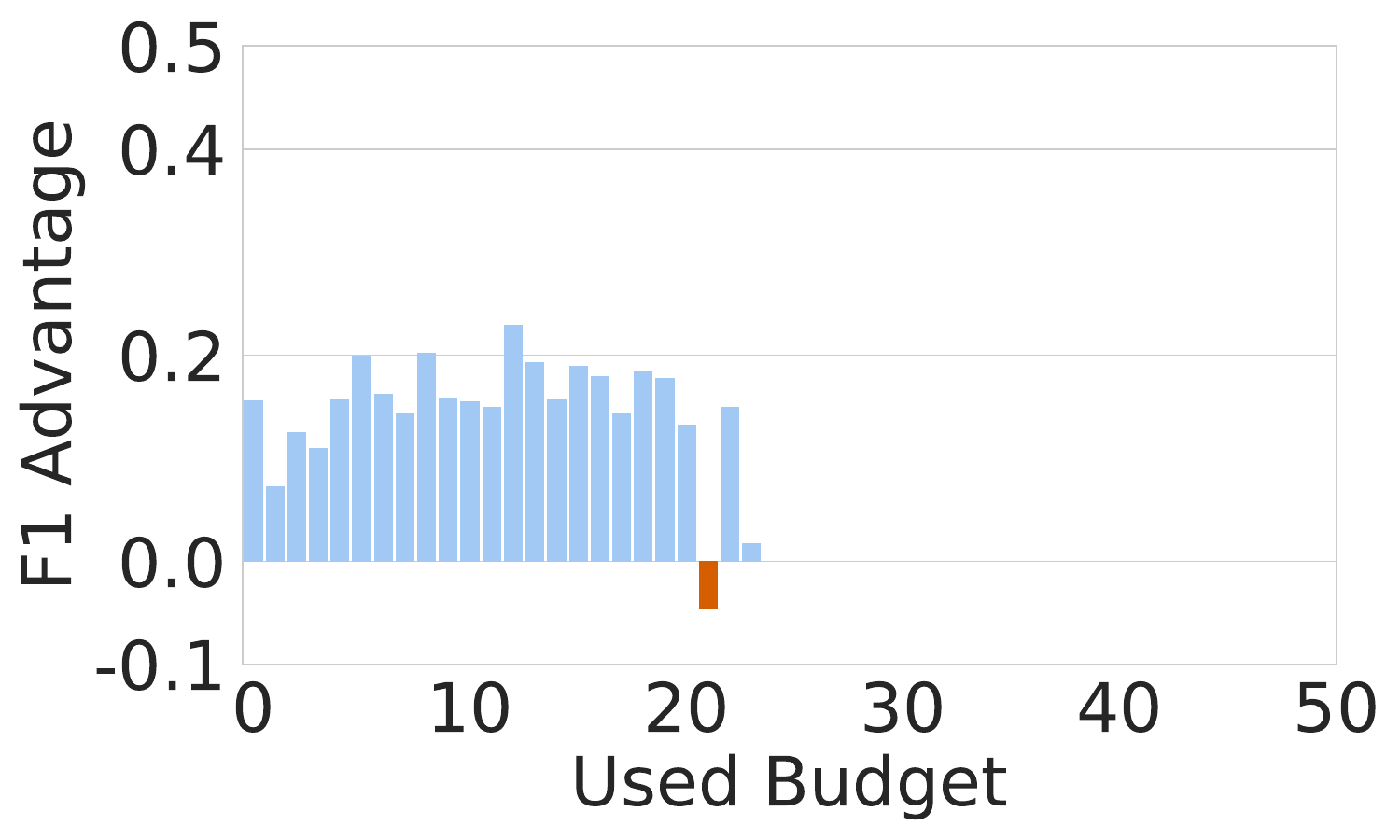}
    \end{subfigure}
    
    \vspace{-0.1em}

        \raisebox{1.2\height}{\rotatebox{90}{\textbf{Churn}}}\hspace{0.3em}%
    \begin{subfigure}{0.24\textwidth}
        \includegraphics[width=\linewidth]{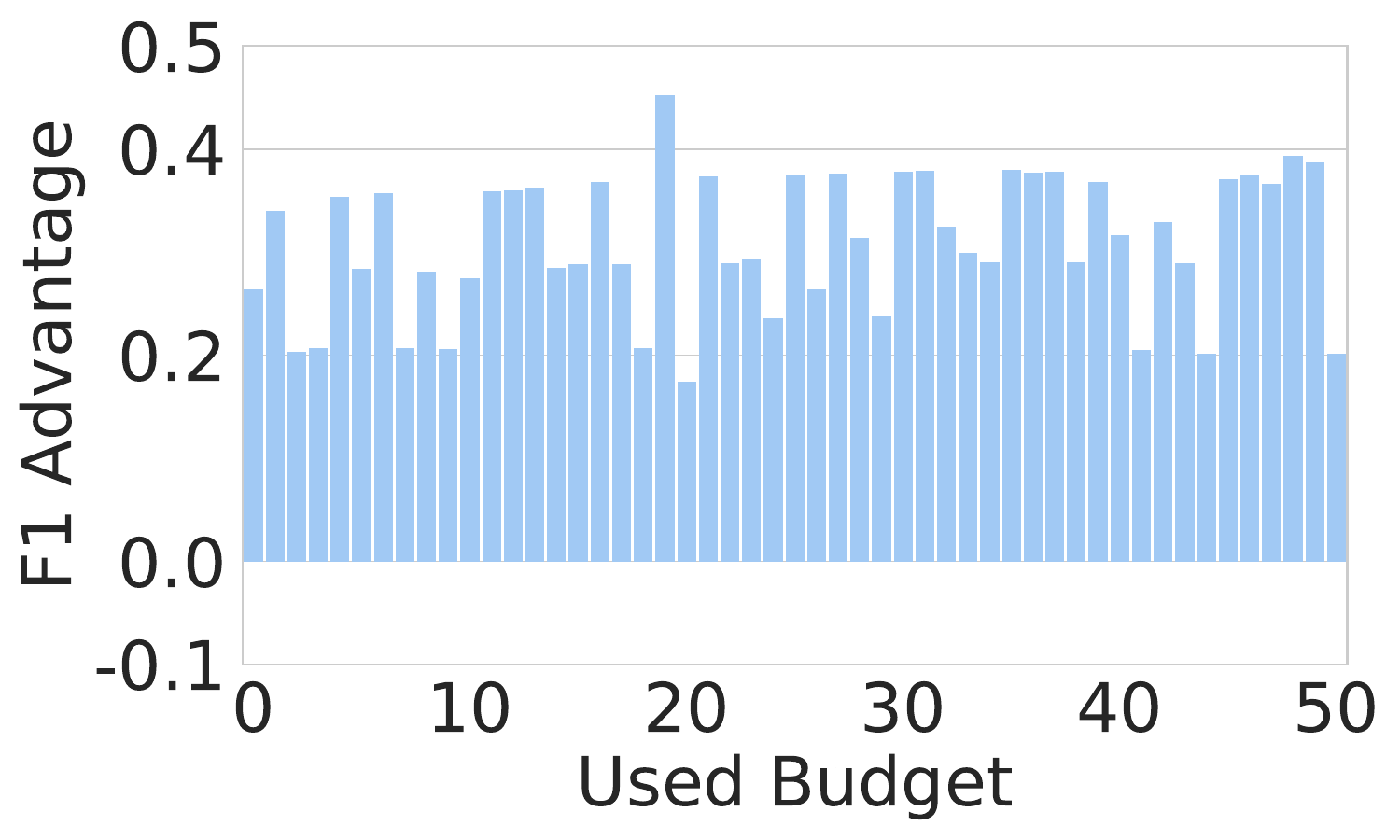}
    \end{subfigure}\hfill
    \begin{subfigure}{0.24\textwidth}
        \includegraphics[width=\linewidth]{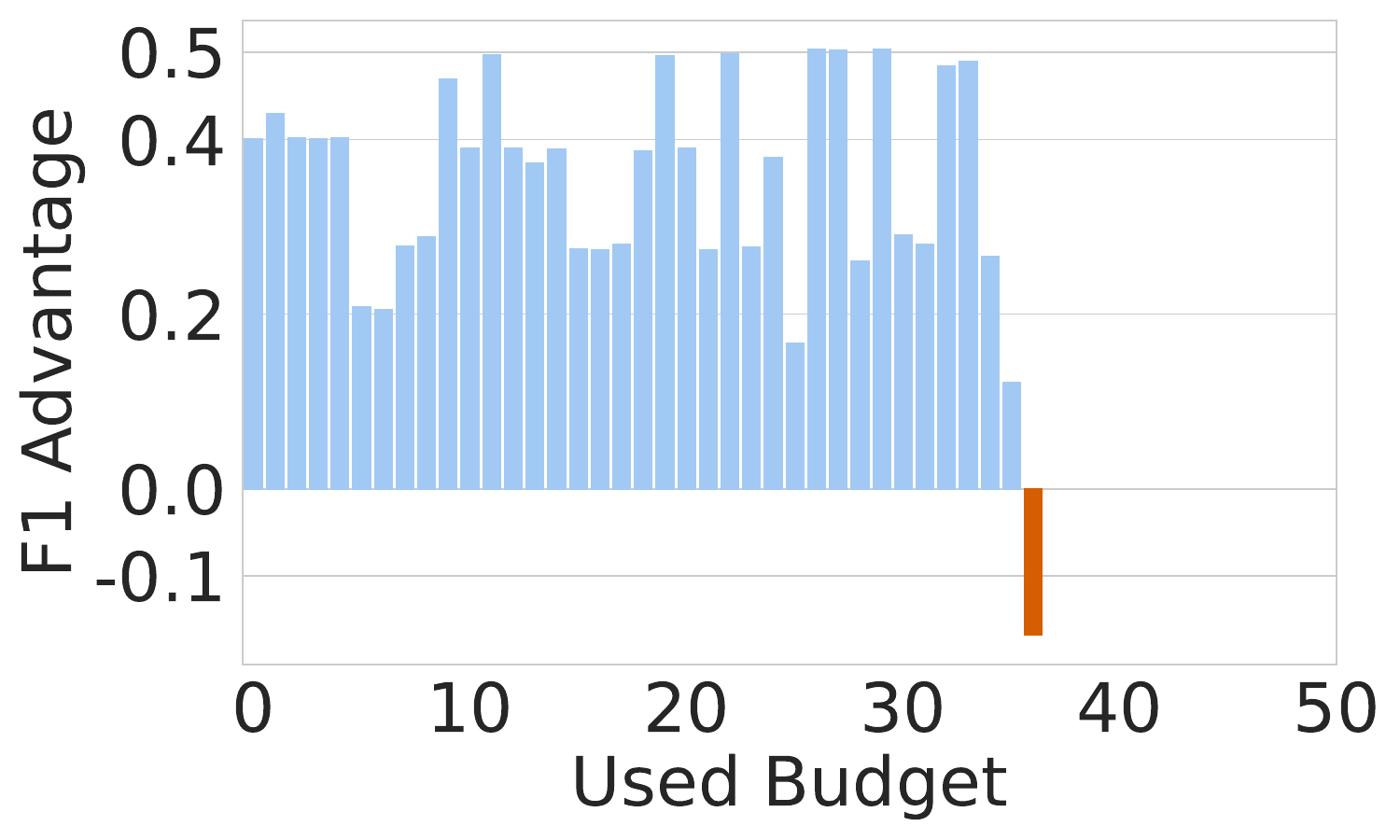}
    \end{subfigure}\hfill
    \begin{subfigure}{0.24\textwidth}
        \includegraphics[width=\linewidth]{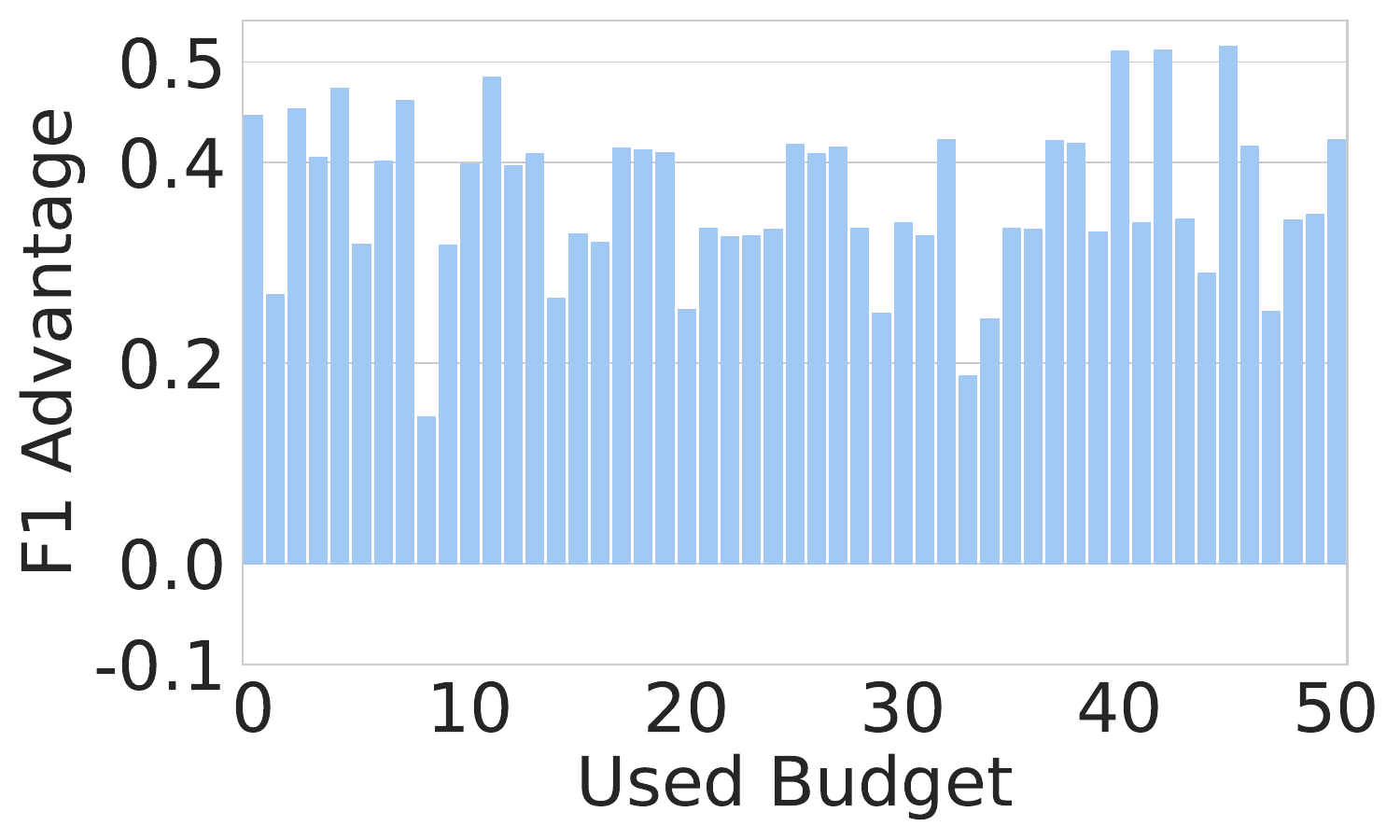}
    \end{subfigure}\hfill
    \begin{subfigure}{0.24\textwidth}
        \includegraphics[width=\linewidth]{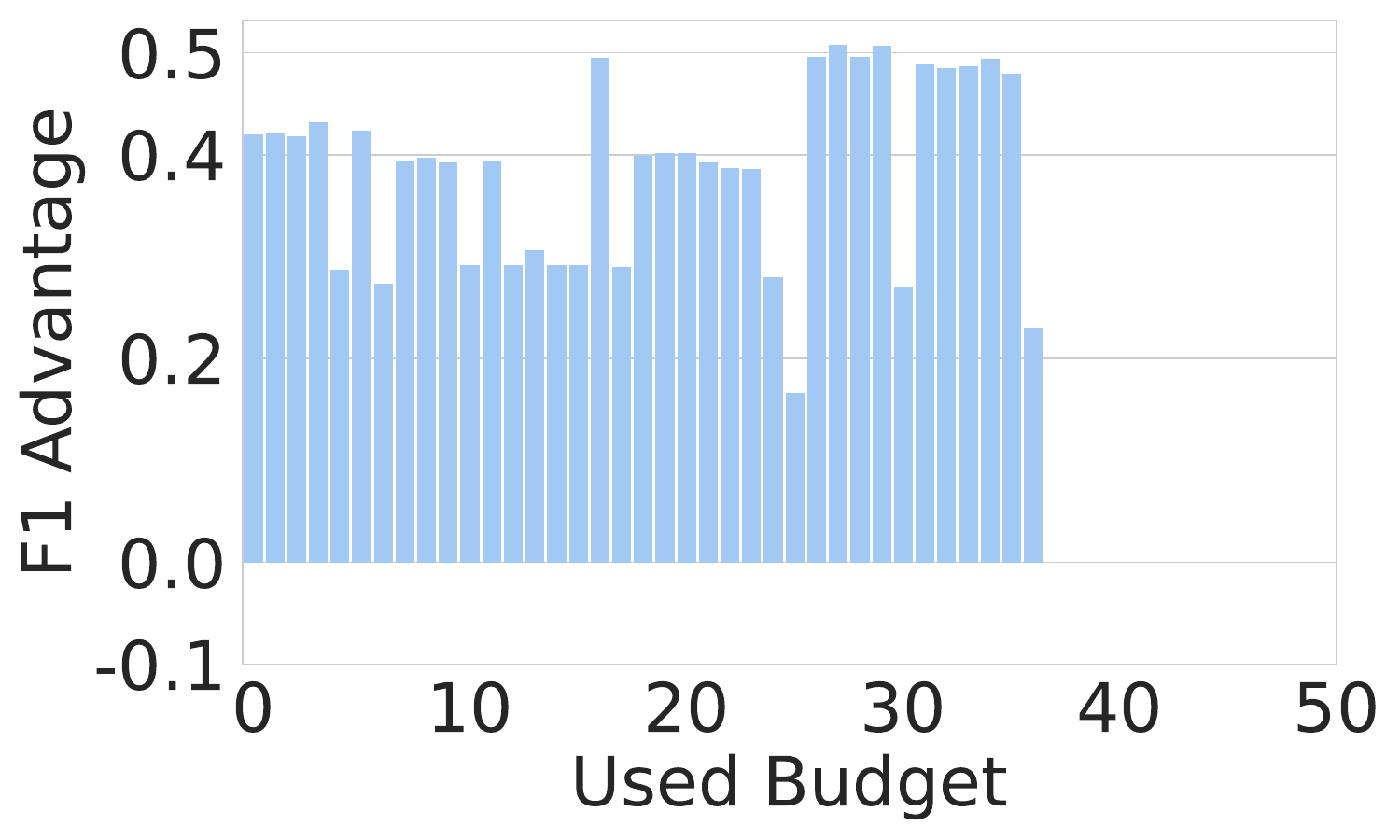}
    \end{subfigure}
    
    \vspace{-0.1em}

    \raisebox{2.\height}{\rotatebox{90}{\textbf{EEG}}}\hspace{0.3em}%
    \begin{subfigure}{0.24\textwidth}
        \centering\raisebox{3.85\height}{\parbox{0.75\linewidth}{\texttt{EEG only contains numerical features.}}}
    \end{subfigure}\hfill
    \begin{subfigure}{0.24\textwidth}
        \includegraphics[width=\linewidth]{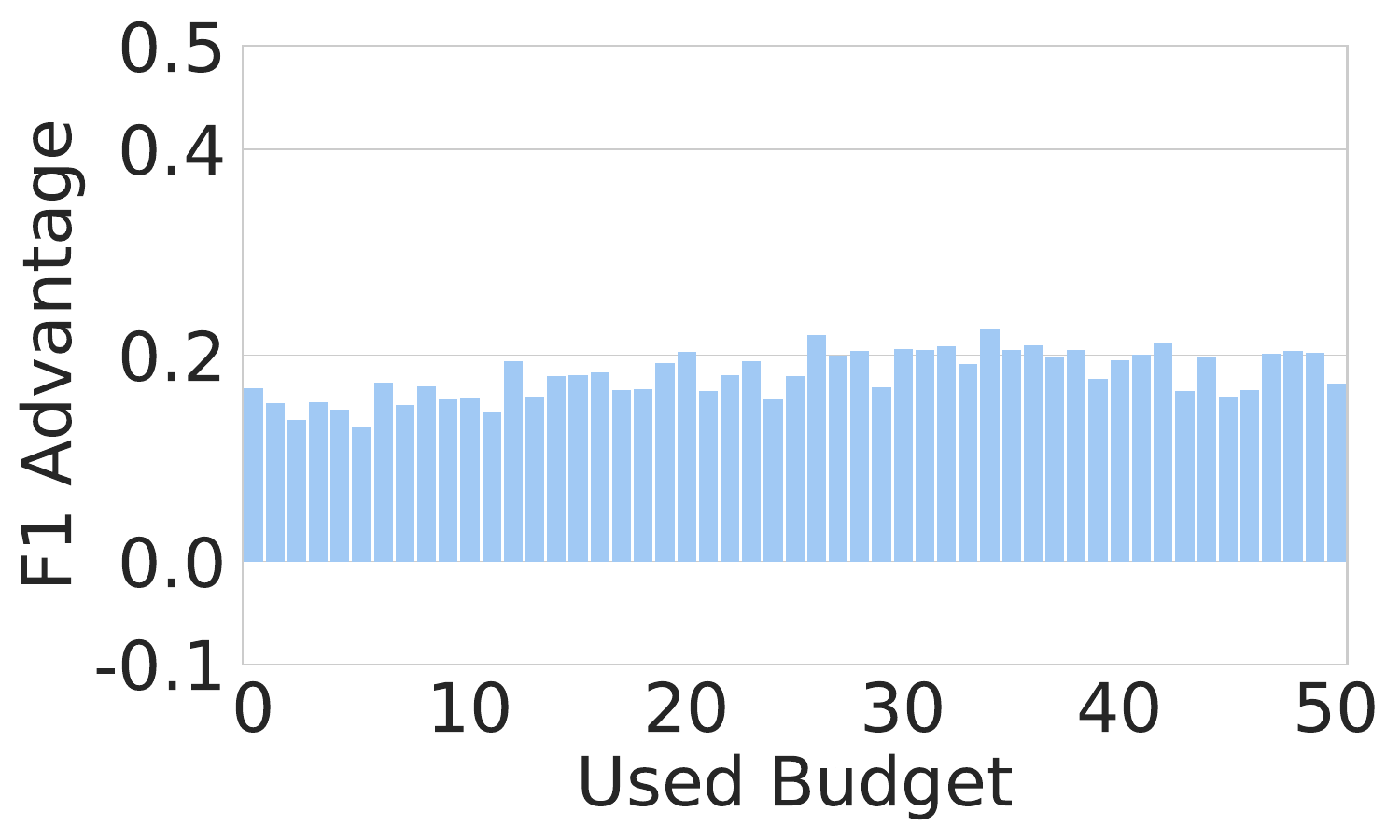}
    \end{subfigure}\hfill
    \begin{subfigure}{0.24\textwidth}
        \includegraphics[width=\linewidth]{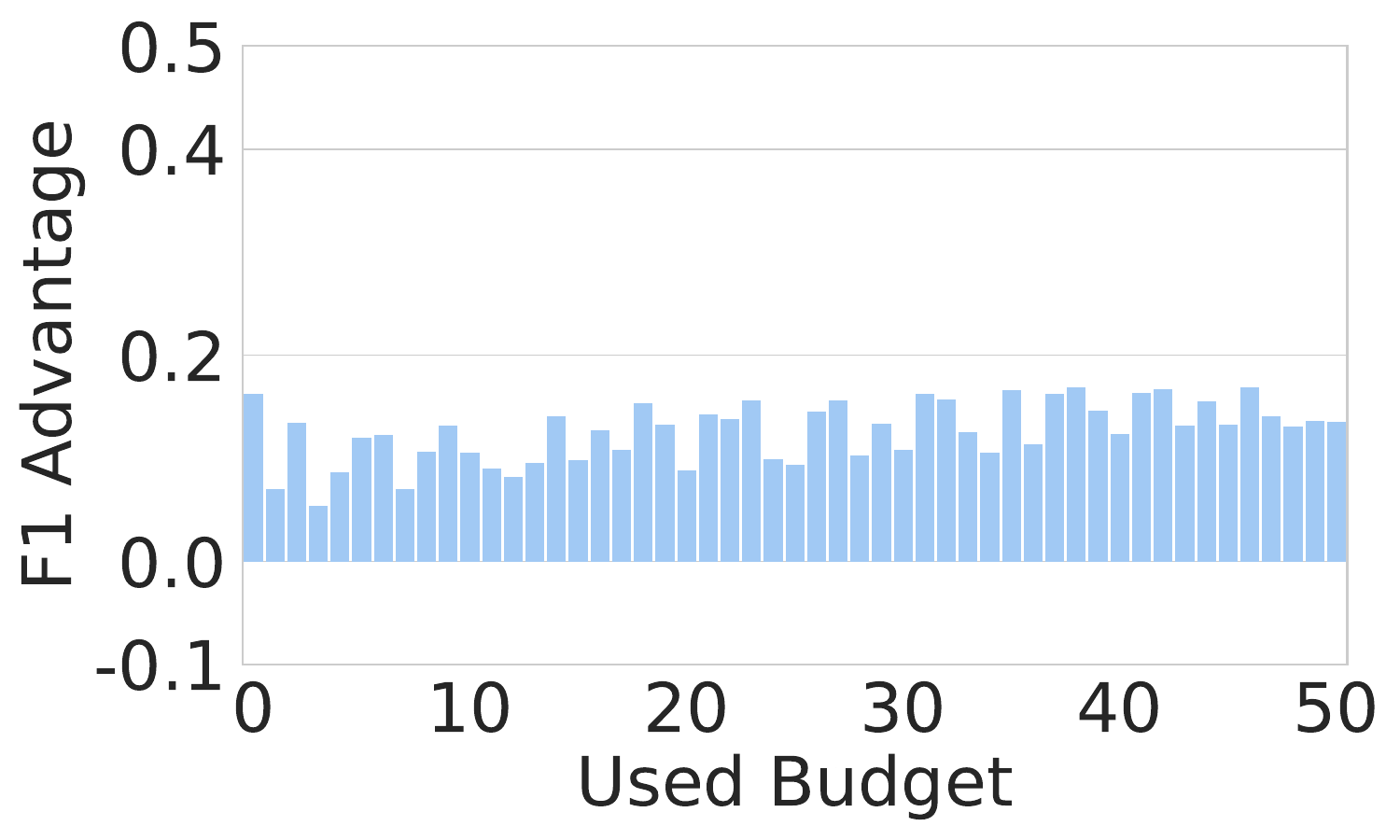}
    \end{subfigure}\hfill
    \begin{subfigure}{0.24\textwidth}
        \includegraphics[width=\linewidth]{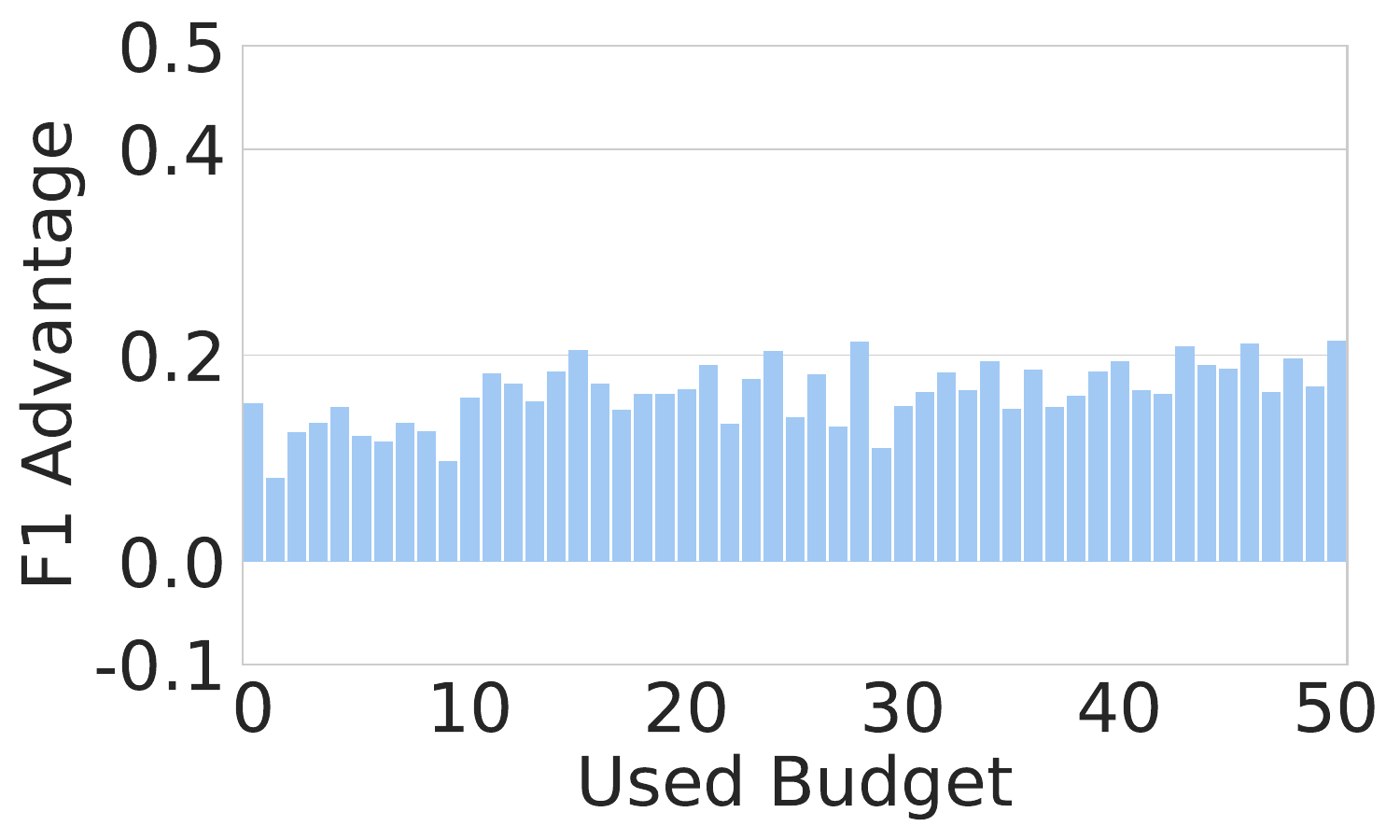}
    \end{subfigure}
    
    \vspace{-0.1em}
    
    \raisebox{1.2\height}{\rotatebox{90}{\textbf{S-Credit}}}\hspace{0.3em}%
    \begin{subfigure}{0.24\textwidth}
        \includegraphics[width=\linewidth]{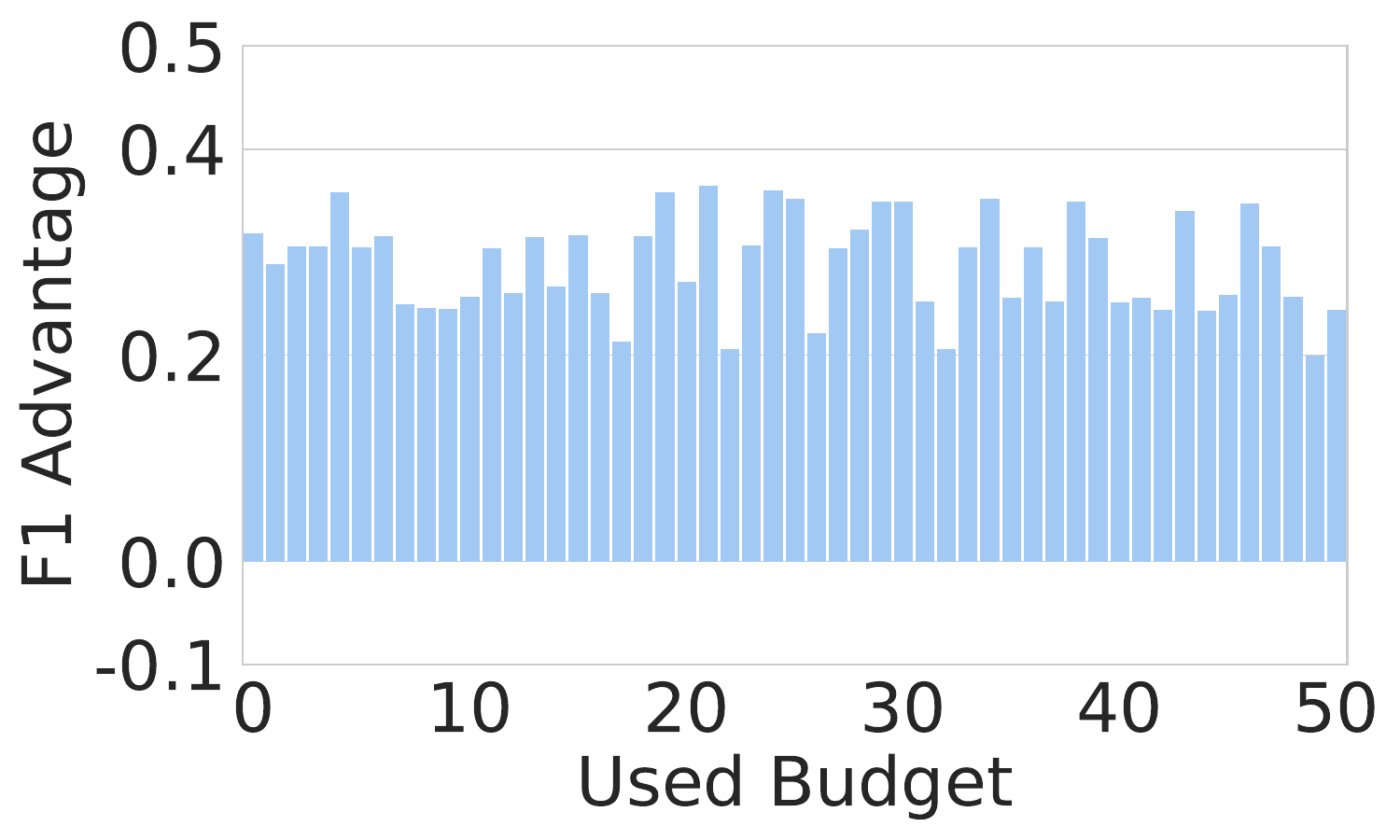}
        \caption{Categorical Shift}
    \end{subfigure}\hfill
    \begin{subfigure}{0.24\textwidth}
        \includegraphics[width=\linewidth]{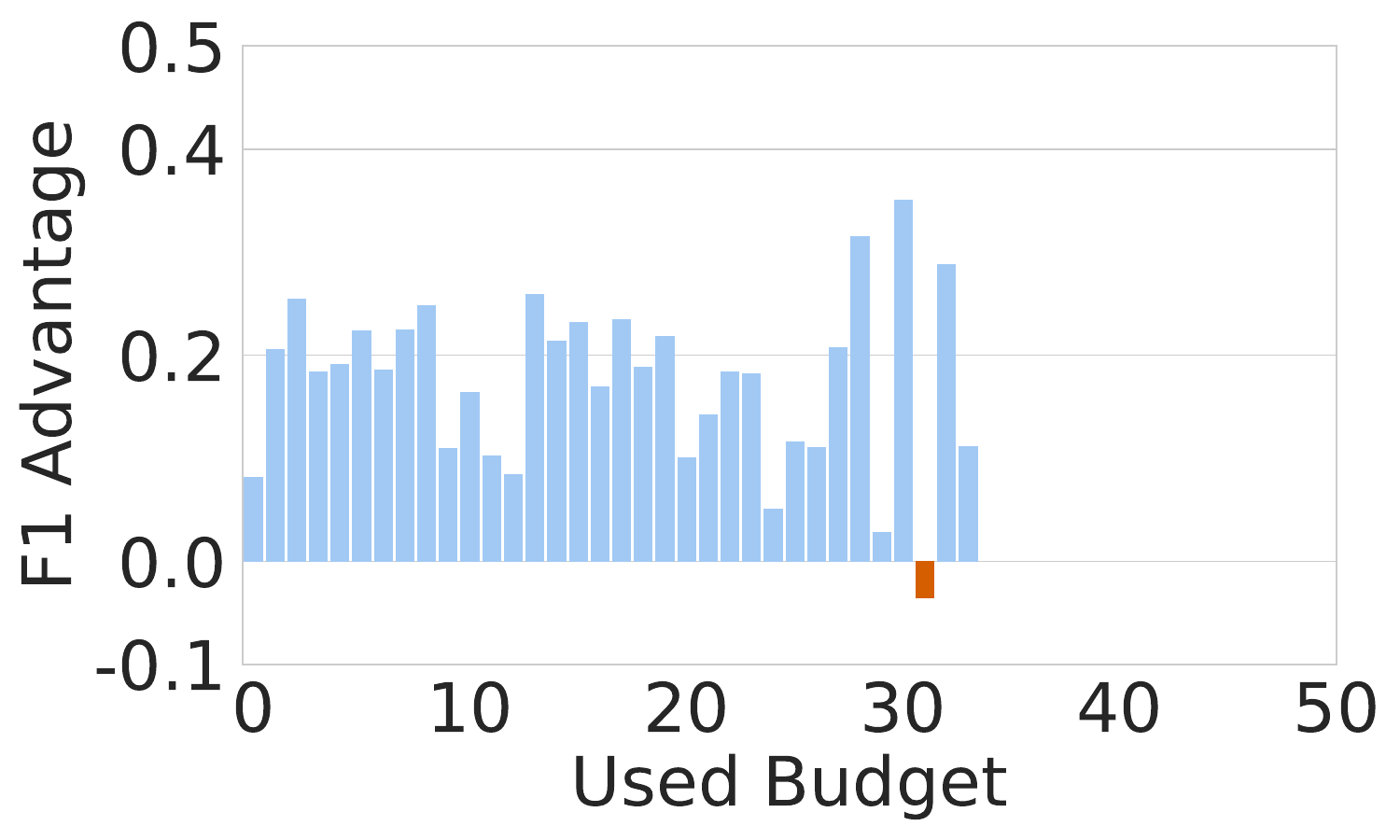}
        \caption{Gaussian Noise}
    \end{subfigure}\hfill
    \begin{subfigure}{0.24\textwidth}
        \includegraphics[width=\linewidth]{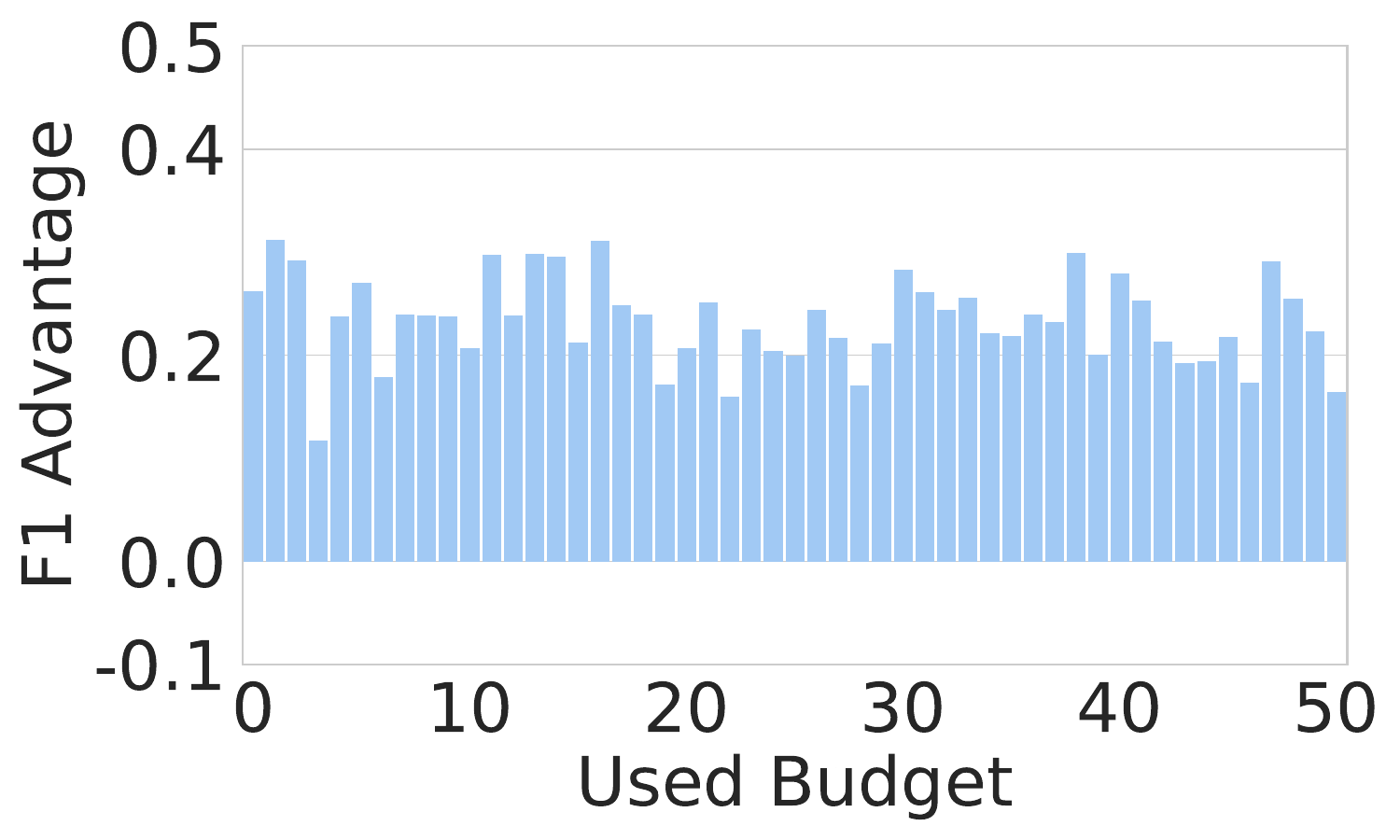}
        \caption{Missing Values}
    \end{subfigure}\hfill
    \begin{subfigure}{0.24\textwidth}
        \includegraphics[width=\linewidth]{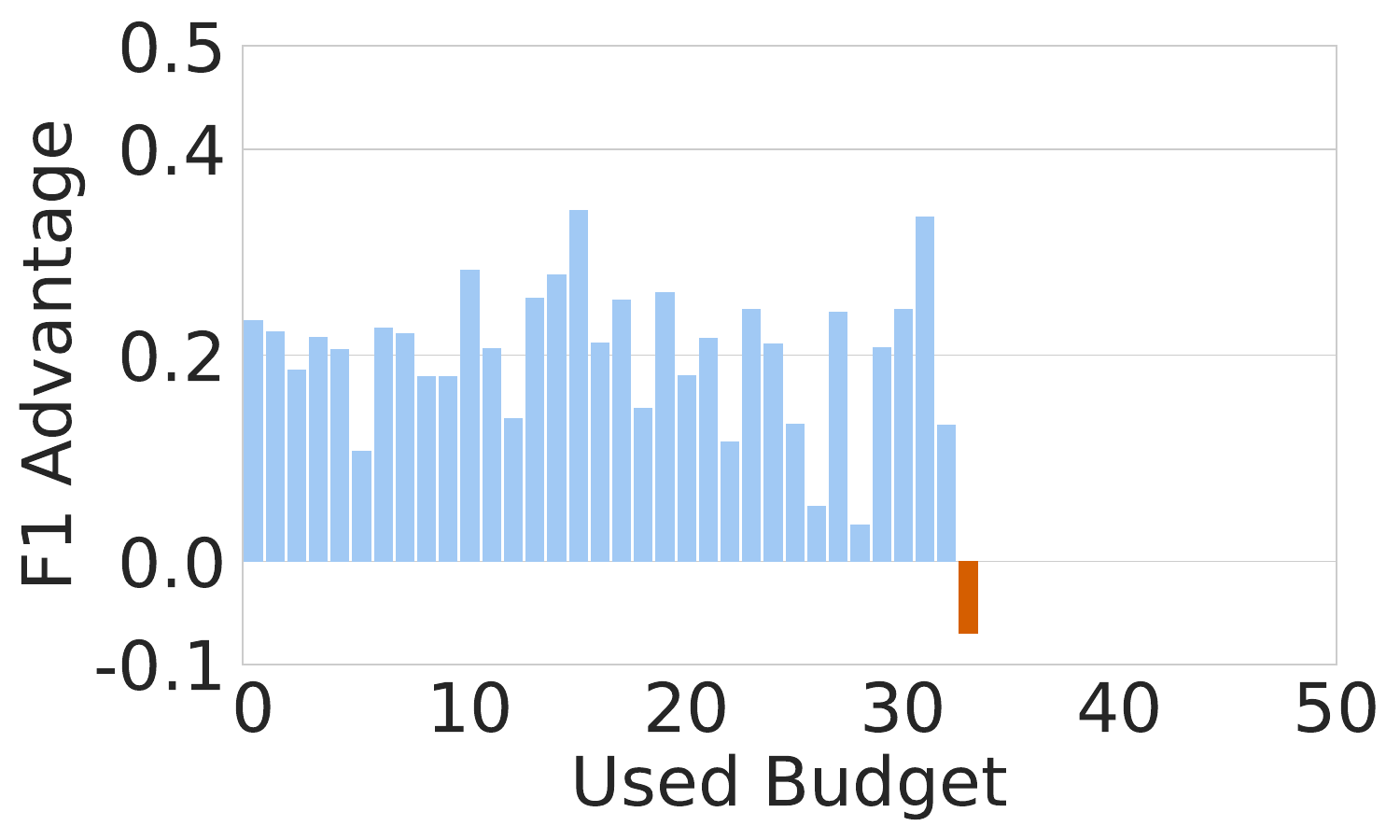}
        \caption{Scaling}
    \end{subfigure}
    \caption{Comparison of~\systemname with AC for LIR across error types.}
    \label{fig:agg_ac_results_lir}
\end{figure*}

\begin{figure*}[h!]
    \centering
    \begin{subfigure}{0.24\textwidth}
        \includegraphics[width=\linewidth]{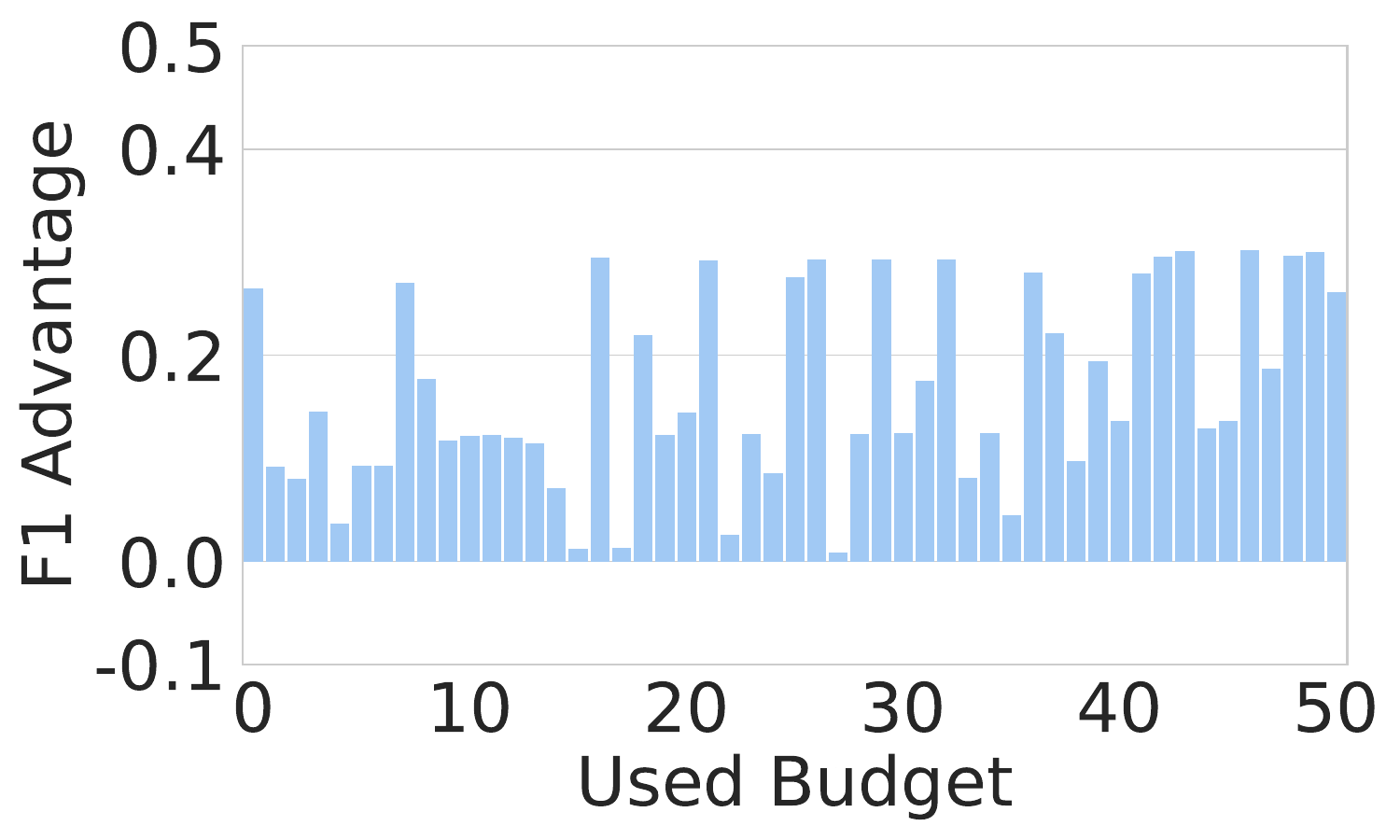}
        \caption{Airbnb - Scaling}
    \end{subfigure}
    \begin{subfigure}{0.24\textwidth}
        \includegraphics[width=\linewidth]{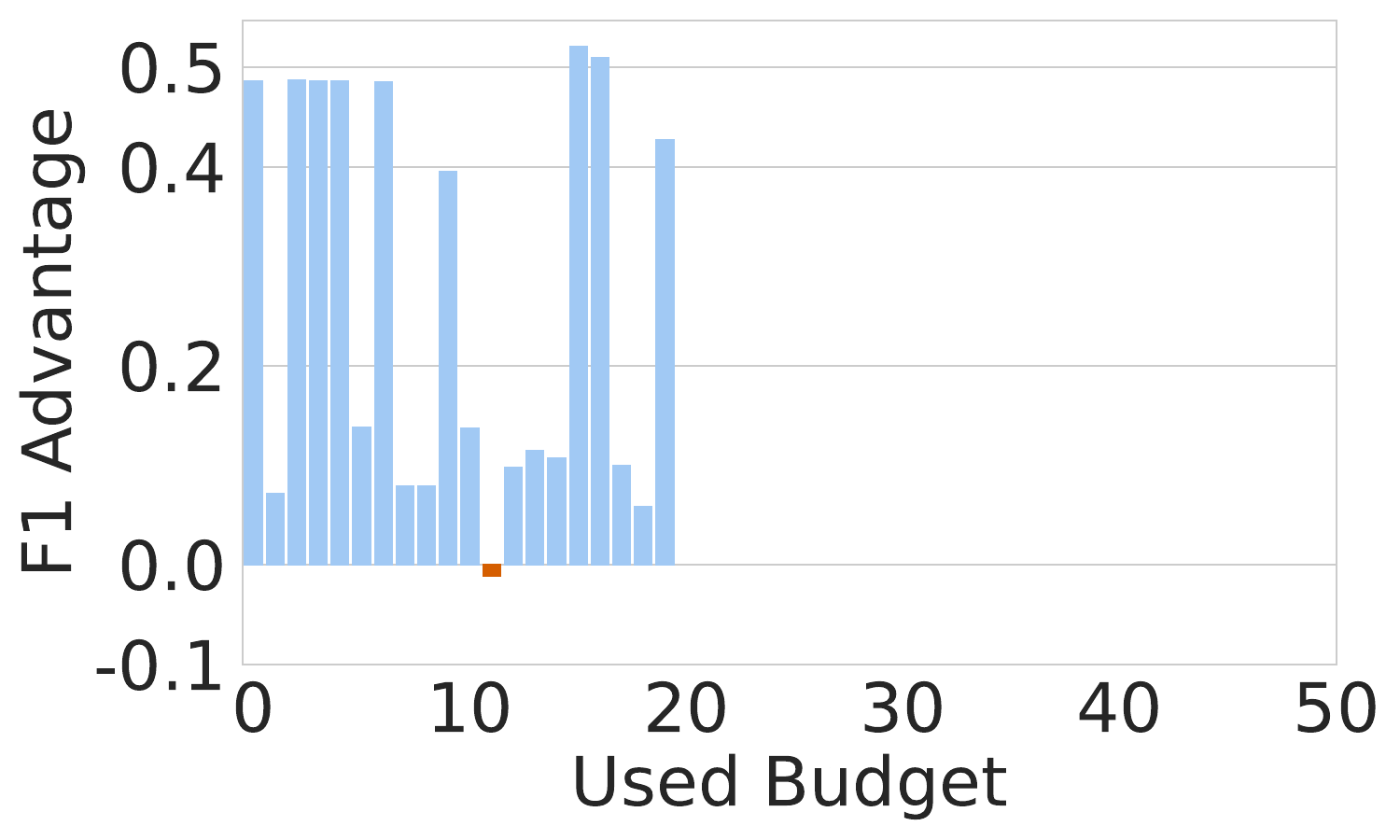}
        \caption{Credit - Scaling}
    \end{subfigure}
    \begin{subfigure}{0.24\textwidth}
        \includegraphics[width=\linewidth]{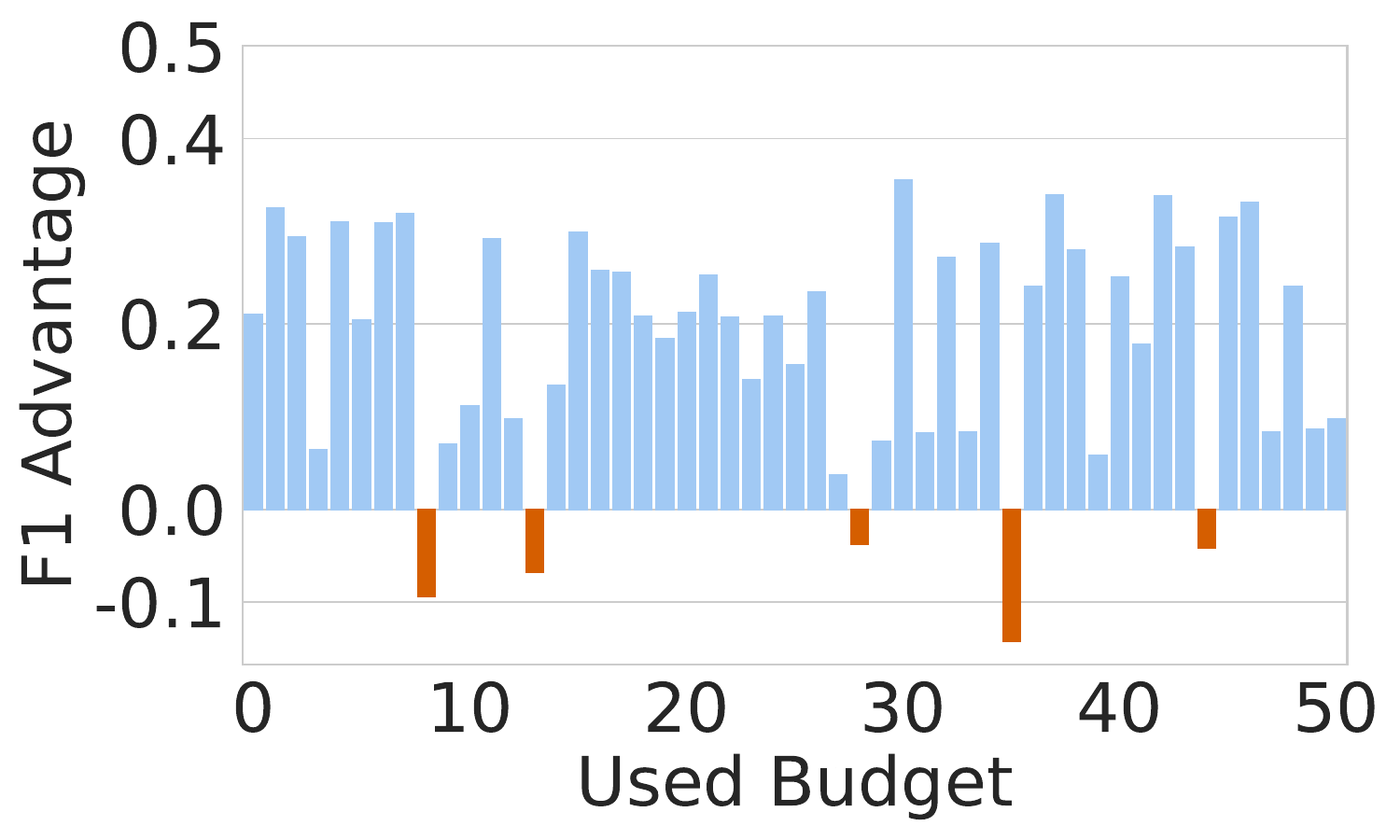}
        \caption{Titanic - Missing Values}
    \end{subfigure}
    \caption{Comparison of~\systemname with AC for LIR across error types, for datasets from CleanML.}
    \label{fig:agg_ac_results_lir2}
\end{figure*}

\begin{figure*}[h!]
    \centering
    \raisebox{1.4\height}{\rotatebox{90}{\textbf{CMC}}}\hspace{0.3em}%
    \begin{subfigure}{0.24\textwidth}
        \includegraphics[width=\linewidth]{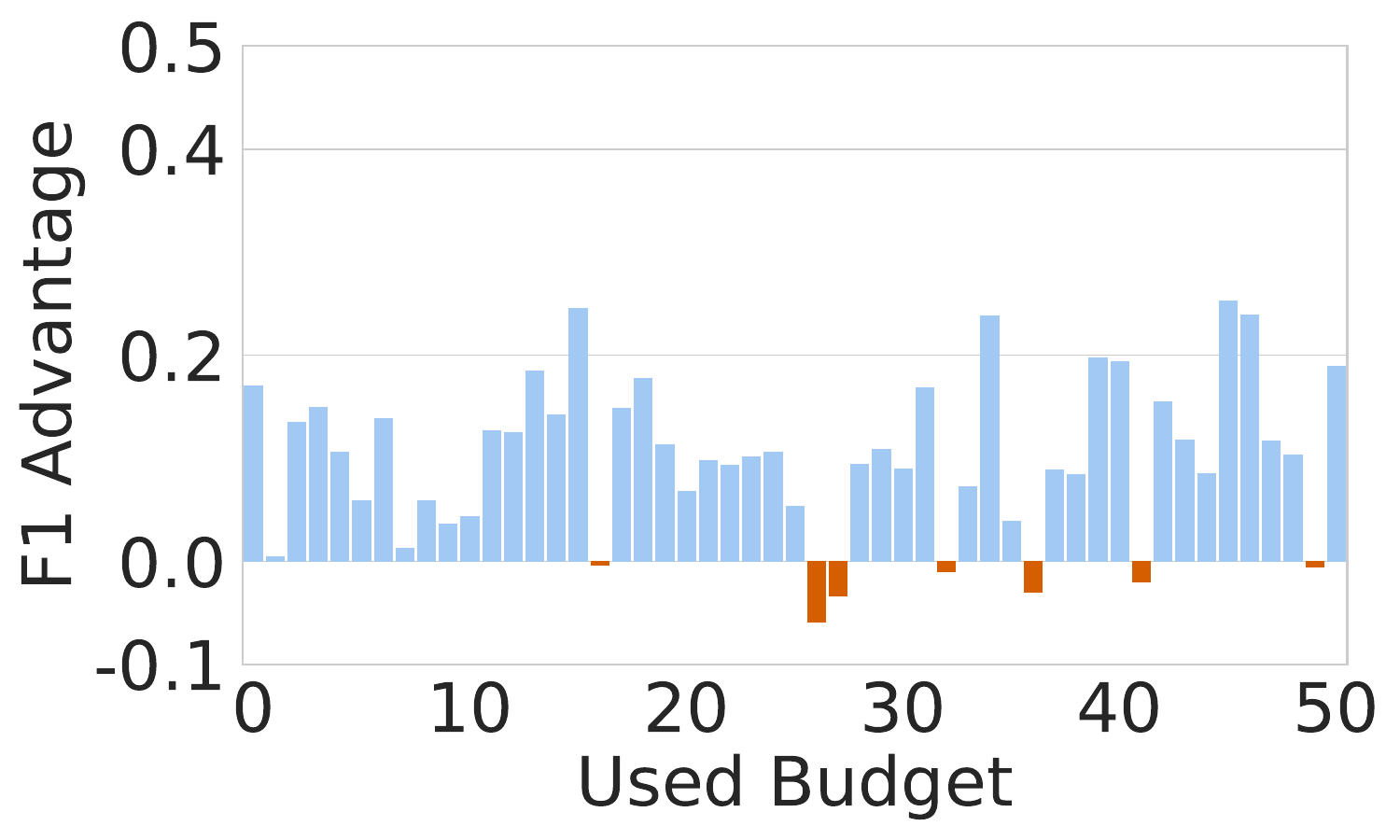}
    \end{subfigure}\hfill
    \begin{subfigure}{0.24\textwidth}
        \includegraphics[width=\linewidth]{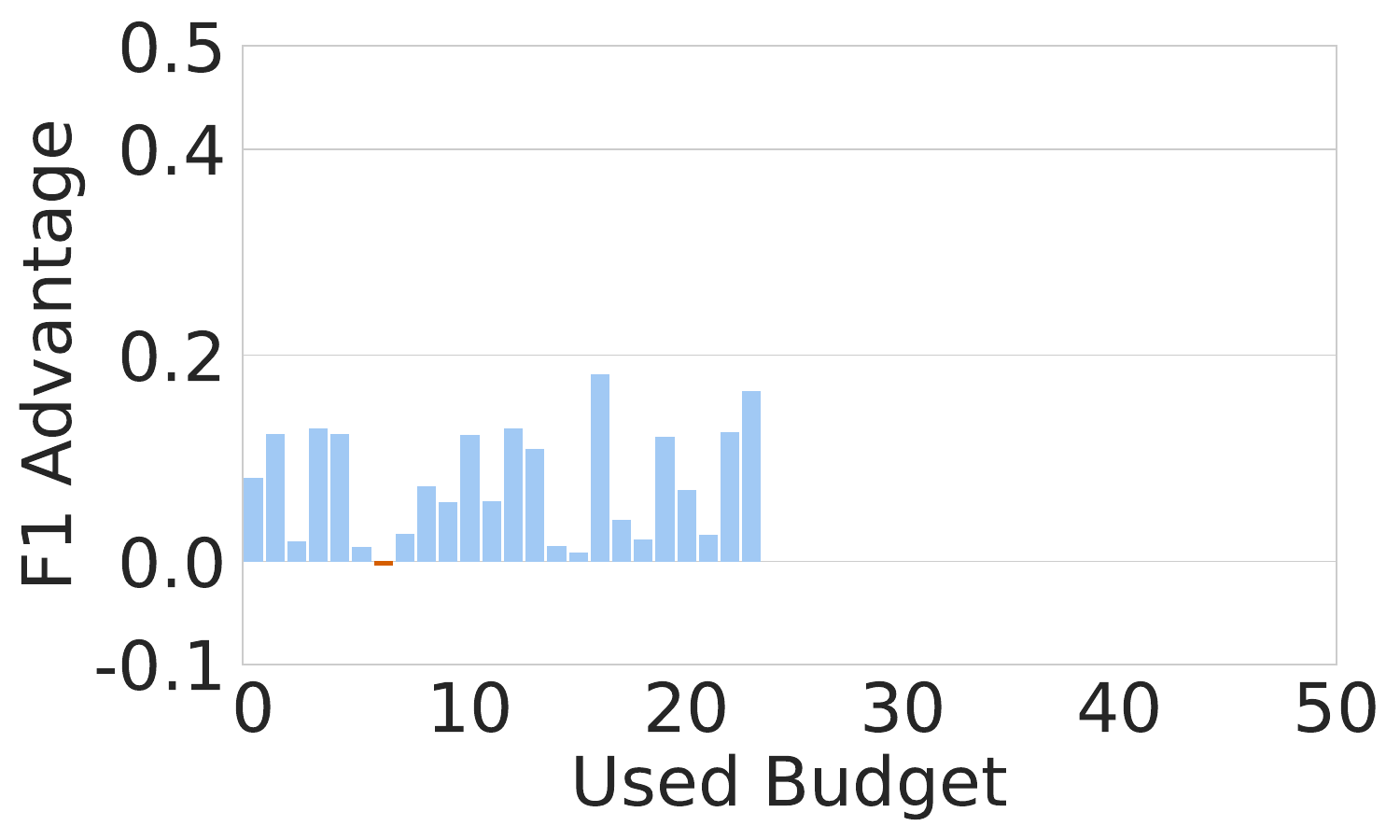}
    \end{subfigure}\hfill
    \begin{subfigure}{0.24\textwidth}
        \includegraphics[width=\linewidth]{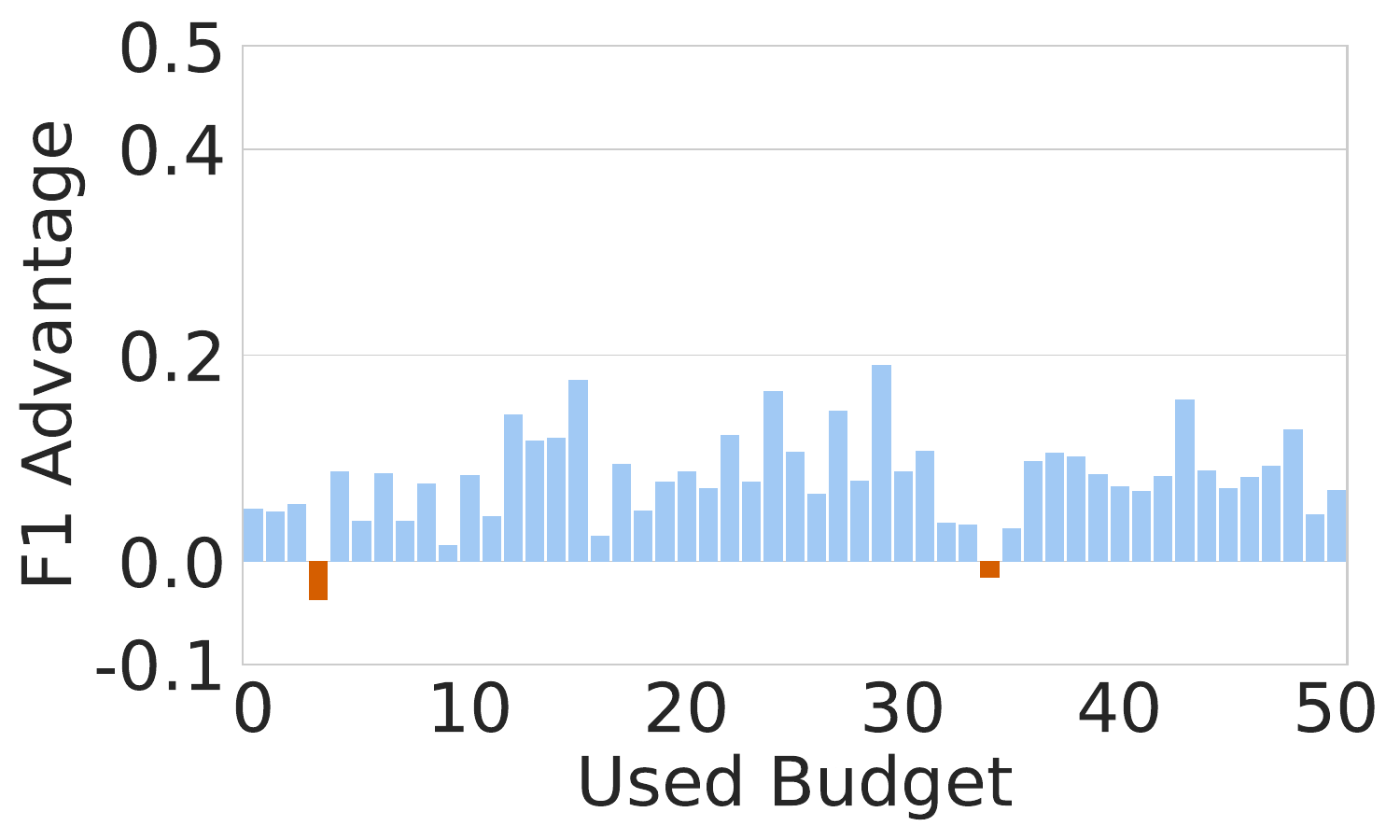}
    \end{subfigure}\hfill
    \begin{subfigure}{0.24\textwidth}
        \includegraphics[width=\linewidth]{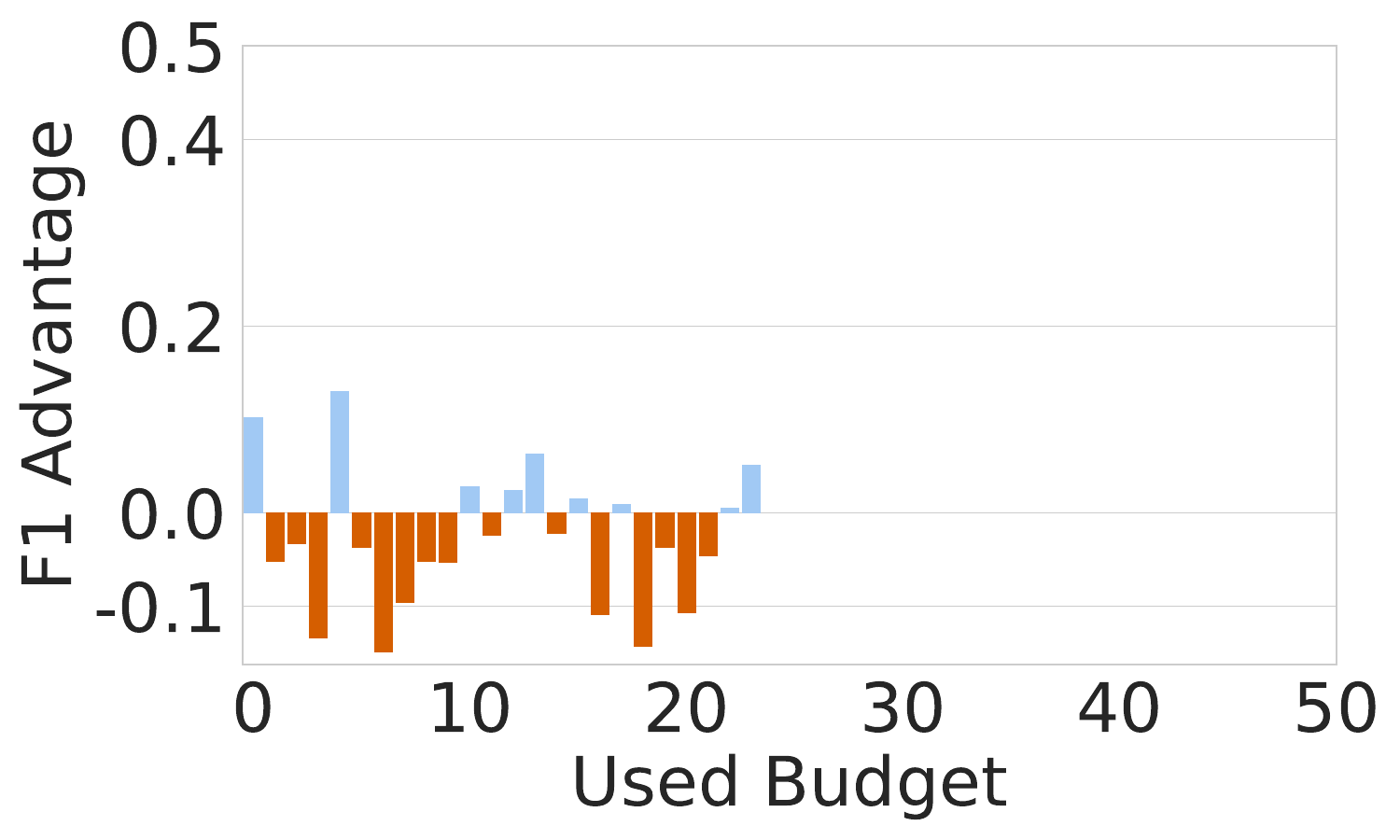}
    \end{subfigure}
    
    \vspace{-0.1em}

        \raisebox{1.2\height}{\rotatebox{90}{\textbf{Churn}}}\hspace{0.3em}%
    \begin{subfigure}{0.24\textwidth}
        \includegraphics[width=\linewidth]{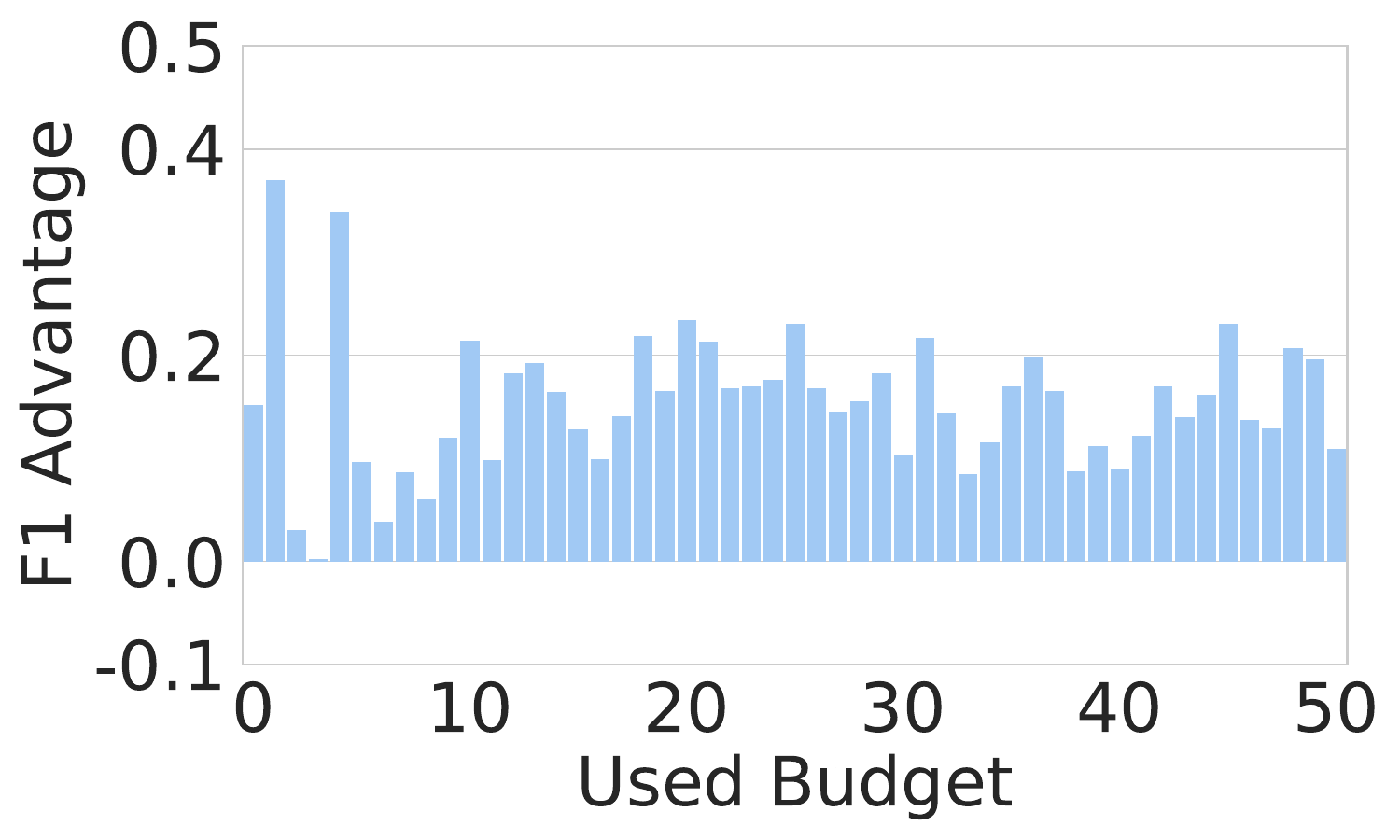}
    \end{subfigure}\hfill
    \begin{subfigure}{0.24\textwidth}
        \includegraphics[width=\linewidth]{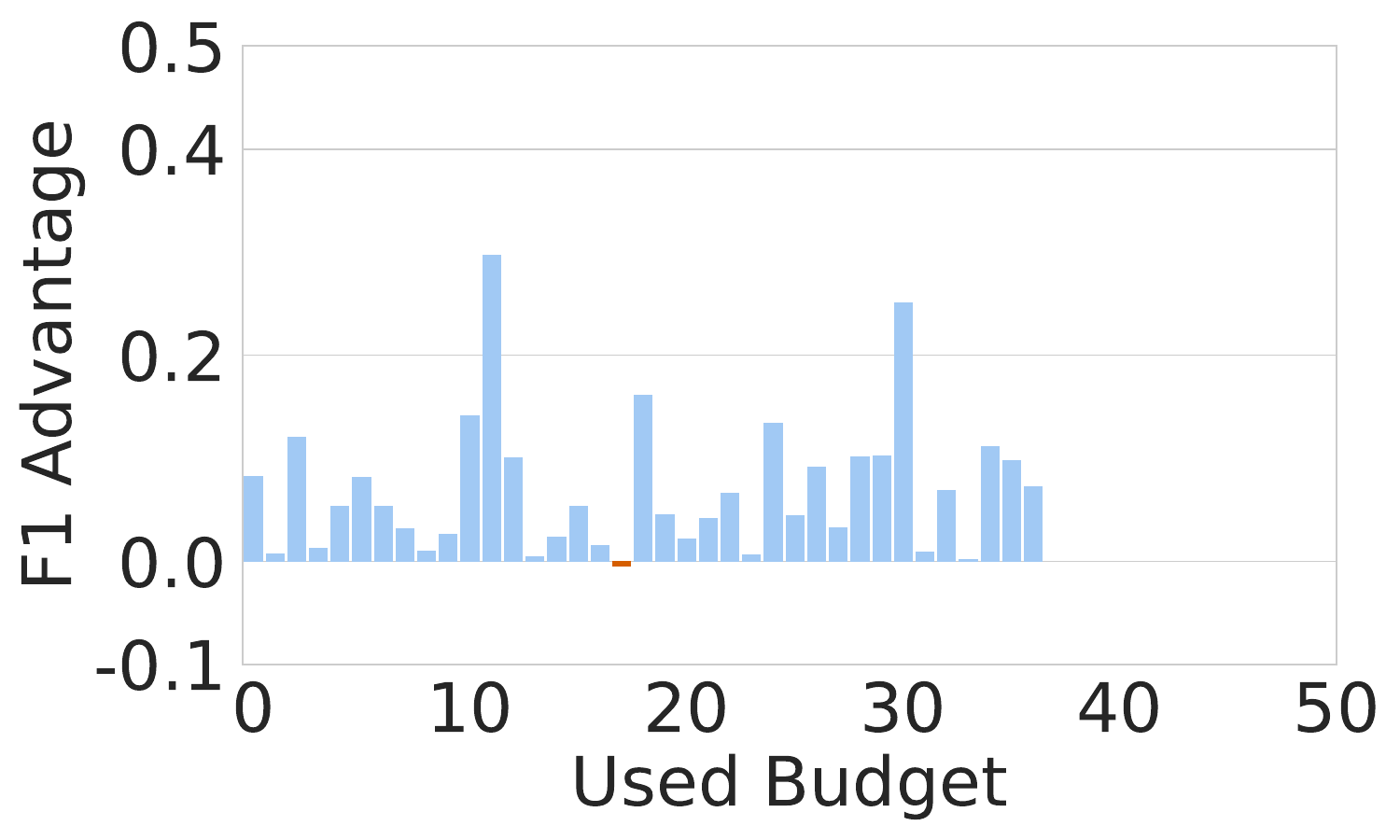}
    \end{subfigure}\hfill
    \begin{subfigure}{0.24\textwidth}
        \includegraphics[width=\linewidth]{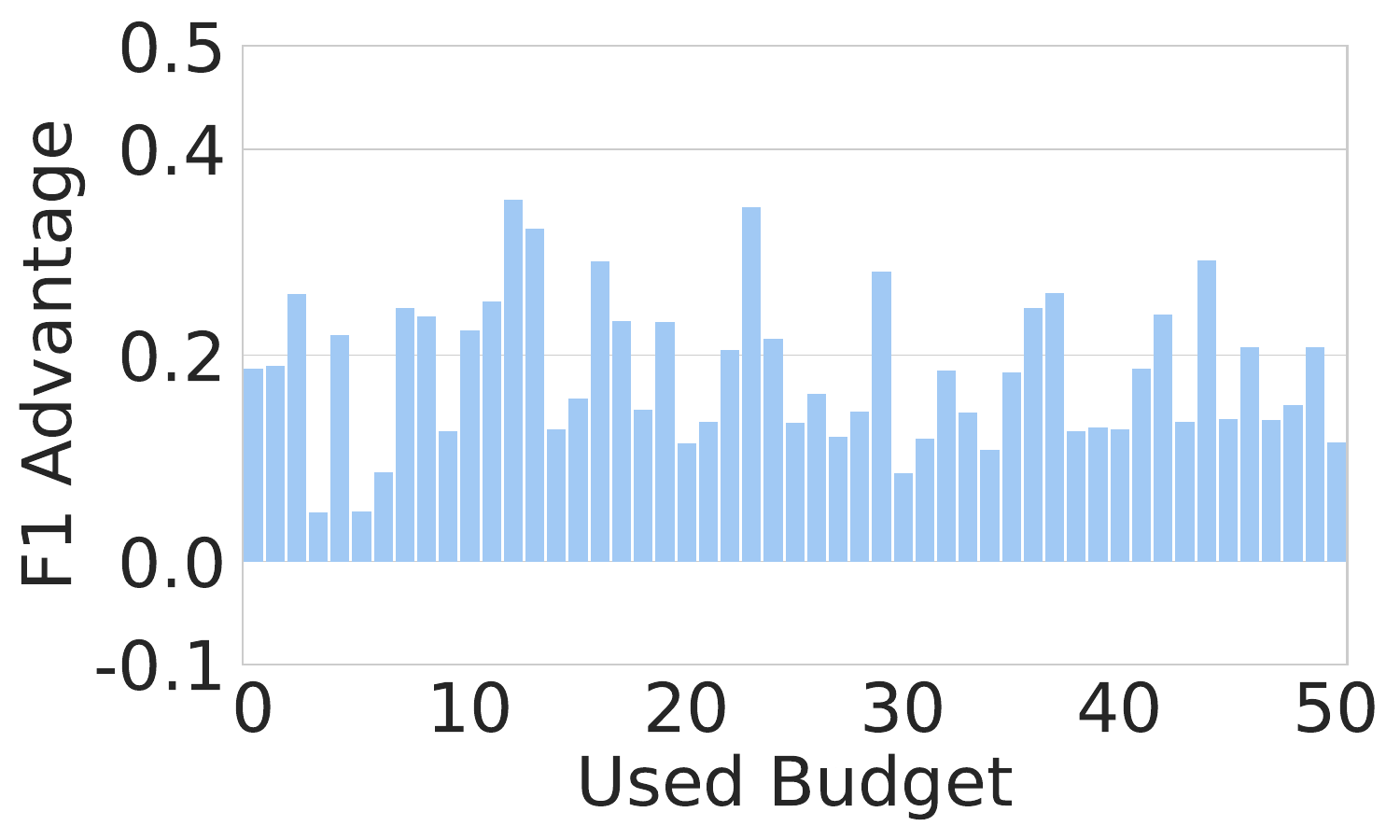}
    \end{subfigure}\hfill
    \begin{subfigure}{0.24\textwidth}
        \includegraphics[width=\linewidth]{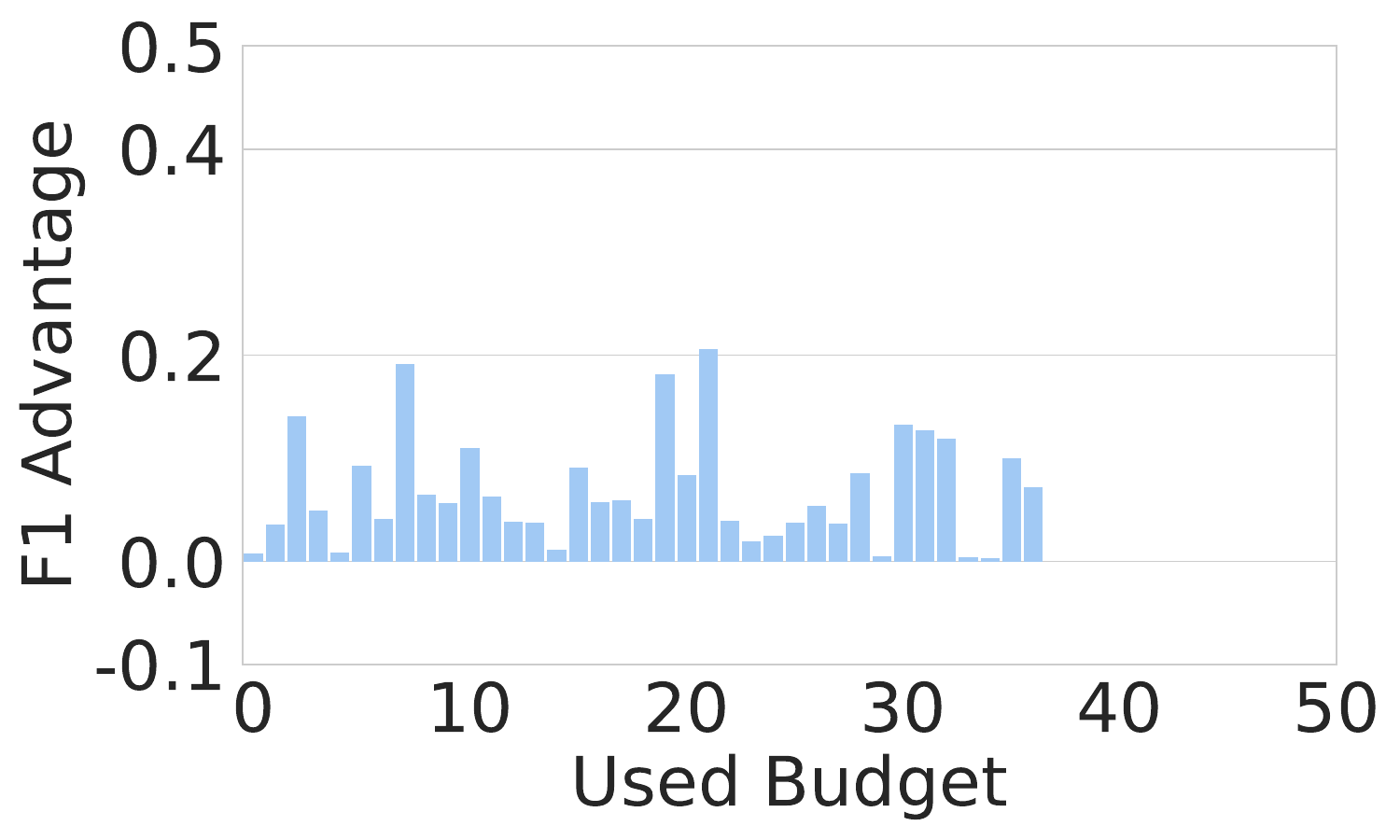}
    \end{subfigure}
    
    \vspace{-0.1em}

    \raisebox{2.\height}{\rotatebox{90}{\textbf{EEG}}}\hspace{0.3em}%
    \begin{subfigure}{0.24\textwidth}
        \centering\raisebox{3.85\height}{\parbox{0.75\linewidth}{\texttt{EEG only contains numerical features.}}}
    \end{subfigure}\hfill
    \begin{subfigure}{0.24\textwidth}
        \includegraphics[width=\linewidth]{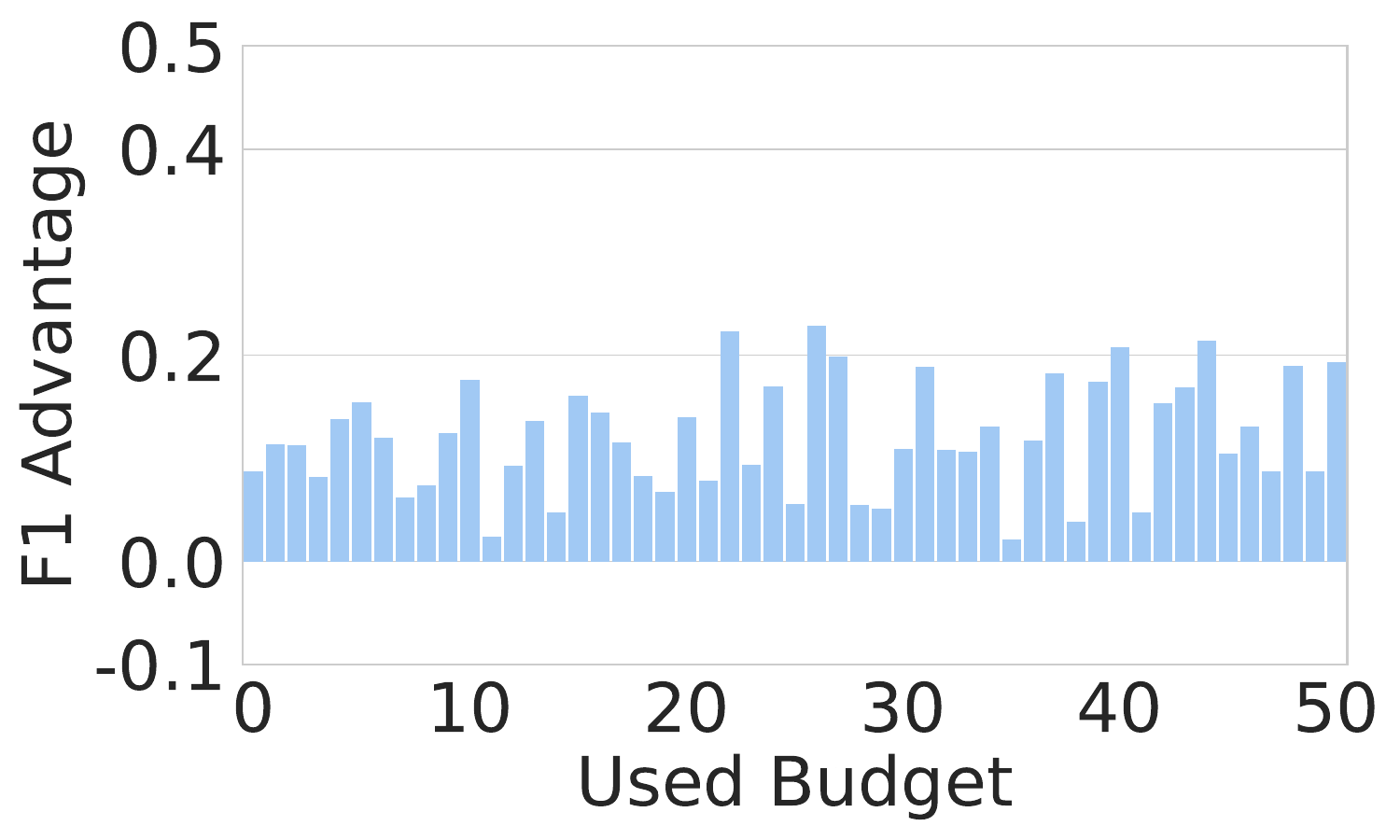}
    \end{subfigure}\hfill
    \begin{subfigure}{0.24\textwidth}
        \includegraphics[width=\linewidth]{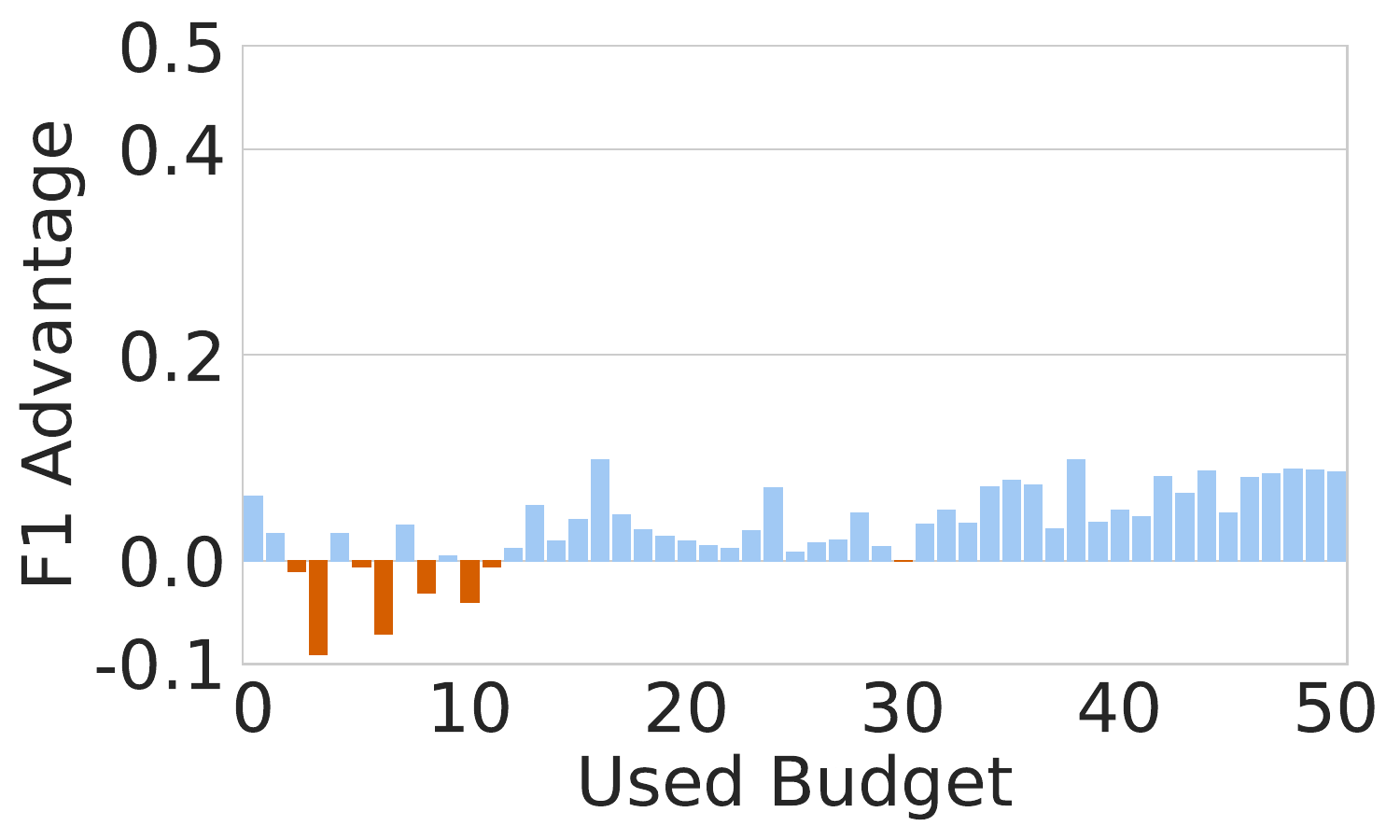}
    \end{subfigure}\hfill
    \begin{subfigure}{0.24\textwidth}
        \includegraphics[width=\linewidth]{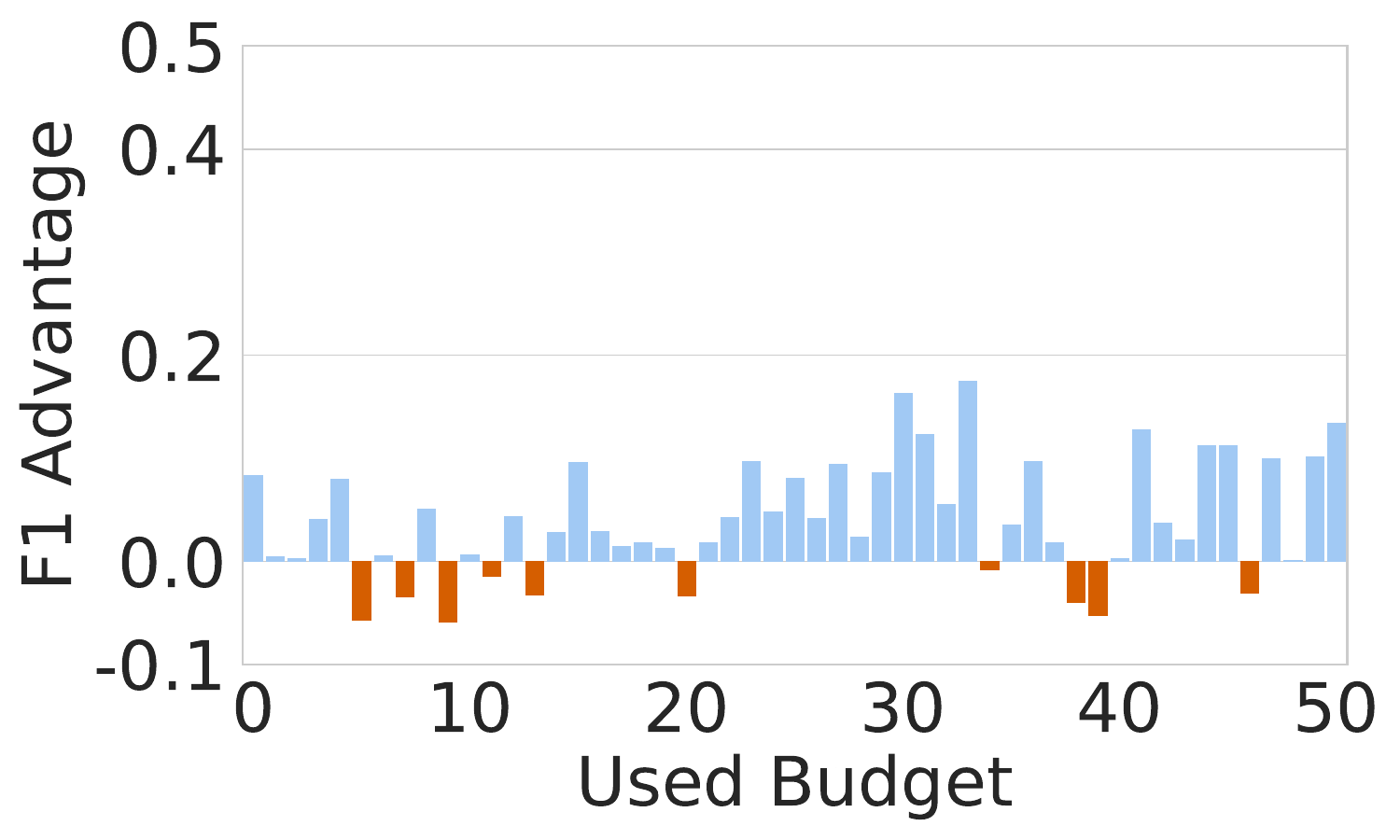}
    \end{subfigure}
    
    \vspace{-0.1em}
    
    \raisebox{1.2\height}{\rotatebox{90}{\textbf{S-Credit}}}\hspace{0.3em}%
    \begin{subfigure}{0.24\textwidth}
        \includegraphics[width=\linewidth]{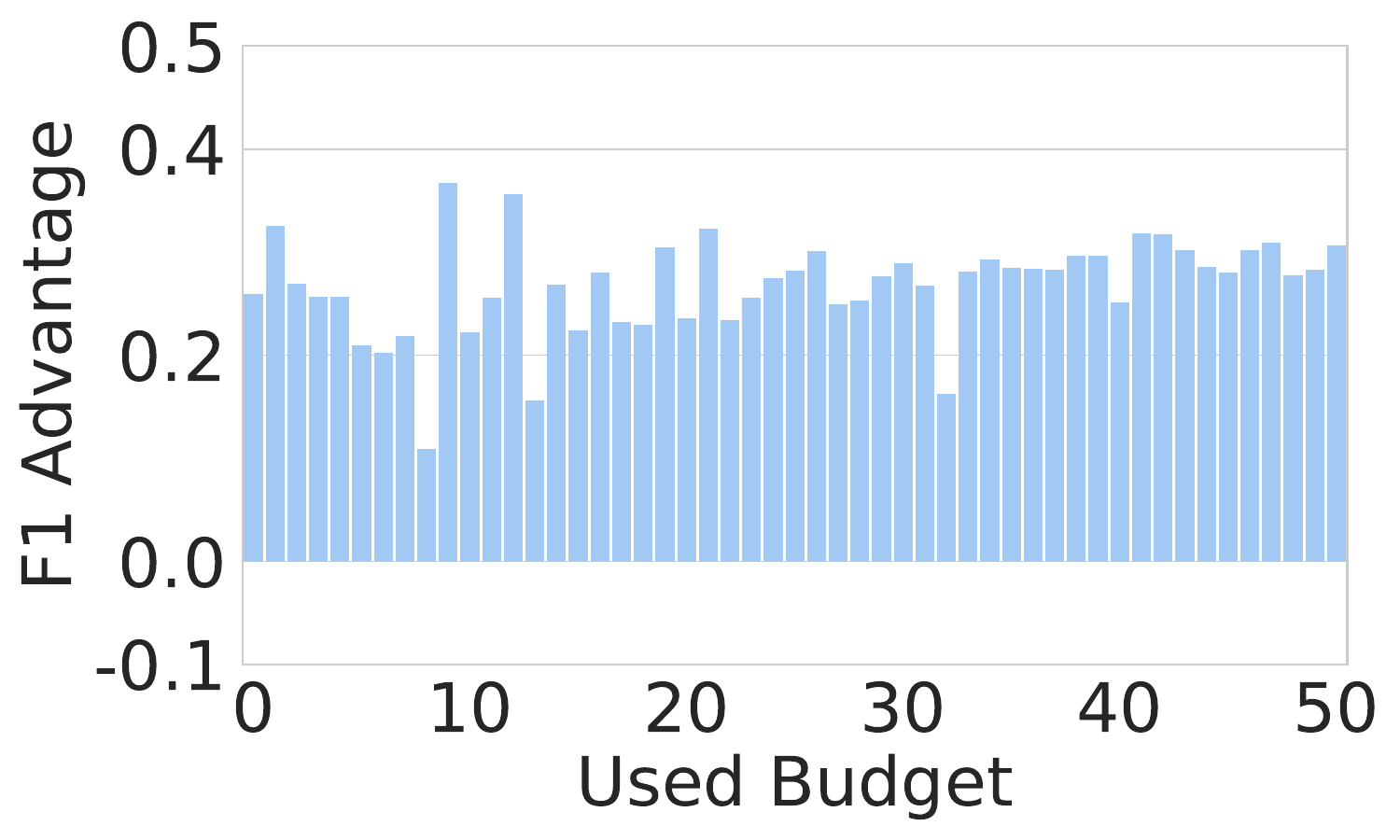}
        \caption{Categorical Shift}
    \end{subfigure}\hfill
    \begin{subfigure}{0.24\textwidth}
        \includegraphics[width=\linewidth]{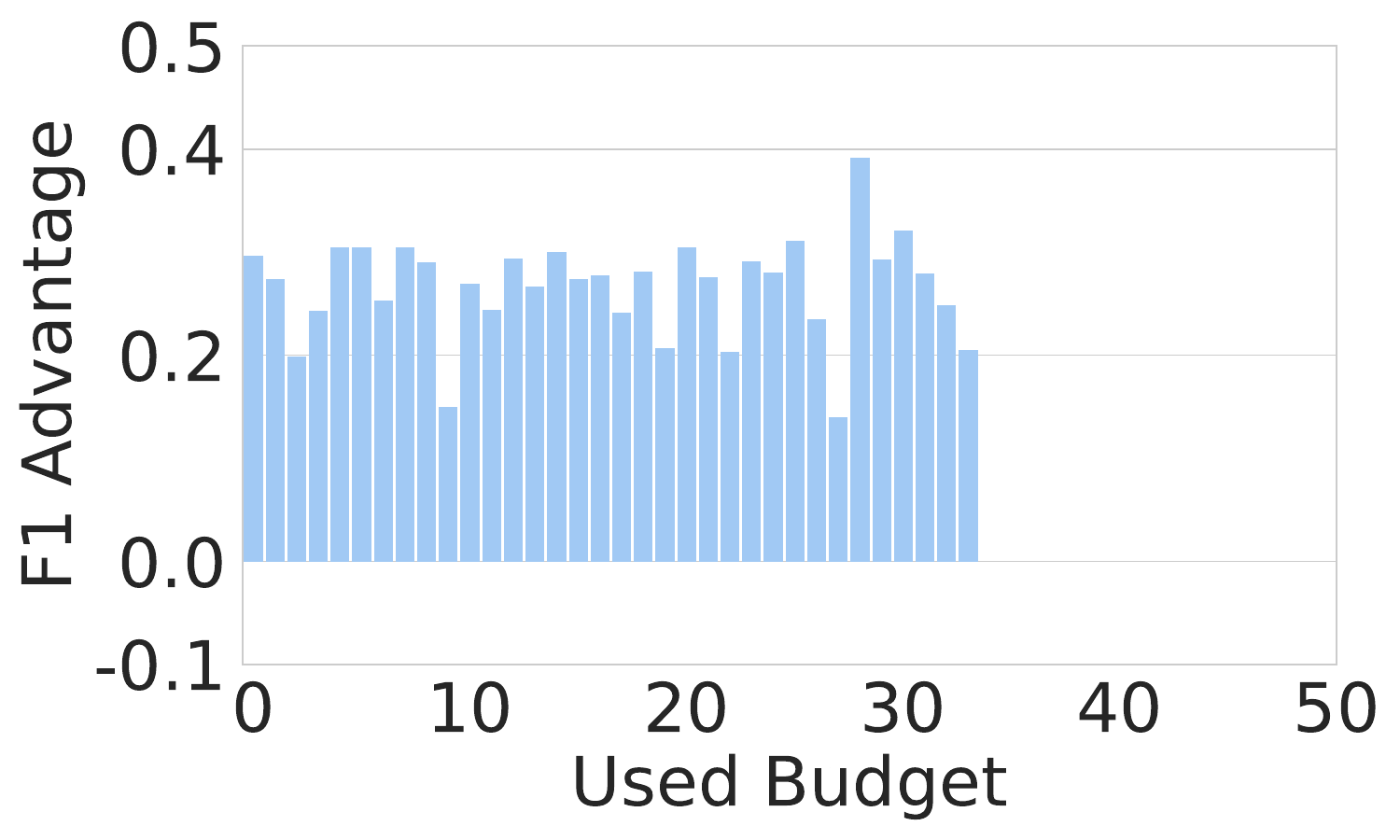}
        \caption{Gaussian Noise}
    \end{subfigure}\hfill
    \begin{subfigure}{0.24\textwidth}
        \includegraphics[width=\linewidth]{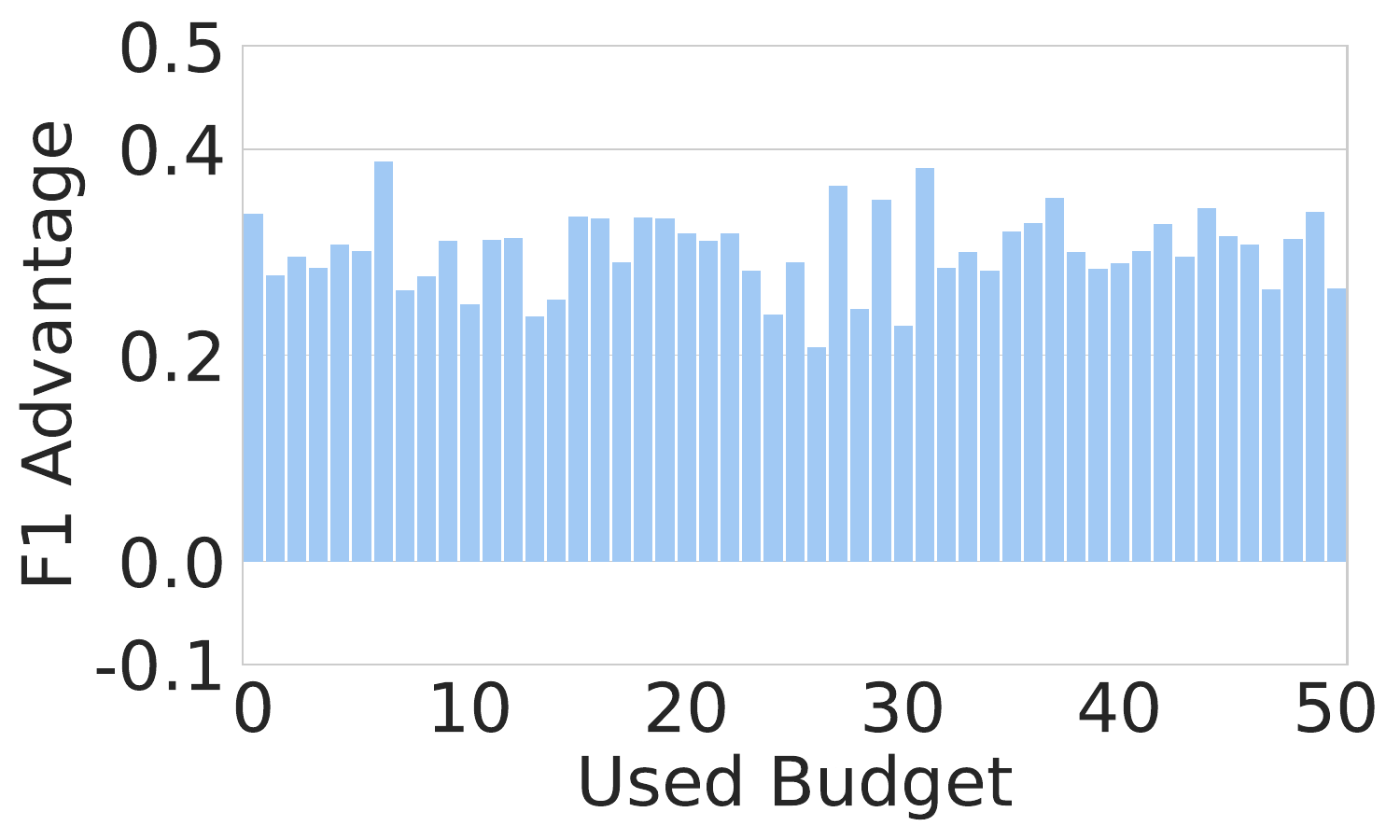}
        \caption{Missing Values}
    \end{subfigure}\hfill
    \begin{subfigure}{0.24\textwidth}
        \includegraphics[width=\linewidth]{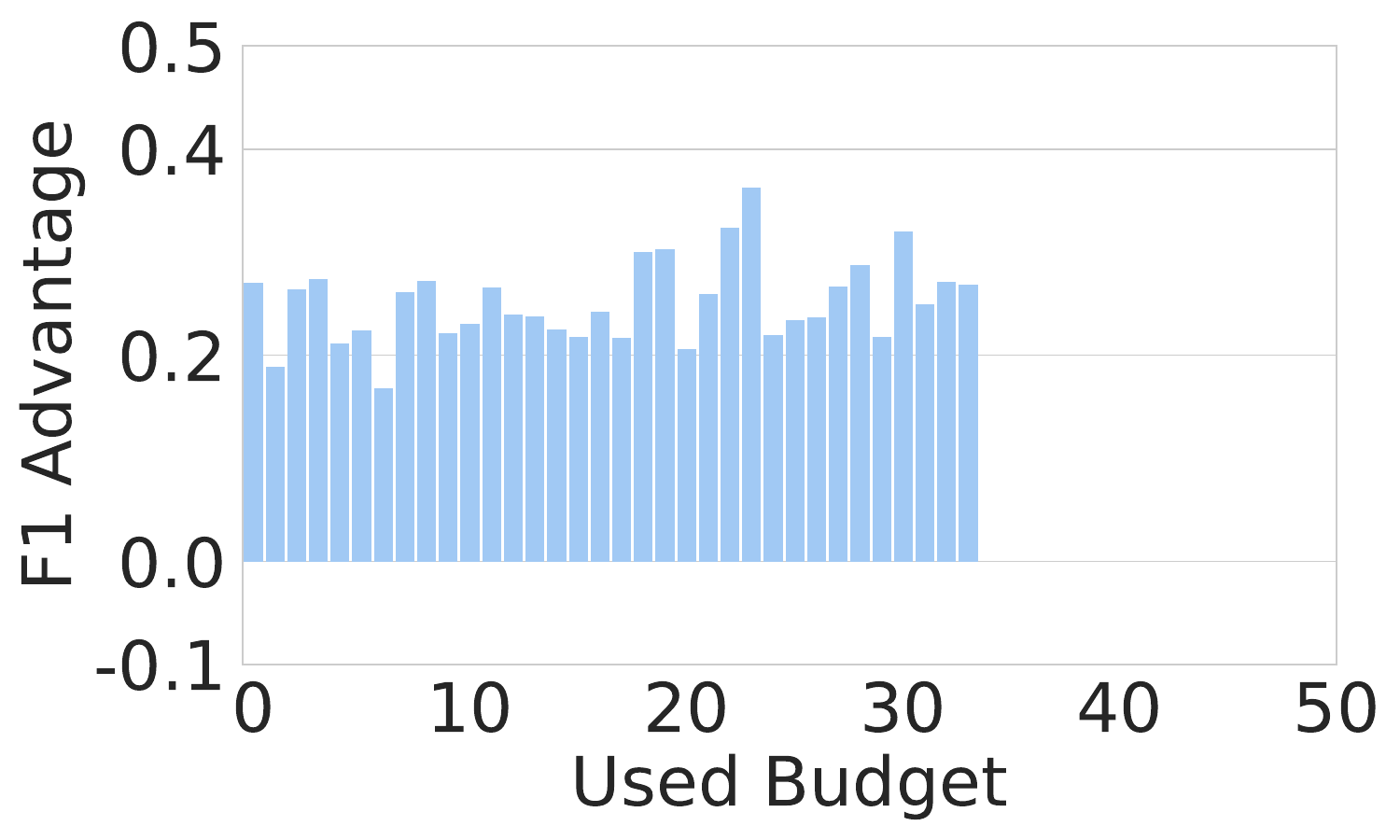}
        \caption{Scaling}
    \end{subfigure}
    \caption{Comparison of~\systemname with AC for LOR across error types.}
    \label{fig:agg_ac_results_lor}
\end{figure*}

\begin{figure*}[h!]
    \centering
    \begin{subfigure}{0.24\textwidth}
        \includegraphics[width=\linewidth]{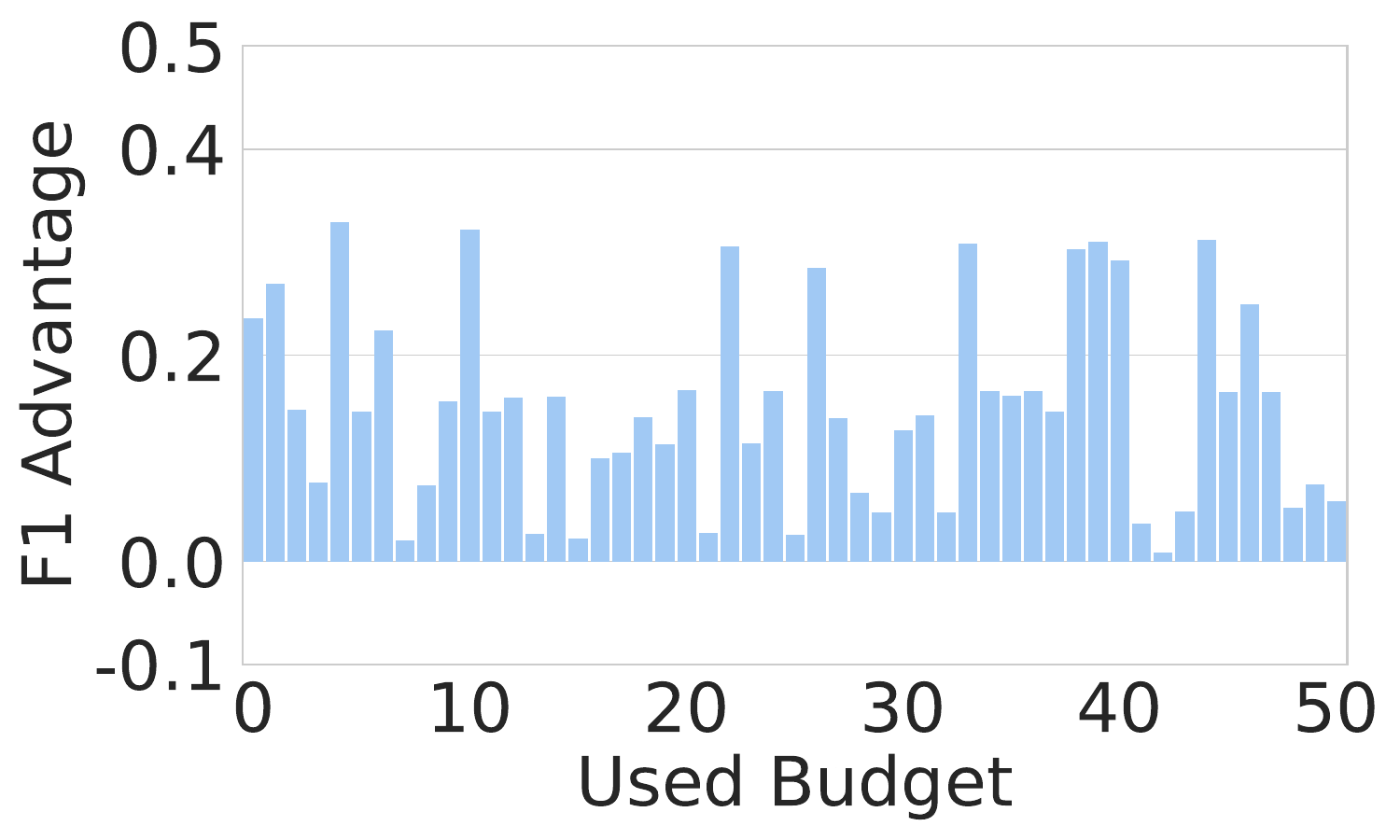}
        \caption{Airbnb - Scaling}
    \end{subfigure}
    \begin{subfigure}{0.24\textwidth}
        \includegraphics[width=\linewidth]{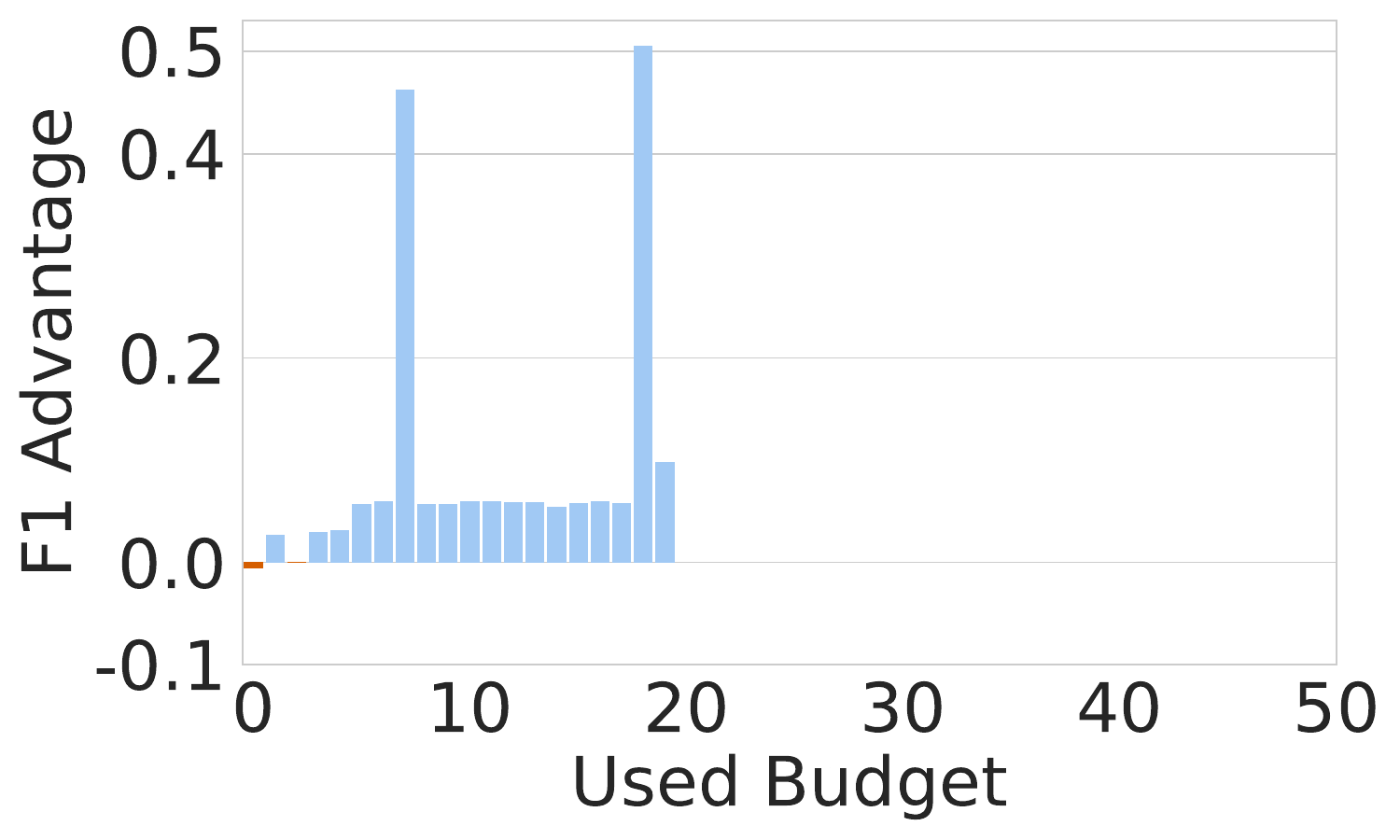}
        \caption{Credit - Scaling}
    \end{subfigure}
    \begin{subfigure}{0.24\textwidth}
        \includegraphics[width=\linewidth]{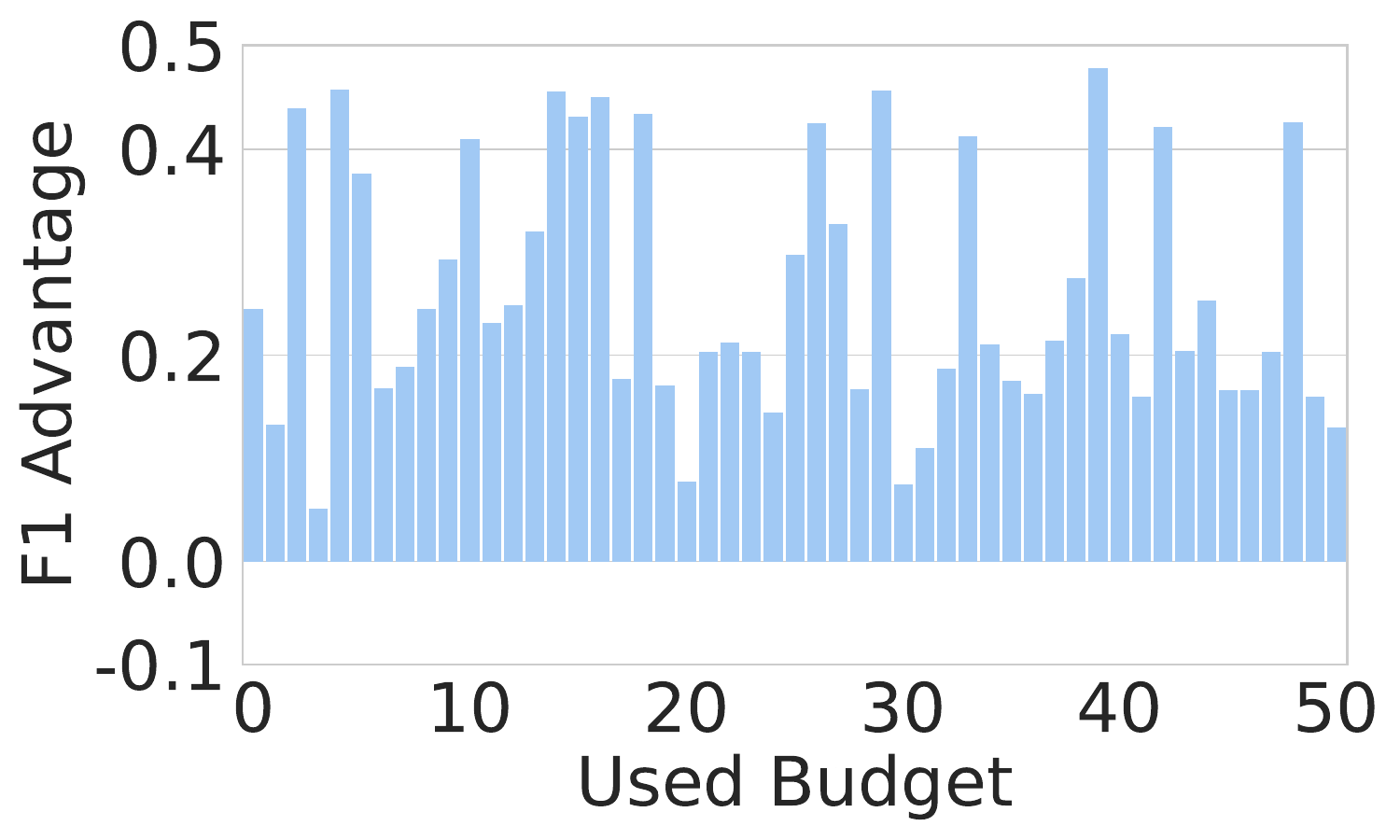}
        \caption{Titanic - Missing Values}
    \end{subfigure}
    \caption{Comparison of~\systemname with AC for LOR across error types, for datasets from CleanML.}
    \label{fig:agg_ac_results_lor2}
\end{figure*}

\end{document}